\documentclass[12pt]{article}
\usepackage{a4wide}
\usepackage{amssymb,amsmath,bm}
\usepackage{graphicx}
\usepackage{mathdots}
\usepackage[vcentermath]{youngtab}
\usepackage{tikz,pbox}
\usepackage{cite,hyperref,mcite}
\usetikzlibrary{calc,decorations.markings}

\makeatletter
\@addtoreset{equation}{section}
\makeatother

\newcommand{\bea}{\begin{eqnarray}} 
\newcommand{\eea}{\end{eqnarray}}

\newcommand{\la}{\left\langle}
\newcommand{\ra}{\right\rangle}
\newcommand{\trVN}[1]{\tr_{V_N^{\otimes #1}}}

\newcommand{\mytikz}[1]{
	\pbox{\textwidth}{\begin{tikzpicture}[>=stealth,decoration={
    markings,
    mark=at position 0.5 with {\arrow{>}}}]	
		#1
	\end{tikzpicture}}
}

\def\mP{\mathbb{P}} 
\def\mC{\mathbb{C}}

\def\mZ{\mathbb{Z}} 
\def\mT{\mathbb{T}} 
\def\mV{\mathbb{V}} 
\def\mI{\mathbb{I}}
\def\mR{\mathbb{R}}

\def\br{\bm{r}}
\def\bs{\bm{s}}
\def\bV{\bm{V}} 
\def\bR{\bm{R}}
\def\bL{\bm{L}}

\def\bK{\bm{K}}
\def\bn{\bm{n}}
\def\bl{\bm{l}}
\def\bPhi{\bm{\Phi}}
\def\bsig{\bm{\sigma}}
\def\bgam{\bm{\gamma}}

\def\bnu{\bm{\nu}}
\def\btau{\bm{\tau}}
\def\brho{\bm{\rho}}

\def\bS{ {\bm{S}} }

\def\tj{\tilde{j}} 
\def\ti{\tilde{i}}

\def\tQ{\tilde{Q}}

\def\cO{\mathcal{O}} 
\def\cC{\mathcal{C}} 
 
\def\cN{\mathcal{N}} 
\def\cP{\mathcal{P}}
\def\cH{\mathcal{H}}

\def\cF{\mathcal{F}} 
 
\def\cP{\mathcal{P}}

\def\cS{\mathcal{S}} 

\def\det{{\rm det}}

\def\tr{{\rm tr}}
\def\Sym{{\rm Sym}}
\def\Dim{{\rm Dim}}
\def\Adj{{\rm Adj}}

\def\g{\gamma} 
 
\def\a{\alpha}
\def\tl{\tilde}
\def\s{\sigma} 
\def\L{\Lambda} 

\def\C3Z2{\mC^3/\mZ_2}
\def\dP0{dP_0}

\def\makeatletter{\catcode`\@=11}% 11:letter
\makeatletter
\def\mathbox#1{\hbox{$\m@th#1$}}%
\def\math@ccstyles#1#2#3#4#5#6#7{{\leavevmode
      \setbox0\mathbox{#6#7}%
      \setbox2\mathbox{#4#5}%
      \dimen@ #3%
      \baselineskip\z@\lineskiplimit#1\lineskip\z@
      \vbox{\ialign{##\crcr
             \hfil \kern #2\box2 \hfil\crcr
             \noalign{\kern\dimen@}%
             \hfil\box0\hfil\crcr}}}}
\def\mathaccstyles{\math@ccstyles\maxdimen}
\def\maththroughstyles{\math@ccstyles{-\maxdimen}}
\def\unity%
 {\maththroughstyles{.45\ht0}\z@\displaystyle {\mathchar"006C}\displaystyle 1}
\begin{document}
\rightline{QMUL-PH-12-17}

\vskip 1cm

\centerline{{\LARGE \bf  Quivers as Calculators: }}
\centerline{{\LARGE \bf  Counting, Correlators and Riemann Surfaces  }}
 
\medskip

\vspace{.4cm}

\centerline{   {\large \bf Jurgis Pasukonis}\footnote{j.pasukonis@qmul.ac.uk}
{ \bf  and }  {\large \bf Sanjaye Ramgoolam}\footnote{s.ramgoolam@qmul.ac.uk}  } 
\vspace*{4.0ex}
\begin{center}
{\large School of Physics and Astronomy\\
Queen Mary, University of London\\
Mile End Road\\
London E1 4NS UK\\
}
\end{center}

\vspace*{5.0ex}

\centerline{\bf Abstract} \bigskip
The spectrum of chiral operators in supersymmetric  quiver gauge theories 
is typically much larger in the free limit, where the superpotential terms vanish. 
 We find that the finite $N$  counting of operators in  any free quiver theory, with a product of unitary gauge groups,  
 can be described by associating Young diagrams and Littlewood-Richardson multiplicities to a simple modification of the quiver, which we 
call the split-node quiver. The large $N$ limit leads to a surprisingly simple 
infinite product formula for counting gauge invariant operators, 
valid for any quiver with bifundamental fields.  
An orthogonal basis for the operators, in the finite $N$ CFT inner product, is given in terms of 
{ \it quiver characters}. These are constructed by inserting permutations in the split-node quivers and 
intepreting the resulting diagrams in terms of symmetric group matrix elements and branching coefficients. 
The fusion coefficients in the chiral ring  - valid both in the UV and in the IR - are computed at finite $N$. The derivation follows  simple 
 diagrammatic moves on the quiver. The large $N$ counting and correlators  are expressed in terms of topological field theories on Riemann surfaces obtained by thickening the quiver.
  The TFTs are based on symmetric groups and defect observables associated with subgroups play an important role. 
We outline the application of the free field results to the  construction of BPS operators in the case of 
non-zero super-potential.

\newpage

\tableofcontents

\newpage

\setcounter{footnote}{0}

%%%%%%%%%%%%%%%%%%%%%%%%%%%%%%%%%%%%%%%%%%%%%%%%%%%%%%%%%%%%%%%%%%%%%
%%%%%%%%%%%%%%%%%%%%%%%%%%%%%%%%%%%%%%%%%%%%%%%%%%%%%%%%%%%%%%%%%%%%%

\section{Introduction and Summary } 

In the AdS/CFT correspondence  \cite{malda,gkp,witten}, the Hamiltonian for translations of the global time 
in the AdS side corresponds to the scaling operator on the CFT side. Classifying  the states 
of a given energy and computing their interactions   allows comparisons between the two sides. 
States in CFT are related to local operators through radial quantization.  
This paper is primarily concerned with the counting of states and computation 
of correlators in a large class of 
free theories, parametrized by quivers. A quiver is a directed graph used to describe the gauge group 
and matter content of the theory \cite{DM96}. 
 The nodes correspond to gauge groups which we will take to be unitary groups, 
so that the gauge group is $\prod_a U(N_a) $ where $a$  is an index running over labels $ \{ 1, 2, \cdots \}  $ for the nodes. 
Each directed edge starting from $a$ and ending at $b$ correspond to a bifundamental field $( N_a , \bar N_b )$ transforming 
in the fundamental of $U(N_a)$ and anti-fundamental of $U(N_b)$. 
 Our results show that the quiver diagram itself
 becomes a powerful computational tool. Counting of gauge invariant operators can be expressed using 
 the operation of splitting each node into a pair called the plus (or incoming) node 
 and the minus (or outgoing) node. The plus node has all the incoming lines of the original quiver 
 and the minus node has all the outgoing line. A new line is introduced for each pair, going from plus to minus. 
 This modified quiver is called the {\it split-node } quiver.  In going from counting of operators to their correlators,  
 the split-node quiver is used to define  { \it quiver characters} which
  encode representation theory data associated with permutation groups and their representations.
These are  generalizations of symmetric group characters, parametrized by quivers, and obey analogous identities. 
They are constructed by inserting permutations in the split-node quiver 
and intepreting the resulting diagram in terms of matrix elements  of permutations 
and branching coefficients for symmetric group reductions. This reprises the theme that there is a close 
connection between the counting and construction of operators, when we use the right group 
theoretic framework \cite{pasram,dMRDC}.   
The quiver diagram thus gives elegant expressions 
for the counting of chiral operators, the two point functions between chiral and anti-chiral operators,
the chiral ring fusion coefficients, both for finite rank $N_a$  as well as at large $N_a$.
The combinatorial data related to counting and correlators is also shown to have 
an interpretation in terms topological field theories on surfaces which are obtained 
by thickening the  quiver.  At large $N_a$, the combinatorics can be expressed in terms 
of the counting of covering spaces of these surfaces. 

Before explaining some of these results in more detail, we will describe some of the background to 
this work, with particular attention to the significance of finite $N$ results  in AdS/CFT.  The canonical example of AdS/CFT is 
the duality between type IIB on $ AdS_5 \times S^5$  and $\cN=4$ SYM with $U(N)$ gauge group. 
The half-BPS sector of gauge theory operators contains duals to a rich variety of space-time 
objects including perturbative Kaluza-Klein states, giant gravitons and LLM geometries \cite{witten,gkp,mst,bbns,cjr,llm}.
Thanks to non-renormalization theorems (see the review  \cite{dhokfreed} for the references on this)
 the counting and extremal correlators of BPS states 
do not change from the zero coupling answer. The lowest weights of the half-BPS representations are 
holomorphic traces and products of traces of a complex matrix, such as $tr Z , tr Z^2 , (tr Z )^2 $. 
 The two-point correlators between holomorphic and anti-holomorphic operators is diagonalized by 
 Schur Polynomial operators \cite{cjr}
 \bea\label{schurops}  
 \chi_{ R } ( Z ) = { 1 \over n! } \sum_{ \sigma \in S_n } \chi_R ( \sigma ) Z^{i_1}_{ i_{\s(1)} } \cdots Z^{i_n }_{ i_{\s(n)} } 
 \eea
so that 
\bea\label{schurorthog}  
\langle  \chi_{ R } ( Z )  \chi_S ( Z^{\dagger} ) \rangle = \delta_{RS} f_R  
\eea
where $R$ is a Young diagram with $n$ boxes, $\chi_R ( \sigma) $ is a character of the  $S_n$ group element 
$\sigma$ in the irreducible representation (irrep)  $R$ of $S_n$ , $f_R$ is a polynomial in $N$. In the leading large $N$ 
limit, the trace basis is also an orthogonal basis - this is large $N$ factorization - but this does not hold at finite $N$. 
Finite $N$ effects are nicely encoded in the Young diagram $R$, which does not have more than $N$ rows. 
Giant gravitons are particularly interesting since their semiclassical properties 
are sensitive to finite $N$ cutoffs \cite{mst}.  The three-point functions of the Schur Polynomial operators 
are computed in terms of Littlewood-Richardson coefficients \cite{fulhar}. They have recently been tested
using     semiclassical methods in spacetime \cite{bkyz,cdz,lin3pt}.   The Young diagram description 
of operators dual to giants forms the starting point for modifications of the operators which correspond to strings 
attached to giants  \cite{OSgiant1,*OSgiant2,*OSgiant3,*OSgiant4,*OSgiant5,*OSgiant6,*OSgiant7,*OSgiant8}.

There has been a lot of  work on the extension of the dictionary between giants and 
operators, to the case of quarter and eighth BPS giants. The story is substantially more complicated 
in this case. The spectrum of BPS operators now jumps in going from zero coupling to weak coupling
and is conjectured to remain unchanged from weak to strong coupling \cite{KMMR05,BGLM07}.
At zero coupling, we have holomorphic  operators  constructed from two complex matrices 
$X , Y  $ of size $N$ for the quarter BPS sector and three complex matrices $ X , Y , Z $ 
for the eighth-BPS sector (There are also additional 
eighth-BPS operators where the lowest weights are constructed 
with fermions \cite{BDHO}, but they will not be our concern here). 
Diagonal bases for the free field CFT inner product on these spaces of 
multi-matrix gauge invariant operators at finite $N$ have been constructed 
\cite{BHR1,BHR2,KR1,db1}.  Not all of these operators are annihilated by the 
one-loop dilatation operator \cite{BeisertCOL}. These operators which get 
anomalous dimensions are desdendants and contain commutators e.g. 
$[ X , Y ] $ \cite{ryzhov,DHHR03}. 
  
The correct BPS operators at weak coupling, annihilated by the 1-loop dilatation operator, 
can also be characterized as those that are orthogonal to the descendants in the zero-coupling inner product. 
This illustrates the usefulness of the zero coupling inner product for physics at weak coupling, 
Another remarkable example of the power of zero coupling, is that bases constructed to diagonalize 
the free field inner product by exploiting the enhanced symmetries of this limit \cite{KR1,KRT1,*KRT2,EHS}, notably Brauer 
algebra symmetries, have been shown to give  a large subset of quarter BPS operators 
to all orders in $1/N$ \cite{KimuraQBPS1,*KimuraQBPS2} with a proposed matching to states from LLM geometries \cite{KimLin}.

The limit of zero coupling is of intrinsic interest, beyond the application to semiclassical giants 
at strong coupling. In this limit, there is a huge amount of data from the gauge theory
Ideas for the dual string theory can be tested. Aspects related to higher spin symmetries have been 
explored in \cite{BBMS04}. One  approach to the construction of 
the dual string theories for the free limit is to follow the example of low dimensional example of 
two dimensional Yang Mills theory \cite{grta}. 
 In this solvable model, the large N expansion can be computed 
in terms of symmetric groups and a topological string model can be derived. Much the same 
strategy can be applied to study the combinatoric aspects of correlators in the free limit of CFTs in any dimension.  
For two and three-point functions, the space-time dependence is determined by conformal invariance, 
so all the non-triviality is in this combinatorics.  In the simplest case of half-BPS operators, it is indeed 
known \cite{tom,dMR,cjr}    that two point functions are related to Belyi maps (holomorphic maps 
with three branch points)  with  sphere as target space.  In this paper, we will find a generalization of 
this fact to any  free quiver gauge theory, where the target space is constructed by starting with a 
thickening of the quiver to a surface, and then cutting the surface to insert some conditions 
on the monodromies of branched covers over the cuts (see section \ref{sec:surfaces}). A version 
of the  connection between 
correlators and  Belyi maps also holds for hermitian Matrix models (involving so-called clean Belyi maps) \cite{dMR}.  This has been used to relate hermitian Matrix Model correlators to the $A$-model topological string 
with $\mP^1$ target \cite{gopak}.

Beyond the standard example of AdS/CFT  which involves the near-horizon geometry of 
branes at a point in  $\mC^3$,  there are several closely related generalizations. 
One infinite class comes from orbifolds \cite{KS98}. An infinite class comes from toric non-compact 
Calabi-Yaus, which may not be orbifolds \cite{BFHMSVW1,*BFHMSVW2}. 
 Among the examples we will use to illustrate the general counting
and correlator formulae, we will use a  $\mC^3/Z_2$ orbifold and a $\mC^3/Z_3$ orbifold. 
As a simple example of toric CY we will use the conifold, where the AdS/CFT dictionary was established in \cite{KW98}. 
The combinatorics of free field correlators in the conifold theory is the same as in  
the ABJM model \cite{ABJM}, where calculations using Young diagram techniques have been studied in 
\cite{ABJMcor1,*ABJMcor2,*ABJMcor3,*ABJMcor4,*ABJMcor5}. 
It should be noted that our results  apply to the free field limit of any quiver, 
even those which are not related to conformal field theories in the infrared. 
If the theory is asymptotically free, the free limit is the same as the deep UV limit, so this is a restriction that 
may be useful to keep in mind. We focus on the correlators of complex scalars, which can exist 
even in non-supersymmetric theories. However, the discussion is particularly meaningful for the case of 
$\cN=1$ supersymmetric theories, where these are the scalar components of a chiral superfield 
and the chiral gauge invariant operators form part of the chiral ring. Our results on the chiral ring
of free gauge theories may  be useful more generally beyond the context of ADS/CFT. 
For example the detailed study of  chiral rings \cite{CDSW02}  
was valuable in understanding connections between  4D dynamics of $\cN=1$ SUSY gauge theory 
and matrix models \cite{DW02}.   While the generic gauge theories have non-zero superpotential, 
the limit of zero superpotential is a special point  of the moduli space with enhanced symmetries,
which can be of higher spin type involving higher derivative currents (e.g as in  \cite{BBMS04})  or of standard type in 
terms of derivatives but involving the matrix structure of fields in a non-trivial way \cite{EHS}).  
Chiral rings give the ring of functions on the vacuum moduli space  and the study of this space for 
vanishing superpotential terms and at finite $N$ should be of interest from a purely gauge theoretic perspective. 

The key qualitative result of this paper is that the quiver diagram, which is initially 
introduced to describe the matter content of a gauge theory with product gauge group, comes to life as a 
powerful tool for the computation of  counting and correlators of chiral operators. 
The explicit formulae in the bulk of the paper covers the cases with any number of $U(N_a)$ gauge 
groups and any number of bifundamentals (which includes adjoints). 
 Section \ref{sec:counting} starts from 
the known counting formula in terms of group integrals to arrive at an expression in terms of 
Young diagrams. Specifically the result is expressed in terms of Littlewood-Richardson (LR) coefficients, which 
are known to given in terms of efficient  combinatoric rules for combining Young diagrams, familiar from 
tensor products of $U(N)$  irreps. The finite $N$ constraints are simply $ l(R_a) \le N_a$, requiring the 
lengths $l(R_a)$ of the first column to be less than $N$.  The form of the LR coefficients can be read off 
by a simple manipulations on the quiver diagram. The general equation is \ref{eq:N_nabalph} and the diagrammatic rules 
are stated after the equation. The rules involve the application of a move we call {\it splitting-the-nodes}, 
the appropriateness of which is immediately visible from an inspection of the Figures 
\ref{fig:quiver_split_C3}, \ref{fig:quiver_split_con}, \ref{fig:quiver_split_C3Z2}.
The quiver obtained by thus splitting the nodes of the quiver defining  the gauge theory,
is called the {\it split-node quiver}.   When there are multiple flavours $M_{ab}$  of fields for 
the same initial and end-points of the quiver, 
we can organize the counting in terms of representations these flavour groups $ U(M_{ab} ) $. 
This covariant counting is given in eqn. (\ref{eq:N_Lab}).  In addition to LR coefficients, it involves the 
Kronecker product coefficients, which are multiplicities depending on a triple of Young diagrams 
all with  the same number of boxes.  In section \ref{sec:infiniteproduct} we turn to the simplifications 
which arise when we consider operators containing a total number of fields which is less than the $N_a$. 
This allows us to  derive an infinite product generating function \ref{Infprod}, of somewhat sirprising simplicity, 
containining terms which have a simple description in terms of loops in the quiver. 

In section \ref{sec:constops} we show that the effectiveness  of the quiver diagrams 
continues when we consider the two point functions in the quiver theory. In particular 
we compute the 2-point functions involving gauge-invariants constructed from 
holomorphic functions of  the  chiral matter fields inserted at one point, and anti-holomorphic 
operators at another point. By taking one point to zero and the other to infinity, 
this defines an inner product for the operators. We find,  for a general quiver $Q$, the analogs 
of the equations (\ref{schurops})  and (\ref{schurorthog}). 

To motivate our strategy for arriving at the quiver analogs of these,
we note that  permutation group characters appearing in (\ref{schurops})
obey some orthogonality and invariance properties  e.g. 
which are useful in considering the correlators of  the half-BPS sector in $\cN=4$ SYM
\begin{equation}\begin{split}\label{snchareqs}  
 \chi_R ( \sigma  )  &=  \chi_R (\alpha\sigma \alpha^{-1} ) \\  
\sum_{ \sigma \in S_n } \chi_R ( \sigma ) \chi_S ( \sigma )  &=    n! \delta_{ RS } \\  
\end{split}\end{equation}
A more complete list of the identities is in Appendix  \ref{app:identities_characters}. 
For a general quiver $Q$, we   choose   integers $n_{ab} \ge 0$  for each directed edge of the quiver,  
which determine integers $n_a$ for each node according to  $ n_a = \sum_b n_{ba} = \sum_b n_{ba} $. 
In addition we choose  irreducible representation labels $R_a$ of  $  S_{n_a} $ for each node, 
i.e Young diagrams with $n_a$ boxes (restricted to $ l( R_a) \le N ) $, i.e no more than $N_a$ rows).
We choose  irreps $r_{ab} $ of $S_{n_{ab}}  $ for each edge. 
Finally $ \nu^+_a  $ is a choice from the multiplicity of the irrep   $ \otimes_b  r_{ba} $ of  $  \times_b S_{n_{ba} }  $ 
in  the restriction  of irrep  $R_a $ of  $ S_{n_a}  $ to the subgroup  $  \times_b S_{n_{ba} }  $  ; 
 $\nu^-_a $ is  a choice from the multiplicity of the irrep   $ \otimes_b  r_{ab} $ of  $  \times_b S_{n_{ab } }  $  in  the
restriction  of irreps  $R_a$ of  $ S_{n_a}  $ to the subgroup  $  \times_b S_{n_{ab} }  $.  
These multiplicities are given by Littlewood-Richardson coefficients. 
We use the label $ \bL $  for the whole set $ \{ R_a , r_{ab}   , \nu^-_{a}  , \nu^+_{a}  \} $
of representation theoretic labels.  Given this data, we define { \it quiver characters} , 
\begin{equation}\begin{split} 
\chi_Q ( \bL ,  \sigma_a ) 
\end{split}\end{equation}
which obey analogs of the above \ref{snchareqs}.  There is one permutation $\sigma_a$ for each node. 
These properties are stated in equations \ref{eq:chiQ_inv}\ref{eq:chiQ_L_sum}\ref{eq:CQ_sigsig_sum}
and proved in Appendix  \ref{appsec:proofs-gen-char-ids}.   
The standard symmetric group identities can be viewed as a special case of 
theese quiver character identities, when  the quiver consists of one edge 
connecting a node to itself. This simple quiver is the one  relevant to the half-BPS sector of $N=4$ SYM.

The quiver  characters are written in terms of matrix elements of the permutations 
$\sigma_a$  in irreps $R_a$, contracted with branching coefficients of symmetric groups. 
In terms of split-node quiver, the formulae for the quiver chracters can be written 
down by inserting the $\sigma_a$ in the lines introduced in  the splitting of the nodes, 
which join the $\nu^+_a$ node to the $\nu^-_a$ node. Branching coefficients 
are associated with these nodes. The contractions of these branching coefficients and matrix elements are most 
clearly understood by looking at a few examples. We   recommend to the reader a casual look 
at Figures  \ref{eq:chiQ_C3_diag},\ref{conifoldchar} \ref{eq:chiL_C3Z2}which are relevant for 
$\mC^3 , \cC , \mC^3/Z_2$ respectively,  before delving into 
the detailed formulae for the quiver characters.
The precise rules for associating formulae to these diagrams are explained in 
Section \ref{sec:constops} and Appendix \ref{appsec:symgpform}.

Bases diagonalizing the CFT inner product for chiral operators are  not unique. 
This is well known already in studies of the eighth-BPS sector of $\cN=4$ SYM. 
We have the restricted Schur basis, where there are Young diagram labels for 
each type of chiral field. This basis is not covariant under the global symmetries
mixing the different types of arrows with the same start and end  points. 
The basis described above, labelled by $\bL$ is the generalization to any 
quiver of the restricted Schur basis. We will, not surprisingly, call it the restricted 
Schur basis for general quivers. We also develop the covariant basis for general 
quivers. There are again generalized characters for any quiver $Q$, with representation labels $\bK$. 
Analogous character identities are derived and used to prove the orthogonality of 
the corresponding operators.

Section \ref{sec:chir-ring} gives the structure constants of the chiral ring 
both in the restricted Schur basis and the covariant basis. 
The Littlewood-Richardson coefficients 
\begin{equation}\begin{split} 
g ( R_1 , R_2 , R_3 ) =  { 1 \over n_1 ! n_2! } \sum_{\s_1 \in S_{n_1}  } \sum_{ \s_2 \in S_{n_2} } \chi_{R_1} ( \s_1  ) \chi_{R_2}  ( \s_2  ) \chi_{ R_3} ( \s_1 \circ \s_2 )    
\end{split}\end{equation}
have a generalization 
\begin{equation}\begin{split} 
g^Q ( \bL^{(1)}  , \bL^{(2)}  , \bL^{(3)}  ) = 
\sum_{ \bsig_1 , \bsig_2 } \chi_Q ( \bL^{(1)}   , \bsig^{(1)}  ) \chi_Q (\bL^{(2)}  ,  \bsig^{(2)}  ) \chi^{Q} ( \bL^{(3)}  , 
\bsig^{(1)}  \circ \bsig^{(2)} )  
\end{split}\end{equation}
These are the chiral ring structure constants for the free quiver theories. 
By studying these structure constants, we obtain selection rules for the 
$ r_{ab} , R_a $ in the restricted Schur basis, as stated in equation 
(\ref{RSBsel}).  
The result is expressed in a factorized form : there is  a product over the gauge groups, 
and for each gauge group there is product with the  $\nu^+ $ and $\nu^-$ multiplicity labels  
appearing in separate factors (see equation \ref{eq:GLLL_result}).  
Analogous selection rules are derived for the covariant basis (\ref{CBsel}). 
 There is a factorization over the gauge groups with $\nu^-$ labels again separated 
from $\nu^+$ factors, while there are also factors for the directed edges (\ref{eq:GKKK_result}).  
Even for the case of $\mC^3$, all these selection rules controlling the 
chiral ring structure constants at finite $N$ have not been made explicit before.

Section \ref{sec:surfaces} observes that the counting and correlators of the gauge invariant operators
can be interpreted in terms of observables in topological field theory on 
Riemann surfaces, with $S_n$ gauge group. The integer $n$ depends on the $n_{ab;\a} $.
The Riemann surface is obtained by thickening the quiver. See Figures 
\ref{fig:c3gluing},  \ref{fig:congluing}, \ref{fig:C3Z2surface} for the Riemann surfaces arising in the case of 
$\mC^3 , \cC , \mC^3/Z_2$ respectively. The $S_n$ topological field theory 
on the thickened quiver is related to counting of covers of this Riemann surface. 
The covering spaces can be interpreted as string worldsheets following an analogous logic 
which lead to the development of the string theory of large $N$ two dimensional Yang Mills
\cite{grta}. It will be interesting to clarify the role of this thickened quiver Riemann surface 
in the context of Sasaki-Einstein geometries arising in AdS/CFT for quiver gauge theories
in the toric cases  \cite{BFHMSVW1,*BFHMSVW2}.

Section \ref{sec:int-chir-ring} explains how the results on free chiral 
 operators developed here can 
be used to approach the construction of the chiral ring when a non-zero superpotential is 
turned on. This allows us to make some comments  on our original motivating interest, 
the connection between giant gravitons and operators.  Section \ref{sec:fundmatt} starts the discussion of how to extend the results for general theories with bi-fundamental fields (which
may include adjoints), to the case where there are fundamentals or anti-fundamentals.  For the case of SQCD,  we describe counting formulae in terms 
of Young diagrams, making contact with recent literature, and we give a corresponding orthognal 
basis of operators. 
Restricting for concreteness to the conifold case, 
Section \ref{sec:IRfxdpoint} recalls the difference between the UV and IR fixed points 
(both in the limit of  zero super-potential) and explains 
  the fact that the chiral ring structure constants calculated at the 
free UV fixed point are the same as at the IR fixed point with vanishing 
superpotential.  This section concludes with some avenues for future research.

%%%%%%%%%%%%%%%%%%%%%%%%%%%%%%%%%%%%%%%%%%%%%%%%%%%%%%%%%%%%%%%%%%%%%
%%%%%%%%%%%%%%%%%%%%%%%%%%%%%%%%%%%%%%%%%%%%%%%%%%%%%%%%%%%%%%%%%%%%%

\section{Counting operators}\label{sec:counting} 

In this section we derive counting formulas for chiral gauge invariant operators in a general quiver gauge theory. We find that counting is neatly expressed in terms of the \emph{split-node quiver}, which is a simple modification of the quiver diagram, with Young diagram labels on the edges, and Littlewood-Richardson multiplicities associated with the nodes. In the case of the covariant basis, we will also need Kronecker product multiplicities for the symmetric groups. 

An $\cN=1$ supersymmetric quiver gauge theory is defined by a directed graph, called quiver, a gauge group factor associated to each quiver node, and a superpotential. For most of this paper we consider a free theory, with vanishing superpotential. We  take the gauge group to be $\prod_{a=1}^G U(N_a)$, where $a$ runs over $G$ nodes. Each arrow in the quiver between nodes $a$ and $b$ denotes a chiral multiplet transforming as $(N_a, \bar N_b)$. We denote the number of directed arrows from  $a$ to $b$ by $M_{ab}$. The free theory has a global symmetry $\prod_{a,b} U(M_{ab})$. The full matter content is denoted by
\begin{equation}
\bPhi = \{ \Phi_{ab;\a} : \;\alpha \in \{ 1, \ldots, M_{ab} \} \; \}
\end{equation}
An example that we will often use is the quiver for $\C3Z2$ theory, with a gauge group 
generalized to  $U(N_1)\times U(N_2)$ shown in Figure~\ref{fig:quiverC3Z2}. It is rich enough to demonstrate the different ingredients we will need to deal with the most general quiver.

\begin{figure}[h]
\centering
\mytikz{
	\node (s1) at (-1.5,0) [circle,draw] {$1$};
	\node (s2) at (1.5,0) [circle,draw] {$2$};
	\draw [postaction={decorate}] (s1.-120) .. controls +(-120:0.5) and +(0,-0.6) .. (-2.5,0) node[left]{$\Phi_{11}$} .. controls +(0,0.6) and +(120:0.5) .. (s1.120);
	\draw [postaction={decorate}] (s2.60) .. controls +(60:0.5) and +(0,0.6) .. (2.5,0) node[right]{$\Phi_{22}$} .. controls +(0,-0.6) and +(-60:0.5) .. (s2.-60);		
	\draw [postaction={decorate}] (s1) to [bend right=40] node[above]{$\Phi_{12;1}$} (s2);
	\draw [postaction={decorate}] (s1) to [bend right=60] node[below]{$\Phi_{12;2}$} (s2);
	\draw [postaction={decorate}] (s2) to [bend right=40] node[below]{$\Phi_{21;1}$} (s1);
	\draw [postaction={decorate}] (s2) to [bend right=60] node[above]{$\Phi_{21;2}$} (s1);	
}
\caption{$\C3Z2$ quiver}
\label{fig:quiverC3Z2}
\end{figure}

Here we consider counting of chiral gauge invariant operators, such as, for the $\C3Z2$ example:
\begin{equation}
	\tr(\Phi_{11}\Phi_{11}), \; \tr(\Phi_{12;1} \Phi_{21;2}), \; \tr(\Phi_{11}\Phi_{12;1}\Phi_{22}\Phi_{21;1}), \; \ldots
\end{equation}
graded by the number of times $\{ n_{ab;\a} \}$ each field appears in the operator.  The numbers $n_{ab;\a}$ determine the numbers of indices in the fundamental and anti-fundamental of each gauge group $U(N_a)$.  These have to be equal by gauge invariance and they are denoted by $n_a$
\bea 
n_a = \sum_{ b }\sum_{\alpha=1}^{M_{ba}} n_{ba;\a} = \sum_{ b } \sum_{\alpha=1}^{M_{ab}} \ n_{ab;\a}
\eea

Note that in the limit $N_a \rightarrow \infty$ gauge invariant operators are in one-to-one correspondence with closed cycles in the quiver, but for finite $N_a$ there are non-trivial identifications between operators. In Section~\ref{sec:groupintegral} we use group integral formula to directly derive finite $N_a$ results, which will be our main focus in this paper. Furthermore, in Section~\ref{sec:infiniteproduct} we also show how in the $N_a \rightarrow \infty$ limit our results lead to particularly nice formulas for counting closed cycles in a directed graph.

\subsection{The group integral formula} 
\label{sec:groupintegral}

There is a group integral formula for the counting of gauge-invariant 
operators \cite{sundborg,ammpv,GRR,kim}. 
It has  been useful in the context of computation of indices recently. 
We will use the group integral formula to show that the finite $N$ counting 
can be expressed in terms of Young diagrams $R_a$ at the nodes with $n_a$ boxes (i.e. $R_a \vdash n_a$),  
$r_{ab;\a}  \vdash n_{ab;\a} $ at the edges and Littlewood-Richardson coefficients 
$ \prod_{ a } g ( \cup_{ b ,\alpha} r_{ab;\a} ; R_a ) g ( \cup_{ b ,\alpha} r_{ba;\a} ; R_a ) $ 
at the edges. The index $\alpha $ always appears on symbols carrying subscripts $a,b$ which run over 
the pairs of gauge groups and range over $ 1 \le \alpha \le M_{ab}  $.  When $M_{ab} =0$, all 
symbols  carrying the corresponding $\alpha$ are dropped from the formulae. 

The partition function for counting operators in any quiver is:
\begin{equation} 
\begin{split} 
\cN (\{    t_{ab;\a} \}; \{  N_a \} ,  \{  M_{ab} \}  ) 
 =   \int \prod_a dU_a  ~~~  e^{ \sum_n  \sum_{a, b ,\alpha} {  ( t_{ab;\a} )^n   \over n } \tr\, U_a^n \tr ( U_b^{\dagger} )^n }
\end{split} 
\end{equation}
where $t_{ab;\a}$ are fugacities associated with $n_{ab;\a}$. That is, if $\cN (\{    n_{ab;\a} \}; \{  N_a \} ,  \{  M_{ab} \}  ) $ is the number of operators with charges $\{ n_{ab;\a} \}$ then the partition function is
\begin{equation}
\label{eq:Nn_to_Nt}
\cN (\{ t_{ab;\a} \}; \{N_a\} , \{M_{ab}\}  ) \equiv
\sum_{ \{n_{ab;\a}\} }  \left(\prod_{a , b ,\alpha}\, 
  ( t_{ab;\a} )^{ n_{ab;\a} } \right)
 \cN (\{ n_{ab;\a} \}; \{N_a\} , \{M_{ab}\})
\end{equation}

We will henceforth write $ \int $ for  $\int \prod_a dU_a$. Writing the exponential as a product and expanding in series 
\begin{equation} 
\begin{split} 
&   \cN ( \{    t_{ab;\a} \} ; \{  N_a \} ,  \{  M_{ab} \}  )     \\
& =  \sum_{ \{ k_{ab ; \a}^{(n)}\}   =0   }^{\infty}  ~~  \int \prod_{ a , b ,\alpha  , n } (t_{ab;\a} )^{ n k_{ab;\a}^{(n)} }
 ~~ {   ( \tr\, U_a^n )^{k_{ab;\a}^{(n)}  }  ( \tr\, U_b^{\dagger ~ n } )^{k_{ab;\a}^{(n)}  } \over n^{k_{ab;\a}^{(n)} } k_{ab;\a}^{(n)}!  }   \\
 & = \int  \sum_{ \{ k_{ab ; \a}^{(n)} \}  =0 }^{\infty} ~ 
 \prod_{ a , b ,\alpha}  ( t_{ab;\a}  )^{\sum_{ n } n k_{ab;\a}^{(n)}  } 
 ~ \prod_n \prod_{ a , b ,\alpha} {   ( \tr\, U_a^n )^{k_{ab;\a}^{(n)}  }  ( \tr\, U_b^{\dagger ~ n } )^{k_{ab;\a}^{(n)}  } \over n^{k_{ab;\a}^{(n)} } k_{ab;\a}^{(n)}!  } \\
 & = \int \sum_{ \{ n_{ab;\a} \} = 0 }^{ \infty } ~  \prod_{a , b ,\alpha} ~ 
 { ( t_{ab;\a} )^{ n_{ab;\a} }    \over n_{ab;\a} ! }  
 \\ & \qquad
\times \sum_{ \sigma_{ab;\a}  \in S_{n_{ab;\a} }}
\prod_{a} \sum_{ R_a \vdash n_a } \chi_{ R_a} ( \cup_{b ,\alpha} \sigma_{ab;\a} ) \chi_{ R_a} ( U_a ) 
 \sum_{ S_a \vdash n_a } \chi_{ S_a} ( \cup_{ b ,\alpha} \sigma_{ ba;\a}  ) \chi_{ S_a} ( U_a^{\dagger} ) 
\end{split} 
\end{equation} 
We have factored the powers  $  ( t_{ab;\a} )^{ n_{ab;\a} } $, recognized that 
for fixed $ n_{ab;\a}$, the sums over $k_{ab;\a}^{(n)} $ run over partitions 
of   $ n_{ab;\a}$, which correspond to conjugacy classes in $ S_{n_{ab;\a} } $. 
We observe that 
\begin{equation} 
\begin{split} 
\prod_n \prod_{ a , b ,\alpha}   ( \tr U_a^n )^{k_{ab;\a}^{(n)}  }  
= \sum_{ \substack{ R_a \vdash n_a \\ l(R_a) \le N_a } }\chi_{R_a} ( \cup_{  a , b  ,\alpha} \s_{ab;\a} )   \chi_{R_a} ( U_a ) 
\end{split} 
\end{equation} 
for $ \s_{ab;\a}  $ being a permutation in the conjugacy class  of  $ n_{ab;\a}$ 
specified by $k_{ab;\a}^{(n)} $. 
Since the number of permutations in the specified conjugacy class
is precisely 
\bea 
{ n_{ab;\a} ! \over  \prod_n n^{k_{ab;\a}^{(n)} } k_{ab;\a}^{(n)}  ! } 
\eea 
 we have converted the sums over partitions to sums over permutations. We have also recognized that the
 traces can be expanded in terms of Schur Polynomials with coefficients given by the characters of 
 these permutations. 
Note, crucially, the height of the Young diagram $R_a$ is at most $N_a$, this fully captures the effect of finite $N_a$.
Using the orthogonality of the Schur Polynomials under group integration 
\begin{equation}
\int dU_a \; \chi_{R_a}(U_a) \chi_{S_a}(U_a^\dagger) = \delta_{R_a S_a}
\end{equation}
we can expand characters in irreps $R_a$ of $S_{n_a} $ 
 into characters of $ \prod_{ b ,\alpha} r_{ab;\a} $  with expansion coefficients which are Littlewood-Richardson 
 numbers
\begin{equation}
 \chi_{ R_a} ( \cup_{b ,\alpha} \sigma_{ab;\a} ) = 
 \sum_{ r_{ab;\a} \vdash n_{ab;\a} } 
 g(\cup_{b,\alpha} r_{ab;\alpha};R_a)
 \prod_{ b ,\alpha} \chi_{ r_{ab;\a} }  ( \sigma_{ab;\a}  ) 
\end{equation}
 This leads to 
\begin{equation}
\label{eq:N_tabalph}
\begin{split} 
 &  \cN (\{ t_{ab;\a} \}; \{N_a\} , \{M_{ab}\}  ) \\ 
 & = \sum_{ \{n_{ab;\a}\} }  \prod_{a , b ,\alpha} ~ 
 { ( t_{ab;\a} )^{ n_{ab;\a} }    \over n_{ab;\a} ! } 
 \sum_{ \sigma_{ab;\a}  \in S_{n_{ab;\a} }}   
 \sum_{ \substack{ R_a \vdash n_a \\ l(R_a) \le N_a } }
 \sum_{ r_{ab;\a}  \vdash n_{ab;\a} } 
 \sum_{ s_{ab;\a}  \vdash n_{ab;\a} } 
  \\ 
 &~~~~\qquad    
 \prod_a 
 g ( \cup_{ b ,\alpha} r_{ab;\a} ; R_a ) 
  g ( \cup_{b ,\alpha} s_{ba;\a} ; R_a )
 \prod_{ a, b ,\alpha}  
 \chi_{r_{ab;\a}  } (  \sigma_{ab;\a}  ) 
 \chi_{s_{ab;\a}  } (  \sigma_{ab;\a}  )  
 \\
&=  
 \sum_{ \{ n_{ab;\a} \} }  \prod_{a , b ,\alpha} \;
 { ( t_{ab;\a} )^{ n_{ab;\a} }   }   
 \sum_{ \substack{ R_a \vdash n_a \\ l(R_a) \le N_a } }
 \sum_{ r_{ab;\a}  \vdash n_{ab;\a} } 
 \prod_a g ( \cup_{ b ,\alpha} r_{ab;\a} ; R_a ) g ( \cup_{b ,\alpha} r_{ba;\a} ; R_a )  
\end{split} 
\end{equation}
In the second line we used orthogonality of characters $\sum_{\sigma} \chi_r(\sigma) \chi_s(\sigma) = n! \delta_{rs}$.
This form of the partition function, comparing with (\ref{eq:Nn_to_Nt}),
gives explicit counting for each choice of charges $\{ n_{ab;\a} \}$
\begin{equation}
\label{eq:N_nabalph}
\boxed{ 
\cN (\{ n_{ab;\a} \}; \{N_a\} , \{M_{ab}\}  ) =
 \sum_{ \substack{ R_a \vdash n_a \\ l(R_a) \le N_a } }
 \sum_{ r_{ab;\a}  \vdash n_{ab;\a} } 
 \prod_a  g ( \cup_{ b ,\alpha} r_{ab;\a} ; R_a ) g ( \cup_{b ,\alpha} r_{ba;\a} ; R_a )  
}
\end{equation}

There is a  simple diagrammatic description of this formula, deriving directly from the quiver itself: 

{ \bf Diagrammatic Rules for counting local operators in the quiver theory  }
 
\begin{itemize}

\item Choose integers $n_{ab;\a}\ge 0 $ for all the edges of the quiver $Q$, subject to $n_a = \sum_{ b } n_{ba}$. 

\item Replace each node with a pair of nodes, joined by  a line labelled by a Young diagram $R_a $ with $n_a$ boxes. 
One of these two nodes, called the plus node,  has all incoming lines and the other, called the minus node,  has all outgoing lines. The resulting diagram is the {\it split-mode quiver}. 

\item To all the previously existing edges, attach  Young diagrams $r_{ab;\a}$ with $n_{ab;\a}$ boxes.  

\item To  each minus  node attach a Littlewood-Richardson multiplicity   $g ( \bigcup_b  \bigcup_{\alpha =1 }^{M_{ab} } r_{ab;\a} ; R_a ) $ which couples all the incoming lines to $R_a$. To each plus node  attach the LR multiplicity
$ g ( \bigcup_b  \bigcup_{\alpha =1 }^{M_{ba} } r_{ba;\a} ; R_a ) $

\item Take the product of LR-coefficients over all the nodes. This is the counting of free chiral operators 
with numbers $\{ n_{ba;\a}  \} $ of fields of type $\alpha$ transforming as $(N_a , \bar N_b )$. 

\end{itemize}

\begin{figure}[h] %%% C3 %%%%
\centering
$
\mytikz{ 
	\node (s1) at (0,0) [circle,draw] {$1$};							
	\draw [postaction={decorate}] (s1.-60) .. controls +(-60:0.5) and +(0,-0.6) .. (1,0)
		.. controls +(0,0.6) and +(60:0.5) .. (s1.60);
	\draw [postaction={decorate}] (s1.-75) .. controls +(-75:0.7) and +(0,-0.9) .. (1.3,0)
		.. controls +(0,0.9) and +(75:0.7) .. (s1.75);
	\draw [postaction={decorate}] (s1.-90) .. controls +(-90:0.9) and +(0,-1.2) .. (1.6,0)
		.. controls +(0,1.2) and +(90:0.9) .. (s1.90);		
}
\quad \rightarrow \quad
\mytikz{	
	\node (m) at (0,0.7) [circle,draw,inner sep=0.5mm] {};
	\node (n) at (0,-0.7) [circle,draw,inner sep=0.5mm] {};
	\draw[postaction={decorate}] (m) to node[left]{$R$} (n);
	\draw[postaction={decorate}] (n) .. controls +(-0:0.5) and +(0,-0.5) .. (0.8,0) node[left]{$r_1$}
	.. controls +(0,0.5) and +(0:0.5) .. (m);
	\draw[postaction={decorate}] (n) .. controls +(-30:0.5) and +(0,-0.7) .. (1.3,0) node{$\quad r_2$}
	.. controls +(0,0.7) and +(30:0.5) .. (m);	
	\draw[postaction={decorate}] (n) .. controls +(-60:0.5) and +(0,-1.0) .. (1.8,0) node[right]{$r_3$}
	.. controls +(0,1.0) and +(60:0.5) .. (m);		
}
$
\caption{Split-node quiver for $\mC^3$.}
\label{fig:quiver_split_C3}
\end{figure}

\begin{figure}[h]
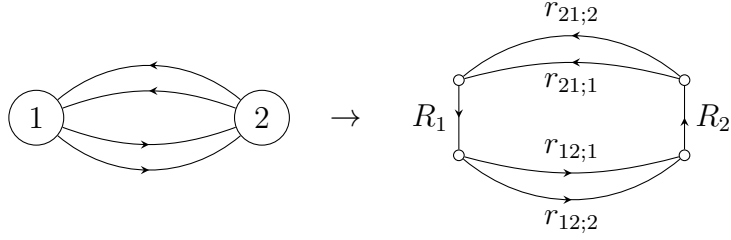
 %%% conifold %%%%
\centering
$
\mytikz{
	\node (s1) at (-1.5,0) [circle,draw] {$1$};
	\node (s2) at (1.5,0) [circle,draw] {$2$};
	\draw [postaction={decorate}] (s1) to [bend right=20] (s2);
	\draw [postaction={decorate}] (s1) to [bend right=40] (s2);
	\draw [postaction={decorate}] (s2) to [bend right=20] (s1);
	\draw [postaction={decorate}] (s2) to [bend right=40] (s1);
}
\quad \rightarrow \quad
\mytikz{	
	\node (m1) at (-1.5,0.5) [circle,draw,inner sep=0.5mm] {};
	\node (n1) at (-1.5,-0.5) [circle,draw,inner sep=0.5mm] {};
	\node (m2) at (1.5,-0.5) [circle,draw,inner sep=0.5mm] {};
	\node (n2) at (1.5,0.5) [circle,draw,inner sep=0.5mm] {};	
	\draw[postaction={decorate}] (m1) to node[left]{$R_1$} (n1);
	\draw[postaction={decorate}] (m2) to node[right]{$R_2$} (n2);
	\draw [postaction={decorate}] (n1) to [bend right=15] node[above]{$r_{12;1}$} (m2);
	\draw [postaction={decorate}] (n1) to [bend right=40] node[below]{$r_{12;2}$} (m2);
	\draw [postaction={decorate}] (n2) to [bend right=15] node[below]{$r_{21;1}$} (m1);
	\draw [postaction={decorate}] (n2) to [bend right=40] node[above]{$r_{21;2}$} (m1);			
}
$
\caption{Split-node quiver for the conifold.}
\label{fig:quiver_split_con}
\end{figure}

\begin{figure}[h] %%% C3/Z2 %%%%
\centering
$
\mytikz{
	\node (s1) at (-1.5,0) [circle,draw] {$1$};
	\node (s2) at (1.5,0) [circle,draw] {$2$};
	\draw [postaction={decorate}] (s1.-120) .. controls +(-120:0.5) and +(0,-0.6) .. (-2.5,0)
		.. controls +(0,0.6) and +(120:0.5) .. (s1.120);
	\draw [postaction={decorate}] (s2.60) .. controls +(60:0.5) and +(0,0.6) .. (2.5,0)
		.. controls +(0,-0.6) and +(-60:0.5) .. (s2.-60);		
	\draw [postaction={decorate}] (s1) to [bend right=20] (s2);
	\draw [postaction={decorate}] (s1) to [bend right=40] (s2);
	\draw [postaction={decorate}] (s2) to [bend right=20] (s1);
	\draw [postaction={decorate}] (s2) to [bend right=40] (s1);	
}
\quad \rightarrow \quad
\mytikz{	
	\node (m1) at (-1.5,0.7) [circle,draw,inner sep=0.5mm] {};
	\node (n1) at (-1.5,-0.7) [circle,draw,inner sep=0.5mm] {};
	\node (m2) at (1.5,-0.7) [circle,draw,inner sep=0.5mm] {};
	\node (n2) at (1.5,0.7) [circle,draw,inner sep=0.5mm] {};
	\draw[postaction={decorate}] (m1) to node[right]{$R_1$} (n1);
	\draw[postaction={decorate}] (m2) to node[left]{$R_2$} (n2);
	\draw [postaction={decorate}] (n1) to [bend right=15] node[above]{$r_{12;1}$} (m2);
	\draw [postaction={decorate}] (n1) to [bend right=40] node[below]{$r_{12;2}$} (m2);
	\draw [postaction={decorate}] (n2) to [bend right=15] node[below]{$r_{21;1}$} (m1);
	\draw [postaction={decorate}] (n2) to [bend right=40] node[above]{$r_{21;2}$} (m1);	
	\draw[postaction={decorate}] (n1) .. controls +(-180:0.5) and +(0,-0.5) .. (-2.3,0) node[left]{$r_{11}$}
	.. controls +(0,0.5) and +(180:0.5) .. (m1);		
	\draw[postaction={decorate}] (n2) .. controls +(0:0.5) and +(0,0.5) .. (2.3,0) node[right]{$r_{22}$}
	.. controls +(0,-0.5) and +(-0:0.5) .. (m2);		
}
$
\caption{Split-node quiver for $\C3Z2$.}
\label{fig:quiver_split_C3Z2}
\end{figure}

These steps are illustrated for $\mC^3$ in Figure~\ref{fig:quiver_split_C3}. We have suppressed the $a,b $ indices labeling 
the nodes of the quiver, since there is only one node in this case. 
\begin{equation}
\cN_{\mC^3}( n_{1} , n_{2} , n_{3} ; N)  = \sum_{\substack{ R  \vdash n \\ l ( R ) \le N}   } 
  g ( r_{1} , r_{2} , r_{3} ; R  ) \,  g ( r_{1} , r_{2} , r_{3} ; R ) 
\end{equation}
This equation was given in \cite{db1,collins}. 
For $\cC $, we read off the counting from (\ref{eq:N_nabalph}) or by following the steps in 
Figure~\ref{fig:quiver_split_con}. 
\begin{equation}
\label{eq:counting_con}
\begin{split}
& \cN_{\cC} ( n_{12;1}  , n_{12;2}  ,n_{21;1}  , n_{21;2} ; N_1 , N_2     ) 
 = \sum_{\substack { R_1 \vdash n_1 \\ l(R_1 ) \le N_1  } }  \sum_{\substack { R_2 \vdash n_2 \\ l(R_2 ) \le N_2  } } 
\sum_{r_{12;1} \vdash n_{12;1}  }  \sum_{r_{12;2} \vdash n_{12;2}  }   \sum_{r_{21;1} \vdash n_{21;1}  }   \sum_{r_{21;2} \vdash n_{21;2}  }  \\
& \qquad \qquad \qquad 
 g( r_{12;1} ,  r_{12;2} ; R_1 ) \, g( r_{12;1} ,  r_{12;2} ; R_2 ) \,  g( r_{21;1} ;  r_{21;2} , R_1 ) \,  g( r_{21;1} ,  r_{21;2} ; R_2 )
\end{split}
\end{equation}
This counting for the free conifold operators has not been given before. 
For $ \mC^3/\mZ_2 $, again following the steps above shown in Figure~\ref{fig:quiver_split_C3Z2} or specializing  (\ref{eq:N_nabalph}), we get 
\begin{equation}
\label{eq:counting_c3z2}
\begin{split}
& \cN_{\C3Z2} (n_{11} , n_{22} ,  n_{12;1}  , n_{12;2}  ,   n_{21;1}  , n_{21;2} ; N_1 , N_2     ) 
\\
& =  \sum_{\substack { R_1 \vdash n_1 \\ l(R_1 ) \le N_1  } }  \sum_{\substack { R_2 \vdash n_2 \\ l(R_2 ) \le N_2  } } 
\sum_{ r_{11} \vdash n_{11} } \sum_{ r_{22} \vdash n_{22} } 
\sum_{r_{12;1} \vdash n_{12;1}  }  \sum_{r_{12;2} \vdash n_{12;2}  }   \sum_{r_{21;1} \vdash n_{21;1}  }   \sum_{r_{21;2} \vdash n_{21;2}  }  \\ 
& \quad
 g(r_{11} ,  r_{12;1} ,  r_{12;2} , R_1 ) \, g(r_{22} ,  r_{12;1} ,  r_{12;2} , R_2 ) \,  g(r_{11}  ,  r_{21;1} ,  r_{21;2} , R_1 ) \,  g( r_{22} ,  r_{21;1} ,  r_{21;2} , R_2 )
\end{split}
\end{equation}

There is another useful form of the counting formula  where we do not specify $\{ n_{ab;\a} \}$ but only $\{ n_{ab} \}$
\begin{equation}
 n_{ab} = \sum_{\alpha} n_{ab;\a}
\end{equation}
that is, the total number of fields transforming under $U(M_{ab})$ global symmetry group. This will be related to the covariant basis, where we can count states according to representations of the global symmetry group $\prod_{ab} U(M_{ab})$. We group together representations $\cup_{\alpha} r_{ab;\a}$ corresponding to the same pair $(a,b)$, and expand the multiplicities in (\ref{eq:N_nabalph}) as
\begin{equation}
\begin{split}
	g(\cup_{b,\alpha} r_{ab;\a};R_a) &= \sum_{ \{ s_{ab}^- \} } g(\cup_b s_{ab}^-;R_a) \prod_b g(\cup_{\alpha} r_{ab;\a};s_{ab}^-)
	\\
	g(\cup_{b,\alpha} r_{ba;\a};R_a) &= \sum_{ \{ s_{ba}^+ \} } g(\cup_b s_{ba}^+;R_a) \prod_b g(\cup_{\alpha} r_{ba;\a};s_{ba}^+)
\end{split}
\end{equation}
$s_{ab}^\pm$ are intermediate representations in the reductions $R_a \rightarrow \{ \cup_b s_{ab}^- \} \rightarrow \{ \cup_{b,\alpha} r_{ab;\a} \}$ and $R_a \rightarrow \{ \cup_b s_{ba}^+ \} \rightarrow \{ \cup_{b,\alpha} r_{ba;\a} \}$. Next, we apply (\ref{eq:gg_Cg}) for fixed $(a,b)$:
\begin{equation}
	\sum_{ \{ r_{ab;\a} \} } 
	g( \cup_\alpha r_{ab;\a} ; s_{ab}^+)
	g( \cup_\alpha r_{ab;\a} ; s_{ab}^-)
	=
	\sum_{\Lambda_{ab}} C(s_{ab}^+,s_{ab}^-,\Lambda_{ab}) 
	g(\cup_\alpha [n_{ab;\a}];\Lambda_{ab})
\end{equation}
where $\cup_\alpha [n_{ab;\a}]$ is the irrep of $ \times_{\alpha}  S_{n_{ab;\alpha}}$
consisting of the single row symmetric irreps $[n_{ab;\a}]$ for each factor.
We find  
\begin{equation}
\label{eq:N_abalph_cov}
\begin{split}
\cN (\{ n_{ab;\a} \}; \{N_a\} , \{M_{ab}\}  ) &=
 \sum_{ \substack{ R_a \vdash n_a \\ l(R_a) \le N_a } }
 \sum_{ s_{ab}^+ \vdash n_{ab} } 
 \sum_{ s_{ab}^- \vdash n_{ab} }
 \sum_{ \substack{ \Lambda_{ab} \vdash n_{ab} \\ l(\Lambda_{ab}) \le M_{ab} } }
 \prod_a    
 g ( \cup_{b} s_{ab}^{-} ; R_a ) 
 g ( \cup_{b} s_{ba}^{+} ; R_a ) 
 \\
 & ~~~~~~~~~~~~~~~~~~~~~~~~~~~~~
 \times \prod_{a,b} C(s_{ab}^+,s_{ab}^-,\Lambda_{ab}) 
 g(\cup_{\alpha} [n_{ab;\a}]; \Lambda_{ab})
\end{split}  
\end{equation}
The new labels $\Lambda_{ab}$ are precisely the $U(M_{ab})$ representations. (\ref{eq:N_abalph_cov}) can be understood by 
noting that the  number of states in the irrep $\Lambda_{ab} $, a Young diagram of $U(M_{ab} ) $ with 
$n_{ab}$ boxes,  with specified charges 
$n_{ab;\a}$ under the diagonal $U(1)^{M_{ab}} $, is given by the Littlewood-Richardson 
multiplicity
\bea 
g ( \cup_{\alpha} [ n_{ab;\a} ]  ; \Lambda_{ab} ) 
=  \frac{1}{\prod_{a,b,\alpha} n_{ab;\a} ! }  \sum_{ \sigma_{ab;\a} \in S_{n_{ab;\a}  }}  
  \chi_{\Lambda_{ab}  }  ( \cup_{\alpha}  \sigma_{ab;\a} ) 
\eea
Thus if we do not refine by $n_{ab;\a}$, but count all the states with fixed $\{n_{ab}\}$, we count the total number of states in the representation
\begin{equation}
\label{eq:N_nab}
\begin{split}
\cN (\{ n_{ab} \}; \{N_a\} , \{M_{ab}\}  ) &=
 \sum_{ \substack{ R_a \vdash n_a \\ l(R_a) \le N_a } }
 \sum_{ s_{ab}^+ \vdash n_{ab} } 
 \sum_{ s_{ab}^- \vdash n_{ab} }
 \sum_{ \substack{ \Lambda_{ab} \vdash n_{ab} \\ l(\Lambda_{ab}) \le M_{ab} } }
 \prod_a    
 g ( \cup_{b} s_{ab}^{-} ; R_a ) 
 g ( \cup_{b} s_{ba}^{+} ; R_a ) 
 \\
 & ~~~~~~~~~~~~~~~~~~~~~~~~~~~~~
 \times \prod_{a,b} C(s_{ab}^+,s_{ab}^-,\Lambda_{ab}) 
 \Dim(\Lambda_{ab})
\end{split}  
\end{equation}
where $\Dim(\Lambda_{ab})$ is the size of $U(M_{ab})$ irrep $\Lambda_{ab}$.
We can also, instead of counting individual states, count how many times a particular global symmetry representation $\otimes_{ab} \Lambda_{ab}$ appears
\begin{equation}
\label{eq:N_Lab}
\boxed{
\begin{split}
\cN (\{ \Lambda_{ab} \}; \{N_a\} , \{M_{ab}\}  ) &=
 \sum_{ \substack{ R_a \vdash n_a \\ l(R_a) \le N_a } }
 \sum_{\substack{ s_{ab}^+ \vdash n_{ab} \\ s_{ab}^- \vdash n_{ab} }} 
 \prod_a    
 g ( \cup_{b} s_{ab}^{-} ; R_a ) 
 g ( \cup_{b} s_{ba}^{+} ; R_a ) 
\prod_{a,b}
 C(s_{ab}^+,s_{ab}^-,\Lambda_{ab})  
\end{split}  
}
\end{equation}

The following figures illustrate the structure of this formula to the case of $\mC^3 , \cC $ and $\mC^3/\mZ_2$ quivers. The white nodes again represent LR multiplicities and the new black nodes represent Kronecker product multiplicities $C(s^+_{ab},s^-_{ab},\Lambda_{ab})$.

\begin{figure}[h] %%% C3 %%%%
\centering
$
\mytikz{ 
	\node (s1) at (0,0) [circle,draw] {$1$};							
	\draw [postaction={decorate}] (s1.-60) .. controls +(-60:0.5) and +(0,-0.6) .. (1,0)
		.. controls +(0,0.6) and +(60:0.5) .. (s1.60);
	\draw [postaction={decorate}] (s1.-75) .. controls +(-75:0.7) and +(0,-0.9) .. (1.3,0)
		.. controls +(0,0.9) and +(75:0.7) .. (s1.75);
	\draw [postaction={decorate}] (s1.-90) .. controls +(-90:0.9) and +(0,-1.2) .. (1.6,0)
		.. controls +(0,1.2) and +(90:0.9) .. (s1.90);		
}
\quad \rightarrow \quad
\mytikz{	
	\node (m) at (0,0.7) [circle,draw,inner sep=0.5mm] {};
	\node (n) at (0,-0.7) [circle,draw,inner sep=0.5mm] {};
	\node (t) at (1.3,0) [circle,fill,inner sep=0.5mm] {};
	\node (b) at (2.0,0) {$\Lambda$};
	\draw[postaction={decorate}] (m) to node[left]{$R$} (n);
	\draw[postaction={decorate}] (n) to [bend right=60] node[below]{$R$} (t);
	\draw[postaction={decorate}] (t) to [bend right=60] node[above]{$R$} (m);
	\draw[-] (t) to (b);
}
$
\caption{Covariant quiver for $\mC^3$.}
\label{fig:quiver_split_cov_C3}
\end{figure}

\begin{figure}[h]
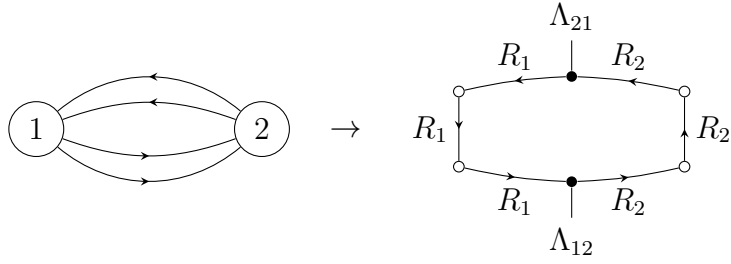
 %%% conifold %%%%
\centering
$
\mytikz{
	\node (s1) at (-1.5,0) [circle,draw] {$1$};
	\node (s2) at (1.5,0) [circle,draw] {$2$};
	\draw [postaction={decorate}] (s1) to [bend right=20] (s2);
	\draw [postaction={decorate}] (s1) to [bend right=40] (s2);
	\draw [postaction={decorate}] (s2) to [bend right=20] (s1);
	\draw [postaction={decorate}] (s2) to [bend right=40] (s1);
}
\quad \rightarrow \quad
\mytikz{	
	\node (m1) at (-1.5,0.5) [circle,draw,inner sep=0.5mm] {};
	\node (n1) at (-1.5,-0.5) [circle,draw,inner sep=0.5mm] {};
	\node (m2) at (1.5,-0.5) [circle,draw,inner sep=0.5mm] {};
	\node (n2) at (1.5,0.5) [circle,draw,inner sep=0.5mm] {};	
	\node (t12) at (0,-0.7) [circle,fill,inner sep=0.5mm] {};	
	\node (b12) at (0,-1.5) {$\Lambda_{12}$};
	\node (t21) at (0,0.7) [circle,fill,inner sep=0.5mm] {};
	\node (b21) at (0,1.5) {$\Lambda_{21}$};
	\draw[postaction={decorate}] (m1) to node[left]{$R_1$} (n1);
	\draw[postaction={decorate}] (m2) to node[right]{$R_2$} (n2);
	\draw [postaction={decorate}] (n1) to [bend right=5] node[below]{$R_1$} (t12);
	\draw [postaction={decorate}] (t12) to [bend right=5] node[below]{$R_2$} (m2);
	\draw [postaction={decorate}] (n2) to [bend right=5] node[above]{$R_2$} (t21);
	\draw [postaction={decorate}] (t21) to [bend right=5] node[above]{$R_1$} (m1);
	\draw[-] (t12) to (b12);
	\draw[-] (t21) to (b21);
}
$
\caption{Covariant quiver for the conifold.}
\label{fig:quiver_split_cov_con}
\end{figure}

\begin{figure}[h] %%% C3/Z2 %%%%
\centering
$
\mytikz{
	\node (s1) at (-1.5,0) [circle,draw] {$1$};
	\node (s2) at (1.5,0) [circle,draw] {$2$};
	\draw [postaction={decorate}] (s1.-120) .. controls +(-120:0.5) and +(0,-0.6) .. (-2.5,0)
		.. controls +(0,0.6) and +(120:0.5) .. (s1.120);
	\draw [postaction={decorate}] (s2.60) .. controls +(60:0.5) and +(0,0.6) .. (2.5,0)
		.. controls +(0,-0.6) and +(-60:0.5) .. (s2.-60);		
	\draw [postaction={decorate}] (s1) to [bend right=20] (s2);
	\draw [postaction={decorate}] (s1) to [bend right=40] (s2);
	\draw [postaction={decorate}] (s2) to [bend right=20] (s1);
	\draw [postaction={decorate}] (s2) to [bend right=40] (s1);	
}
\quad \rightarrow \quad
\mytikz{	
	\node (m1) at (-1.5,0.7) [circle,draw,inner sep=0.5mm] {};
	\node (n1) at (-1.5,-0.7) [circle,draw,inner sep=0.5mm] {};
	\node (m2) at (1.5,-0.7) [circle,draw,inner sep=0.5mm] {};
	\node (n2) at (1.5,0.7) [circle,draw,inner sep=0.5mm] {};
	\node (t12) at (0,-1) [circle,fill,inner sep=0.5mm] {};	
	\node (b12) at (0,-2) {$\Lambda_{12}$};
	\node (t21) at (0,1) [circle,fill,inner sep=0.5mm] {};
	\node (b21) at (0,2) {$\Lambda_{21}$};
	\draw[postaction={decorate}] (m1) to node[right]{$R_1$} (n1);
	\draw[postaction={decorate}] (m2) to node[left]{$R_2$} (n2);
	\draw [postaction={decorate}] (n1) to [bend right=10] node[below]{$s_{12}^-$} (t12);
	\draw [postaction={decorate}] (t12) to [bend right=10] node[below]{$s_{12}^+$} (m2);
	\draw [postaction={decorate}] (n2) to [bend right=10] node[above]{$s_{21}^-$} (t21);
	\draw [postaction={decorate}] (t21) to [bend right=10] node[above]{$s_{21}^+$} (m1);
	\draw[-] (t12) to (b12);
	\draw[-] (t21) to (b21);
	\draw[postaction={decorate}] (n1) .. controls +(-180:0.5) and +(0,-0.5) .. (-2.3,0) node[left]{$s_{11}$}
	.. controls +(0,0.5) and +(180:0.5) .. (m1);		
	\draw[postaction={decorate}] (n2) .. controls +(0:0.5) and +(0,0.5) .. (2.3,0) node[right]{$s_{22}$}
	.. controls +(0,-0.5) and +(-0:0.5) .. (m2);		
}
$
\caption{Covariant quiver for $\C3Z2$.}
\label{fig:quiver_split_cov_C3Z2}
\end{figure}

The corresponding formula for $\mC^3$ according to Figure~\ref{fig:quiver_split_cov_C3}
\begin{equation}
\cN_{\mC^3}(\Lambda;N) = \sum_{ \substack{ R \vdash n \\ l(R) \le N } } C(R,R,\Lambda)
\end{equation}
It was first obtained in \cite{Dolan0704} and the matching construction of orthogonal operators given in \cite{BHR1}. 
Since there is only single incoming and outgoing arrow from the white branching nodes in Figure~\ref{fig:quiver_split_cov_C3}, there is no actual branching, and the labels on both sides are $R$. That is, compared to general formula (\ref{eq:N_Lab}) we have $s^+=s^-=R$.

For conifold we have Figure~\ref{fig:quiver_split_cov_con}
\begin{equation}
\cN_{\cC}(\Lambda_{12},\Lambda_{21};N) = \sum_{ \substack{ R_1 \vdash n \\ l(R_1) \le N } } \sum_{ \substack{ R_2 \vdash n \\ l(R_2) \le N } } C(R_1,R_2,\Lambda_{12}) C(R_2,R_1,\Lambda_{21})
\end{equation}
Again the white node multiplicities are trivial, setting $s_{ab}^\pm$ to $R_a$.

For $\C3Z2$ we find non-trivial branching multiplicities, following the diagram Figure~\ref{fig:quiver_split_cov_C3Z2}:
\begin{equation}
\begin{split}
&\cN_{\C3Z2}(\Lambda_{12},\Lambda_{21},n_{11},n_{22};N) = 
\sum_{ \substack{ R_1 \vdash n_1 \\ l(R_1) \le N } } 
\sum_{ \substack{ R_2 \vdash n_2 \\ l(R_2) \le N } }
\sum_{ \substack{ s_{12}^- \vdash n_{12} } }
\sum_{ \substack{ s_{12}^+ \vdash n_{12} } }
\sum_{ \substack{ s_{21}^- \vdash n_{12} } }
\sum_{ \substack{ s_{21}^+ \vdash n_{12} } }
\sum_{ \substack{ s_{11} \vdash n_{11} } }
\sum_{ \substack{ s_{22} \vdash n_{22} } }
\\ & \quad
g(s_{11},s_{12}^-;R_1) g(s_{11},s_{21}^+;R_1)
g(s_{22},s_{21}^-;R_2) g(s_{22},s_{12}^+;R_2) 
C(s_{12}^-,s_{12}^+,\Lambda_{12}) C(s_{21}^-,s_{21}^+,\Lambda_{21})
\end{split}
\end{equation}
The only simplification compared to the generic formula (\ref{eq:N_Lab}) is that $s_{11}^+ = s_{11}^- \equiv s_{11}$ and $s_{22}^+ = s_{22}^- \equiv s_{22}$, since the original quiver has $M_{11}=M_{22}=1$, the corresponding global symmetry factor is abelian, and so $\Lambda_{11}=[n_{11}],\Lambda_{22}=[n_{22}]$ are trivial.

\subsection{Infinite product generating functions}\label{sec:infiniteproduct}

In this section we will use the covariant basis counting (\ref{eq:N_nab}) to 
derive a simple infinite product formula valid when the numbers of fields
are less than the ranks $N_a$. In this case counting gauge invariant operators is the same as counting closed loops in the quiver.

Counting the gauge invariant local operators for fixed ranks $N_a$, numbers  $ M_{ab}$ 
of fields transforming in $(N_a,\bar N_b)$  in the theory, and numbers $n_{ab}$ for the total number of fields of type $(N_a,\bar N_b)$
we have (\ref{eq:N_nab})
\begin{equation}
\label{eq:N_nab_2}
\begin{split}
\cN (\{ n_{ab} \}; \{N_a\} , \{M_{ab}\}  ) &=
 \sum_{ \substack{ R_a \vdash n_a \\ l(R_a) \le N_a } }
 \sum_{ s_{ab}^+ \vdash n_{ab} } 
 \sum_{ s_{ab}^- \vdash n_{ab} }
 \sum_{ \substack{ \Lambda_{ab} \vdash n_{ab} \\ l(\Lambda_{ab}) \le M_{ab} } }
 \prod_a    
 g ( \cup_{b} s_{ab}^{-} ; R_a ) 
 g ( \cup_{b} s_{ba}^{+} ; R_a ) 
 \\
 & ~~~~~~~~~~~~~~~~~~~~~~~~~~~~~
 \times \prod_{a,b} C(s_{ab}^+,s_{ab}^-,\Lambda_{ab}) 
 \Dim(\Lambda_{ab})
\end{split}  
\end{equation}
The finite $N$ constraints are encoded in the requirement that the Young 
diagrams $R_a$ have no more than $N_a$ rows. 

Let us convert it to a partition function with fugacities $\{ t_{ab;\a} \}$ for numbers $\{ n_{ab;\a} \}$. The contribution from a single irrep $\Lambda_{ab}$ is
\begin{equation}
\chi_{ \L_{ab} } ( \mT_{ab} )
\end{equation}
where $ \mT_{ab}$ is a  square matrix of size $M_{ab}$ with  entries $ t_{ab;\a}$ along the diagonal. Thus we can replace $\Dim(\Lambda_{ab})$ with $\chi_{ \L_{ab} } ( \mT_{ab} )$ in (\ref{eq:N_nab_2}) and sum over all representations without restriction on the number of boxes, to get the full partition function: 
\begin{equation}
\label{covcountgen}  
\begin{split}
 \cN (\{   t_{ab;\a} \}; \{ N_a \}  , \{  M_{ab}  \}   )  
&= \sum_{ \substack{ R_a \\  l( R_a ) \le N_a   } } 
 \sum_{ s_{ab}^+ \vdash n_{ab} } 
 \sum_{ s_{ab}^- \vdash n_{ab} }
\sum_{ \substack{  \L_{ab} \vdash n_{ab}   \\  l ( \Lambda_{ab})   \le M_{ab}  }} 
 \prod_{ a } g ( \cup_{ b } s^+_{ab} ; R_a )  g ( \cup_b s^-_{ab} ; R_a )    
 \\
 & ~~~~~~~~~~~~~~~~~~~~~~~~~~~~~
\times \prod_{a,b} C ( s_{ab}^+ , s_{ab}^- , \L_{ab}  ) \chi_{ \L_{ab} } ( \mT_{ab} )
\end{split}
\end{equation}
Note this is the same partition function as in the derivation in the previous section (\ref{eq:N_tabalph}), but now using the covariant basis we can conveniently package $(t_{ab;\a})^{n_{ab;\a}}$ into $\chi_{ \L_{ab} } ( \mT_{ab} )$.

The counting formula (\ref{covcountgen}) can be used to derive an elegant 
infinite product formula for large $N_a$. If we assume $n_a \le N_a$ so sums over $R_a$ are unconstrained,  we can do the sums over  
 $R_a , \L_{ab}, s_{ab}^\pm$ to end up with a product of delta functions over the groups
\begin{equation}
 \label{deltamuaforcount} 
\cN(\{t_{ab;\a}\};  \{ M_{ab} \}  )  =  \sum_{ \{\g_a\} } \sum_{ \{\sigma_{ab}\}  } \prod_{ a  }   \delta_{ S_{n_a} }  \left (     \left ( \prod_{b }^{ \circ}  \sigma_{ba} \right )  \g_a  \left ( \prod_{ b }^{\circ}  \sigma_{ab} \right )  \g_a^{-1}  \right ) 
          \prod_{a,b}   \tr_{ n_{ab} } (  \mT_{ab} \sigma_{ab}  ) 
\end{equation}
where
\begin{equation}
	\cN(\{t_{ab;\a}\}; \{ M_{ab} \}  ) \, \equiv \, \cN (\{   t_{ab;\a} \}; \{ N_a =\infty\}  , \{  M_{ab}  \}   )  
\end{equation}
The limit $N_a = \infty$ holds as long as $n_a \le N_a$.

The derivation is described in more detail in Appendix (\ref{App:dercount}). 
The sum is over permutations $ \gamma_1 , \gamma_2 ,  \cdots \gamma_G$, 
one for each node (or  group), with $\gamma_a \in S_{n_a} $ ; as well as 
a sum over permutations $ \s_{ab}$, one for every pair $(a, b ) $ of nodes of the quiver which have a non-zero number $M_{ab}$ 
of arrows from $a$ to $b$. The $\s_{ab}$ are permutations in $S_{n_{ab}} $. Note that $ \prod_{b }^{ \circ}  \sigma_{ba} $
is an outer product of permutations, e.g if there are $3$ values of $b$ for which $n_{ba}$ is non-zero, say $b=1,2,3$, then 
the product gives a permutation $ \sigma_{11} \circ \sigma_{ 21 } \circ \sigma_{ 31} $ which lives in the $S_{n_{1a} } \times S_{ n_{ 2a} } \times S_{ n_{3a} } $ 
subgroup of $ S_{n_a}  = S_{ n_{1a} + n_{2a} + n_{3a} }  $. 

Consider cycles of length $i$. Let $\s_{ab}$ have $p_{ab}^{(i)} $ cycles of this length. 
The delta functions associated with each node lead to the condition $ \sum_{ b } p_{ab}^{(i)} = \sum_b p_{ba}^{(i)} $. Given 
any $\gamma_a , \sigma_{ab} $ which solve the delta function, we can generate the other solutions for the same $ \sigma_{ab}$,  
 by considering by multiplying $\gamma_a$ on the right with 
permutations $\gamma_a$ in the stabilizer of  $\left ( \prod_{ b }^{\circ}  \sigma_{ab} \right ) $. This generates a multiplicity of  
\bea\label{sumparts}  
 \prod_i  \prod_a   \left( \sum_b  p_{ab}^{(i)} \right)  !  \,   i^{ \sum_b   p_{ab}^{(i)} }    
\eea
We can see that the sums over $\gamma_a$  in (\ref{deltamuaforcount}) only depends on the conjugacy class of $ \s_{ab}$ in $S_{n_{ab}} $, 
since conjugating $ \s_{ab}$ by elements of $S_{n_{ab}} $ can be absorbed in $\gamma_a \in S_{n_a} $  the summations by exploiting  the 
invariance of these sums under left or right multiplication by elements of the $S_{n_{ab}} $ subgroups of $S_{n_a}$. 
This means that the sums over $ \s_{ab}$ can be converted into sums over $p_{ab}^{(i)}  $. There is a 
multiplicity 
\bea 
\prod_i \prod_{a,b} \frac{ n_{ab} !  }{ i^{p_{ab}^{(i)} }  (p_{ab}^{(i)} )  !   } 
\eea

Combining these facts we arrive at
\begin{equation}
\label{prodformcount} 
\cN (\{t_{ab;\a}\} ; \{M_{ab}\}   ) = 
\prod_{i=1}^\infty \left[ 
\sum_{ \{p_{ab}^{(i)}\} = 0 }^{\infty }   \prod_{a}  \delta \left ( \sum_{b } p_{ba}^{(i)}  - \sum_b p_{ab}^{(i)}    \right )  \left ( \sum_{ b }   p_{ab}^{(i)} \right ) !    \prod_{a,b}{   \left( \sum_{\alpha} (t_{ab;\a})^i \right)^{ p_{ab}^{(i)}   }  \over      p_{ab}^{(i)} !  }   
\right]
\end{equation}
For each $i$ we need to do a sum of the form 
\begin{equation} 
\cS ( \{ t_{ab} \}  ) =   \sum_{ \{ p_{ab} \}  = 0 }^{\infty } \prod_{a}  \delta \left ( \sum_{b } p_{ba} - \sum_b p_{ab}  \right ) \left  ( \sum_{ b }   p_{ab}  \right ) !  
  \prod_{a,b}{   \left  (  t_{ab}  \right  )^{ p_{ab} }  \over  p_{ab} !  }   
\end{equation}
It is convenient to write the Kronecker delta as a contour integral, using 
\begin{equation}
\delta ( p ) = \oint { dz \over 2 \pi i z } z^p  
\end{equation}
which gives
\begin{equation}
\begin{split}
& \cS ( \{ t_{ab} \}  ) =    \sum_{ \{ p_{ab} \}  = 0 }^{\infty } \prod_{a}  
( \sum_{ b }   p_{ab}  ) !
\oint { dz_a \over 2 \pi i z_a } z_a^{ \sum_b p_{ba} - \sum_{b} p_{ab} }     
  \prod_{a,b}{   \left (  t_{ab}  \right  )^{ p_{ab} }  \over      p_{ab} !  }   
\\ 
& =     \oint \left(\prod_{a}{  dz_a \over 2 \pi i z_a }\right) \sum_{ \{ p_{ab} \}  = 0 }^{\infty }   \prod_{a} ( \sum_{ b }   p_{ab}  ) !  
\prod_{a,b}{   \left ( z_a^{-1} z_b  t_{ab}  \right  )^{ p_{ab} }  \over      p_{ab} !  } \\ 
& =   \oint \left(\prod_{a}{  dz_a \over 2 \pi i z_a }\right)   \prod_{a}   { 1 \over 1 - \sum_{ b } z_a^{-1} z_b  t_{ab} } 
\end{split}
\end{equation}
We can obtain the desired sum by calculating residues. 

We find that the result can be expressed in an elegant and intuitive form.
Let $\bV $ be the set $ \{ 1 , 2, \cdots  G  \} $ of nodes in the quiver. 
We will let $ \mV$ be any subset of $\bV$, and define $\Sym ( \mV ) $ to be 
the group of all permutations of the elements in $\mV$.  For each permutation 
$\sigma$ we will define a monomial $T_{\sigma} (  \{ t_{ab} \} )$ built from the set $\{ t_{ab} \} $. 
Any permutation $\sigma$ is a product of cycles $ \sigma = \prod_{j} \sigma^{(j)} $. 
The monomial $T_{\sigma} (  \{ t_{ab} \} )$  is a product over these cycles. 
\bea 
T_{\sigma}  (  \{ t_{ab} \}   )   =    \prod_{j  } (-1) ~ T_{ \sigma^{(j)} }  (  \{ t_{ab} \}   )   
\eea
For a cycle, such as $(a_1,a_2, \cdots a_ k )$ with integers $a_1 , \cdots a_k $ chosen from $ \{ 1 , \cdots  , G \}$,  the factor is 
\bea 
T_{ ( a_1 , a_2  \cdots , a_k )}  (  \{ t_{ab} \}   )   
= t_{a_1 a_2} t_{a_2 a_3} \cdots   t_{a_{k-1} a_k }  t_{a_k a_1} 
\eea
We find that 
\begin{equation}\label{Infprod} 
\fbox{ 
$\displaystyle{  \mathcal S }( \{ t_{ab} \} ) = { 1 \over ( 1 - \sum_{ \mV \subset \bV } \sum_{ \sigma \in Sym ( \mV ) }  T_{\sigma} (  \{ t_{ab} \} ) }
$ }
\end{equation}  
The sign of each term is $(-1)^{C_{\sigma}}$ where $C_{\sigma}$ is  the number of cycles in the corresponding permutation.
Each cycle $\sigma^{(i)}  $ corresponds to an elementary  closed loop in the quiver, elementary in the sense that 
it does not involve visiting any node more than once.  The  permutation $\sigma$ corresponds to 
a product of disjoint elementaty loops. For example, for a quiver with three nodes, this becomes 
\begin{equation}
\begin{split}
& \cS ( t_{11} , t_{22} , t_{33} , t_{12} , t_{13} , t_{23} )
\\
& \qquad =
(  1 - t_{11}  - t_{22} - t_{33} + t_{11}  t_{22}  - t_{ 12} t_{21} + t_{22} t_{33} - t_{23}t_{32} + t_{11} t_{33} - t_{13}t_{31} \\
& \qquad - t_{11}t_{22} t_{33} + t_{12}t_{21} t_{33} +  t_{13}t_{31} t_{22} + t_{11} t_{23} t_{32} - t_{12}t_{23}t_{31} 
- t_{13} t_{32} t_{21}  )^{-1} 
\end{split}
\end{equation}
The first three terms  after $1$ come from the $3$ 1-element subsets of $\bV = \{ 1 , 2, 3 \}$. 
The next three pairs come from the $3$ two-element subsets of $ \bV $. 
The first of each pair comes from the identity permuttaion of the subset, the second from 
the swop. The last  line comes from  permutations of $\mV = \bV $. 

The  large $N$ counting function  can then  be written as 
\begin{equation}\label{largeNcount}  
\begin{split}
\cN (\{ t_{ab;\a} \}  ; \{  M_{ab} \}   ) =  \prod_{i=1}^{\infty}  \cS (  \{ t_{ab} \rightarrow   \sum_{\alpha =1}^{M_{ab} }  (t_{ab;\a})^i \}  ) 
\end{split} 
\end{equation} 
In this equation, we have the counting for a quiver with $G$ nodes and any number of arrows 
for any specified pair of start and end points. When there are no arrows between 
a specified start and end point, we set the corresponding $t_{ab}$ variable to zero. 

Let us now explain how to specialize the above formula for some specific cases. 
Take the half-BPS sector of $\cN=4$ SYM. This is described by one node and one arrow
starting and ending at that node. The set $\bV$ has one element $\{ 1 \}$ and there is one $t_{11}$ 
parameter. There are two subsets, $\mV = \emptyset $ or $\mV = \bV$. In calculating 
$\cS ( t_{11} ) $,  the monomial 
coming from the emptyset is $1$. The monomial from $\mV = \bV$ is $- t_{11} $.  So 
\bea 
\cN_{\mC }(t_{11}) = \prod_{i=1}^{\infty} { 1 \over  1 - t_{11}^i  } 
\eea
For the one-node quiver with three lines starting and ending at the node, 
$\bV = \{ 1 \}$.  The set of t-variables (``fugacities'') is $ \{ t_{11; 1 } ,   t_{11; 2 } , t_{11;3} \} $.  
\bea
\cS_{\mC^3}( t_{11} )   = ( 1 - t_{11} )^{-1}  
\eea
The counting function is 
\bea
\cN_{\mC^3 } (  \{ t_{11 ;\alpha} \}  ) = \prod_{ i =1}^{\infty} { 1 \over  1 - t^i_{11;1}  - t^i_{11;2} - t^i_{11; 3 }  } 
\eea
This formula was  written down in \cite{BDHO}. 

Beyond these examples, the analogous formulae have not been previously written down. 
For the conifold, we have $\bV = \{ 1,2 \} $. The $\cS $ function is 
\bea 
\cS_{ \cC }(t_{12},t_{21}) = ( 1- t_{12}t_{21} )^{-1}
\eea
The variables $t_{11} , t_{22} $ are set to zero, since there are no 
arrows joining any node to itself. The $1$ comes as usual from the empty set, the second term from 
the permutation $(12)$ in  $\Sym( \mV ) $ for $\mV = \bV $. All other terms are zero due to the 
vanishing of $t_{11}, t_{22}$. 
Since there is a multiplicity $2$ for 
the arrows going from $1$ to $2$  and conversely from $2$ to $1$, 
we have variables $t_{12;1} , t_{12;2} , t_{21;1} , t_{21;2} $ 
and the counting function 
\begin{equation}
\label{eq:counting_con_inf}
\begin{split}
\cN_{\cC }(\{t_{12;\alpha},t_{21;\alpha}\}) &=  \prod_{i=1}^{\infty } { 1 \over  1 -  ( t_{12 ; 1 }^i + t_{12; 2 }^i ) (    t_{21 ; 1}^i +  t_{21 ; 2 }^i  )   }   \\
 & = \prod_{i=1}^{\infty } { 1 \over  1 - t_{12 ; 1 }^i t_{21 ; 1 }^i  -    t_{12 ; 2 }^i t_{21 ; 2 }^i  -  t_{12 ; 1 }^i t_{21 ; 2 }^i  -  t_{12 ; 2 }^i t_{21 ; 1 }^i   } 
\end{split}
\end{equation}
For the example of $ \mC^3 / Z_2 $, the $\cS $  function depends on $t_{11} , t_{22} ,  t_{12 } , t_{21}$, 
The $\cN$   function depends on $t_{11} , t_{22} ,  t_{12 ; 1 } ,   t_{12 ; 1 }  ,  t_{21; 1 } , t_{21 ; 2  } $. 
\begin{equation}
\cS_{\mC^3 / Z_2 }  ( t_{11} , t_{22} ,  t_{12 } , t_{21} )  = ( 1 - t_{11} - t_{22} - t_{12}t_{21} + t_{11} t_{22} )^{-1} 
\end{equation}
Here  $ \bV = \{ 1, 2 \} $. The monomials $t_{11} , t_{22} $ come from choices $\mV = \{ 1 \} $ and $ \mV = \{ 2 \} $. 
The term $t_{12} t_{21} $ comes from permutation $(12)$ in $\Sym ( \mV ) $ for $\mV = \{ 1, 2 \} $. 
The term $t_{11} t_{22} $ comes from permutation $(1) (2)  $  in $\Sym ( \mV ) $ for $\mV = \{ 1, 2 \} $. 
The counting function is 
\begin{equation}
\cN_{\mC^3 / Z_2}(\{t_{11},t_{22},t_{12;\a},t_{21;\a}\}) = \prod_{i=1}^\infty { 1 \over  1 - t^i_{11} - t^i_{22} - ( t^i_{12,1} + t^i_{12,2} ) ( t^i_{21,1} + t^i_{21,2} )  + t^i_{11} t^i_{22}  }
\end{equation}

%%%%%%%%%%%%%%%%%%%%%%%%%%%%%%%%%%%%%%%%%%%%%%%%%%%%%%%%%%%%%%%%%%%%%
%%%%%%%%%%%%%%%%%%%%%%%%%%%%%%%%%%%%%%%%%%%%%%%%%%%%%%%%%%%%%%%%%%%%%

\section{Construction of free orthogonal basis}
\label{sec:constops} 

Motivated by the counting formulae (\ref{eq:N_nabalph}), (\ref{eq:N_Lab}) we proceed in this section with the construction of an explicit operator basis. 
The prescriptions for counting in Figures~\ref{fig:quiver_split_C3},\ref{fig:quiver_split_con},\ref{fig:quiver_split_C3Z2},\ref{fig:quiver_split_cov_C3},\ref{fig:quiver_split_cov_con},\ref{fig:quiver_split_cov_C3Z2} will 
be developed to produce an orthogonal basis of operators (in the free field 
inner product) to match the counting.

\subsection{Review of \texorpdfstring{$\mC^3$}{C3}}

\begin{figure}[h]
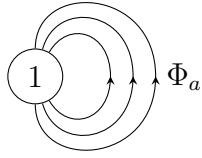

\centering
\mytikz{ 
	\node (s1) at (0,0) [circle,draw] {$1$};							
	\draw [postaction={decorate}] (s1.-60) .. controls +(-60:0.5) and +(0,-0.6) .. (1,0)
		.. controls +(0,0.6) and +(60:0.5) .. (s1.60);
	\draw [postaction={decorate}] (s1.-75) .. controls +(-75:0.7) and +(0,-0.9) .. (1.3,0)
		.. controls +(0,0.9) and +(75:0.7) .. (s1.75);
	\draw [postaction={decorate}] (s1.-90) .. controls +(-90:0.9) and +(0,-1.2) .. (1.6,0) node[right]{$\Phi_a$}
		.. controls +(0,1.2) and +(90:0.9) .. (s1.90);		
}
\caption{Quiver for $\mC^3$, arrows correspond to three chiral multiplets $\Phi_1,\Phi_2,\Phi_3$.}
\label{fig:quiverC3}
\end{figure}

Let us first review $\cN=4$ $U(N)$ SYM, for which the orthogonal basis of free chiral operators has been constructed before \cite{BHR1,db1}. We can view $\cN=4$ as a special case of $\cN=1$ quiver gauge theory with the quiver shown in Figure~\ref{fig:quiverC3}.
Theory contains three $\cN=1$ chiral multiplets $\Phi_a$ transforming in the adjoint of $U(N)$. There is a global $U(3)$ flavor symmetry. The chiral gauge invariant operators are built from the chiral adjoint scalars $\Phi_a$, so we have single traces
\begin{equation}
	\tr(\Phi_{a_1}\Phi_{a_2} \ldots \Phi_{a_n})
\end{equation}
and products of such traces. We will be interested in cases where $N$ is finite and the operators involve more than $N$ fields. In that case we need to take care of relationships between products of traces, arising from the fact that $\Phi_a$ are $N$-by-$N$ matrices.

Consider all possible multitrace operators with $U(1)^3 \subset U(3)$ charges $\bn=(n_1,n_2,n_3)$ and bare dimension $n=n_1+n_2+n_3$.  A natural way to label the operators is by using a permutation $\sigma \in S_n$:
\begin{equation}
\label{eq:Osig_C3_1}
	\cO(\bn,\sigma) = 
	\prod_{k=1}^{n_1} (\Phi_1)^{i_k}_{i_{\sigma(k)}} 
	\prod_{k=n_1+1}^{n_1+n_2} (\Phi_2)^{i_k}_{i_{\sigma(k)}}
	\prod_{k=n_1+n_2+1}^{n_1+n_2+n_3} (\Phi_3)^{i_k}_{i_{\sigma(k)}}
\end{equation}
That is, the operator involves a product of fields $(\Phi_1)^{n_1}(\Phi_2)^{n_2}(\Phi_3)^{n_3}$ and the permutation $\sigma$ indicates that $k$'th upper index is contracted with $\sigma(k)$'th lower index. Each cycle in $\sigma$ corresponds to a single trace. 

At this point let us introduce some convenient notation. $(\Phi_a)^i_j$ is a matrix, which can be thought of as linear operator acting on $N$-dimensional vector space $V_N$. Then the object:  
\begin{equation}
\label{eq:Phi_tensor_n}
	\left(\Phi_1^{\otimes n_1}\otimes\Phi_2^{\otimes n_2}\otimes\Phi_3^{\otimes n_3}\right)^{i_1\ldots i_n}_{j_1 \ldots j_n} \equiv 
	\prod_{k=1}^{n_1} (\Phi_1)^{i_k}_{j_k} 
	\prod_{k=n_1+1}^{n_1+n_2} (\Phi_2)^{i_k}_{j_k}
	\prod_{k=n_1+n_2+1}^{n_1+n_2+n_3} (\Phi_3)^{i_k}_{j_k}
\end{equation}
is a linear operator acting on the $N^n$-dimensional vector space $V_N^{\otimes n}$. Permutations $\sigma$ are also  linear operators in $V_N^{\otimes n}$ 
which acts by permuting the $V_N$ factors of the tensor product : 
\begin{equation}
\label{eq:sigma_tensor_n}
	(\sigma)^{i_1 i_2 \ldots i_n}_{j_1 j_2 \ldots j_n} \equiv \delta^{i_1}_{j_{\sigma(1)}} \delta^{i_2}_{j_{\sigma(2)}} \ldots \delta^{i_n}_{j_{\sigma(n)}}
\end{equation}
Then (\ref{eq:Osig_C3_1}) can be expressed as 
\begin{equation}
\label{eq:Osig_C3_2}
	\cO(\bn,\sigma) = \trVN{n}\left(\sigma \, \Phi_1^{\otimes n_1}\otimes\Phi_2^{\otimes n_2}\otimes\Phi_3^{\otimes n_3} \right)
\end{equation}
where the product of operators and the trace is over $V_N^{\otimes n}$, which means contracted indices of (\ref{eq:Phi_tensor_n}) and (\ref{eq:Phi_tensor_n}).

Let us also introduce diagrammatic notation for matrix multiplication and traces. 
\begin{equation}
A^i_j = \;
	\mytikz{
		\node (a) at (0,0) [rectangle,draw] {$A$};
		\draw [postaction={decorate}] ($(a)+(0,1)$) to node[right]{$i$} (a);
		\draw [postaction={decorate}] (a) to node[right]{$j$} +(0,-1);
	}
\quad\quad
(AB)^i_j = \;
	\mytikz{
		\node (a) at (0,0) [rectangle,draw] {$AB$};
		\draw [postaction={decorate}] ($(a)+(0,1)$) to node[right]{$i$} (a);
		\draw [postaction={decorate}] (a) to node[right]{$j$} +(0,-1.0);
	}
\; = A^i_k B^k_j = \;
	\mytikz{
		\node (a) at (0,0) [rectangle,draw] {$A$};
		\node (b) at (0,-1) [rectangle,draw] {$B$};
		\draw [postaction={decorate}] ($(a)+(0,1)$) to node[right]{$i$} (a);
		\draw [postaction={decorate}] (b) to node[right]{$j$} +(0,-1.0);
		\draw [postaction={decorate}] (a) to (b);
	}
\quad\quad \tr(A) = \;
	\mytikz{
		\node (a) at (0,0) [rectangle,draw] {$A$};		
		\draw [postaction={decorate}] ($(a)+(0,1)$) to (a);
		\draw [postaction={decorate}] (a) to +(0,-1);
		\draw [-] (-0.2,1) to (0.2,1);
		\draw [-] (-0.2,-1) to (0.2,-1);
	}
\end{equation}
Incoming and outgoing arrows represent upper and lower indices respectively. Since in matrix multiplication conventionally lower index is contracted with upper, then in the diagram matrices are multiplied in the direction following arrows. When matrices are laid out vertically, the multiplication conventionally flows from top to bottom, and we can omit the arrows. The indices can, of course, belong to the vector space $V^{\otimes n}_N$, in which case lines represent the whole set $\{i_1 \ldots i_n\}$ of contracted indices. Using this, we get a nice expression for the operator (\ref{eq:Osig_C3_2})
\begin{equation}
\label{eq:Osig_C3_diag}
	\cO(\bn,\sigma) = \; \mytikz{
		\node (s) at (0,0) [rectangle,draw] {$\quad\sigma\quad$};
		\node (p) at (0,-1) [rectangle,draw] {$\Phi_1^{\otimes n_1}\otimes\Phi_2^{\otimes n_2}\otimes \Phi_3^{\otimes n_3}$};						
		\draw [-] (s) to ++(0,1) to ++(-0.2,0) to ++(0.4,0);
		\draw [-] (s) to (p);
		\draw [-] (p) to ++(0,-1) to ++(-0.2,0) to ++(0.4,0);
	}
\end{equation}

Note an operator is not labelled by a unique $\sigma$. $\cO(\bn,\sigma)$ does not change if we conjugate $\sigma$ by the subgroup:
\begin{equation}
\label{eq:Osig_C3_conj}
	\cO(\bn,\gamma \sigma \gamma^{-1}) = \cO(\bn,\sigma), \quad \gamma \in S_{n_1} \times S_{n_2} \times S_{n_3}
\end{equation} 
This can be seen from (\ref{eq:Osig_C3_1}), where the conjugation can be brought from $\sigma$ to act on $\Phi_1^{\otimes n_1}\otimes\Phi_2^{\otimes n_2}\otimes\Phi_3^{\otimes n_3}$, which is invariant. Furthermore, we still have the problem of finite $N$ relationships.

One complete basis for the gauge invariant operators at finite $N$ was constructed in \cite{db1}, and is called ``Restricted Schur'' basis: 
\begin{equation}
\label{eq:OL_C3}
	\begin{split}
		\cO(\bL) &= \frac{1}{n_1! n_2! n_3!} \sum_{\s\in S_n} 
			\chi_{R\rightarrow\br}^{\nu^-,\nu^+}\left( \s \right) \, \cO(\bn, \s)		
	\end{split}
\end{equation}
The operators are uniquely specified by the set of group theoretic labels
\begin{equation}
\label{eq:L_C3}
	\bm{L} = \{ R, r_1, r_2, r_3, \nu^-, \nu^+ \}
\end{equation}
$R, r_1, r_2, r_3$ are Young diagrams
\begin{equation}
	R \vdash n, \quad r_1 \vdash n_1, \quad r_2 \vdash n_2, \quad r_3 \vdash n_3
\end{equation}
$R$ labels the representation of $S_n$ and $\br=(r_1,r_2,r_3)$ labels the representation of the subgroup $S_{n_1}\times S_{n_2}\times S_{n_3} \subset S_n$ which appears in the decomposition of $R$ in terms of subgroup irreps
\begin{equation}
	R \rightarrow (r_1, r_2, r_3)
\end{equation}
In case $\br$ appears in the decomposition more than once, the two numbers $\nu^{\pm}$ each label runs over the multiplicity given by Littlewood-Richardson coefficient $1 \le \nu^\pm \le g(r_1,r_2,r_3;R)$. For a summary of the facts about subgroup decomposition and branching coefficients see Appendix~\ref{app:branching}. 
The finite $N$ constraint appears simply as a cutoff on the number of rows in $R$:
\begin{equation}
	l(R) \le N
\end{equation}
and there are no further relationships between the operators.

The key ingredient in (\ref{eq:OL_C3}) is the coefficient $\chi_{R\rightarrow\br}^{\nu^-,\nu^+}\left( \s \right)$ called ``restricted character''. It is a generalization of the usual character $\chi_R(\sigma) = \tr(D^R(\sigma))$ and defined as
\begin{equation}
\label{eq:chi_C3_1}
	\chi_{R\rightarrow\br}^{\nu^-,\nu^+}\left( \s \right) = 
	\tr\left( P^{\nu^-,\nu^+}_{R\rightarrow \br} D^R(\sigma) \right)
\end{equation}
$P^{\nu^-,\nu^+}_{R\rightarrow \br}$ is a  projector-like operator \footnote{
If $\nu^-=\nu^+\equiv\nu$, then $P^{\nu,\nu}_{R\rightarrow \br}$ is precisely the projector to $(\br,\nu)$ in $R$. But the ``off-diagonal'' ones with $\nu^-\neq\nu^+$ are not strictly projectors, they are intertwining operators mapping between 
different copies of the same irrep $\br$ in $R$}
\begin{equation}
\label{eq:P_nu_Rr}
	P^{\nu^-,\nu^+}_{R\rightarrow \br }  =  \sum_{ l_1,l_2,l_3=1 }^{d_r} | R ; \br, \nu^-, \bl \rangle \langle R ; \br, \nu^+ , \bl  |	
\end{equation}
or in terms of Branching coefficients  (see (\ref{eq:Bil_matrix}))
\begin{equation}
\label{eq:Pij_Bil}
	(P^{\nu^-,\nu^+}_{R\rightarrow \br })_{ij} = \sum_{\bl} B^{R\rightarrow \br,\nu^-}_{i\rightarrow \bl} B^{R\rightarrow \br,\nu^+}_{j \rightarrow \bl}
\end{equation}
Using diagramatic notation (\ref{eq:Bil_diagram}) we can represent the restricted character 
\begin{equation}
\label{eq:chiQ_C3_diag}
	\chi_{R\rightarrow\br}^{\nu^-,\nu^+}\left( \s \right) = \;
	\mytikz{	
		\node (s) at (0,0) [rectangle,draw] {$\sigma$};		
		\node (m) at (1,1) [circle,draw,inner sep=0.5mm,label=above:$\nu^+$] {};
		\node (n) at (1,-1) [circle,draw,inner sep=0.5mm,label=below:$\nu^-$] {};
		\draw [postaction={decorate}] (m) to [bend right=45] node[left]{$R$} (s);
		\draw [postaction={decorate}] (s) to [bend right=45] (n);
		\draw [postaction={decorate}] (n) to [bend right=0] node[left]{$r_1$} (m);		
		\draw [postaction={decorate}] (n) to [bend right=45] node[left]{$r_2$} (m);
		\draw [postaction={decorate}] (n) .. controls +(0:1) and +(0:1) .. node[right]{$r_3$} (m);
	}
\end{equation}
The edges now correspond to contracted indices in irreducible representations $R, r_1, r_2, r_3$, as labelled.

The basis (\ref{eq:OL_C3}) is not only complete, it is, in fact, orthogonal in the free field Zamolodchikov metric obtained from the two point function
\begin{equation}
	\langle (\Phi_a)^i_j (\Phi_b^\dagger)^k_l \rangle = \delta_{ab} \delta^i_l \delta^k_j
\end{equation}
Then 
\begin{equation}
\label{eq:OL2pt_C3}
	\langle O(R,\br,\nu^-,\nu^+) O(\tl R,\tl \br, \tl\nu^-, \tl\nu^+) \rangle =
	\frac{h(R) f_N(R)}{h(r_1) h(r_2) h(r_3)} \delta_{R\tl R} \delta_{r_1\tl r_1} \delta_{r_2\tl r_2} \delta_{r_3\tl r_3} \delta_{\nu^+\tl\nu^+} \delta_{\nu^-\tl\nu^-}
\end{equation}
$h(R)$ is the product of hooks of the Young diagram, and $f_N(R)$ is the weight of the diagram in $U(N)$. That is the only place that $N$ dependence comes in, and it nicely captures the cutoff, because if the height of $R$ exceeds $N$, then $f_N(R)=0$, which means the operator is 0.

There is another complete orthogonal basis found in \cite{BHR1}, where operators are organized into irreducible representation of the global symmetry $U(3)$. We will refer to it as ``covariant basis'', since operators transform covariantly with the global symmetry group. The operators are 
\begin{equation}
\label{eq:OK_C3}
	\cO(\bK) = \frac{1}{n!}\sum_{\sigma\in S_n} B^{\Lambda\rightarrow [\bn],\beta}_{m} S^{R R,\Lambda \tau}_{\,i\,j,\,m} D^R_{ij}(\sigma) \, \cO(\bn, \sigma)
\end{equation}
The group theory labels in this case are
\begin{equation}
	\bK = \{ R, \Lambda, \tau, \bn, \beta \}
\end{equation}
where $R, \Lambda \vdash n$ are Young diagrams with $n=n_1+n_2+n_3$ boxes. $R$ is the same as before, with a cutoff of at most $N$ rows, and $\Lambda$ is an irrep of $U(3)$ with at most 3 rows. $\tau$ is the multiplicity label for the Kronecker product of $S_n$ irreps 
\begin{equation}
R \otimes R \rightarrow \Lambda
\end{equation}
and $S^{R R,\Lambda \tau}_{\,i\,j,\,m}$ is the associated Clebsch-Gordan coefficient. For the review of the facts about Kronecker product and Clebsch-Gordan coefficients see Appendix~\ref{app:clebsch}. $\bn=(n_1,n_2,n_3)$ specifies how many fields of each flavor there are (note in $\bL$ this information was contained in $\br$). $B^{\Lambda\rightarrow [\bn],\beta}_{m}$ is the branching coefficient for the reduction from $S_n$ irrep $\Lambda$ to the trivial one-dimensional irrep $[n_1,n_2,n_3]$ of $S_{n_1}\times S_{n_2}\times S_{n_3}$, and $\beta$ is the multiplicity label. In other words, $\beta$ labels the invariants of $\Lambda$ under $S_{n_1}\times S_{n_2}\times S_{n_3}$, and $B^{\Lambda\rightarrow [\bn],\beta}_{m}$ are the invariant vectors.
Note, compared with the usual branching coefficient notation $B^{\Lambda\rightarrow [\bn],\beta}_{m\rightarrow i}$, we suppress the index $i$ since $[\bn]$ is one-dimensional.

Again it will be useful to have a diagrammatic notation for the basis. Define
\begin{equation}
	\chi(\bK,\sigma) = B^{\Lambda\rightarrow [\bn],\beta}_{m} S^{R R,\Lambda \tau}_{\,i\,j,\,m} D^R_{ij}(\sigma)
\end{equation}
so that
\begin{equation}
	\cO(\bK) = \frac{1}{n!}\sum_{\sigma\in S_n} \chi(\bK,\sigma) \, \cO(\bn, \sigma)
\end{equation}
The coefficient $\chi(\bK,\sigma)$ can be expressed, using the diagrammatic notation (\ref{eq:CG_diag}) for the Clebsch-Gordan coefficient, as
\begin{equation}
\label{eq:chiK_diag_C3}
	\chi(\bK,\sigma) = 
	\mytikz{	
		\node (s) at (0,0) [rectangle,draw] {$\sigma$};		
		\node (t) at (1.5,0) [circle,fill,inner sep=0.5mm,label=above:$\tau$] {};
		\node (b) at (2.3,0) [circle,draw,inner sep=0.5mm,label=above:$\beta$] {};
		\node (l) at (3.2,0)  {};	
		\draw [postaction={decorate}] (s.-90) to [bend right=45] node[below]{$R$} (t);
		\draw [postaction={decorate}] (t) to [bend right=45] node[above]{$R$} (s.90);
		\draw [-] (t) to node[below]{$\Lambda$} (b);
		\draw [-] (b) to node[below]{$[\bn]$} (l);
	}
\end{equation}
The open line, which normally has an associated state label, corresponds to the unique  $i=1$ basis state of $[\bn]$ in the branching $B^{\Lambda\rightarrow [\bn],\beta}_{m\rightarrow i}$.

The two-point function between the operators is
\begin{equation}
	\langle \cO(\bK) \cO(\tl\bK)^\dagger \rangle = 
	\frac{n_1!n_2!n_3!\Dim_N(R)}{d_R^2} \delta_{R\tl R} \delta_{\Lambda\tl\Lambda} \delta_{\tau\tl\tau} \delta_{\bn\tl\bn} \delta_{\beta\tl\beta}
\end{equation}

%%%%%%%%%%%%%%%%%%%%%%%%%%%%%%%%%%%%%%%%%%%%%%%%%%%%%%%%%%%%%%%%%%%%%

\subsection{Generalized restricted Schur basis}\label{sec:constmethod} 

Let us assume we have a general quiver $Q$. We will often use $\mC^3/\mZ_2$ as an example, see Figure~\ref{fig:quiverC3Z2}. The goal in this section is to derive a free orthogonal basis $\cO_Q(\bL)$ for arbitrary quiver, analogous to the restricted Schur basis (\ref{eq:OL_C3}) in $\mC^3$. We extend this to covariant basis $\cO_Q(\bK)$ in the next section.

In order to build a gauge-invariant operator\footnote{We restrict to the mesonic sector, or, in other words, $\prod_a U(N_a)$ gauge group, not $\prod_a SU(N_a)$.} we contract the incoming and outgoing fields at each group node. In a more complicated quiver such as $\C3Z2$ there are different ``paths'' that an operator can take. We can build, for example:
\begin{equation}
	\tr(\Phi_{11}\Phi_{11}), \; \tr(\Phi_{12;1} \Phi_{21;2}), \; \tr(\Phi_{11}\Phi_{12;1}\Phi_{22}\Phi_{21;1}), \; \ldots
\end{equation}
It is possible to capture all the different possibilities by fixing the number of times $n_{ab;\a}$ each field appears, and then contracting the indices corresponding to each group according to a permutation $\sigma_a$. This defines an operator which, in correspondence with (\ref{eq:Osig_C3_diag}), diagrammatically looks like:
\begin{equation}\label{quivertrace} 
	\cO_{\C3Z2}( \bn,\bsig) = \;
	%\cO_{\C3Z2}(\{\sigma_a\}, \{n_{ab;\a}\}, \{\Phi_{ab;\a}\}) =
	\mytikz{
	%
	% Quiver as trace
	%
		\node (s1) 		at (2,0) [rectangle,draw,minimum height=0.8cm,minimum width=6cm] {$\sigma_1$};
		\node (p11) 	at (0,-1.5) [rectangle,draw,minimum height=0.8cm] {$\Phi_{11}^{\otimes n_{11}}$};
		\node (p121) 	at (2,-1.5) [rectangle,draw,minimum height=0.8cm] {$\Phi_{12;1}^{\otimes n_{12;1}}$};
		\node (p122) 	at (4,-1.5) [rectangle,draw,minimum height=0.8cm] {$\Phi_{12;2}^{\otimes n_{12;2}}$};
		\node (s2) 		at (4,-3) [rectangle,draw,minimum height=0.8cm,minimum width=6cm] {$\sigma_2$};		
		\node (p211) 	at (2,-4.5) [rectangle,draw,minimum height=0.8cm] {$\Phi_{21;1}^{\otimes n_{21;1}}$};
		\node (p212) 	at (4,-4.5) [rectangle,draw,minimum height=0.8cm] {$\Phi_{21;2}^{\otimes n_{21;2}}$};
		\node (p22) 	at (6,-4.5) [rectangle,draw,minimum height=0.8cm] {$\Phi_{22}^{\otimes n_{22}}$};
		\draw [-] (-1,0.75) to (7,0.75);
		\draw [-] (-1,-5.25) to (7,-5.25);
		\draw [postaction={decorate}] (0,0.75) to +(0,-0.35);
		\draw [postaction={decorate}] (2,0.75) to +(0,-0.35);
		\draw [postaction={decorate}] (4,0.75) to +(0,-0.35);
		\draw [postaction={decorate}] (6,0.75) to (6,-2.6);
		\draw [postaction={decorate}] (0,-0.4) to +(0,-0.7);
		\draw [postaction={decorate}] (2,-0.4) to +(0,-0.7);
		\draw [postaction={decorate}] (4,-0.4) to +(0,-0.7);
		\draw [postaction={decorate}] (0,-1.9) to (0,-5.25);		
		\draw [postaction={decorate}] (2,-1.9) to +(0,-0.7);
		\draw [postaction={decorate}] (4,-1.9) to +(0,-0.7);
		\draw [postaction={decorate}] (2,-3.4) to +(0,-0.7);
		\draw [postaction={decorate}] (4,-3.4) to +(0,-0.7);
		\draw [postaction={decorate}] (6,-3.4) to +(0,-0.7);
		\draw [postaction={decorate}] (2,-4.9) to +(0,-0.35);
		\draw [postaction={decorate}] (4,-4.9) to +(0,-0.35);
		\draw [postaction={decorate}] (6,-4.9) to +(0,-0.35);
	}
\end{equation}
The lines represent indices in $V_N^{\otimes n_{ab;\a}}$. Note that if $n_{11} \neq n_{22}$, permutations $\sigma_1, \sigma_2$ are elements of symmetric 
 groups of different size
\begin{equation}
\begin{split}
	\sigma_1 \in S_{n_1}, \quad n_1 &\equiv n_{11} + n_{12;1} + n_{12;2} \\
	\sigma_2 \in S_{n_2}, \quad n_2 &\equiv n_{22} + n_{12;1} + n_{12;2}
\end{split}
\end{equation}
acting as operators in $V_{N_1}^{\otimes n_1} $ and  $V_{N_2}^{\otimes n_2} $.
 If we rearrange the above diagram we get just the quiver itself with a permutation $\sigma_a$ at each group node and an operator $(\Phi_{ab;\a})^{\otimes n_{ab;\a}}$ on each field line
\begin{equation}
\label{eq:Osig_C3Z2}
	\cO_{\C3Z2}(\bn, \bsig) = \;
	\mytikz{
	%
	% Quiver with Phi boxes
	%
		\node (s1) at (0,0) [rectangle,draw] {$\quad\sigma_1\quad$};				
		\node (s2) at (4,0) [rectangle,draw] {$\quad\sigma_2\quad$};
		\node (p11) at (-2,0) [rectangle,draw] {$\Phi_{11}^{\otimes n_{11}}$};
		\node (p22) at (6,0) [rectangle,draw] {$\Phi_{22}^{\otimes n_{22}}$};
		\node (p121) at (2,-0.5) [rectangle,draw] {$\Phi_{12;1}^{\otimes n_{12;1}}$};
		\node (p122) at (2,-1.5) [rectangle,draw] {$\Phi_{12;2}^{\otimes n_{12;2}}$};
		\node (p211) at (2,0.5) [rectangle,draw] {$\Phi_{21;1}^{\otimes n_{21;1}}$};
		\node (p212) at (2,1.5) [rectangle,draw] {$\Phi_{21;2}^{\otimes n_{21;2}}$};		
		\draw [postaction={decorate}] (s1.-120) to [bend left=60] (p11);
		\draw [postaction={decorate}] (p11) to [bend left=60] (s1.120);
		\draw [postaction={decorate}] (s1.-60) to [bend right=10] (p121.180);
		\draw [postaction={decorate}] (p121.0) to [bend right=10] (s2.-120);
		\draw [postaction={decorate}] (s1.-90) to [bend right=30] (p122.180);
		\draw [postaction={decorate}] (p122.0) to [bend right=30] (s2.-90);
		\draw [postaction={decorate}] (p22) to [bend left=60] (s2.-60);
		\draw [postaction={decorate}] (s2.60) to [bend left=60] (p22);		
		\draw [postaction={decorate}] (s2.120) to [bend right=10] (p211);
		\draw [postaction={decorate}] (p211) to [bend right=10] (s1.60);
		\draw [postaction={decorate}] (s2.90) to [bend right=30] (p212);
		\draw [postaction={decorate}] (p212) to [bend right=30] (s1.90);
	}
\end{equation}
It is clear that we can define $\cO_Q(\bn,\bsig)$ in such a way for any quiver $Q$: it is a generalization of (\ref{eq:Osig_C3_2}), but instead of contractions performed sequentially in a single trace, now the operators $\sigma_a$ and $(\Phi_{ab;\a})^{\otimes n_{ab;\a}}$ are contracted along $Q$. With  the diagrammatic 
representation of linear operators using boxes and lines, we are inserting the
boxes for  $(\Phi_{ab;\a})^{\otimes n_{ab;\a}}$ along the edge of the split-node quiver labelled $\alpha$ going from $a$ to $b$, and we are inserting $\sigma_a$ 
 in the $a$'th line joining the $a$'th plus and minus nodes. 
 Explicitly we can write:
\begin{equation}
\label{eq:OQ_sig_defn}
\boxed{
	\cO_Q (\bn,\bsig) = 
	\prod_{a,b} \prod_{\alpha=1}^{M_{ab}} \left( \Phi_{ab;\a}^{\otimes n_{ab;\a}} \right)^{\bm{I}_{ab;\a}}_{\bm{J}_{ab;\a}}
	\prod_a \left( \sigma_a \right)^{ \bigcup_{b,\alpha} \bm{J}_{ba;\a}}_{ \bigcup_{b,\alpha} \bm{I}_{ab;\a}}
}
\end{equation}
The indices $a,b$ run over all group nodes, and it is understood that we skip the terms where $M_{ab}=0$.
$\bm{I}_{ab;\a}$ and $\bm{J}_{ab;\a}$ are indices in the vector space $V_{N_a}^{\otimes n_{ab;\a}}$ and $\check V_{N_b}^{\otimes n_{ab;\a}}$, 
i.e $ \bm{I}_{ab;\a} = \{ i_1 , \cdots , i_{ n_{ab;\alpha } } \}  $ 
and $ \bm{J}_{ab;\a} = \{ j_1 , \cdots , j_{ n_{ab;\alpha } } \}  $ with the $i_1 , i_2 \cdots $ each living in $V_{N_a}$ and $j_1 , j_2 , \cdots $ each in $V_{N_b}$.   $(\Phi_{ab;\a})^{\otimes n_{ab;\a}}$ are linear maps $V_{N_a}^{\otimes n_{ab;\a}} \rightarrow V_{N_b}^{\otimes n_{ab;\a}  } $, and $\sigma_a$ are linear operators on $V_{N_a}^{\otimes n_{a}}$ where
\begin{equation}
	n_a = \sum_{b,\alpha} n_{ab;\a} = \sum_{b,\alpha} n_{ba;\a}
\end{equation}
%\begin{align}
%	(\Phi^{\otimes n})^{\bm{I}}_{\bm{J}} &\equiv \Phi^{I_1}_{J_1} \Phi^{I_2}_{J_2} \ldots \Phi^{I_n}_{J_n} \\
%	(\sigma)^{\bm{I}}_{\bm{J}} &\equiv \delta^{I_1}_{J_{\sigma(1)}} \delta^{I_2}_{J_{\sigma(2)}} \ldots \delta^{I_n}_{J_{\sigma(n)}}
%\end{align}
The indices of $\sigma_a$ are unions $\bigcup_{b,\alpha} \bm{J}_{ba;\a}$ and $\bigcup_{b,\alpha} \bm{I}_{ab;\a}$, meaning that upper indices of $\sigma_a$ are contracted with lower indices of all fields $\Phi_{ba;\a}$ that enter node $a$, and lower indices of $\sigma_a$ are contracted with upper indices of all fields $\Phi_{ab;\a}$ that leave node $a$.

As a basic example consider an operator in $\C3Z2$ with
\begin{equation}
	\bn = \{ n_{11}, n_{22}, n_{12;1}, n_{12;2}, n_{21;1}, n_{21;2} \} = \{ 1, 1, 1, 0, 1, 0 \}
\end{equation}
that is, build from fields $(\Phi_{11}, \Phi_{22}, \Phi_{12;1}, \Phi_{21;1})$. We have
\begin{equation}
	\cO_{\C3Z2}(\bn, \sigma_1,\sigma_2) = 
	(\Phi_{11})^{i_1}_{j_1} \, (\Phi_{22})^{i_2}_{j_2} \, (\Phi_{12;1})^{i_3}_{j_3} \, (\Phi_{21;1})^{i_4}_{j_4} \, (\sigma_1)^{j_1 j_4}_{i_1 i_3} (\sigma_2)^{j_2 j_3}_{i_2 i_4}
\end{equation}
with $\sigma_1,\sigma_2 \in S_2$. For different combinations of $\sigma_a$ we get
\begin{equation}
\begin{split}
	\cO(\mI, \mI ) &= \tr(\Phi_{11}) \tr(\Phi_{22}) \tr(\Phi_{12;1} \Phi_{21;1}) \\
	\cO((12), \mI ) &= \tr(\Phi_{11}\Phi_{12;1} \Phi_{21;1}) \tr(\Phi_{22}) \\
	\cO(\mI, (12) ) &= \tr(\Phi_{11}) \tr(\Phi_{22}\Phi_{21;1}\Phi_{12;1}) \\
	\cO((12), (12) ) &= \tr(\Phi_{11}\Phi_{12;1}\Phi_{22}\Phi_{21;1})
\end{split}
\end{equation}

As  in the previous section for the case of $\mC^3$, the operators $\cO_Q (\bn,\bsig)$ are not uniquely labelled by $\bsig$, that is, the basis is overcomplete and different $\bsig$ can correspond to the same operator. Specifically, we have an identification
\begin{equation}
\label{eq:OQ_adj}
	\cO_Q (\bn,\bsig) = \cO_Q (\bn, \Adj_{\bgam}(\bsig))
\end{equation}
where 
\begin{equation}
\label{eq:gamma_subgroup}
	\bgam = \{ \gamma_{ab;\a} \} \; \in \bigotimes_{a,b,\alpha} S_{n_{ab;\a}}
\end{equation}
and the adjoint action is defined as
\begin{equation}
	\Adj_{\bgam}(\bsig) = \left\{ (\otimes_{b,\alpha} \gamma_{ba;\a} ) \, \sigma_{a} \, (\otimes_{b,\alpha} \gamma_{ab;\a}^{-1}) \right\}
\end{equation}
This is easily seen from the definition (\ref{eq:OQ_sig_defn}) and the fact that each $n_{ab;\a}$ block of identical fields is invariant under permutations
\begin{equation}
	\left( \Phi_{ab;\a}^{\otimes n_{ab;\a}} \right) = 
	\gamma^{-1} \left( \Phi_{ab;\a}^{\otimes n_{ab;\a}} \right) \gamma
\end{equation}
These permutations can then be moved to act on $\bsig$.

It is shown in \cite{BHR1,BHR2} that for $\mC^3$ the complete orthogonal bases (\ref{eq:OL_C3}) and (\ref{eq:OK_C3}) can be derived by essentially ``solving'' the invariance (\ref{eq:Osig_C3_conj}). We will use the same method here to find generalized bases $\cO_Q(\bL)$ and $\cO_Q(\bK)$ for any quiver $Q$. As an illustration let us take the simplest example of half-BPS operators \cite{cjr}. The idea is that the invariance
\begin{equation}
	\cO_\mC(\sigma) = \frac{1}{n!} \sum_{\gamma\in S_n} \cO_\mC(\gamma^{-1} \sigma \gamma)
\end{equation}
can be rewritten as
\begin{equation}
	\cO_\mC(\sigma) 
	= \sum_\tau \left( \frac{1}{n!} \sum_{\gamma} \delta(\gamma \sigma \gamma^{-1} \tau^{-1}) \right) \cO_\mC(\tau)
	= \sum_\tau \left( 	\frac{1}{n!}\sum_{R \vdash n} \chi_R(\sigma) \chi_R(\tau) \right) \cO_\mC(\tau)
\end{equation}
which looks like a projector to a lower-dimensional space labelled by Young diagram $R$. This motivates the Schur polynomial basis
\begin{equation}
	\cO_\mC(R) = \frac{1}{n!} \sum_\tau \chi_R(\sigma) \cO_\mC(\sigma)
\end{equation}
which indeed turns out to be complete and orthogonal. For $\mC^3$ we have similarly (\ref{eq:Osig_C3_conj}) leading to
\begin{equation}
	\cO_{\mC^3}(\bn, \sigma) \sim \sum_{\tau} \left( \sum_{R,\br,\nu^-,\nu^+} \chi_{R\rightarrow\br}^{\nu^-,\nu^+}(\sigma) \chi_{R\rightarrow\br}^{\nu^-,\nu^+}(\tau) \right) \cO_{\mC^3}(\bn,\tau)
\end{equation}
which suggests the basis (\ref{eq:OL_C3}). In order to generalize this to arbitrary quiver, we define ``quiver characters'' $\chi_Q(\bL, \bsig)$ obeying, schematically
\begin{equation}
	\sum_{\bL} \chi_Q(\bL, \bsig) \chi_Q(\bL, \btau) \sim \sum_{\bgam} \delta(\bsig, \Adj_{\bgam}(\btau) )
\end{equation}
where $\bL$ is a generalized set of group theory labels. With a help of quiver characters we can analogously express invariance (\ref{eq:OQ_adj}) as
\begin{equation}
\label{eq:inv_projection_approx}
	\cO_Q(\bn,\bsig) \sim \sum_{\btau} \left( \sum_{\bL} \chi_Q(\bL, \bsig) \chi_Q(\bL, \btau) \right) \cO_Q(\bn,\btau)
\end{equation}
leading to define a basis 
\begin{equation}
	\cO_Q(\bL) \sim \sum_{\bsig} \chi_Q(\bL, \bsig) \cO_Q(\bn, \bsig)
\end{equation}

The details of the derivation can be found in Appendix~\ref{app:basis_from_invariance}, the result is that we can define restricted quiver characters as
\begin{equation}
\label{eq:chiQ_defn}
\boxed{
\begin{split}
	\chi_Q(\bL,\bsig) &= 
	\prod_a 
	D^{R_a}_{i_a j_a} (\sigma_a)
	B^{R_a \rightarrow \bigcup_{b,\alpha} r_{ab;\a}, \nu^-_a}_{j_a \rightarrow \bigcup_{b,\alpha} l_{ab;\a}}
	B^{R_a \rightarrow \bigcup_{b,\alpha} r_{ba;\a}, \nu^+_a}_{i_a \rightarrow \bigcup_{b,\alpha} l_{ba;\a}}
 \\
	\bL &\equiv \{ R_a, r_{ab;\a}, \nu^-_a, \nu^+_a \}
\end{split}
}
\end{equation}
They obey the required invariance and orthogonality properties, listed in Appendix~\ref{app:identities_restricted}, which are analogous to those of symmetric group characters.
The complete basis of operators with a convenient normalization can then be defined as:
\begin{equation}
\label{eq:OL_defn}
\boxed{
	\cO_Q(\bL) = \frac{1}{\prod n_a!} \sqrt{\frac{\prod d(R_a)}{\prod d(r_{ab;\a})}} \; \sum_{\bsig} \chi_Q(\bL,\bsig) \cO_Q(\bn,\bsig)
}
\end{equation}
The group theory labels $\bL$ are:
\begin{itemize}
\item $r_{ab;\a}$:
a Young diagram with $n_{ab;\a}$ boxes for each set of fields $\Phi_{ab;\a}$.
\item $R_a$: 
a Young diagram for each group factor, labelling representation of $S_{n_a}$, where $n_a=\sum_{b,\alpha} n_{ba;\a} = \sum_{b,\alpha} n_{ab;\a}$ is the number of incoming and outgoing fields.
\item $\nu^-_a$:
multiplicity index for outgoing field reduction $R_a \rightarrow \bigcup_{b,\alpha} r_{ab;\a}$.
\item $\nu^+_a$:
multiplicity index for incoming field reduction $R_a \rightarrow \bigcup_{b,\alpha} r_{ba;\a}$.
\end{itemize}
The structure can most easily be seen with a diagram, which is the split-node quiver with permutations $\sigma_a$ inserted 
\begin{equation}
\label{eq:chiL_C3Z2}
	\chi_{\C3Z2}(\bL,\bsig) = \;
	\mytikz{
	%
	% Quiver with R,r,mu,nu labels
	%
		\node (s1) at (0,0) [rectangle,draw] {$\sigma_1$};		
		\node (m1) at (0,1) [circle,draw,inner sep=0.5mm,label=above:$\nu^+_1$] {};
		\node (n1) at (0,-1) [circle,draw,inner sep=0.5mm,label=below:$\nu^-_1$] {};
		\node (s2) at (3,0) [rectangle,draw] {$\sigma_2$};
		\node (n2) at (3,1) [circle,draw,inner sep=0.5mm,label=above:$\nu^-_2$] {};
		\node (m2) at (3,-1) [circle,draw,inner sep=0.5mm,label=below:$\nu^+_2$] {};
		\draw [postaction={decorate}] (m1) to node[auto]{$R_1$} (s1);
		\draw [postaction={decorate}] (s1) to node[auto]{} (n1);
		\draw [postaction={decorate}] (n1) to [bend left=90] node[auto]{$r_{11}$} (m1);
		\draw [postaction={decorate}] (n1) to [bend left=10] node[above]{$r_{12;1}$} (m2);
		\draw [postaction={decorate}] (n1) to [bend right=10] node[below]{$r_{12;2}$} (m2);
		\draw [postaction={decorate}] (m2) to node[auto]{$R_2$} (s2);
		\draw [postaction={decorate}] (s2) to node[auto]{} (n2);
		\draw [postaction={decorate}] (n2) to [bend left=90] node[auto]{$r_{22}$} (m2);
		\draw [postaction={decorate}] (n2) to [bend left=10] node[below]{$r_{21;1}$} (m1);
		\draw [postaction={decorate}] (n2) to [bend right=10] node[above]{$r_{21;2}$} (m1);		
	}
\end{equation}
Each group node carries a permutation in representation $R_a$ (denoted by a box), which is then contracted via branching coefficients (denoted by white nodes) to representations $r_{ab;\a}$ associated with fields. There are multiplicities $\nu^\pm_a$ associated to each branching coefficient node. The lines denote contracted matrix indices $i_a,j_a,l_{ab;\a}$. Note that $\chi_Q(\bL,\bsig)$ reduces precisely to (\ref{eq:chiQ_C3_diag}) for the $\mC^3$ quiver! Also, for the trivial quiver $\mC$ consisting of one node and one field $\Phi_{11}$, corresponding to the half-BPS sector, we get $R_{1} = r_{11}$, all the branching coefficients are unit matrices, and the quiver character \emph{is} the usual symmetric group character.

Using the orthogonality properties of quiver characters we can write the inverse of the basis change (\ref{eq:OL_defn}):
\begin{equation}
\label{eq:OL_inverse}
	\cO_Q(\bn,\bsig) = \sum_{\bL}
	\sqrt{\frac{\prod d(R_a)}{\prod d(r_{ab;\a})}} \;  \chi_Q(\bL,\bsig) \cO_Q(\bL)
\end{equation}

%%%%%%%%%%%%%%%%%%%%%%%%%%%%%%%%%%%%%%%%%%%%%%%%%%%%%%%%%%%%%%%%%%%%%

\subsection{Two-point function}
\label{sec:twopoint}

We will show here that the general basis (\ref{eq:OL_defn}) is orthogonal in free field metric for any quiver $Q$
\begin{equation}
\label{eq:OL2pt}
\boxed{
	\la \cO_Q(\bL) \cO_Q(\tl\bL)^\dagger \ra =
	\delta_{\bL\tl\bL}
	\frac{\prod n_{ab;\a} !}{\prod n_a !}
	\prod_a f_{N_a}(R_a)	
}
\end{equation}
$f_{N_a}(R_a)$ is the product of weights of a $U(N_a)$ diagram $R_a$. We can see it is a straightforward generalization of the result (\ref{eq:OL2pt_C3}) for $\mC^3$, except with a different normalization, due to different normalization of the operators (\ref{eq:OL_defn}), compared to (\ref{eq:OL_C3}). It is important to note, that again $N$-dependence is in the factors $f_{N_a}(R_a)$ which vanish if the height of $R_a$ exceeds $N_a$. So at finite $N$ the Hilbert space consists of operators $\cO_Q(\bL)$ where the height of all $R_a$ does not exceed $N_a$
\begin{equation}
	\cH = \{ \cO_Q(\bL) \; | \; \forall_a l(R_a) \le N_a \}
\end{equation}

The derivation of (\ref{eq:OL2pt}) is similar to that of (\ref{eq:OL2pt_C3}) in \cite{db1}, but now using analogous properties of quiver characters $\chi_Q(\bL,\bsig)$ from Appendix~\ref{app:identities_restricted}. We have the free field metric
\begin{equation}
\label{eq:free_metric}
	\la (\Phi_{ab;\a})^i_j (\Phi_{cd;\beta}^{\dagger})^k_l \ra = 
	\delta_{ac} \delta_{bd} \delta_{\alpha\beta} \delta^i_l \delta^k_j
\end{equation}
Then the two point function of $\cO_Q(\bn,\bsig)$ operators is
\begin{equation}
\label{eq:OQ_sig_2pt}
	\la \cO_Q(\bn,\bsig)\cO_Q(\bn,\tl\bsig)^\dagger \ra =
	\sum_{\bgam} \prod_a \tr_{V_{N_a}} {n}(\Adj_{\bgam}(\sigma_a) \tl\sigma_a^{-1})	
\end{equation}
The sum is over $\bgam \equiv \{ \gamma_{ab;\a} \in S_{n_{ab;\a}} \}$ -- Wick contractions arising from each set of fields. For the derivation of (\ref{eq:OQ_sig_2pt}) see Appendix~\ref{appsec:2ptfunction}. Next, we apply (\ref{eq:OQ_sig_2pt}) to the definition of $\cO_Q(\bL)$ (\ref{eq:OL_defn}):
\begin{equation}
	\la \cO_Q(\bL) \cO_Q(\tl\bL)^\dagger \ra = 
	c_{\bL} c_{\tl \bL}
	\sum_{\bsig,\tl\bsig,\bgam} \chi_Q(\bL,\bsig) \chi_Q(\tl\bL,\tl\bsig) \prod_a \tr_{V_{N_a}}^{n_a}  (\Adj_{\bgam}(\sigma_a) \tl\sigma_a^{-1})	
\end{equation}
where $c_{\bL}=\frac{1}{\prod n_a! }\sqrt{\frac{\prod d(R_a)}{\prod d(r_{ab;\a})}}$ is the normalization coefficient appearing in front of the sum in (\ref{eq:OL_defn}).
Note that $\chi_Q(\bL,\bsig)$ is a real number, so we drop complex conjugation. Now redefining $\sigma_a \rightarrow  \Adj_{\bgam}(\sigma_a)$ and using invariance property (\ref{eq:chiQ_inv}) the dependence on $\bgam$ drops out
\begin{equation}
	\la \cO_Q(\bL) \cO_Q(\tl\bL)^\dagger \ra = 
	\left( c_{\bL} c_{\tl \bL} \prod n_{ab;\a} ! \right)
	\sum_{\bsig,\tl\bsig} \chi_Q(\bL,\bsig) \chi_Q(\tl\bL,\tl\bsig) \prod_a 
\tr_{V_{N_a}}^{n_a} (\sigma_a \tl\sigma_a^{-1})	
\end{equation}
Next, applying (\ref{eq:CQ_sigsig_sum})
\begin{equation}
\begin{split}
	\la \cO_Q(\bL) \cO_Q(\tl\bL)^\dagger \ra &= 
	\delta_{\bR\tl\bR}\delta_{\br\tl\br}\delta_{\bnu^-\tl\bnu^-} 
	\left({c_{\bL}}^2 \prod n_{ab;\a} ! \right)
	\\ & \times
	\sum_{\bsig} \prod_a \frac{n_a!}{d(R_a)} \tr\left(D^{R_a}(\sigma_a) P^{\nu^+_a\tl\nu^+_a}_{R_a\rightarrow \bigcup_{b,\alpha} r_{ab;\a}} \right) \tr_{V_{N_a}}^{n_a} (\sigma_a)	
\end{split}
\end{equation}
Finally (\ref{eq:DP_tr_sum}) gives
\begin{equation}
\begin{split}
	\la \cO_Q(\bL) \cO_Q(\tl\bL)^\dagger \ra &= 
	\delta_{\bL \tl\bL}
	\,
	{c_{\bL}}^2	\;
	\frac{\prod n_{ab;\a} ! \prod n_a! \prod d(r_{ab;\a})}{\prod d(R_a)}	\prod_a f_{N_a}(R_a)
\\
	&= \delta_{\bL\tl\bL}
	\frac{\prod n_{ab;\a} !}{\prod n_a !}
	\prod_a f_{N_a}(R_a) 	
\end{split}
\end{equation}
proving (\ref{eq:OL2pt}) .

%%%%%%%%%%%%%%%%%%%%%%%%%%%%%%%%%%%%%%%%%%%%%%%%%%%%%%%%%%%%%%%%%%%%%

\subsection{Covariant basis}
\label{sec:covbasis}

We can define another complete, free orthogonal basis, which is a generalization of (\ref{eq:OK_C3})
\begin{equation}
\label{eq:OK_defn}
\boxed{
	\cO_Q(\bK) = \frac{\sqrt{\prod d(R_a)}}{\prod n_a!} \;  \sum_{\bsig} \chi_Q(\bK,\bsig) \, \cO_Q(\bn,\bsig)
}
\end{equation}
We refer to it as the  \emph{covariant basis}, because the labels $\bK$ include representations of the global symmetry group $\prod_{a,b} U(M_{ab})$. The basis arises from the possibility to ``solve the invariance'' as in (\ref{eq:inv_projection_approx}) using \emph{covariant} quiver characters:
\begin{equation}
\label{eq:chiK_defn}
\begin{split}
	\chi_Q(\bK,\bsig) &= 
	\left(
	\prod_a 
	D^{R_a}_{i_a j_a} (\sigma_a)
	B^{R_a \rightarrow \bigcup_{b} s_{ab}^-, \nu^-_a}_{j_a \rightarrow \bigcup_{b} l_{ab}^-}
	B^{R_a \rightarrow \bigcup_{b} s_{ba}^+, \nu^+_a}_{i_a \rightarrow \bigcup_{b} l_{ba}^+}
	\right)
\left(
	\prod_{a,b} 
	B^{\Lambda_{ab} \rightarrow [\bn_{ab}],\beta_{ab}}_{l_{ab}} 
	S^{\; s_{ab}^+ \; s_{ab}^-,\, \Lambda_{ab}\tau_{ab}}_{\;l_{ab}^+\;\tl l_{ab}^-,\,l_{ab}}
\right)
\end{split}
\end{equation}
with a different set of labels
\begin{equation}
\label{eq:K_defn}
	\bK = \{ R_a, s_{ab}^+, s_{ab}^-, \nu_a^+, \nu_a^-, \Lambda_{ab}, \tau_{ab}, n_{ab;\a}, \beta_{ab} \}
\end{equation}
The covariant quiver characters $\chi_Q(\bK,\bsig)$ also obey an analogous set of character orthogonality identities, listed in Appendix~\ref{app:identities_covariant}.
For the details of the derivation of the basis and how the two options $\chi_Q(\bL,\bsig)$ and $\chi_Q(\bK,\bsig)$ arise see Appendix~\ref{app:basis_from_invariance}.

The covariant quiver characters are again most neatly expressed diagrammatically, as a modification of the original quiver. For $\C3Z2$ (\ref{eq:chiK_defn}) becomes
\begin{equation}
\label{eq:chiK_diag}
	\chi_{\C3Z2}(\bK,\bsig) = \;
	\mytikz{
	%
	% Quiver with R,r,mu,nu labels
	%
		\node (s1) at (0,0) [rectangle,draw] {$\sigma_1$};		
		\node (m1) at (0,1) [circle,draw,inner sep=0.5mm,label=above:$\nu^+_1$] {};
		\node (n1) at (0,-1) [circle,draw,inner sep=0.5mm,label=below:$\nu^-_1$] {};
		\node (s2) at (3,0) [rectangle,draw] {$\sigma_2$};
		\node (n2) at (3,1) [circle,draw,inner sep=0.5mm,label=above:$\nu^-_2$] {};
		\node (m2) at (3,-1) [circle,draw,inner sep=0.5mm,label=below:$\nu^+_2$] {};
		\node (t12) at (1.5,-1) [circle,fill,inner sep=0.5mm,label=above:$\tau_{12}$] {};
		\node (t21) at (1.5,1) [circle,fill,inner sep=0.5mm,label=below:$\tau_{21}$] {};
		\node (t11) at (-1,0) [circle,fill,inner sep=0.5mm] {};
		\node (t22) at (4,0) [circle,fill,inner sep=0.5mm] {};
		\node (n12) at (1.5,-2.5) {};
		\node (n21) at (1.5,2.5) {};
		\node (b11) at (-2.5,0) {$\Lambda_{11}=[n_{11}]$};
		\node (b22) at (5.5,0) {$\Lambda_{22}=[n_{22}]$};		
		\node (b12) at (1.5,-2) [circle,draw,inner sep=0.5mm,label=left:$\beta_{12}$] {};
		\node (b21) at (1.5,2) [circle,draw,inner sep=0.5mm,label=right:$\beta_{21}$] {};
		\draw [postaction={decorate}] (m1) to node[left]{$R_1$} (s1);
		\draw [postaction={decorate}] (s1) to node[left]{} (n1);
		\draw [postaction={decorate}] (n1) to [bend left=45] node[left]{$s_{11}$} (t11);
		\draw [postaction={decorate}] (t11) to [bend left=45] node[left]{$s_{11}$} (m1);
		\draw [postaction={decorate}] (n1) to node[above]{$s_{12}^-$} (t12);
		\draw [postaction={decorate}] (t12) to node[above]{$s_{12}^+$} (m2);
		\draw [-] (t12) to node[right]{$\Lambda_{12}$} (b12);
		\draw [-] (b12) to node[right]{$\bn_{12}$} (n12);
		\draw [postaction={decorate}] (m2) to node[right]{$R_2$} (s2);
		\draw [postaction={decorate}] (s2) to node[right]{} (n2);
		\draw [postaction={decorate}] (n2) to [bend left=45] node[right]{$s_{22}$} (t22);
		\draw [postaction={decorate}] (t22) to [bend left=45] node[right]{$s_{22}$} (m2);
		\draw [postaction={decorate}] (n2) to node[below]{$s_{21}^-$} (t21);
		\draw [postaction={decorate}] (t21) to node[below]{$s_{21}^+$} (m1);		
		\draw [-] (t21) to node[left]{$\Lambda_{21}$} (b21);
		\draw [-] (b21) to node[left]{$\bn_{12}$} (n21);
		\draw [-] (t11) to node[below]{} (b11);
		\draw [-] (t22) to node[below]{} (b22);		
	}
\end{equation}
The labels involved are:
\begin{itemize}
\item $R_a \vdash n_a$ diagram associated to each group node factor is the same as before, with finite $N$ cutoff $l(R_a) \le N_a$.
\item Each set of $M_{ab}$ arrows between given pair of nodes is collapsed into one, and there is an associated diagram $\Lambda_{ab} \vdash n_{ab}$, 
where $n_{ab} = \sum_{ \alpha } n_{ab;\alpha} $.  It labels a representation of the global symmetry $U(M_{ab})$, and so $l(\Lambda_{ab}) \le M_{ab}$. Since in $\C3Z2$ we have $M_{11}=M_{22}=1$, the associated $\Lambda_{11},\Lambda_{22}$ are fixed to be single-row diagrams, one-dimensional irreps.
\item There are two additional diagrams $s_{ab}^\pm \vdash n_{ab}$ associated to each line. In case $M_{ab}=1$ they are equal $s_{ab}^+=s_{ab}^-$ and the same as $r_{ab}$ in the restricted basis.
\item As in the restricted basis, we have branching at the white nodes $R_a \rightarrow \cup_b s_{ba}^+$ and $R_a \rightarrow \cup_b s_{ab}^-$ and the associated Littlewood-Richardson multiplicity labels $\nu_a^\pm$.
\item There is a black node on each field line denoting Kronecker product $s_{ab}^+ \otimes s_{ab}^- \rightarrow \Lambda_{ab}$ and the associated Clebsch-Gordan multiplicity label $\tau_{ab}$.
\item The extra labels $\beta_{ab}$, together with charges $\bn_{ab} \equiv \{ n_{ab;\alpha}\} $, identify a state in $U(M_{ab})$ irrep $\Lambda_{ab}$. That is equivalent to specifying a branching multiplicity label for $\Lambda_{ab} \rightarrow \cup_\alpha [n_{ab;\a}]$ reduction (see e.g.  \cite{fulhar} for this fact). 
\end{itemize}

Let us also note, that in the case of the trivial $\Lambda_{11},\Lambda_{22}$ the corresponding Clebsch-Gordan coefficient still has to be included in (\ref{eq:chiK_defn})
\begin{equation}
S^{s^+ s^- , ~ \Lambda=[n]}_{\,i\;\;j\;\;\;,~ 1} = \delta_{s^+ s^-} \frac{\delta_{ij}}{\sqrt{d(s^+)}}
\end{equation} 
It forces $s^+ = s^-$, and is itself proportional to a delta function, but it includes the coefficient $\frac{1}{\sqrt{d(s)}}$. Diagrammatically 
\begin{equation}
\mytikz{
	\node (t) at (0,0) [circle,fill,inner sep=0.5mm] {};
	\node (l) at (0,1) {$\Lambda=[n]$};
	\node (dummy) at (0,-1) {};
	\draw [postaction={decorate}] (t) to node[above]{$s$} +(1,0);
	\draw [postaction={decorate}] ($(t)+(-1,0)$) to node[above]{$s$} (t);
	\draw [-] (t) to (l);
}
\; = \;
\frac{1}{\sqrt{d(s)}} \quad
\mytikz{
	\draw [postaction={decorate}] (0,0) to node[above]{$s$} +(2,0);	
}
\end{equation}

The key property of this basis is that the transformations under global symmetry group $\prod_{a,b} U(M_{ab})$ are made explicit
\begin{itemize}
\item $\{ \Lambda_{ab} \}$ labels pick the representation of $\prod_{a,b} U(M_{ab})$
\item $\{ R_a, s_{ab}^+, s_{ab}^-, \nu_a^+, \nu_a^-, \tau_{ab} \}$ then distinguish different multiplets transforming under $\{ \Lambda_{ab} \}$
\item $\{ \bn_{ab}, \beta_{ab} \}$ label a state in $\{ \Lambda_{ab} \}$.
\end{itemize}

The free two-point function in the covariant basis can be calculated in analogous way as in the previous section, now using the properties of covariant characters in Appendix~\ref{app:identities_covariant}. With our normalization the result is exactly the same as (\ref{eq:OL2pt}):
\begin{equation}
	\la \cO_Q(\bK) \cO_Q(\tl\bK)^\dagger \ra = 
	\delta_{\bK\tl\bK}
	\frac{\prod n_{ab;\a} !}{\prod n_a !}
	\prod_a f_{N_a}(R_a) 	
\end{equation}

Finally, the inverse basis transformation is:
\begin{equation}
	\cO_Q(\bn,\bsig) = \sum_{\bK} \sqrt{\prod d(R_a)} \; \chi_Q(\bK,\bsig) \, \cO_Q(\bK)
\end{equation}

%%%%%%%%%%%%%%%%%%%%%%%%%%%%%%%%%%%%%%%%%%%%%%%%%%%%%%%%%%%%%%%%%%%%%

\subsection{Examples}

Let us go over a few specific examples of quivers, to illustrate our general methods.

\subsubsection{Conifold}

\begin{figure}[h]
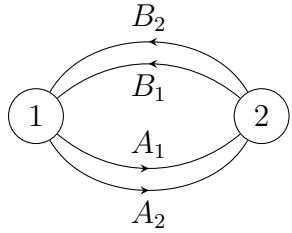

\centering
\mytikz{
	\node (s1) at (-1.5,0) [circle,draw] {$1$};
	\node (s2) at (1.5,0) [circle,draw] {$2$};	
	\draw [postaction={decorate}] (s1) to [bend right=40] node[above]{$A_1$} (s2);
	\draw [postaction={decorate}] (s1) to [bend right=60] node[below]{$A_2$} (s2);
	\draw [postaction={decorate}] (s2) to [bend right=40] node[below]{$B_1$} (s1);
	\draw [postaction={decorate}] (s2) to [bend right=60] node[above]{$B_2$} (s1);	
}
\caption{Quiver for the conifold theory.}
\label{fig:quivercon}
\end{figure}

The quiver for the conifold theory is shown in Figure~\ref{fig:quivercon}.
The gauge group is $U(N_1) \times U(N_2)$ and we have bifundamental fields
\begin{equation}
	A_1, A_2, B_1, B_2
\end{equation}
transforming in a global $U(2) \times U(2)$ flavor symmetry. Note according to the labelling in the previous section the fields correspond to $A_1=\Phi_{12;1}, A_2=\Phi_{12;2}, B_1=\Phi_{21;1}, B_2=\Phi_{21;2}$.

The gauge invariant mesonic operators are traces of alternating products $\tr(A_{i_1} B_{j_1} A_{i_2} B_{j_2} \ldots)$. According to the general prescription (\ref{eq:OQ_sig_defn}), a general gauge invariant operator can be specified by charges and two permutations
\begin{equation}
	\cO_\cC(\bn, \{\sigma_1,\sigma_2\}) = 
	\trVN{n}\left(\sigma_1 (A_1^{\otimes n_1}\otimes A_2^{\otimes n_2}) \sigma_2 (B_1^{\otimes m_1}\otimes B_2^{\otimes m_2}) \right)
\end{equation}
or diagrammatically
\begin{equation}
	\cO_{\cC}( \bn,\{\sigma_1,\sigma_2\}) = \;	
	\mytikz{	
		\node (s1) 		at (3,0) [rectangle,draw,minimum height=0.8cm,minimum width=4cm] {$\sigma_1$};		
		\node (p121) 	at (2,-1.5) [rectangle,draw,minimum height=0.8cm] {$A_1^{\otimes n_1}$};
		\node (p122) 	at (4,-1.5) [rectangle,draw,minimum height=0.8cm] {$A_2^{\otimes n_2}$};
		\node (s2) 		at (3,-3) [rectangle,draw,minimum height=0.8cm,minimum width=4cm] {$\sigma_2$};		
		\node (p211) 	at (2,-4.5) [rectangle,draw,minimum height=0.8cm] {$B_1^{\otimes m_1}$};
		\node (p212) 	at (4,-4.5) [rectangle,draw,minimum height=0.8cm] {$B_2^{\otimes m_2}$};		
		\draw [-] (1,0.75) to (5,0.75);
		\draw [-] (1,-5.25) to (5,-5.25);
		\draw [postaction={decorate}] (2,0.75) to +(0,-0.35);
		\draw [postaction={decorate}] (4,0.75) to +(0,-0.35);				
		\draw [postaction={decorate}] (2,-0.4) to +(0,-0.7);
		\draw [postaction={decorate}] (4,-0.4) to +(0,-0.7);		
		\draw [postaction={decorate}] (2,-1.9) to +(0,-0.7);
		\draw [postaction={decorate}] (4,-1.9) to +(0,-0.7);		
		\draw [postaction={decorate}] (2,-3.4) to +(0,-0.7);
		\draw [postaction={decorate}] (4,-3.4) to +(0,-0.7);				
		\draw [postaction={decorate}] (2,-4.9) to +(0,-0.35);
		\draw [postaction={decorate}] (4,-4.9) to +(0,-0.35);		
	}
\end{equation}
Here we denote $n=n_1+n_2=m_1+m_2$ the total number of $A$'s or $B$'s, which has to be equal.

The counting is given by the split-node quiver, which was shown in Figure~\ref{fig:quiver_split_con} and (\ref{eq:counting_con}). Now the restricted quiver characters obtained by inserting $(\s_1 , \s_2 )$ in the    the same split-node quiver are
\begin{equation}\label{conifoldchar}
	\chi_\cC(\bL,\{\sigma_1,\sigma_2\}) =
	\mytikz{
		\node (s1) at (0,0) [rectangle,draw] {$\sigma_1$};		
		\node (m1) at (0,0.8) [circle,draw,inner sep=0.5mm,label=above:$\nu^+_1$] {};
		\node (n1) at (0,-0.8) [circle,draw,inner sep=0.5mm,label=below:$\nu^-_1$] {};
		\node (s2) at (3,0) [rectangle,draw] {$\sigma_2$};
		\node (n2) at (3,0.8) [circle,draw,inner sep=0.5mm,label=above:$\nu^-_2$] {};
		\node (m2) at (3,-0.8) [circle,draw,inner sep=0.5mm,label=below:$\nu^+_2$] {};
		\draw [postaction={decorate}] (m1) to node[left]{$R_1$} (s1);
		\draw [postaction={decorate}] (s1) to (n1);		
		\draw [postaction={decorate}] (n1) to [bend left=10] node[above]{$r_{A_1}$} (m2);
		\draw [postaction={decorate}] (n1) to [bend right=10] node[below]{$r_{A_2}$} (m2);
		\draw [postaction={decorate}] (m2) to node[right]{$R_2$} (s2);
		\draw [postaction={decorate}] (s2) to (n2);		
		\draw [postaction={decorate}] (n2) to [bend left=10] node[below]{$r_{B_1}$} (m1);
		\draw [postaction={decorate}] (n2) to [bend right=10] node[above]{$r_{B_2}$} (m1);		
	}	
\end{equation}
leading to the restricted Schur basis operators (\ref{eq:OL_defn}):
\begin{equation}
	\cO_\cC(\bL) = \frac{1}{(n!)^2} \sqrt{\frac{d(R_1)d(R_2)}{d(r_{A_1})d(r_{A_2})d(r_{B_1})d(r_{B_2})}} \sum_{\sigma_1,\sigma_2}	
	\chi_\cC(\bL,\{\sigma_1,\sigma_2\}) \,
	\cO_\cC(\bn, \{\sigma_1,\sigma_2\})
\end{equation}
The labels are
\begin{equation}
	\bL = \{ R_1, R_2, r_{A_1}, r_{A_2}, r_{B_1}, r_{B_2}, \nu_1^\pm, \nu_2^\pm \}
\end{equation}
where $R_1, R_2 \vdash n$ are Young diagrams associated with each of the group factors, limited to at most $N_1,N_2$ rows, $r_{A_1}, r_{A_2}, r_{B_1}, r_{B_2}$ are Young diagrams associated with each field type. They are constrained such that $R_1,R_2$ appear in the Littlewood-Richardson products
\begin{equation}
\begin{split}
	r_{A_1} \otimes r_{A_2} &\rightarrow R_1 \\
	r_{A_1} \otimes r_{A_2} &\rightarrow R_2 \\
	r_{B_1} \otimes r_{B_2} &\rightarrow R_1 \\
	r_{B_1} \otimes r_{B_2} &\rightarrow R_2
\end{split}
\end{equation}
and $\nu_1^\pm$, $\nu_2^\pm$ are the associated multiplicity labels, when $R_1,R_2$ appears more than once in the product.

In this case, as  in (\ref{eq:chi_C3_1}) for $\mC^3$, we can write the restricted quiver character $\chi_\cC(\bL,\bsig)$ as a sort of restricted trace. Define a projector
\begin{equation}
	(P^{\nu^-,\nu^+}_{R\rightarrow \br \leftarrow S })_{ij}  =  
		\sum_l
		B^{R \rightarrow \br}_{i \rightarrow l}
		B^{S \rightarrow \br}_{j \rightarrow l}
\end{equation}
which projects from two different representations $R,S$ of $S_n$ into the same irrep $\br=(r_1,r_2)$ of the subgroup $S_{n_1}\times S_{n_2}$. Then we can write the quiver character as
\begin{equation}
	\chi_\cC(\bL,\{\sigma_1,\sigma_2\}) = \tr\left( 
		D^{R_1}(\sigma_1) 
		P^{\nu_1^-,\nu_2^+}_{R_1\rightarrow \br_A \leftarrow R_2 }
		D^{R_2}(\sigma_2) 
		P^{\nu_2^-,\nu_1^+}_{R_2\rightarrow \br_B \leftarrow R_2 }
	\right)
\end{equation}
The Restricted Schur basis operators are, explicitly:
\begin{equation}
\label{eq:OL_con}
\begin{split}
	\cO_\cC(\bL) &= 
		c_{\bL} \sum_{\sigma_1,\sigma_2} 
		\tr\left( 
		D^{R_1}(\sigma_1) 
		P^{\nu_1^-,\nu_2^+}_{R_1\rightarrow \br_A \leftarrow R_2 }
		D^{R_2}(\sigma_2) 
		P^{\nu_2^-,\nu_1^+}_{R_2\rightarrow \br_B \leftarrow R_2 }
	\right)
		\cO_\cC(\bn,\{\sigma_1,\sigma_2\})
\end{split}
\end{equation}

Let us demonstrate the simplest example, with the charges $\bn=\{1,1,1,1\}$, that is, each field occurs once. The only choice for $r$ diagrams is
\begin{equation}
 	r_{A_1} = r_{A_2} = r_{B_1} = r_{B_2} = \yng(1)
\end{equation}
Littlewood-Richardson product is
\begin{equation}
	\yng(1) \otimes \yng(1) \rightarrow \yng(2) \oplus \yng(1,1)
\end{equation}
each diagram appearing once, so there is no multiplicity. We can choose each $R_1,R_2$ independently to be either of the diagrams, giving 4 operators
\begin{equation}
\begin{split}
	&\cO(\yng(2),\yng(2)) \\& \quad = 
	\frac{1}{4} \left(	\tr(A_1 B_1)\tr(A_2 B_2) + \tr(A_1 B_2)\tr(A_2 B_1)
		+ \tr(A_1 B_1 A_2 B_2) + \tr(A_1 B_2 A_2 B_1) \right)
\\		
	&\cO(\yng(1,1),\yng(1,1)) \\& \quad = 
	\frac{1}{4} \left(	\tr(A_1 B_1)\tr(A_2 B_2) + \tr(A_1 B_2)\tr(A_2 B_1)
		- \tr(A_1 B_1 A_2 B_2) - \tr(A_1 B_2 A_2 B_1) \right)
\\		
	&\cO(\yng(2),\yng(1,1)) \\& \quad = 
	\frac{1}{4} \left(	\tr(A_1 B_1)\tr(A_2 B_2) - \tr(A_1 B_2)\tr(A_2 B_1)
		+ \tr(A_1 B_1 A_2 B_2) - \tr(A_1 B_2 A_2 B_1)	\right)
\\		
	&\cO(\yng(1,1),\yng(2)) \\& \quad= 
	\frac{1}{4} \left(	\tr(A_1 B_1)\tr(A_2 B_2) - \tr(A_1 B_2)\tr(A_2 B_1)
		- \tr(A_1 B_1 A_2 B_2) + \tr(A_1 B_2 A_2 B_1)	\right)			
\end{split}
\end{equation}
It can be checked that they are orthogonal in the free field metric. These operators are particularly easy to evaluate, since all the representations are one-dimensional, and so all branching coefficients are equal to 1. The only dependence comes from $D^{R_1}(\sigma_1)$, $D^{R_2}(\sigma_2)$. Note also the way this basis captures finite-$N$ cutoff: if $N=1$ the height of $R_1,R_2$ is limited to 1, so the only operator that survives is $\cO(\yng(2),\yng(2))$. It is easy to see that the others are 0 if the fields are replaced by scalar values.

Covariant basis operators (\ref{eq:OK_defn}) for conifold are
\begin{equation}
	\cO_\cC(\bK) = \frac{\sqrt{d(R_1)d(R_2)}}{(n!)^2} \sum_{\sigma_1,\sigma_2}
	\chi_\cC(\bK,\{\sigma_1,\sigma_2\}) \cO_\cC(\bn,\{\sigma_1,\sigma_2\})
\end{equation}
\begin{equation}
\label{eq:chiK_diag_con}
	\chi_\cC(\bK,\{\sigma_1,\sigma_2\}) = 
	\mytikz{	
		\node (s1) at (-1.5,0) [rectangle,draw] {$\sigma_1$};		
		\node (s2) at (1.5,0) [rectangle,draw] {$\sigma_2$};		
		\node (ta) at (0,-0.8) [circle,fill,inner sep=0.5mm,label=above:$\tau_A$] {};
		\node (tb) at (0,0.8) [circle,fill,inner sep=0.5mm,label=below:$\tau_B$] {};
		\node (beta) at (0,-1.5) [circle,draw,inner sep=0.5mm,label=left:$\beta_A$] {};
		\node (betb) at (0,1.5) [circle,draw,inner sep=0.5mm,label=left:$\beta_B$] {};
		\node (na) at (0,-2.2)  {};	
		\node (nb) at (0,2.2)  {};	
		\draw [postaction={decorate}] (s1) to [bend right=20] node[above]{$R_1$} (ta);
		\draw [postaction={decorate}] (ta) to [bend right=20] node[above]{$R_2$} (s2);
		\draw [postaction={decorate}] (s2) to [bend right=20] node[below]{$R_2$} (tb);
		\draw [postaction={decorate}] (tb) to [bend right=20] node[below]{$R_1$} (s1);		
		\draw [-] (ta) to node[right]{$\Lambda_A$} (beta);
		\draw [-] (beta) to node[right]{$[n_1,n_2]$} (na);
		\draw [-] (tb) to node[right]{$\Lambda_B$} (betb);
		\draw [-] (betb) to node[right]{$[m_1,m_2]$} (nb);
	}
\end{equation}
with the labels
\begin{equation}
\label{eq:K_con}
\begin{split}
	\bK &= \{ R_1, R_2, \Lambda_A, \Lambda_B, \tau_A, \tau_B, \bn, \beta_A, \beta_B \}	
\end{split}
\end{equation}
The $R_1,R_2 \vdash n$ are Young diagrams associated to the group nodes like before. But now, instead of $r_{A_i}, r_{B_i}$ we have global symmetry representation labels $\Lambda_A, \Lambda_B \vdash n$. They are constrained to appear in the irrep decomposition of the $S_n$ Kronecker product
\begin{equation}
\begin{split}
	R_1 \otimes R_2 &\rightarrow \Lambda_A \\
	R_1 \otimes R_2 &\rightarrow \Lambda_B
\end{split}	
\end{equation}
If $\Lambda_A, \Lambda_B$ appear multiple times in the decomposition, $\tau_A, \tau_B$ is the multiplicity label. The remaining labels $\{ n_A, n_B, \beta_A, \beta_B \}$ then label a specific state in the $U(M) \times U(M)$ irrep $(\Lambda_A, \Lambda_B)$. Note, compared to the general case (\ref{eq:K_defn}), we do not need additional labels $s_A^\pm, s_B^\pm, \nu_1^\pm, \nu_2^\pm$. This is because there is no ``branching'' in the quiver -- all arrows outgoing from node 1 go to node 2 and vice-versa, which enforces $R_1 = s_A^- = s_B^+$ and $R_2 = s_A^+ = s_B^-$.

Let us again work out the example with $n=2$, that is 2 $A$ fields and 2 $B$ fields. Like with Restricted Schur basis, we have 4 choices for $R_1,R_2$. In this simple case $\Lambda_A,\Lambda_B$ are uniquely determined by the choice of $R_1,R_2$, since
\begin{equation}
\begin{split}
	\yng(2) \otimes \yng(2) &\rightarrow \yng(2) \\
	\yng(2) \otimes \yng(1,1) &\rightarrow \yng(1,1) \\
	\yng(1,1) \otimes \yng(1,1) &\rightarrow \yng(2)
\end{split}
\end{equation}
that is, only one irrep appears in the product, so $\Lambda_A=\Lambda_B = R_1 \otimes R_2$. For each choice of $R_1,R_2,\Lambda_A,\Lambda_B$ we list the highest-weight state in $(\Lambda_A,\Lambda_B)$:
\begin{equation}
\begin{split}
	& \cO^{\rm hw}(R_1=\yng(2),\,R_2=\yng(2),\,\Lambda_A=\Lambda_B=\yng(2)) 
	= \frac{1}{2} \tr(A_1 B_1) \tr(A_1 B_1) + \frac{1}{2} \tr(A_1 B_1 A_1 B_1) \\
	& \cO^{\rm hw}(R_1=\yng(1,1),\,R_2=\yng(1,1),\,\Lambda_A=\Lambda_B=\yng(2))
	= \frac{1}{2} \tr(A_1 B_1) \tr(A_1 B_1) - \frac{1}{2} \tr(A_1 B_1 A_1 B_1) \\
	& \cO^{\rm hw}(R_1=\yng(2),\,R_2=\yng(1,1),\,\Lambda_A=\Lambda_B=\yng(1,1)) \\
	& \quad = \frac{1}{4}\left( \tr(A_1 B_1)\tr(A_2 B_2) - \tr(A_1 B_2)\tr(A_2 B_1)
		+ \tr(A_1 B_1 A_2 B_2) - \tr(A_1 B_2 A_2 B_1)	\right) \\
	& \cO^{\rm hw}(R_1=\yng(1,1),\,R_2=\yng(2),\,\Lambda_A=\Lambda_B=\yng(1,1)) \\
	& \quad = \frac{1}{4}\left( \tr(A_1 B_1)\tr(A_2 B_2) - \tr(A_1 B_2)\tr(A_2 B_1)
		- \tr(A_1 B_1 A_2 B_2) + \tr(A_1 B_2 A_2 B_1)	\right)
\end{split}
\end{equation}

\subsubsection{\texorpdfstring{$\C3Z2$}{C3/Z2}}

We have used the theory of $D3$ branes on a $\C3Z2$ singularity throughout, so here we just collect the references. 

The quiver and the split-node quiver is displayed in Figure~\ref{fig:quiver_split_C3Z2}. The gauge symmetry is $U(N_1)\times U(N_2)$ and the global symmetry in the free limit is $U(2) \times U(2)$. The split-node quiver leads to counting (\ref{eq:counting_c3z2}). The restricted characters $\chi_{\C3Z2}(\bL,\bsig)$ that give an explicit implementation of the counting  are shown in (\ref{eq:chiL_C3Z2}). Combining with the operators $\cO_{\C3Z2}(\bn,\bsig)$ shown in (\ref{eq:Osig_C3Z2}) we get the basis $\cO_{\C3Z2}(\bL)$ (\ref{eq:OL_defn}). The labels are
\begin{equation}
\bL = \{ R_1,R_2,r_{11},r_{22},r_{12;1},r_{12;2},r_{21;1},r_{21;2},\nu_1^\pm,\nu_2^\pm \}
\end{equation}

The covariant basis $\cO_{\C3Z2}(\bK)$ is built with covariant characters shown in (\ref{eq:chiK_diag}). 

\subsubsection{\texorpdfstring{$dP_0$}{dP0}}

\begin{figure}[h]
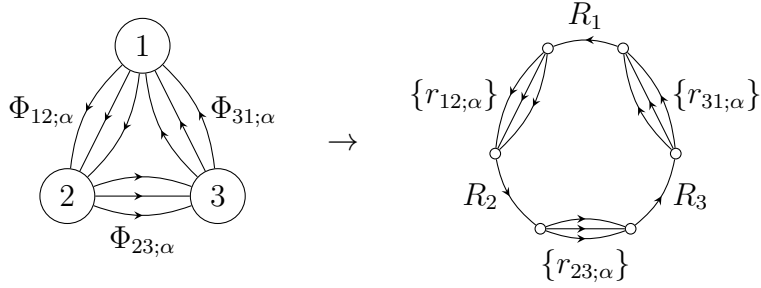

\centering
$$
	\mytikz{
		\node (s1) at (0,0) [circle,draw] {$1$};				
		\node (s2) at (-1,-2) [circle,draw] {$2$};
		\node (s3) at (1,-2) [circle,draw] {$3$};
		\draw [postaction={decorate}] (s1) to [bend right=20] node[left]{$\Phi_{12;\a}$} (s2);
		\draw [postaction={decorate}] (s1) to [bend right=0] (s2);
		\draw [postaction={decorate}] (s1) to [bend right=-20] (s2);
		\draw [postaction={decorate}] (s2) to [bend right=20] node[below]{$\Phi_{23;\a}$} (s3);
		\draw [postaction={decorate}] (s2) to [bend right=0] (s3);
		\draw [postaction={decorate}] (s2) to [bend right=-20] (s3);
		\draw [postaction={decorate}] (s3) to [bend right=20] node[right]{$\Phi_{31;\a}$} (s1);
		\draw [postaction={decorate}] (s3) to [bend right=0] (s1);
		\draw [postaction={decorate}] (s3) to [bend right=-20] (s1);
	}	
\quad \rightarrow \quad
	\mytikz{				
		\node (n1) at (-0.5,-1.0) [circle,draw,inner sep=0.5mm] {};
		\node (m2) at (-1.2,-2.4) [circle,draw,inner sep=0.5mm] {};
		\node (n2) at (-0.6,-3.4) [circle,draw,inner sep=0.5mm] {};
		\node (m3) at (0.6,-3.4) [circle,draw,inner sep=0.5mm] {};
		\node (n3) at (1.2,-2.4) [circle,draw,inner sep=0.5mm] {};				
		\node (m1) at (0.5,-1.0) [circle,draw,inner sep=0.5mm] {};
		\draw [postaction={decorate}] (m1) to [bend right=20] node[above]{$R_1$} (n1);		
		\draw [postaction={decorate}] (n1) to [bend right=20] node[left]{$\{r_{12;\a}\}$} (m2);
		\draw [postaction={decorate}] (n1) to [bend right=0] (m2);		
		\draw [postaction={decorate}] (n1) to [bend right=-20] (m2);		
		\draw [postaction={decorate}] (m2) to [bend right=20] node[left]{$R_2$} (n2);
		\draw [postaction={decorate}] (n2) to [bend right=20] node[below]{$\{r_{23;\a}\}$} (m3);
		\draw [postaction={decorate}] (n2) to [bend right=0] (m3);		
		\draw [postaction={decorate}] (n2) to [bend right=-20] (m3);
		\draw [postaction={decorate}] (m3) to [bend right=20] node[right]{$R_3$} (n3);
		\draw [postaction={decorate}] (n3) to [bend right=20] node[right]{$\{r_{31;\a}\}$} (m1);
		\draw [postaction={decorate}] (n3) to [bend right=0] (m1);		
		\draw [postaction={decorate}] (n3) to [bend right=-20] (m1);						
	}	
$$
\caption{Quiver for $dP_0$ theory, and the split-node quiver for operator counting.}
\label{fig:quiverdP0}
\end{figure}

The theory of $D3$ branes on $\mC^3/\mZ^3$ singularity \cite{Douglas:1997de}, also known as $dP_0$, has a quiver shown in Figure~\ref{fig:quiverdP0}.
The gauge group is $U(N_1)\times U(N_2) \times U(N_3)$, and we have a total of 9 bifundamental chiral multiplets
\begin{equation}
	\{ \Phi_{12;\a}, \Phi_{23;\a}, \Phi_{31;\a} \}, \quad \alpha \in \{1,2,3\}
\end{equation}
There is a global flavor symmetry group $U(3)\times U(3) \times U(3)$. The counting of finite-$N$ gauge invariant operators following (\ref{eq:N_nabalph}) is given by the labelled split-node quiver, also in Figure~\ref{fig:quiverdP0}:
\begin{equation}
\begin{split}
&\cN_{dP_0}(\{n_{ab;\a}\};N_1,N_2,N_3) = 
\sum_{\substack{R_1 \vdash n \\ l(R_1)\le N_1}}
\sum_{\substack{R_2 \vdash n \\ l(R_1)\le N_1}}
\sum_{\substack{R_3 \vdash n \\ l(R_1)\le N_1}}
\sum_{\{ r_{12;\a} \}} \sum_{\{ r_{23;\a} \}} \sum_{\{ r_{31;\a} \}} 
\\ & 
g(\{r_{31;\a}\};R_1) \, g(\{r_{12;\a}\};R_1) \,
g(\{r_{12;\a}\};R_2) \, g(\{r_{23;\a}\};R_2) \,
g(\{r_{23;\a}\};R_3) \, g(\{r_{31;\a}\};R_3)
\end{split}
\end{equation}

The gauge invariant mesonic operators are traces of products going around the quiver $\tr(\Phi_{12;\alpha_1} \Phi_{23;\alpha_2} \Phi_{31;\alpha_3} \Phi_{12;\alpha_4} \ldots)$. According to the general prescription (\ref{eq:OQ_sig_defn}), a general gauge invariant operator can be specified by charges and three permutations
\begin{equation}
	\cO_{\dP0}(\bn, \{\sigma_1,\sigma_2,\sigma_3\}) = 
	\trVN{n}\left(
		\sigma_1 \, (\Phi_{12;\a})^{\otimes \{n_{12;\a}\} }
		\, \sigma_2 \, (\Phi_{23;\a})^{\otimes \{n_{23;\a}\} }
		\, \sigma_3 \, (\Phi_{31;\a})^{\otimes \{n_{31;\a}\} }
	\right)
\end{equation}
Here $n=\sum_{\alpha} n_{12;\a}=\sum_{\alpha} n_{23;\a}=\sum_{\alpha} n_{31;\a}$ is the total number of $\Phi_{12;\a}$'s or $\Phi_{23;\a}$'s or $\Phi_{31;\a}$'s . Since the $\dP0$ quiver is ``linear'', without any branchings like in $\C3Z2$, we can think of the operators $\cO_{\dP0}(\bn,\bsig)$ as traces in $V_N^{\otimes n}$.

Restricted Schur basis operators (\ref{eq:OL_defn}) are:
\begin{equation}
	\cO_{\dP0}(\bL) = c_{\bL} \sum_{\sigma_1,\sigma_2,\sigma_3}	
	\chi_{\dP0}(\bL,\{\sigma_1,\sigma_2,\sigma_3\}) \,
	\cO_{\dP0}(\bn, \{\sigma_1,\sigma_2,\sigma_3\})
\end{equation}
with the restricted quiver character as a further decorated split-node quiver:
\begin{equation}
	\chi_{\dP0}(\bL,\{\sigma_1,\sigma_2,\sigma_3\}) =
	\mytikz{		
		\node (s1) at (0,0) [rectangle,draw] {$\sigma_1$};				
		\node (n1) at (-0.5,-1.0) [circle,draw,inner sep=0.5mm,label=right:$\nu^-_1$] {};
		\node (m2) at (-1.2,-2.4) [circle,draw,inner sep=0.5mm,label=right:$\nu^+_2$] {};
		\node (s2) at (-1.7,-3.4) [rectangle,draw] {$\sigma_2$};
		\node (n2) at (-0.6,-3.4) [circle,draw,inner sep=0.5mm,label=above:$\nu^-_2$] {};
		\node (m3) at (0.6,-3.4) [circle,draw,inner sep=0.5mm,label=above:$\nu^+_3$] {};
		\node (s3) at (1.7,-3.4) [rectangle,draw] {$\sigma_3$};
		\node (n3) at (1.2,-2.4) [circle,draw,inner sep=0.5mm,label=left:$\nu^-_3$] {};				
		\node (m1) at (0.5,-1.0) [circle,draw,inner sep=0.5mm,label=right:$\nu^+_1$] {};
		\draw [postaction={decorate}] (m1) to (s1);
		\draw [postaction={decorate}] (s1) to node[left]{$R_1$} (n1);		
		\draw [postaction={decorate}] (n1) to [bend right=20] node[left]{$\{r_{12;\a}\}$} (m2);
		\draw [postaction={decorate}] (n1) to [bend right=0] (m2);		
		\draw [postaction={decorate}] (n1) to [bend right=-20] (m2);		
		\draw [postaction={decorate}] (m2) to (s2);
		\draw [postaction={decorate}] (s2) to node[below]{$R_2$} (n2);
		\draw [postaction={decorate}] (n2) to [bend right=20] node[below]{$\{r_{23;\a}\}$} (m3);
		\draw [postaction={decorate}] (n2) to [bend right=0] (m3);		
		\draw [postaction={decorate}] (n2) to [bend right=-20] (m3);
		\draw [postaction={decorate}] (m3) to (s3);
		\draw [postaction={decorate}] (s3) to node[right]{$R_3$} (n3);
		\draw [postaction={decorate}] (n3) to [bend right=20] node[right]{$\{r_{31;\a}\}$} (m1);
		\draw [postaction={decorate}] (n3) to [bend right=0] (m1);		
		\draw [postaction={decorate}] (n3) to [bend right=-20] (m1);						
	}	
\end{equation}
The labels are
\begin{equation}
	\bL = \{ R_1, R_2, R_3, r_{12;\a}, r_{23;\a}, r_{31;\a}, \nu_1^\pm, \nu_2^\pm, \nu_3^\pm \}
\end{equation}

Covariant basis operators (\ref{eq:OK_defn}) for $\dP0$ are
\begin{equation}
	\cO_{\dP0}(\bK) = \frac{\sqrt{d(R_1)d(R_2)d(R_3)}}{(n!)^3} \sum_{\sigma_1,\sigma_2,\sigma_3}
	\chi_{\dP0}(\bK,\{\sigma_1,\sigma_2,\sigma_3\}) \cO_{\dP0}(\bn,\{\sigma_1,\sigma_2,\sigma_3\})
\end{equation}
\begin{equation}
\label{eq:chiK_diag_dP0}
	\chi_{\dP0}(\bK,\{\sigma_1,\sigma_2,\sigma_3\}) = 
	\mytikz{	
		\node (s1) at (0,0) [rectangle,draw] {$\sigma_1$};		
		\node (s2) at (-1.5,-3.0) [rectangle,draw] {$\sigma_2$};		
		\node (s3) at (1.5,-3.0) [rectangle,draw] {$\sigma_3$};		
		\node (t12) at (-0.75,-1.5) [circle,fill,inner sep=0.5mm,label=right:$\tau_{12}$] {};
		\node (t31) at (0.75,-1.5) [circle,fill,inner sep=0.5mm,label=left:$\tau_{31}$] {};
		\node (t23) at (0,-3.0) [circle,fill,inner sep=0.5mm,label=above:$\tau_{23}$] {};		
		\node (b12) at (-1.55,-1.1) [circle,draw,inner sep=0.5mm,label=below:$\beta_{12}$] {};		
		\node (b31) at (1.55,-1.1) [circle,draw,inner sep=0.5mm,label=right:$\beta_{31}$] {};
		\node (b23) at (0,-3.7) [circle,draw,inner sep=0.5mm,label=left:$\beta_{23}$] {};
		\node (n12) at (-2.35,-0.7)  {};			
		\node (n31) at (2.35,-0.7)  {};	
		\node (n23) at (0,-4.4)  {};	
		\draw [postaction={decorate}] (s1) to node[above]{$R_1$} (t12);
		\draw [postaction={decorate}] (t12) to node[left]{$R_2$} (s2);
		\draw [postaction={decorate}] (s2) to node[above]{$R_2$} (t23);
		\draw [postaction={decorate}] (t23) to node[above]{$R_3$} (s3);		
		\draw [postaction={decorate}] (s3) to node[right]{$R_3$} (t31);		
		\draw [postaction={decorate}] (t31) to node[above]{$R_1$} (s1);		
		\draw [-] (t12) to node[above]{$\Lambda_{12}$} (b12);
		\draw [-] (b12) to node[above]{$[n_{12;\a}]$} (n12);
		\draw [-] (t31) to node[above]{$\Lambda_{31}$} (b31);
		\draw [-] (b31) to node[above]{$[n_{31;\a}]$} (n31);
		\draw [-] (t23) to node[right]{$\Lambda_{23}$} (b23);
		\draw [-] (b23) to node[right]{$[n_{23;\a}]$} (n23);
	}
\end{equation}
with the labels
\begin{equation}
\begin{split}
	\bK &= \{ R_1, R_2, R_3, \Lambda_{12}, \Lambda_{23}, \Lambda_{31}, \tau_{ab}, n_{ab;\a}, \beta_{ab} \}	
\end{split}
\end{equation}
That is, an operator $U(M)^3$ multiplet is defined by the global symmetry irrep $(\Lambda_{12}, \Lambda_{23}, \Lambda_{31})$, the diagrams $R_1,R_2,R_3 \vdash n$ for each gauge group factor and 3 multiplicity labels $\tau_{ab}$ for Clebsch-Gordan decompositions
\begin{equation}
\begin{split}
	R_1 \otimes R_2 &\rightarrow \Lambda_{12} \\
	R_2 \otimes R_3 &\rightarrow \Lambda_{23} \\
	R_3 \otimes R_1 &\rightarrow \Lambda_{31}
\end{split}
\end{equation}

\subsubsection{\texorpdfstring{$\mC^2/\mZ_n \times \mC$}{C2/Zn x C}}

As a final example let us take the quiver of the $\mC^2/\mZ_n \times \mC$ theory \cite{DM96}, Figure~\ref{fig:quiverC3Zn}. In $\cN=1$ language it is a circular quiver with $n$ nodes and fields $\Phi_{a,a+1}, \Phi_{a,a-1}, \Phi_{a,a}$.
\begin{figure}[h]
\centering
\mytikz{
		\node (s1) at (0,1.5) [circle,draw] {$1$};
		\draw [postaction={decorate}] (s1.150) .. controls +(150:0.5) and +(180:0.6) .. (0,2.5) node[above]{$\Phi_{11}$} .. controls +(0:0.6) and +(30:0.5) .. (s1.30);
	\begin{scope}[rotate=-72]
		\node (s2) at (0,1.5) [circle,draw] {$2$};
		\draw [postaction={decorate}] (s2.78) .. controls +(150:0.5) and +(180:0.6) .. (0,2.5) node[right]{$\Phi_{22}$} .. controls +(0:0.6) and +(30:0.5) .. (s2.-42);
	\end{scope}	
	\begin{scope}[rotate=-144]
		\node (s3) at (0,1.5) [circle,draw] {$3$};
		\draw [postaction={decorate}] (s3.6) .. controls +(150:0.5) and +(180:0.6) .. (0,2.5) node[right]{$\Phi_{33}$} .. controls +(0:0.6) and +(30:0.5) .. (s3.-114);
	\end{scope}	
	\begin{scope}[rotate=-216]
		\node (s4) at (0,1.5) {$\cdots$};		
	\end{scope}	
	\begin{scope}[rotate=72]
		\node (s5) at (0,1.5) [circle,draw] {$n$};
		\draw [postaction={decorate}] (s5.222) .. controls +(150:0.5) and +(180:0.6) .. (0,2.5) node[left]{$\Phi_{nn}$} .. controls +(0:0.6) and +(30:0.5) .. (s5.102);
	\end{scope}	
	\draw [postaction={decorate}] (s1) to [bend left=10] node[above]{$\Phi_{12}$} (s2);	
	\draw [postaction={decorate}] (s2) to [bend left=10] node[below]{$\Phi_{21}$} (s1);	
	\draw [postaction={decorate}] (s2) to [bend left=10] node[right]{$\Phi_{23}$} (s3);	
	\draw [postaction={decorate}] (s3) to [bend left=10] node[left]{$\Phi_{32}$} (s2);	
	\draw [postaction={decorate}] (s3) to [bend left=10] node[below]{} (s4);	
	\draw [postaction={decorate}] (s4) to [bend left=10] node[above]{} (s3);
	\draw [postaction={decorate}] (s4) to [bend left=10] node[left]{} (s5);	
	\draw [postaction={decorate}] (s5) to [bend left=10] node[right]{} (s4);	
	\draw [postaction={decorate}] (s5) to [bend left=10] node[above]{$\Phi_{n1}$} (s1);	
	\draw [postaction={decorate}] (s1) to [bend left=10] node[below]{$\Phi_{1n}$} (s5);		
}
\caption{$\mC^2/\mZ_n \times \mC$ quiver}
\label{fig:quiverC3Zn}
\end{figure}

The corresponding split-node quiver is shown in Figure~\ref{fig:quiverC3Zn_split}.
\begin{figure}[h]
\centering
\mytikz{
	\def\y{1}
	\def\x{2}
	\def\xx{4}
	\node (n0m) at (-\xx,\y) [circle,draw,inner sep=0.5mm,label=left:$\cdots$] {};
	\node (n0p) at (-\xx,-\y) [circle,draw,inner sep=0.5mm,label=left:$\cdots$] {};	
	\node (n1p) at (-\x,\y) [circle,draw,inner sep=0.5mm,label=above:$\,$] {};
	\node (n1m) at (-\x,-\y) [circle,draw,inner sep=0.5mm,label=below:$\,$] {};		
	\node (n2m) at (0,\y) [circle,draw,inner sep=0.5mm,label=above:$\,$] {};
	\node (n2p) at (0,-\y) [circle,draw,inner sep=0.5mm,label=below:$\,$] {};
	\node (n3p) at (\x,\y) [circle,draw,inner sep=0.5mm,label=above:$\,$] {};
	\node (n3m) at (\x,-\y) [circle,draw,inner sep=0.5mm,label=below:$\,$] {};		
	\node (n4m) at (\xx,\y) [circle,draw,inner sep=0.5mm,label=right:$\cdots$] {};
	\node (n4p) at (\xx,-\y) [circle,draw,inner sep=0.5mm,label=right:$\cdots$] {};		
	\draw [postaction={decorate}] (n1p) to node[left]{$R_1$} (n1m);
	\draw [postaction={decorate}] (n2p) to node[left]{$R_2$} (n2m);
	\draw [postaction={decorate}] (n3p) to node[left]{$R_3$} (n3m);		
	\draw [postaction={decorate}] (n0m) to node[above]{$r_{n1}$} (n1p);
	\draw [postaction={decorate}] (n1m) to node[below]{$r_{12}$} (n2p);
	\draw [postaction={decorate}] (n1m) to node[below]{$r_{1n}$} (n0p);
	\draw [postaction={decorate}] (n2m) to node[above]{$r_{21}$} (n1p);
	\draw [postaction={decorate}] (n2m) to node[above]{$r_{23}$} (n3p);
	\draw [postaction={decorate}] (n3m) to node[below]{$r_{32}$} (n2p);
	\draw [postaction={decorate}] (n3m) to node[below]{$r_{34}$} (n4p);			
	\draw [postaction={decorate}] (n4m) to node[above]{$r_{43}$} (n3p);	
	\draw [postaction={decorate}] (n1m) to [bend right=45] node[right]{$r_{11}$} (n1p);	
	\draw [postaction={decorate}] (n2m) to [bend left=45] node[right]{$r_{22}$} (n2p);
	\draw [postaction={decorate}] (n3m) to [bend right=45] node[right]{$r_{33}$} (n3p);		
	}
\caption{Split-node quiver for $\mC^2/\mZ_n \times \mC$}
\label{fig:quiverC3Zn_split}
\end{figure}
This leads to finite-$N$ counting of operators
\begin{equation}
\begin{split}
\cN_{\mC^2/\mZ_n\times \mC}(\{n_{ab}\},\{ N_a \}) &=
\sum_{\substack{\{ R_a \vdash n_a \} \\ l(R_a) \le N_a }}
\sum_{ \{ r_{a,a+1} \} } \sum_{ \{ r_{a,a-1} \} } \sum_{ \{ r_{a,a} \} }
\\ &
\prod_a 
	g(r_{a,a},r_{a,a-1},r_{a,a+1};R_a) \;
	g(r_{a,a},r_{a-1,a},r_{a+1,a};R_a)
\end{split}	
\end{equation}

The restricted Schur basis $\cO_Q(\bL)$ can be constructed by writing down quiver characters according to the split-node quiver, with the multiplicity labels $\nu_a^\pm$.

%%%%%%%%%%%%%%%%%%%%%%%%%%%%%%%%%%%%%%%%%%%%%%%%%%%%%%%%%%%%%%%%%%%%%
%%%%%%%%%%%%%%%%%%%%%%%%%%%%%%%%%%%%%%%%%%%%%%%%%%%%%%%%%%%%%%%%%%%%%

\section{Chiral ring structure constants}
\label{sec:chir-ring}

In this section we obtain general expressions for the chiral ring structure constants. In Section~\ref{sec:restschurfus} we work out the result for the restricted Schur basis
\begin{equation}
\cO_Q(\bL^{(1)}) \cO_Q(\bL^{(2)}) \equiv \sum_{\bL^{(3)}} G ( \bL^{(1)} , \bL^{(2)} ; \bL^{(3) } )  \cO_Q ( \bL^{(3)} )
\end{equation}
and in Section~\ref{sec:covariantfus} we deal with the covariant basis
\begin{equation}
\cO_Q(\bK^{(1)}) \cO_Q(\bK^{(2)}) \equiv \sum_{\bK^{(3)}} G ( \bK^{(1)} , \bK^{(2)} ; \bK^{(3) } )  \cO_Q ( \bK^{(3)} )
\end{equation}
We find that $G ( \bL^{(1)} , \bL^{(2)} ; \bL^{(3) } )$ and $G ( \bK^{(1)} , \bK^{(2)} ; \bK^{(3) } )$ can be written diagrammatically. They involve two types of 
vertices --- solid vertices for inner products; and white vertices for outer products. 

The main result is that all Young diagram labels combine according to the Littlewood-Richardson rule. For the restricted Schur basis the resulting diagram (\ref{eq:GLLL_diag}) involves the branching coefficients for $R_a^{(3)} \rightarrow (R_a^{(1)}, R_a^{(2)})$ and $r_{ab;\a}^{(3)} \rightarrow (r_{ab;\a}^{(1)}, r_{ab;\a}^{(2)})$ 
\begin{equation}
G ( \bL^{(1)} , \bL^{(2)} ; \bL^{(3) } ) 
\; \sim \;
\prod_a
\mytikz{
	\node (n) at (0,0) [circle,draw,inner sep=0.5mm] {};
	\draw [postaction={decorate}] (0,1) to node[right]{$R_a^{(3)}$} (n);
	\draw [postaction={decorate}] (n) to node[left]{$R_a^{(1)}$} (-0.5,-1);
	\draw [postaction={decorate}] (n) to node[right]{$R_a^{(2)}$} (0.5,-1);
}
\;
\prod_{a,b,\a}
\mytikz{
	\node (n) at (0,0) [circle,draw,inner sep=0.5mm] {};
	\draw [postaction={decorate}] (0,1) to node[right]{$r_{ab;\a}^{(3)}$} (n);
	\draw [postaction={decorate}] (n) to node[left]{$r_{ab;\a}^{(1)}$} (-0.5,-1);
	\draw [postaction={decorate}] (n) to node[right]{$r_{ab;\a}^{(2)}$} (0.5,-1);
}
\end{equation}
and so the chiral ring structure constant vanishes unless the Littlewood-Richardson coefficients $g ( R_a^{(1)}  , R_a^{(2)}  ; R_{a}^{(3)})$ and $g ( r_{ab;\a}^{(1)}  , r_{ab;\a}^{(2)}  ; r_{ab;\a}^{(3)}  )$ are all non-zero.

Analogously, the covariant basis structure constants (\ref{eq:GKKK_diag}) involve the branching coefficients for $R_a^{(3)} \rightarrow (R_a^{(1)}, R_a^{(2)})$, $\L_{ab}^{(3)} \rightarrow (\L_{ab}^{(1)}, \L_{ab}^{(2)})$ and $s_{ab}^{(3)\pm} \rightarrow (s_{ab}^{(1)\pm}, s_{ab}^{(2)\pm})$ 
\begin{equation}
G ( \bK^{(1)} , \bK^{(2)} ; \bK^{(3) } ) 
\; \sim \;
\prod_a
\mytikz{
	\node (n) at (0,0) [circle,draw,inner sep=0.5mm] {};
	\draw [postaction={decorate}] (0,1) to node[right]{$R_a^{(3)}$} (n);
	\draw [postaction={decorate}] (n) to node[left]{$R_a^{(1)}$} (-0.5,-1);
	\draw [postaction={decorate}] (n) to node[right]{$R_a^{(2)}$} (0.5,-1);
}
\;
\prod_{a,b}
\mytikz{
	\node (n) at (0,0) [circle,draw,inner sep=0.5mm] {};
	\draw [postaction={decorate}] (0,1) to node[right]{$\L_{ab}^{(3)}$} (n);
	\draw [postaction={decorate}] (n) to node[left]{$\L_{ab}^{(1)}$} (-0.5,-1);
	\draw [postaction={decorate}] (n) to node[right]{$\L_{ab}^{(2)}$} (0.5,-1);
}
\mytikz{
	\node (n) at (0,0) [circle,draw,inner sep=0.5mm] {};
	\draw [postaction={decorate}] (0,1) to node[right]{$s_{ab}^{(3)+}$} (n);
	\draw [postaction={decorate}] (n) to node[left]{$s_{ab}^{(1)+}$} (-0.5,-1);
	\draw [postaction={decorate}] (n) to node[right]{$s_{ab}^{(2)+}$} (0.5,-1);
}
\mytikz{
	\node (n) at (0,0) [circle,draw,inner sep=0.5mm] {};
	\draw [postaction={decorate}] (0,1) to node[right]{$s_{ab}^{(3)-}$} (n);
	\draw [postaction={decorate}] (n) to node[left]{$s_{ab}^{(1)-}$} (-0.5,-1);
	\draw [postaction={decorate}] (n) to node[right]{$s_{ab}^{(2)-}$} (0.5,-1);
}
\end{equation}
and thus vanish unless $g ( R_a^{(1)}  , R_a^{(2)}  ; R_{a}^{(3)})$,  $g ( \L_{ab}^{(1)}  , \L_{ab}^{(2)}  ; \L_{ab}^{(3)}  )$, $g ( s_{ab}^{(1)\pm}  , s_{ab}^{(2)\pm}  ; s_{ab}^{(3)\pm}  )$ are all non-zero.

Note that, if we consider correlators of $n$ holomorphic 
operators and one anti-holomorphic, the coefficient would involve the appropriate Littlewood-Richardson coefficient for the branching $R_a^{(n+1)} \rightarrow (R_a^{(1)},R_a^{(2)}, \ldots, R_a^{(n)})$ and so on for other labels. We leave it as an exercise for the reader to write out the 
explicit formulae for that case, following the analogous expressions we present for $n=2$, i.e two holomorphic operators fusing into one. 

%%%%%%%%%%%%%%%%%%%%%%%%%%%%%%%%%%%%%%%%%%%%%%%%%%%%%%%%%%%%%%%%%%%%%

\subsection{Restricted Schur basis}\label{sec:restschurfus}

Consider the product of operators (\ref{eq:OL_defn})
\begin{equation}\label{prodops} 
\begin{split}
& \cO_{Q} ( \bL^{(1)}  ) \cO_{Q} ( \bL^{(2)}  )  \\
& =  
\frac{1}{\prod n_a^{(1)}!} \frac{1}{\prod n_a^{(2)}!}
\sum_{   \bsig^{(1)} }  \sum_{ \bsig^{(2)} }  \hat\chi_Q (\bL^{(1)},  \bsig^{ (1) }  )  \hat\chi_Q ( \bL^{(2)} ,  \bsig^{(2)}   ) \cO_{Q} (  \bsig^{(1)}  ) \cO_Q (   \bsig^{(2)}    )    \\ 
& =  \frac{1}{\prod n_a^{(1)}!} \frac{1}{\prod n_a^{(2)}!} \sum_{   \bsig^{(1)} }  \sum_{ \bsig^{(2)} }   \hat\chi_Q (\bL^{(1)},  \bsig^{ (1) }  )  \hat\chi_Q ( \bL^{(2)} ,  \bsig^{(2)}   ) \cO_{Q} (  \bsig^{(1)}  \circ  \bsig^{(2)}  ) 
\end{split}
\end{equation}
Here we use a conveniently normalized quiver character
\begin{equation}
	\hat\chi_Q(\bL,\bsig) \equiv \sqrt{\frac{\prod d(R_a)}{\prod d(r_{ab;\a})}} \chi_Q(\bL,\bsig)
\end{equation}
The outer product    $ \bsig^{(1)}  \circ  \bsig^{(2)} $  consists of pairs of  permutations $      \sigma_{ a }^{(1)} \circ    \sigma_{ a }^{(2)} $ 
in $  S_{n_a^{(1)} } \times S_{n_a^{(2)} } \subset S_{n_a^{(1)} + n_a^{(2)} } $. 
We can expand the permutation-basis operators as a sum of Fourier basis operators using (\ref{eq:OL_inverse}) to get 
\begin{equation}
\begin{split}
& \cO_{Q} ( \bL^{(1)} ) \cO_{Q} ( \bL^{(2)} )  \\ 
& =   \frac{1}{\prod n_a^{(1)}!} \frac{1}{\prod n_a^{(2)}!} \sum_{   \bsig^{(1)} }  \sum_{ \bsig^{(2)} }   \sum_{ \bL^{(3) } } \hat\chi_Q (\bL^{(1)},  \bsig^{ (1) }  )  \hat\chi_Q ( \bL^{(2)} ,  \bsig^{(2)}   ) \hat\chi_Q ( \bL^{(3)}  ,  \bsig^{(1)}  \circ  \bsig^{(2)}   )  \cO_Q ( \bL^{(3)} )
\\
& \equiv \sum_{\bL^{(3)}} G ( \bL^{(1)} , \bL^{(1)} ; \bL^{(3) } )  \cO_Q ( \bL^{(3)} )
\end{split}
\end{equation}
The sum $\bL^{(3)}$ runs over labels with $\bn^{(3)} = \bn^{(1)}+\bn^{(2)}$. This leads to the expression for the chiral ring structure constants 
\begin{equation}\begin{split}\label{formula-chiring} 
& G ( \bL^{(1)} , \bL^{(1)} ; \bL^{(3) } ) = \frac{1}{\prod n_a^{(1)}!} \frac{1}{\prod n_a^{(2)}!} \sum_{   \bsig^{(1)} }  \sum_{ \bsig^{(2)} } \hat\chi_Q (\bL^{(1)},  \bsig^{ (1) }  )  \hat\chi_Q ( \bL^{(2)} ,  \bsig^{(2)}   ) \hat\chi_Q ( \bL^{(3)}  ,  \bsig^{(1)}  \circ  \bsig^{(2)}   )
\end{split}\end{equation}
which can be evaluating by doing the sums over $ \bsig^{(1)} , \bsig^{(2)} $. 

Let us first describe the answer and then sketch the steps in the derivation. The final result is, diagrammatically:
\begin{equation}
\label{eq:GLLL_diag}
\begin{split}
	&G(\bL^{(1)},\bL^{(2)};\bL^{(3)}) = 
		f_{\bL^{(1)}\bL^{(2)}}^{\bL^{(3)}}
	\sum_{\{\mu_a\}} \sum_{\{\mu_{ab;\a}\}} 
\\
	& \prod_a \left(
	\mytikz{		
		\node (np1) at (-1.5,0) [circle,draw,inner sep=0.5mm,label=left:$\nu_a^{(1)-}$] {};
		\node (np2) at (0,0) [circle,draw,inner sep=0.5mm,label=right:$\nu_a^{(2)-}$] {};
		\node (np3) at (1.5,0) [circle,draw,inner sep=0.5mm,label=right:$\nu_a^{(3)-}$] {};
		\node (n) at (0,1.5) [circle,draw,inner sep=0.5mm,label=above:$\mu_a$] {};
		\node (n1) at (-1,-1.5) [circle,draw,inner sep=0.5mm] {};
		\node (n2) at (0,-1.5) [circle,draw,inner sep=0.5mm,label=below:$\bigcup_{b,\alpha} \mu_{ab;\a}$] {};
		\node (n3) at (1,-1.5) [circle,draw,inner sep=0.5mm] {};
		\draw [postaction={decorate}] (n) to node[above left]{$R_a^{(1)}$} (np1);		
		\draw [postaction={decorate}] (n) to node{$R_a^{(2)}$} (np2);		
		\draw [postaction={decorate}] (np3) to node[above right]{$R_a^{(3)}$} (n);		
		\draw [postaction={decorate}] (np1) to node[left]{$\bigcup_{b,\alpha} r_{ab;\a}^{(1)}$} (n1);		
		\draw [postaction={decorate}] (np1) to (n2);		
		\draw [postaction={decorate}] (np1) to (n3);		
		\draw [postaction={decorate}] (np2) to (n1);		
		\draw [postaction={decorate}] (np2) to (n2);		
		\draw [postaction={decorate}] (np2) to (n3);		
		\draw [postaction={decorate}] (n1) to (np3);		
		\draw [postaction={decorate}] (n2) to (np3);		
		\draw [postaction={decorate}] (n3) to node[right]{$\bigcup_{b,\alpha} r_{ab;\a}^{(3)}$} (np3);		
	}	
	\mytikz{		
		\node (np1) at (-1.5,0) [circle,draw,inner sep=0.5mm,label=left:$\nu_a^{(1)+}$] {};
		\node (np2) at (0,0) [circle,draw,inner sep=0.5mm,label=right:$\nu_a^{(2)+}$] {};
		\node (np3) at (1.5,0) [circle,draw,inner sep=0.5mm,label=right:$\nu_a^{(3)+}$] {};
		\node (n) at (0,-1.5) [circle,draw,inner sep=0.5mm,label=below:$\mu_a$] {};
		\node (n1) at (-1,1.5) [circle,draw,inner sep=0.5mm] {};
		\node (n2) at (0,1.5) [circle,draw,inner sep=0.5mm,label=above:$\bigcup_{b,\alpha}\mu_{ba;\a}$] {};
		\node (n3) at (1,1.5) [circle,draw,inner sep=0.5mm] {};
		\draw [postaction={decorate}] (np1) to node[below left]{$R_a^{(1)}$} (n);		
		\draw [postaction={decorate}] (np2) to node{$R_a^{(2)}$} (n);		
		\draw [postaction={decorate}] (n) to node[below right]{$R_a^{(3)}$} (np3);		
		\draw [postaction={decorate}] (n1) to node[left]{$\bigcup_{b,\alpha} r_{ba;\a}^{(1)}$} (np1);		
		\draw [postaction={decorate}] (n2) to (np1);		
		\draw [postaction={decorate}] (n3) to (np1);		
		\draw [postaction={decorate}] (n1) to (np2);		
		\draw [postaction={decorate}] (n2) to (np2);		
		\draw [postaction={decorate}] (n3) to (np2);		
		\draw [postaction={decorate}] (np3) to (n1);		
		\draw [postaction={decorate}] (np3) to (n2);		
		\draw [postaction={decorate}] (np3) to node[right]{$\bigcup_{b,\alpha} r_{ba;\a}^{(3)}$} (n3);		
	}			
	\right)
\end{split}
\end{equation}
with the constant prefactor
\begin{equation}
	f_{\bL^{(1)}\bL^{(2)}}^{\bL^{(3)}} = 
	\sqrt{\frac{\prod_a d(R^{(1)}_a) d(R^{(2)}_a) d(R^{(3)}_a)}{\prod_{a,b,\alpha} d(r_{ab;\a}^{(1)}) d(r_{ab;\a}^{(2)}) d(r_{ab;\a}^{(3)})}}
	\frac{1}{\prod_a d(R^{(1)}_a) d(R^{(2)}_a)}
	\frac{1}{\prod_{a,b,\alpha} d(r_{ab;\a}^{(1)}) d(r_{ab;\a}^{(2)})}
\end{equation}
For illustration purposes in (\ref{eq:GLLL_diag}) we draw three outgoing arrows $r_{ab;\alpha}$ from each branching node $\nu_a^-$ and three incoming arrows $r_{ba;\alpha}$ to each branching node $\nu_a^+$. The precise structure depends on the quiver (on the other hand, the lines and nodes labelled by $^{(1),(2),(3)}$ are always three, associated with the three operators). 

The explicit expression corresponding to (\ref{eq:GLLL_diag}) is
\begin{equation}
\label{eq:GLLL_result}
\begin{split}
 G ( \bL^{(1)} , \bL^{(2)} ; \bL^{(3)} ) &= 
	f_{\bL^{(1)}\bL^{(2)}}^{\bL^{(3)}}
	\sum_{\{\mu_a\}} \sum_{\{\mu_{ab;\a}\}} 
\\ &	
\prod_a
\cF\left( \cup_I R_a^{(I)} , \{ \cup_{I, b,\a} r_{ab;\a}^{(I)} \} , \cup_I \nu_a^{(I)-} ; \mu_a, \{ \cup_{b,\a} \mu_{ab;\a} \} \right) \; \\ 
& \times
\cF\left(\cup_I  R_a^{(I)} , \{ \cup_{I, b,\a} r_{ba;\a}^{(I)} \} , \cup_I \nu_a^{(I)+} ; \mu_a, \{ \cup_{b,\a} \mu_{ba;\a} \} \right)
\end{split}
\end{equation}
with the object $\cF$ equal to the single connected piece in (\ref{eq:GLLL_diag}):
\begin{equation}
\label{eq:Frrr_defn}
\begin{split}
	&\cF\left( \cup_I R^{(I)} , \{ \cup_{I, \alpha} r_{\a}^{(I)} \} , \cup_I \nu^{(I)} ; \mu, \{ \cup_{\a} \mu_{\a} \} \right)
\\ 
&=	
	B^{R^{(1)} \rightarrow \cup_{\a} r_{\a}^{(1)}; \nu^{(1)+}}_{~  i^{(1)} \rightarrow \cup_{\a} l_{\a}^{(1)}  }      	
	B^{R^{(2)} \rightarrow \cup_{\a} r_{\a}^{(2)}; \nu^{(2)+}}_{~  i^{(2)} \rightarrow \cup_{\a} l_{\a}^{(2)}  }      
	B^{R^{(3)} \rightarrow \cup_{\a} r_{\a}^{(3)}; \nu^{(3)+}}_{~  i^{(3)} \rightarrow \cup_{\a} l_{\a}^{(3)}  }      		
	B^{ R^{(3)} \rightarrow R^{(1)} , R^{(2)}  ; \mu }_{i^{(3)} \rightarrow i^{(1)}  , i^{(2)} } 	
	\prod_{\a} B^{r_{\a}^{(3)} \rightarrow r_{\a}^{(1)} , r_{\a}^{(2)}  ; \mu_{\a}}_{ l_{\a}^{(3)}  \rightarrow l_{\a}^{(1)}, l_{\a}^{(2)} }
\end{split}\end{equation}
The two pieces $\cF( \cup_I R_a^{(I)} , \{ \cup_{I, b,\a} r_{ab;\a}^{(I)} \} , \cup_I \nu_a^{(I)-} ; \mu_a, \{ \cup_{b,\a} \mu_{ab;\a} \})$ and \break 
$\cF( \cup_I R_a^{(I)} , \{ \cup_{I, b,\a} r_{ba;\a}^{(I)} \} ,\cup_I  \nu_a^{(I)+} ; \mu_a, \{ \cup_{b,\a} \mu_{ba;\a} \} )$ originally appear with reversed arrows, but have the same expression (\ref{eq:Frrr_defn}) due to reality of branching coefficients.

The key feature of (\ref{eq:GLLL_diag}) is that sums $\mu_a$ are over multiplicity $g(R^{(1)}_a,R^{(2)}_a;R^{(3)}_a)$ and $\mu_{ab;\a}$ are over $g(r_{ab;\a}^{(1)},r_{ab;\a}^{(2)};r_{ab;\a}^{(3)})$, and so the structure constant vanishes, unless all diagrams of $\bL^{(3)}$ appear in the respective Littlewood-Richardson products
\begin{equation}\label{RSBsel}
\boxed{ 
\begin{aligned}
	R^{(1)}_a \otimes R^{(2)}_a &\rightarrow R^{(3)}_a \\ 
	r_{ab;\a}^{(1)} \otimes r_{ab;\a}^{(2)} &\rightarrow r_{ab;\a}^{(3)}
\end{aligned}
}
\end{equation} 
The branching coefficients in $(\ref{eq:GLLL_diag})$ are contracted in the natural way, given these selection rules. For each group node $a$ there are two terms -- one for $\nu^+$ and one for $\nu^-$. Within each term, the branching coefficients arising from each operator $B^{R_a^{(I)} \rightarrow \cup_{b,\alpha} r_{ab;\a}^{(I)} ;  \nu_{a}^{(I) \pm }}$ are coupled via extra branching coefficients: $B^{ R^{(3)}_a \rightarrow R^{(1)}_a , R^{(2)}_a  ; \mu_a }$ for $R_a$'s, and 	
	$B^{  r_{ab;\a}^{(3)} \rightarrow r_{ab;\a}^{(1)} , r_{ab;\a}^{(2)}  ; \mu_{ab;\a}}$ for $r_{ab;\a}$'s.

Let us take as an example the structure constants of $\mC^3$, which were  discussed
 in the restricted Schur basis in \cite{db2}. The operators are defined via quiver characters (\ref{eq:chiQ_C3_diag}), and for a single-node quiver (\ref{eq:GLLL_diag}) reduces to:
\begin{equation}
\begin{split}
	&G_{\mC^3}(\bL^{(1)},\bL^{(2)};\bL^{(3)})
	\\		
	&= 
	f_{\bL^{(1)}\bL^{(2)}}^{\bL^{(3)}} \sum_{\substack{\mu, \\ \mu_1,\mu_2,\mu_3}}		
	\;
	\mytikz{		
		\node (np1) at (-1.5,0) [circle,draw,inner sep=0.5mm,label=left:$\nu^{(1)+}$] {};
		\node (np2) at (0,0) [circle,draw,inner sep=0.5mm,label=right:$\nu^{(2)+}$] {};
		\node (np3) at (1.5,0) [circle,draw,inner sep=0.5mm,label=right:$\nu^{(3)+}$] {};
		\node (n) at (0,1.5) [circle,draw,inner sep=0.5mm,label=above:$\mu$] {};
		\node (n1) at (-1,-1.5) [circle,draw,inner sep=0.5mm,label=below:$\mu_1$] {};
		\node (n2) at (0,-1.5) [circle,draw,inner sep=0.5mm,label=below:$\mu_2$] {};
		\node (n3) at (1,-1.5) [circle,draw,inner sep=0.5mm,label=below:$\mu_3$] {};
		\draw [postaction={decorate}] (np1) to node[above left]{$R^{(1)}$} (n);		
		\draw [postaction={decorate}] (np2) to node{$R^{(2)}$} (n);		
		\draw [postaction={decorate}] (n) to node[above right]{$R^{(3)}$} (np3);		
		\draw [postaction={decorate}] (n1) to node[left]{$r^{(1)}_\alpha$} (np1);		
		\draw [postaction={decorate}] (n2) to (np1);		
		\draw [postaction={decorate}] (n3) to (np1);		
		\draw [postaction={decorate}] (n1) to node[above right]{$r^{(2)}_\alpha$} (np2);		
		\draw [postaction={decorate}] (n2) to (np2);		
		\draw [postaction={decorate}] (n3) to (np2);		
		\draw [postaction={decorate}] (np3) to (n1);		
		\draw [postaction={decorate}] (np3) to (n2);		
		\draw [postaction={decorate}] (np3) to node[right]{$r^{(3)}_\alpha$} (n3);		
	}		
	\mytikz{		
		\node (np1) at (-1.5,0) [circle,draw,inner sep=0.5mm,label=left:$\nu^{(1)-}$] {};
		\node (np2) at (0,0) [circle,draw,inner sep=0.5mm,label=right:$\nu^{(2)-}$] {};
		\node (np3) at (1.5,0) [circle,draw,inner sep=0.5mm,label=right:$\nu^{(3)-}$] {};
		\node (n) at (0,1.5) [circle,draw,inner sep=0.5mm,label=above:$\mu$] {};
		\node (n1) at (-1,-1.5) [circle,draw,inner sep=0.5mm,label=below:$\mu_1$] {};
		\node (n2) at (0,-1.5) [circle,draw,inner sep=0.5mm,label=below:$\mu_2$] {};
		\node (n3) at (1,-1.5) [circle,draw,inner sep=0.5mm,label=below:$\mu_3$] {};
		\draw [postaction={decorate}] (n) to node[above left]{$R^{(1)}$} (np1);		
		\draw [postaction={decorate}] (n) to node{$R^{(2)}$} (np2);		
		\draw [postaction={decorate}] (np3) to node[above right]{$R^{(3)}$} (n);		
		\draw [postaction={decorate}] (np1) to node[left]{$r^{(1)}_\alpha$} (n1);		
		\draw [postaction={decorate}] (np1) to (n2);		
		\draw [postaction={decorate}] (np1) to (n3);		
		\draw [postaction={decorate}] (np2) to node[above right]{$r^{(2)}_\alpha$} (n1);		
		\draw [postaction={decorate}] (np2) to (n2);		
		\draw [postaction={decorate}] (np2) to (n3);		
		\draw [postaction={decorate}] (n1) to (np3);		
		\draw [postaction={decorate}] (n2) to (np3);		
		\draw [postaction={decorate}] (n3) to node[right]{$r^{(3)}_\alpha$} (np3);		
	}	
\end{split}	
\end{equation}

Let us now go over the steps in the derivation of (\ref{eq:GLLL_diag}). For clarity we mostly use diagrammatic notation. The starting point is the sum (\ref{formula-chiring}), which we 
write out as a trace of a product of  group factors 
\begin{equation}
\label{eq:GLLL_deriv2}
	G ( \bL^{(1)} , \bL^{(2)} ; \bL^{(3) } ) = \tl f_{\bL^{(1)}\bL^{(2)}}^{\bL^{(3)}}
	{ \bf tr } \prod_a \left(
	\frac{1}{n_a^{(1)}!n_a^{(2)}!} \sum_{\sigma_a^{(1)},\sigma_a^{(2)}}
	\!\!\!\!
	\mytikz{	
		\node (s) at (0,0) [rectangle,draw] {$\sigma_a^{(1)}$};		
		\node (m) at (0,1) [circle,draw,inner sep=0.5mm,label=left:$\nu_a^{(1)+}$] {};
		\node (n) at (0,-1) [circle,draw,inner sep=0.5mm,label=left:$\nu_a^{(1)-}$] {};
		\node (rin) at (0,2) {$\bigcup_{b,\alpha} r_{ba;\a}^{(1)}$};
		\node (rout) at (0,-2) {$\bigcup_{b,\alpha} r_{ab;\a}^{(1)}$};
		\draw [postaction={decorate}] (rin.-45) to (m);
		\draw [postaction={decorate}] (rin.-90) to (m);
		\draw [postaction={decorate}] (rin.-135) to (m);
		\draw [postaction={decorate}] (m) to node[right]{} (s);
		\draw [postaction={decorate}] (s) to node[right]{$R^{(1)}_a$} (n);	
		\draw [postaction={decorate}] (n) to (rout.45);
		\draw [postaction={decorate}] (n) to (rout.90);
		\draw [postaction={decorate}] (n) to (rout.135);	
	}
	\mytikz{	
		\node (s) at (0,0) [rectangle,draw] {$\sigma_a^{(2)}$};		
		\node (m) at (0,1) [circle,draw,inner sep=0.5mm,label=left:$\nu_a^{(2)+}$] {};
		\node (n) at (0,-1) [circle,draw,inner sep=0.5mm,label=left:$\nu_a^{(2)-}$] {};
		\node (rin) at (0,2) {$\bigcup_{b,\alpha} r_{ba;\a}^{(2)}$};
		\node (rout) at (0,-2) {$\bigcup_{b,\alpha} r_{ab;\a}^{(2)}$};
		\draw [postaction={decorate}] (rin.-45) to (m);
		\draw [postaction={decorate}] (rin.-90) to (m);
		\draw [postaction={decorate}] (rin.-135) to (m);
		\draw [postaction={decorate}] (m) to node[right]{} (s);
		\draw [postaction={decorate}] (s) to node[right]{$R^{(2)}_a$} (n);	
		\draw [postaction={decorate}] (n) to (rout.45);
		\draw [postaction={decorate}] (n) to (rout.90);
		\draw [postaction={decorate}] (n) to (rout.135);	
	}
	\mytikz{	
		\node (s) at (0,0) [rectangle,draw] {$(\sigma_a^{(1)}\circ\sigma_a^{(2)})^{-1}$};		
		\node (m) at (0,1) [circle,draw,inner sep=0.5mm,label=left:$\nu_a^{(3)+}$] {};
		\node (n) at (0,-1) [circle,draw,inner sep=0.5mm,label=left:$\nu_a^{(3)-}$] {};
		\node (rin) at (0,2) {$\bigcup_{b,\alpha} r_{ba;\a}^{(3)}$};
		\node (rout) at (0,-2) {$\bigcup_{b,\alpha} r_{ab;\a}^{(3)}$};
		\draw [postaction={decorate}] (m) to (rin.-45);
		\draw [postaction={decorate}] (m) to (rin.-90);
		\draw [postaction={decorate}] (m) to (rin.-135);
		\draw [postaction={decorate}] (s) to node[right]{} (m);
		\draw [postaction={decorate}] (n) to node[right]{$R^{(3)}_a$} (s);	
		\draw [postaction={decorate}] (rout.45) to (n);
		\draw [postaction={decorate}] (rout.90) to (n);
		\draw [postaction={decorate}] (rout.135) to (n);	
	}		
	\right)
\end{equation}
with a prefactor
\begin{equation}
	\tl f_{\bL^{(1)}\bL^{(2)}}^{\bL^{(3)}} = \sqrt{\frac{\prod_a d(R^{(1)}_a) d(R^{(2)}_a) d(R^{(3)}_a)}{\prod_{a,b,\alpha} d(r_{ab;\a}^{(1)}) d(r_{ab;\a}^{(2)}) d(r_{ab;\a}^{(3)})}}
\end{equation}
The trace ${ \bf tr } $ refers to the contraction of the indices associated with 
the  $ \cup_{a,b} r_{ba;\a}^{(I)}$ at the top of the diagram to those of  $ \cup_{a,b} r_{ab;\a}^{(I)}$
at the bottom.  The identification occurs across different group factors,  to make up the quivers for the three quiver characters.
The diagram, with free upper and lower external legs, corresponds to an expression with 
 indices $\{ \cup_{I , a , b ,\alpha} i_{ba;\a}^{(I)} \} $ for the upper legs and  $\{ \cup_{I , a , b ,\alpha} j_{ab;\a}^{(I)} \} $ for the lower legs, each set
 living in $ \otimes_{I , a , b ,\alpha}  r_{ab;\a}^{(I)}$.  The trace operation multiplies with $ \prod_{I , a , b ,\alpha} \delta_{  i_{ab;\a}^{(I)}  ,  j_{ab;\a}^{(I)} } $ and sums over the  indices.  

Applying (\ref{eq:BB_g1g2}) and (\ref{eq:D_sum_sigma}) we have
\begin{equation}
\label{eq:g1g2_factoriz}
	\sum_{\gamma_1,\gamma_2}	
	\;
	\mytikz{	
		\node (s) at (0,0) [rectangle,draw] {$\gamma_1$};		
		\node (m) at (0,1) {};
		\node (n) at (0,-1) {};
		\draw [postaction={decorate}] (m) to node[right]{} (s);
		\draw [postaction={decorate}] (s) to node[right]{$R_1$} (n);	
	}
	\mytikz{	
		\node (s) at (0,0) [rectangle,draw] {$\gamma_2$};		
		\node (m) at (0,1) {};
		\node (n) at (0,-1) {};
		\draw [postaction={decorate}] (m) to node[right]{} (s);
		\draw [postaction={decorate}] (s) to node[right]{$R_2$} (n);	
	}
	\mytikz{	
		\node (s) at (0,0) [rectangle,draw] {$(\gamma_1\circ\gamma_2)^{-1}$};		
		\node (m) at (0,1) {};
		\node (n) at (0,-1) {};
		\draw [postaction={decorate}] (s) to node[right]{} (m);
		\draw [postaction={decorate}] (n) to node[right]{$R_3$} (s);	
	}			
	\; = \;
	\frac{n_1!n_2!}{d(R_1)d(R_2)} \sum_\mu
	\mytikz{
		\node (m1) at (-1,1.5) {};
		\node (m2) at (0,1.5) {};
		\node (m3) at (1.5,1.5) {};
		\node (n1) at (-1,-1.5) {};
		\node (n2) at (0,-1.5) {};
		\node (n3) at (1.5,-1.5) {};
		\node (nn1) at (0,0.5) [circle,draw,inner sep=0.5mm,label=below:$\mu$] {};
		\node (nn2) at (0,-0.5) [circle,draw,inner sep=0.5mm,label=above:$\mu$] {};
		\draw [postaction={decorate}] (m1) to [bend right=20] node[left]{$R_1$} (nn1);
		\draw [postaction={decorate}] (m2) to node[above right]{$R_2$} (nn1);
		\draw [postaction={decorate}] (nn1) to [bend right=30] node[right]{$R_3$} (m3);
		\draw [postaction={decorate}] (nn2) to [bend right=20] node[left]{$R_1$} (n1);
		\draw [postaction={decorate}] (nn2) to node[below right]{$R_2$} (n2);
		\draw [postaction={decorate}] (n3) to [bend right=30] node[right]{$R_3$} (nn2);
	}
\end{equation}
Using this to perform $\sigma_a^{(1)},\sigma_a^{(2)}$ sums we get
\begin{equation}
\label{eq:GLLL_deriv3}
	G ( \bL^{(1)} , \bL^{(2)} ; \bL^{(3) } ) = \tl f_{\bL^{(1)}\bL^{(2)}}^{\bL^{(3)}}
{ \bf tr }  	\prod_a \left(
	\frac{1}{d(R_a^{(1)})d(R_a^{(2)})} \sum_{\mu_a}
	\!\!\!\!
	\mytikz{	
		\node (nna) at (0,0.5) [circle,draw,inner sep=0.5mm,label=below:$\mu_a$] {};
		\node (nnb) at (0,-0.5) [circle,draw,inner sep=0.5mm,label=above:$\mu_a$] {};
		\node (m1) at (-2,1) [circle,draw,inner sep=0.5mm,label=left:$\nu_a^{(1)+}$] {};
		\node (n1) at (-2,-1) [circle,draw,inner sep=0.5mm,label=left:$\nu_a^{(1)-}$] {};
		\node (rin1) at (-2,2) {$\bigcup_{b,\alpha} r_{ba;\a}^{(1)}$};
		\node (rout1) at (-2,-2) {$\bigcup_{b,\alpha} r_{ab;\a}^{(1)}$};
		\draw [postaction={decorate}] (rin1.-45) to (m1);
		\draw [postaction={decorate}] (rin1.-90) to (m1);
		\draw [postaction={decorate}] (rin1.-135) to (m1);
		\draw [postaction={decorate}] (m1) to node[below]{$R^{(1)}_a$} (nna);
		\draw [postaction={decorate}] (nnb) to node[above]{$R^{(1)}_a$} (n1);	
		\draw [postaction={decorate}] (n1) to (rout1.45);
		\draw [postaction={decorate}] (n1) to (rout1.90);
		\draw [postaction={decorate}] (n1) to (rout1.135);	
		\node (m2) at (0,1) [circle,draw,inner sep=0.5mm,label=left:$\nu_a^{(2)+}$] {};
		\node (n2) at (0,-1) [circle,draw,inner sep=0.5mm,label=left:$\nu_a^{(2)-}$] {};
		\node (rin2) at (0,2) {$\bigcup_{b,\alpha} r_{ba;\a}^{(2)}$};
		\node (rout2) at (0,-2) {$\bigcup_{b,\alpha} r_{ab;\a}^{(2)}$};
		\draw [postaction={decorate}] (rin2.-45) to (m2);
		\draw [postaction={decorate}] (rin2.-90) to (m2);
		\draw [postaction={decorate}] (rin2.-135) to (m2);
		\draw [postaction={decorate}] (m2) to node[right]{$R^{(2)}_a$} (nna);
		\draw [postaction={decorate}] (nnb) to node[right]{$R^{(2)}_a$} (n2);	
		\draw [postaction={decorate}] (n2) to (rout2.45);
		\draw [postaction={decorate}] (n2) to (rout2.90);
		\draw [postaction={decorate}] (n2) to (rout2.135);	
		\node (m3) at (2,1) [circle,draw,inner sep=0.5mm,label=right:$\nu_a^{(3)+}$] {};
		\node (n3) at (2,-1) [circle,draw,inner sep=0.5mm,label=right:$\nu_a^{(3)-}$] {};
		\node (rin3) at (2,2) {$\bigcup_{b,\alpha} r_{ba;\a}^{(3)}$};
		\node (rout3) at (2,-2) {$\bigcup_{b,\alpha} r_{ab;\a}^{(3)}$};
		\draw [postaction={decorate}] (m3) to (rin3.-45);
		\draw [postaction={decorate}] (m3) to (rin3.-90);
		\draw [postaction={decorate}] (m3) to (rin3.-135);
		\draw [postaction={decorate}] (nna) to node[below right]{$R^{(3)}_a$} (m3);
		\draw [postaction={decorate}] (n3) to node[above right]{$R^{(3)}_a$} (nnb);	
		\draw [postaction={decorate}] (rout3.45) to (n3);
		\draw [postaction={decorate}] (rout3.90) to (n3);
		\draw [postaction={decorate}] (rout3.135) to (n3);	
	}		
	\right)
\end{equation}
At this point the diagram is still not factorized, because legs are contracted between different factors in $\prod_a$. Next, focus on the lower piece of the diagram, containing $\nu^-$ (equivalently we can pick the upper piece -- they are symmetric). We can insert the following sum over $\gamma_1,\gamma_2$
\begin{equation}
\label{eq:GLLL_deriv1}
\begin{split}
&
	\frac{1}{n_{ab;\a}^{(1)}!n_{ab;\a}^{(2)}!} \sum_{\gamma_1,\gamma_2}
	\mytikz{	
		\node (nna) at (0,-0.5) [circle,draw,inner sep=0.5mm,label=above:$\mu_a$] {};		
		\node (m1) at (-2,-1) [circle,draw,inner sep=0.5mm,label=above:$\nu_a^{(1)-}$] {};		
		\node (g1) at (-2.5,-2) [rectangle,draw] {$\gamma_1$};		
		\draw [postaction={decorate}] (m1) to (g1);		
		\draw [postaction={decorate}] (m1) to +(0,-1.7);
		\draw [postaction={decorate}] (m1) to +(0.4,-1.7);
		\draw [postaction={decorate}] (g1) to +(0,-0.7);
		\draw [postaction={decorate}] (nna) to node[below]{} (m1);		
		\node (m2) at (0,-1) [circle,draw,inner sep=0.5mm,label=left:$\nu_a^{(2)-}$] {};		
		\node (g2) at (-0.5,-2) [rectangle,draw] {$\gamma_2$};	
		\draw [postaction={decorate}] (m2) to (g2);		
		\draw [postaction={decorate}] (m2) to +(0,-1.7);
		\draw [postaction={decorate}] (m2) to +(0.4,-1.7);		
		\draw [postaction={decorate}] (g2) to +(0,-0.7);		
		\draw [postaction={decorate}] (nna) to node[right]{} (m2);		
		\node (m3) at (2,-1) [circle,draw,inner sep=0.5mm,label=above:$\nu_a^{(3)-}$] {};		
		\node (g3) at (1.7,-2.1) [rectangle,draw] {$(\gamma_1\circ\gamma_2)^{-1}$};	
		\draw [postaction={decorate}] (g3) to (m3);		
		\draw [postaction={decorate}] ($(m3)+(1.0,-1.7)$) to ($(m3)+(1.0,-0.7)$) to (m3);
		\draw [postaction={decorate}] ($(m3)+(1.2,-1.7)$) to ($(m3)+(1.2,-0.7)$) to (m3);
		\draw [postaction={decorate}] ($(g3)+(0,-0.7)$) to (g3);	
		\draw [postaction={decorate}] (m3) to node[below right]{} (nna);		
	}	
	\\		
	& ~~~~~~~~~~~~~~~~~~~~~~~~~~~~~~~~~~~ =
	\frac{1}{d(r_{ab;\a}^{(1)})d(r_{ab;\a}^{(2)})} 
	\sum_{\mu_{ab;\a}}
	\mytikz{	
		\node (nna) at (0,-0.5) [circle,draw,inner sep=0.5mm,label=above:$\mu_a$] {};		
		\node (nab1) at (-0.5,-2) [circle,draw,inner sep=0.5mm,label=below:$\mu_{ab;\a}$] {};
		\node (nab2) at (-0.5,-3.1) [circle,draw,inner sep=0.5mm,label=above:$\mu_{ab;\a}$] {};
		\draw [postaction={decorate}] (nab2) to +(-2,-0.2);
		\draw [postaction={decorate}] (nab2) to +(0,-0.5);
		\draw [postaction={decorate}] ($(nab2)+(2,-0.2)$) to (nab2);
		\node (m1) at (-2,-1) [circle,draw,inner sep=0.5mm,label=above:$\nu_a^{(1)-}$] {};				
		\draw [postaction={decorate}] (m1) .. controls +(-135:1) and +(180:1) .. (nab1);		
		\draw [postaction={decorate}] (m1) to +(0,-0.6);
		\draw [postaction={decorate}] (m1) to +(0.4,-0.7);
		\draw [postaction={decorate}] (nna) to node[below]{} (m1);		
		\node (m2) at (0,-1) [circle,draw,inner sep=0.5mm,label=left:$\nu_a^{(2)-}$] {};				
		\draw [postaction={decorate}] (m2) to (nab1);		
		\draw [postaction={decorate}] (m2) to +(0,-0.7);
		\draw [postaction={decorate}] (m2) to +(0.4,-0.7);		
		\draw [postaction={decorate}] (nna) to node[right]{} (m2);		
		\node (m3) at (2,-1) [circle,draw,inner sep=0.5mm,label=above:$\nu_a^{(3)-}$] {};				
		\draw [postaction={decorate}] (nab1) to [bend right=20] (m3);		
		\draw [postaction={decorate}] ($(m3)+(0,-0.7)$) to (m3);
		\draw [postaction={decorate}] ($(m3)+(0.4,-0.7)$) to (m3);	
		\draw [postaction={decorate}] (m3) to node[below right]{} (nna);		
	}			
\end{split}	
\end{equation}

On the left hand side $\gamma_1$ acts on one of the outgoing legs $r_{ab;\a}^{(1)}$ (for some choice of $b,\alpha$), $\gamma_2$ acts on $r_{ab;\a}^{(2)}$, and $(\gamma_1 \circ \gamma_2)^{-1}$ acts on $r_{ab;\a}^{(3)}$. It is equal to the original $\nu^-$ factor in (\ref{eq:GLLL_deriv1}), because we can pull $\gamma$'s through the branching coefficients and cancel. Next we can sum over all $\gamma_1 \circ \gamma_2 \in S_{n_{ab;\a}^{(1)}}\times S_{n_{ab;\a}^{(2)}}$, which allows us to apply (\ref{eq:g1g2_factoriz}) again, resulting in the right hand side. Performing this for each $b,\alpha$ we completely ``cap off'' the outgoing $r_{ab;\a}^{(I)}$ legs, contracting each $r_{ab;\a}^{(1)}\otimes r_{ab;\a}^{(2)} \rightarrow r_{ab;\a}^{(3)}$ respectively, and introducing $\{\mu_{ab;\a}\}$ sums. The leftover branching coefficient with $\mu_{ab;\a}$ (at the bottom of the right hand side) contracts the incoming legs $r_{ba;\a}^{(I)}$ of the respective $\nu^+$ diagram in (\ref{eq:GLLL_deriv1}). Consequently, the diagram completely factorizes, and we get (\ref{eq:GLLL_diag}), with prefactor arising from
\begin{equation}
	f_{\bL^{(1)}\bL^{(2)}}^{\bL^{(3)}} = \frac{\tl f_{\bL^{(1)}\bL^{(2)}}^{\bL^{(3)}}}{ 
	\prod_a d(R^{(1)}_a) d(R^{(2)}_a)
	\prod_{a,b,\alpha} d(r_{ab;\a}^{(1)}) d(r_{ab;\a}^{(2)})
	}
\end{equation}
The equations corresponding to the diagrammatic manipulations above are given 
in Appendix \ref{der:ChirStConst}.

\subsection{Covariant basis}
\label{sec:covariantfus}

Here we calculate the chiral ring structure constants for the covariant basis (\ref{eq:OK_defn}) operators $\cO_Q(\bK)$. As in the previous section, the product is
\begin{equation}
	\cO_Q(\bK^{(1)}) \cO_Q(\bK^{(2)}) = \sum_{\bK^{(3)}} G(\bK^{(1)},\bK^{(2)};\bK^{(3)}) \cO_Q(\bK^{(3)})
\end{equation}
with the structure constants
\begin{equation}
\label{eq:GKKK_defn} 
\begin{split}
& G ( \bK^{(1)} , \bK^{(2)} ; \bK^{(3) } ) 
\\
& \quad = 
\frac{1}{\prod n_a^{(1)}!} \frac{1}{\prod n_a^{(2)}!} 
\sum_{   \bsig^{(1)} }  \sum_{ \bsig^{(2)} } 
\hat\chi_Q (\bK^{(1)},  \bsig^{ (1) }  )  
\hat\chi_Q ( \bK^{(2)} ,  \bsig^{(2)}   ) 
\hat\chi_Q ( \bK^{(3)}  ,  \bsig^{(1)}  \circ  \bsig^{(2)}   )
\end{split}\end{equation}
Here we use conveniently normalized covariant quiver characters
\begin{equation}
	\hat\chi_Q (\bK) \equiv \sqrt{\prod d(R_a)} \chi_Q (\bK) .
\end{equation}

Let us first present the answer and some examples, and sketch the derivation afterwards. Recall from the definition (\ref{eq:chiK_defn}) of the covariant quiver characters, that the labels are $\bK = \{ R_a, s_{ab}^+, s_{ab}^-, \nu_a^+, \nu_a^-, \Lambda_{ab}, \tau_{ab}, n_{ab;\a}, \beta_{ab} \}$, as displayed in (\ref{eq:chiK_diag}). The result of the sum (\ref{eq:GKKK_defn}) is, like in the previous section, that \emph{all} of the Young diagram labels multiply according to the Littlewood-Richardson rule
\begin{equation}\label{CBsel}
\boxed{ 
  \begin{aligned} 
	R_a^{(1)} \otimes R_a^{(2)} &\rightarrow R_a^{(3)} \\
	\Lambda_{ab}^{(1)} \otimes \Lambda_{ab}^{(2)} &\rightarrow \Lambda_{ab}^{(3)} \\
	s_{ab}^{(1)+} \otimes s_{ab}^{(2)+} &\rightarrow s_{ab}^{(3)+} \\
	s_{ab}^{(1)-} \otimes s_{ab}^{(2)-} &\rightarrow s_{ab}^{(3)-}
   \end{aligned} 
}
\end{equation} 
That is, $G ( \bK^{(1)} , \bK^{(2)} ; \bK^{(3)} )$ vanishes unless 
the labels from $\bK^{(3)}$ are contained in the  Littlewood-Richardson tensor product
(also called outer product) of the Young diagrams.
The non-vanishing coefficients are given, similarly as in (\ref{eq:GLLL_diag}), by connecting up all coupled legs via branching coefficients, and summing over the multiplicities for the new branchings. Specifically, we get:
\begin{equation}
\label{eq:GKKK_diag}
\begin{split}
	&G(\bK^{(1)},\bK^{(2)};\bK^{(3)}) = 
		f_{\bK^{(1)}\bK^{(2)}}^{\bK^{(3)}}
	\sum_{\{\mu_a\}} \sum_{\{\mu_{ab}^+\}} \sum_{\{\mu_{ab}^-\}} \sum_{\{\mu_{ab}^\Lambda\}} 
\\
	& \prod_a \left(
%%%%%%%%%%%%%%%%%%%%%%%%%%%%	
	\mytikz{		
		\node (np1) at (-1.5,0) [circle,draw,inner sep=0.5mm,label=left:$\nu_a^{(1)-}$] {};
		\node (np2) at (0,0) [circle,draw,inner sep=0.5mm,label=right:$\nu_a^{(2)-}$] {};
		\node (np3) at (1.5,0) [circle,draw,inner sep=0.5mm,label=right:$\nu_a^{(3)-}$] {};
		\node (n) at (0,1.5) [circle,draw,inner sep=0.5mm,label=above:$\mu_a$] {};
		\node (n1) at (-1,-1.5) [circle,draw,inner sep=0.5mm] {};
		\node (n2) at (0,-1.5) [circle,draw,inner sep=0.5mm,label=below:$\bigcup_{b} \mu_{ab}^-$] {};
		\node (n3) at (1,-1.5) [circle,draw,inner sep=0.5mm] {};
		\draw [postaction={decorate}] (n) to node[above left]{$R_a^{(1)}$} (np1);		
		\draw [postaction={decorate}] (n) to node{$R_a^{(2)}$} (np2);		
		\draw [postaction={decorate}] (np3) to node[above right]{$R_a^{(3)}$} (n);		
		\draw [postaction={decorate}] (np1) to node[left]{$\bigcup_{b} s_{ab}^{(1)-}$} (n1);		
		\draw [postaction={decorate}] (np1) to (n2);		
		\draw [postaction={decorate}] (np1) to (n3);		
		\draw [postaction={decorate}] (np2) to (n1);		
		\draw [postaction={decorate}] (np2) to (n2);		
		\draw [postaction={decorate}] (np2) to (n3);		
		\draw [postaction={decorate}] (n1) to (np3);		
		\draw [postaction={decorate}] (n2) to (np3);		
		\draw [postaction={decorate}] (n3) to node[right]{$\bigcup_{b} s_{ab}^{(3)-}$} (np3);		
	}	
%%%%%%%%%%%%%%%%%%%%%%%%%%%%	
	\mytikz{		
		\node (np1) at (-1.5,0) [circle,draw,inner sep=0.5mm,label=left:$\nu_a^{(1)+}$] {};
		\node (np2) at (0,0) [circle,draw,inner sep=0.5mm,label=right:$\nu_a^{(2)+}$] {};
		\node (np3) at (1.5,0) [circle,draw,inner sep=0.5mm,label=right:$\nu_a^{(3)+}$] {};
		\node (n) at (0,-1.5) [circle,draw,inner sep=0.5mm,label=below:$\mu_a$] {};
		\node (n1) at (-1,1.5) [circle,draw,inner sep=0.5mm] {};
		\node (n2) at (0,1.5) [circle,draw,inner sep=0.5mm,label=above:$\bigcup_{b}\mu_{ba}^+$] {};
		\node (n3) at (1,1.5) [circle,draw,inner sep=0.5mm] {};
		\draw [postaction={decorate}] (np1) to node[below left]{$R_a^{(1)}$} (n);		
		\draw [postaction={decorate}] (np2) to node{$R_a^{(2)}$} (n);		
		\draw [postaction={decorate}] (n) to node[below right]{$R_a^{(3)}$} (np3);		
		\draw [postaction={decorate}] (n1) to node[left]{$\bigcup_{b} s_{ba}^{(1)+}$} (np1);		
		\draw [postaction={decorate}] (n2) to (np1);		
		\draw [postaction={decorate}] (n3) to (np1);		
		\draw [postaction={decorate}] (n1) to (np2);		
		\draw [postaction={decorate}] (n2) to (np2);		
		\draw [postaction={decorate}] (n3) to (np2);		
		\draw [postaction={decorate}] (np3) to (n1);		
		\draw [postaction={decorate}] (np3) to (n2);		
		\draw [postaction={decorate}] (np3) to node[right]{$\bigcup_{b} s_{ba}^{(3)+}$} (n3);		
	}			
%%%%%%%%%%%%%%%%%%%%%%%%%%%%	
	\right)
\\ 
	& \prod_{a,b} \left(
%%%%%%%%%%%%%%%%%%%%%%%%%%%%	
	\mytikz{		
		\node (np1) at (-1.5,0) [circle,fill,inner sep=0.5mm,label=left:$\tau_{ab}^{(1)}$] {};
		\node (np2) at (0,0) [circle,fill,inner sep=0.5mm,label=right:$\tau_{ab}^{(2)}$] {};
		\node (np3) at (1.5,0) [circle,fill,inner sep=0.5mm,label=right:$\tau_{ab}^{(3)}$] {};
		\node (n) at (0,1.5) [circle,draw,inner sep=0.5mm,label=above:$\mu_{ab}^{-}$] {};
		\node (m) at (0,-1.5) [circle,draw,inner sep=0.5mm,label=below:$\mu_{ab}^{+}$] {};	
		\node (nl) at (2.5,1.5) [circle,draw,inner sep=0.5mm,label=above:$\mu_{ab}^{\Lambda}$] {};
		\draw [postaction={decorate}] (n) to node[above left]{$s_{ab}^{(1)-}$} (np1);		
		\draw [postaction={decorate}] (n) to node{$s_{ab}^{(2)-}$} (np2);		
		\draw [postaction={decorate}] (np3) to node[above]{$s_{ab}^{(3)-}$} (n);		
		\draw [postaction={decorate}] (np1) to node[below left]{$s_{ab}^{(1)+}$} (m);		
		\draw [postaction={decorate}] (np2) to node{$s_{ab}^{(2)+}$} (m);		
		\draw [postaction={decorate}] (m) to node[below right]{$s_{ab}^{(3)+}$} (np3);		
		\draw [-] (np1) to (nl);		
		\draw [-] (np2) to (nl);		
		\draw [-] (np3) to node[right]{$\Lambda_{ab}^{(3)}$} (nl);	
		\node (l1) at (1.7,1.6) {$\Lambda_{ab}^{(1)}$};	
		\node (l2) at (1.5,0.8) {$\Lambda_{ab}^{(2)}$};	
	}	
%%%%%%%%%%%%%%%%%%%%%%%%%%%%	
	\mytikz{				
		\node (n) at (0,0) [circle,draw,inner sep=0.5mm,label=above:$\mu_{ab}^{\Lambda}$] {};
		\node (b1) at (2,1.5) [circle,draw,inner sep=0.5mm,label=above:$\beta_{ab}^{(1)}$] {};
		\node (b2) at (2,0) [circle,draw,inner sep=0.5mm,label=above:$\beta_{ab}^{(2)}$] {};
		\node (b3) at (2,-1.5) [circle,draw,inner sep=0.5mm,label=above:$\beta_{ab}^{(3)}$] {};
		\node (n1) at (3,1.5) {$\bn_{ab}^{(1)}$};	
		\node (n2) at (3,0) {$\bn_{ab}^{(2)}$};	
		\node (n3) at (3,-1.5) {$\bn_{ab}^{(3)}$};			
		\draw [-] (n) to node[above]{$\Lambda_{ab}^{(1)}$} (b1);		
		\draw [-] (n) to node[above]{$\Lambda_{ab}^{(2)}$} (b2);		
		\draw [-] (n) to node[above]{$\Lambda_{ab}^{(3)}$} (b3);		
		\draw [-] (b1) to (n1);		
		\draw [-] (b2) to (n2);		
		\draw [-] (b3) to (n3);		
	}		
%%%%%%%%%%%%%%%%%%%%%%%%%%%%
	\right)
\end{split}
\end{equation}
with
\begin{equation}
	f_{\bK^{(1)}\bK^{(2)}}^{\bK^{(3)}} =	
	\frac{\sqrt{\prod_a d(R^{(1)}_a) d(R^{(2)}_a) d(R^{(3)}_a)}}{
		\prod_a d(R^{(1)}_a) d(R^{(2)}_a)
		\prod_{a,b} d(s^{(1)-}_{ab}) d(s^{(2)-}_{ab}) d(s^{(1)+}_{ab}) d(s^{(2)+}_{ab}) d(\Lambda^{(1)}_{ab}) d(\Lambda^{(2)}_{ab})
	}
\end{equation}
As for the  restricted Schur basis, we get two factors of $\cF$ defined in (\ref{eq:Frrr_defn}) for each group node, now $s_{ab}^\pm$ playing the role of $r_{ab;\a}$. In addition to that, for each edge in the quiver we get a factor coupling $\Lambda_{ab}^{(1)} \otimes \Lambda_{ab}^{(2)} \rightarrow \Lambda_{ab}^{(3)}$. Again for illustration we use three outgoing arrows from each branching node $\nu_a^-$  and three incoming arrows to each $\nu_a^+$.
The explicit expression is:
\begin{equation}
\label{eq:GKKK_result}
\begin{split}
	& G ( \bK^{(1)} , \bK^{(2)} ; \bK^{(3)} ) = 
		f_{\bK^{(1)}\bK^{(2)}}^{\bK^{(3)}}
	\sum_{\{\mu_a\}} \sum_{\{\mu_{ab}^+\}} \sum_{\{\mu_{ab}^-\}} \sum_{\{\mu_{ab}^\Lambda\}}  
\\ &
\prod_a
\cF\left( \cup_I R_a^{(I)} , \{ \cup_{I, b} s_{ab}^{(I)-} \} , \cup_I \nu_a^{(I)-} ; \mu_a, \{ \cup_{b} \mu_{ab}^- \} \right) \;
\cF\left(\cup_I  R_a^{(I)} , \{ \cup_{I, b} s_{ba}^{(I)+} \} , \cup_I \nu_a^{(I)+} ; \mu_a, \{ \cup_{b} \mu_{ba}^+ \} \right)	
\\
& \prod_{a,b} \left(
	S^{\; s_{ab}^{(1)+} s_{ab}^{(1)-},\, \Lambda^{(1)}_{ab}\tau^{(1)}_{ab} }_{ \;l^{(1)+}_{ab}\; l^{(1)-}_{ab},\,l^{(1)}_{ab} }
	S^{\; s_{ab}^{(2)+} s_{ab}^{(2)-},\, \Lambda^{(2)}_{ab}\tau^{(2)}_{ab} }_{ \;l^{(2)+}_{ab}\; l^{(2)-}_{ab},\,l^{(2)}_{ab} }
	S^{\; s_{ab}^{(3)+} s_{ab}^{(3)-},\, \Lambda^{(3)}_{ab}\tau^{(3)}_{ab} }_{ \;l^{(3)+}_{ab}\; l^{(3)-}_{ab},\,l^{(3)}_{ab} }		
\right.
\\
&
\left. 
~~~~~~~~~~~~~~~~~~~~~~~~~
\times
B^{  s^{(3) - }_{ab} \rightarrow s_{ab}^{(1) -} , s_{ab}^{(2) -}  ; \mu_{ab}^- }_{ l_{ab}^{(3) -}  \rightarrow l_{ab}^{(1) -}   , l_{ab}^{(2) -  } }
B^{  s^{(3) + }_{ab} \rightarrow s_{ab}^{(1) +} , s_{ab}^{(2) +}  ; \mu_{ab}^+ }_{ l_{ab}^{(3) +}  \rightarrow l_{ab}^{(1) +}   , l_{ab}^{(2) +  } }
B^{ \Lambda^{(3)}_{ab} \rightarrow \Lambda^{(1)}_{ab} , \Lambda^{(2)}_{ab}  ; \mu_{ab}^\Lambda }_{l^{(3)}_{ab} \rightarrow l^{(1)}_{ab}  , l^{(2)}_{ab} } 	
\right)
\\
& \times
\left(
B^{\Lambda^{(1)}_{ab} \rightarrow [\bn^{(1)}_{ab}],\beta^{(1)}_{ab}}_{k^{(1)}_{ab}} 
B^{\Lambda^{(2)}_{ab} \rightarrow [\bn^{(2)}_{ab}],\beta^{(2)}_{ab}}_{k^{(2)}_{ab}} 
B^{\Lambda^{(3)}_{ab} \rightarrow [\bn^{(3)}_{ab}],\beta^{(3)}_{ab}}_{k^{(3)}_{ab}} 
B^{ \Lambda^{(3)}_{ab} \rightarrow \Lambda^{(1)}_{ab} , \Lambda^{(2)}_{ab}  ; \mu_{ab}^\Lambda }_{k^{(3)}_{ab} \rightarrow k^{(1)}_{ab}  , k^{(2)}_{ab} } 
\right)
\end{split}\end{equation}

In its most general form $G(\bK^{(1)},\bK^{(2)};\bK^{(3)})$ looks more complicated than $G(\bL^{(1)},\bL^{(2)};\bL^{(3)})$, because it has to deal with both $s_{ab}^\pm$ and $\Lambda_{ab}$. However, for linear quivers like $\mC^3$ (\ref{eq:chiK_diag_C3}), conifold (\ref{eq:chiK_diag_con}), $dP_0$ (\ref{eq:chiK_diag_dP0}) it simplifies significantly, because $s_{ba}^+ = R_a = s_{ab}^-$, so there are no $s_{ab}^\pm$ or $\nu_a^\pm$ labels at all. In that case the $\cF$ factors reduce to
\begin{equation}
	\cF\left( R_a^{(I)} , R_a^{(I)}, \nu_a^{(I)-}=1 ; \mu_a, \mu_{ab}^- \right)
	\; = \;
	\mytikz{
		\node (n1) at (0,0.7) [circle,draw,inner sep=0.5mm,label=above:$\mu_a$] {};
		\node (n2) at (0,-0.7) [circle,draw,inner sep=0.5mm,label=below:$\mu_{ab}^-$] {};
		\draw[postaction={decorate}] (n1) to [bend right=90] node[left]{$R_a^{(1)}$} (n2);
		\draw[postaction={decorate}] (n1) to [bend right=0] node{$R_a^{(2)}$} (n2);
		\draw[postaction={decorate}] (n2) to [bend right=90] node[right]{$R_a^{(3)}$} (n1);
	}
	\; = \; 
	\delta_{\mu_a \mu_{ab}^-} \, d(R_a^{(1)}) \, d(R_a^{(2)})
\end{equation}
using (\ref{eq:BB_R_delta}). Thus, for example, we can write the chiral ring structure constants for $\mC^3$ as just the term for the single edge in the quiver
\begin{equation}
\label{eq:GKKK_C3}
\begin{split}
	&G_{\mC^3}(\bK^{(1)},\bK^{(2)};\bK^{(3)}) = 
	\frac{\sqrt{d(R^{(1)}) d(R^{(2)}) d(R^{(3)})}}{
		d(R^{(1)}) d(R^{(2)}) d(\Lambda^{(1)}) d(\Lambda^{(2)})
	}	
\\ 
	& ~~~~~~~~~~~~ \times \sum_{\mu} \sum_{\mu^\Lambda}  \left(
%%%%%%%%%%%%%%%%%%%%%%%%%%%%	
	\mytikz{		
		\node (np1) at (-1.5,0) [circle,fill,inner sep=0.5mm,label=left:$\tau^{(1)}$] {};
		\node (np2) at (0,0) [circle,fill,inner sep=0.5mm,label=right:$\tau^{(2)}$] {};
		\node (np3) at (1.5,0) [circle,fill,inner sep=0.5mm,label=right:$\tau^{(3)}$] {};
		\node (n) at (0,1.5) [circle,draw,inner sep=0.5mm,label=above:$\mu$] {};
		\node (m) at (0,-1.5) [circle,draw,inner sep=0.5mm,label=below:$\mu$] {};	
		\node (nl) at (2.5,1.5) [circle,draw,inner sep=0.5mm,label=above:$\mu^{\Lambda}$] {};
		\draw [postaction={decorate}] (n) to node[above left]{$R^{(1)}$} (np1);		
		\draw [postaction={decorate}] (n) to node{$R^{(2)}$} (np2);		
		\draw [postaction={decorate}] (np3) to node[above]{$R^{(3)}$} (n);		
		\draw [postaction={decorate}] (np1) to node[below left]{$R^{(1)}$} (m);		
		\draw [postaction={decorate}] (np2) to node{$R^{(2)}$} (m);		
		\draw [postaction={decorate}] (m) to node[below right]{$R^{(3)}$} (np3);		
		\draw [-] (np1) to (nl);		
		\draw [-] (np2) to (nl);		
		\draw [-] (np3) to node[right]{$\Lambda^{(3)}$} (nl);	
		\node (l1) at (1.7,1.6) {$\Lambda^{(1)}$};	
		\node (l2) at (1.5,0.8) {$\Lambda^{(2)}$};	
	}	
%%%%%%%%%%%%%%%%%%%%%%%%%%%%	
	\mytikz{				
		\node (n) at (0,0) [circle,draw,inner sep=0.5mm,label=above:$\mu^{\Lambda}$] {};
		\node (b1) at (2,1.5) [circle,draw,inner sep=0.5mm,label=above:$\beta^{(1)}$] {};
		\node (b2) at (2,0) [circle,draw,inner sep=0.5mm,label=above:$\beta^{(2)}$] {};
		\node (b3) at (2,-1.5) [circle,draw,inner sep=0.5mm,label=above:$\beta^{(3)}$] {};
		\node (n1) at (3,1.5) {$\bn^{(1)}$};	
		\node (n2) at (3,0) {$\bn^{(2)}$};	
		\node (n3) at (3,-1.5) {$\bn^{(3)}$};			
		\draw [-] (n) to node[above]{$\Lambda^{(1)}$} (b1);		
		\draw [-] (n) to node[above]{$\Lambda^{(2)}$} (b2);		
		\draw [-] (n) to node[above]{$\Lambda^{(3)}$} (b3);		
		\draw [-] (b1) to (n1);		
		\draw [-] (b2) to (n2);		
		\draw [-] (b3) to (n3);		
	}		
%%%%%%%%%%%%%%%%%%%%%%%%%%%%
	\right)
\end{split}
\end{equation}
A diagrammatic form of the fusion coeffieicnt for the $\mC^3$ case, manifestly exhibiting 
the $R^{(1)}  \otimes R^{(2)} \rightarrow R^{(3)} $ LR-selection rule was given in \cite{BHR1}. 
For the conifold we have a product of two terms, one for each edge, using the labeling (\ref{eq:K_con}) $\bK = \{ R_1, R_2, \Lambda_A, \Lambda_B, \tau_A, \tau_B, \bn, \beta_A, \beta_B \}	$:
\begin{equation}
\label{eq:GKKK_con}
\begin{split}
	&G_{\cC}(\bK^{(1)},\bK^{(2)};\bK^{(3)}) = 
	\frac{\sqrt{ d(R_1^{(1)}) d(R_1^{(2)}) d(R_1^{(3)}) d(R_2^{(1)}) d(R_2^{(2)}) d(R_3^{(3)}) }}{
		d(R_1^{(1)}) d(R_1^{(2)}) d(R_2^{(1)}) d(R_2^{(2)}) d(\Lambda_A^{(1)}) d(\Lambda_A^{(2)}) d(\Lambda_B^{(1)}) d(\Lambda_B^{(2)})
	}	
\\ 
	& \times \sum_{\mu_1\mu_2} \sum_{\mu_A^\Lambda\mu_B^\Lambda}  \left(
%%%%%%%%%%%%%%%%%%%%%%%%%%%%	
	\mytikz{		
		\node (np1) at (-1.5,0) [circle,fill,inner sep=0.5mm,label=left:$\tau_A^{(1)}$] {};
		\node (np2) at (0,0) [circle,fill,inner sep=0.5mm,label=right:$\tau_A^{(2)}$] {};
		\node (np3) at (1.5,0) [circle,fill,inner sep=0.5mm,label=right:$\tau_A^{(3)}$] {};
		\node (n) at (0,1.5) [circle,draw,inner sep=0.5mm,label=above:$\mu_1$] {};
		\node (m) at (0,-1.5) [circle,draw,inner sep=0.5mm,label=below:$\mu_2$] {};	
		\node (nl) at (2.5,1.5) [circle,draw,inner sep=0.5mm,label=above:$\mu_A^{\Lambda}$] {};
		\draw [postaction={decorate}] (n) to node[above left]{$R_1^{(1)}$} (np1);		
		\draw [postaction={decorate}] (n) to node{$R_1^{(2)}$} (np2);		
		\draw [postaction={decorate}] (np3) to node[above]{$R_1^{(3)}$} (n);		
		\draw [postaction={decorate}] (np1) to node[below left]{$R_2^{(1)}$} (m);		
		\draw [postaction={decorate}] (np2) to node{$R_2^{(2)}$} (m);		
		\draw [postaction={decorate}] (m) to node[below right]{$R_2^{(3)}$} (np3);		
		\draw [-] (np1) to (nl);		
		\draw [-] (np2) to (nl);		
		\draw [-] (np3) to node[right]{$\Lambda_A^{(3)}$} (nl);	
		\node (l1) at (1.7,1.6) {$\Lambda_A^{(1)}$};	
		\node (l2) at (1.5,0.8) {$\Lambda_A^{(2)}$};	
	}	
%%%%%%%%%%%%%%%%%%%%%%%%%%%%	
	\mytikz{				
		\node (n) at (0,0) [circle,draw,inner sep=0.5mm,label=above:$\mu_A^{\Lambda}$] {};
		\node (b1) at (2,1.5) [circle,draw,inner sep=0.5mm,label=above:$\beta_A^{(1)}$] {};
		\node (b2) at (2,0) [circle,draw,inner sep=0.5mm,label=above:$\beta_A^{(2)}$] {};
		\node (b3) at (2,-1.5) [circle,draw,inner sep=0.5mm,label=above:$\beta_A^{(3)}$] {};
		\node (n1) at (3,1.5) {$\bn_A^{(1)}$};	
		\node (n2) at (3,0) {$\bn_A^{(2)}$};	
		\node (n3) at (3,-1.5) {$\bn_A^{(3)}$};			
		\draw [-] (n) to node[above]{$\Lambda_A^{(1)}$} (b1);		
		\draw [-] (n) to node[above]{$\Lambda_A^{(2)}$} (b2);		
		\draw [-] (n) to node[above]{$\Lambda_A^{(3)}$} (b3);		
		\draw [-] (b1) to (n1);		
		\draw [-] (b2) to (n2);		
		\draw [-] (b3) to (n3);		
	}		
%%%%%%%%%%%%%%%%%%%%%%%%%%%%	
	\right.
	\\ & ~~~~~~~~~~~~~~~~~
	\left.
%%%%%%%%%%%%%%%%%%%%%%%%%%%%	
	\mytikz{		
		\node (np1) at (-1.5,0) [circle,fill,inner sep=0.5mm,label=left:$\tau_B^{(1)}$] {};
		\node (np2) at (0,0) [circle,fill,inner sep=0.5mm,label=right:$\tau_B^{(2)}$] {};
		\node (np3) at (1.5,0) [circle,fill,inner sep=0.5mm,label=right:$\tau_B^{(3)}$] {};
		\node (n) at (0,1.5) [circle,draw,inner sep=0.5mm,label=above:$\mu_2$] {};
		\node (m) at (0,-1.5) [circle,draw,inner sep=0.5mm,label=below:$\mu_1$] {};	
		\node (nl) at (2.5,1.5) [circle,draw,inner sep=0.5mm,label=above:$\mu_B^{\Lambda}$] {};
		\draw [postaction={decorate}] (n) to node[above left]{$R_2^{(1)}$} (np1);		
		\draw [postaction={decorate}] (n) to node{$R_2^{(2)}$} (np2);		
		\draw [postaction={decorate}] (np3) to node[above]{$R_2^{(3)}$} (n);		
		\draw [postaction={decorate}] (np1) to node[below left]{$R_1^{(1)}$} (m);		
		\draw [postaction={decorate}] (np2) to node{$R_1^{(2)}$} (m);		
		\draw [postaction={decorate}] (m) to node[below right]{$R_1^{(3)}$} (np3);		
		\draw [-] (np1) to (nl);		
		\draw [-] (np2) to (nl);		
		\draw [-] (np3) to node[right]{$\Lambda_B^{(3)}$} (nl);	
		\node (l1) at (1.7,1.6) {$\Lambda_B^{(1)}$};	
		\node (l2) at (1.5,0.8) {$\Lambda_B^{(2)}$};	
	}	
%%%%%%%%%%%%%%%%%%%%%%%%%%%%	
	\mytikz{				
		\node (n) at (0,0) [circle,draw,inner sep=0.5mm,label=above:$\mu_B^{\Lambda}$] {};
		\node (b1) at (2,1.5) [circle,draw,inner sep=0.5mm,label=above:$\beta_B^{(1)}$] {};
		\node (b2) at (2,0) [circle,draw,inner sep=0.5mm,label=above:$\beta_B^{(2)}$] {};
		\node (b3) at (2,-1.5) [circle,draw,inner sep=0.5mm,label=above:$\beta_B^{(3)}$] {};
		\node (n1) at (3,1.5) {$\bn_B^{(1)}$};	
		\node (n2) at (3,0) {$\bn_B^{(2)}$};	
		\node (n3) at (3,-1.5) {$\bn_B^{(3)}$};			
		\draw [-] (n) to node[above]{$\Lambda_B^{(1)}$} (b1);		
		\draw [-] (n) to node[above]{$\Lambda_B^{(2)}$} (b2);		
		\draw [-] (n) to node[above]{$\Lambda_B^{(3)}$} (b3);		
		\draw [-] (b1) to (n1);		
		\draw [-] (b2) to (n2);		
		\draw [-] (b3) to (n3);		
	}			
%%%%%%%%%%%%%%%%%%%%%%%%%%%%
	\right)
\end{split}
\end{equation}

The derivation of (\ref{eq:GKKK_diag}) parallels that of the last section, except in addition we have to deal with Clebsch-Gordan coefficient (black) nodes and $\Lambda_{ab}$. The sum over $\bsig^{(1)},\bsig^{(2)}$ in (\ref{eq:GKKK_defn}) is performed the same way as in (\ref{eq:GLLL_deriv2}) and we get analogously to (\ref{eq:GLLL_deriv3}):
\begin{equation}
	G ( \bK^{(1)} , \bK^{(2)} ; \bK^{(3) } ) = \frac{\tl f_{\bK^{(1)}\bK^{(2)}}^{\bK^{(3)}}}
	{ \prod d(R_a^{(1)})d(R_a^{(2)}) }
{ \bf tr } 	\prod_a \left(
	\sum_{\mu_a}
	\!\!\!\!
	\mytikz{	
		\node (nna) at (0,0.5) [circle,draw,inner sep=0.5mm,label=below:$\mu_a$] {};
		\node (nnb) at (0,-0.5) [circle,draw,inner sep=0.5mm,label=above:$\mu_a$] {};
		\node (m1) at (-2,1) [circle,draw,inner sep=0.5mm,label=left:$\nu_a^{(1)+}$] {};
		\node (n1) at (-2,-1) [circle,draw,inner sep=0.5mm,label=left:$\nu_a^{(1)-}$] {};
		\node (rin1) at (-2,2) {$\bigcup_{b} s_{ba}^{(1)+}$};
		\node (la1) at (-2.5,-2.5) [circle,fill,inner sep=0.5mm] {};
		\node (lb1) at (-1.5,-2.5) [circle,fill,inner sep=0.5mm] {};
		\node (rout1) at (-2,-3.5) {$\bigcup_{b} s_{ab}^{(1)+}$};
		\draw [postaction={decorate}] (rin1.-45) to (m1);		
		\draw [postaction={decorate}] (rin1.-135) to (m1);
		\draw [postaction={decorate}] (m1) to node[below]{$R^{(1)}_a$} (nna);
		\draw [postaction={decorate}] (nnb) to node[above]{$R^{(1)}_a$} (n1);	
		\draw [postaction={decorate}] (n1) to node[left]{$s_{ab}^{(1)-}$} (la1);		
		\draw [postaction={decorate}] (n1) to (lb1);		
		\draw [-] (la1) to node[left]{$\Lambda_{ab}^{(1)}$} +(-0.3,0);			
		\draw [-] (lb1) to +(-0.3,0);			
		\draw [postaction={decorate}] (la1) to +(0,-0.6);
		\draw [postaction={decorate}] (lb1) to +(0,-0.6);
		\node (m2) at (0,1) [circle,draw,inner sep=0.5mm,label=left:$\nu_a^{(2)+}$] {};
		\node (n2) at (0,-1) [circle,draw,inner sep=0.5mm,label=left:$\nu_a^{(2)-}$] {};
		\node (rin2) at (0,2) {$\bigcup_{b} s_{ba}^{(2)+}$};
		\node (la2) at (-0.5,-2.5) [circle,fill,inner sep=0.5mm] {};
		\node (lb2) at (0.5,-2.5) [circle,fill,inner sep=0.5mm] {};
		\node (rout2) at (0,-3.5) {$\bigcup_{b} s_{ab}^{(2)-}$};
		\draw [postaction={decorate}] (rin2.-45) to (m2);		
		\draw [postaction={decorate}] (rin2.-135) to (m2);
		\draw [postaction={decorate}] (m2) to node[right]{$R^{(2)}_a$} (nna);
		\draw [postaction={decorate}] (nnb) to node[right]{$R^{(2)}_a$} (n2);	
		\draw [postaction={decorate}] (n2) to node[left]{$s_{ab}^{(2)-}$} (la2);		
		\draw [postaction={decorate}] (n2) to (lb2);		
		\draw [-] (la2) to +(-0.3,0);			
		\draw [-] (lb2) to +(-0.3,0);			
		\draw [postaction={decorate}] (la2) to +(0,-0.6);
		\draw [postaction={decorate}] (lb2) to +(0,-0.6);
		\node (m3) at (2,1) [circle,draw,inner sep=0.5mm,label=right:$\nu_a^{(3)+}$] {};
		\node (n3) at (2,-1) [circle,draw,inner sep=0.5mm,label=right:$\nu_a^{(3)-}$] {};
		\node (rin3) at (2,2) {$\bigcup_{b} s_{ba}^{(3)+}$};
		\node (la3) at (1.5,-2.5) [circle,fill,inner sep=0.5mm] {};
		\node (lb3) at (2.5,-2.5) [circle,fill,inner sep=0.5mm] {};		
		\node (rout3) at (2,-3.5) {$\bigcup_{b} s_{ab}^{(3)-}$};
		\draw [postaction={decorate}] (m3) to (rin3.-45);		
		\draw [postaction={decorate}] (m3) to (rin3.-135);
		\draw [postaction={decorate}] (nna) to node[below right]{$R^{(3)}_a$} (m3);
		\draw [postaction={decorate}] (n3) to node[above right]{$R^{(3)}_a$} (nnb);	
		\draw [postaction={decorate}] (la3) to (n3);		
		\draw [postaction={decorate}] (lb3) to node[right]{$s_{ab}^{(3)-}$} (n3);		
		\draw [-] (la3) to +(0.3,0);			
		\draw [-] (lb3) to node[right]{$\Lambda_{ab}^{(3)}$} +(0.3,0);			
		\draw [postaction={decorate}] ($(la3)+(0,-0.6)$) to (la3);
		\draw [postaction={decorate}] ($(lb3)+(0,-0.6)$) to (lb3);	
	}		
	\right)
\end{equation}
As before, the trace-operation identifies and sums the corresponding  indices from $ \cup_{a,b}  s_{ba}^{(I)} $  at the 
top of the diagram to the indices from  $ \cup_{a,b}  s_{ab}^{(I)} $. 
Now we have extra Clebsch-Gordan nodes between $s_{ab}^{(I)-}$ and $s_{ab}^{(I)+}$. Note the outgoing lines next to $\Lambda_{ab}^{(I)}$ are a shorthand for the whole collection of labels $(\tau_{ab}^{(I)}, \Lambda_{ab}^{(I)}, \beta_{ab}^{(I)}, \bn_{ab}^{(I)})$ like in (\ref{eq:chiK_diag}), including the $\beta_{ab}$ white branching coefficient node.

In order to factorize this diagram we apply (\ref{eq:GLLL_deriv1}) \emph{twice}: both on $s_{ab}^{(I)-}$ legs below $\nu_a^{(I)-}$ nodes, and on $s_{ab}^{(I)+}$ legs above $\nu_a^{(I)+}$. This introduces two sums over new branching coefficients $\mu_{ab}^{+},\mu_{ab}^{-}$ (compared to just one in the last section) and splits the diagram into \emph{three} parts:
\begin{equation}
\label{eq:GKKK_deriv2}
\begin{split}
	&G(\bK^{(1)},\bK^{(2)};\bK^{(3)}) = 
	\frac{\tl f_{\bK^{(1)}\bK^{(2)}}^{\bK^{(3)}}}{ 
		\prod_a d(R_a^{(1)})d(R_a^{(2)}) 
		\prod_{a,b} d(s^{(1)-}_{ab}) d(s^{(2)-}_{ab}) d(s^{(1)+}_{ab}) d(s^{(2)+}_{ab})
	}
\\
	& \times	\sum_{\{\mu_a\}} \sum_{\{\mu_{ab}^+\}} \sum_{\{\mu_{ab}^-\}} \prod_a \left(
%%%%%%%%%%%%%%%%%%%%%%%%%%%%	
	\mytikz{		
		\node (np1) at (-1.5,0) [circle,draw,inner sep=0.5mm,label=left:$\nu_a^{(1)-}$] {};
		\node (np2) at (0,0) [circle,draw,inner sep=0.5mm,label=right:$\nu_a^{(2)-}$] {};
		\node (np3) at (1.5,0) [circle,draw,inner sep=0.5mm,label=right:$\nu_a^{(3)-}$] {};
		\node (n) at (0,1.5) [circle,draw,inner sep=0.5mm,label=above:$\mu_a$] {};
		\node (n1) at (-0.7,-1.5) [circle,draw,inner sep=0.5mm] {};		
		\node (n2) at (0,-2) {$\bigcup_{b}\mu_{ab}^-$}	;
		\node (n3) at (0.7,-1.5) [circle,draw,inner sep=0.5mm] {};
		\draw [postaction={decorate}] (n) to node[above left]{$R_a^{(1)}$} (np1);		
		\draw [postaction={decorate}] (n) to node{$R_a^{(2)}$} (np2);		
		\draw [postaction={decorate}] (np3) to node[above right]{$R_a^{(3)}$} (n);		
		\draw [postaction={decorate}] (np1) to node[left]{$\bigcup_{b} s_{ab}^{(1)-}$} (n1);				
		\draw [postaction={decorate}] (np1) to (n3);		
		\draw [postaction={decorate}] (np2) to (n1);				
		\draw [postaction={decorate}] (np2) to (n3);				
		\draw [postaction={decorate}] (n1) to (np3);				
		\draw [postaction={decorate}] (n3) to node[right]{$\bigcup_{b} s_{ab}^{(3)-}$} (np3);		
	}	
%%%%%%%%%%%%%%%%%%%%%%%%%%%%	
	\mytikz{		
		\node (np1) at (-1.5,0) [circle,draw,inner sep=0.5mm,label=left:$\nu_a^{(1)+}$] {};
		\node (np2) at (0,0) [circle,draw,inner sep=0.5mm,label=right:$\nu_a^{(2)+}$] {};
		\node (np3) at (1.5,0) [circle,draw,inner sep=0.5mm,label=right:$\nu_a^{(3)+}$] {};
		\node (n) at (0,-1.5) [circle,draw,inner sep=0.5mm,label=below:$\mu_a$] {};
		\node (n1) at (-0.7,1.5) [circle,draw,inner sep=0.5mm] {};		
		\node (n2) at (0,2) {$\bigcup_{b}\mu_{ba}^+$}	;
		\node (n3) at (0.7,1.5) [circle,draw,inner sep=0.5mm] {};
		\draw [postaction={decorate}] (np1) to node[below left]{$R_a^{(1)}$} (n);		
		\draw [postaction={decorate}] (np2) to node{$R_a^{(2)}$} (n);		
		\draw [postaction={decorate}] (n) to node[below right]{$R_a^{(3)}$} (np3);		
		\draw [postaction={decorate}] (n1) to node[left]{$\bigcup_{b} s_{ba}^{(1)+}$} (np1);				
		\draw [postaction={decorate}] (n3) to (np1);		
		\draw [postaction={decorate}] (n1) to (np2);				
		\draw [postaction={decorate}] (n3) to (np2);		
		\draw [postaction={decorate}] (np3) to (n1);				
		\draw [postaction={decorate}] (np3) to node[right]{$\bigcup_{b} s_{ba}^{(3)+}$} (n3);		
	}			
%%%%%%%%%%%%%%%%%%%%%%%%%%%%		
	\right.
	\\ & ~~~~~~~~~~~~~~~~~~~~~~~~~~~~~~~~~~~~~~~
	\left.	
%%%%%%%%%%%%%%%%%%%%%%%%%%%%		
	\mytikz{		
		\node (n1) at (-0.7,2) [circle,draw,inner sep=0.5mm] {};		
		\node (n2) at (0,2.5) {$\bigcup_{b}\mu_{ab}^-$}	;
		\node (n3) at (0.7,2) [circle,draw,inner sep=0.5mm] {};
		\node (m1) at (-0.7,-2) [circle,draw,inner sep=0.5mm] {};		
		\node (m2) at (0,-2.5) {$\bigcup_{b}\mu_{ab}^+$}	;
		\node (m3) at (0.7,-2) [circle,draw,inner sep=0.5mm] {};		
		\node (npa1) at (-2.5,0) [circle,fill,inner sep=0.5mm] {};
		\node (npb1) at (-1.5,0) [circle,fill,inner sep=0.5mm] {};
		\node (npa2) at (-0.5,0) [circle,fill,inner sep=0.5mm] {};
		\node (npb2) at (0.5,0) [circle,fill,inner sep=0.5mm] {};
		\node (npa3) at (1.5,0) [circle,fill,inner sep=0.5mm] {};				
		\node (npb3) at (2.5,0) [circle,fill,inner sep=0.5mm] {};				
		\draw [postaction={decorate}] (n1) to node[below]{$s_{ab}^{(1)-}$} (npa1);		
		\draw [postaction={decorate}] (n1) to (npa2);		
		\draw [postaction={decorate}] (npa3) to (n1);		
		\draw [postaction={decorate}] (n3) to (npb1);		
		\draw [postaction={decorate}] (n3) to (npb2);		
		\draw [postaction={decorate}] (npb3) to node[below]{$s_{ab}^{(3)-}$} (n3);		
		\draw [postaction={decorate}] (npa1) to node[above]{$s_{ab}^{(1)+}$} (m1);		
		\draw [postaction={decorate}] (npa2) to (m1);		
		\draw [postaction={decorate}] (m1) to (npa3);				
		\draw [postaction={decorate}] (npb1) to (m3);		
		\draw [postaction={decorate}] (npb2) to (m3);		
		\draw [postaction={decorate}] (m3) to node[above]{$s_{ab}^{(3)+}$} (npb3);		
		\draw [-] (npa1) to node[left]{$\Lambda_{ab}^{(1)}$} +(-0.3,0);			
		\draw [-] (npb1) to +(-0.3,0);
		\draw [-] (npa2) to +(-0.3,0);			
		\draw [-] (npb2) to node[left]{$\Lambda_{ab}^{(2)}$} +(-0.3,0);
		\draw [-] (npa3) to +(0.3,0);			
		\draw [-] (npb3) to node[right]{$\Lambda_{ab}^{(3)}$} +(0.3,0);
	}		
	\right)
%%%%%%%%%%%%%%%%%%%%%%%%%%%%
\end{split}
\end{equation}
%%%%%%%%%%%%%%%%%%%%%%%%%%%%%%%%%%%%%%%%%%%%%%%%%%%%%%%%%%%%%%%%%%%%%%%%%%%%%%%

The diagram involving $\Lambda_{ab}^{(I)}$ factorizes into a piece for each $b$, so we have (now including $\beta_{ab}$ nodes):
\begin{equation}
\label{eq:GKKK_deriv1}
	\prod_{a,b} 
	\mytikz{		
		\node (n) at (0,2) [circle,draw,inner sep=0.5mm,label=above:$\mu_{ab}^-$] {};				
		\node (m) at (0,-2) [circle,draw,inner sep=0.5mm,label=below:$\mu_{ab}^+$] {};				
		\node (l1) at (-2,0) [circle,fill,inner sep=0.5mm,label=right:$\tau_{ab}^{(1)}$] {};		
		\node (l2) at (0,0) [circle,fill,inner sep=0.5mm,label=right:$\tau_{ab}^{(2)}$] {};		
		\node (l3) at (2,0) [circle,fill,inner sep=0.5mm,label=left:$\tau_{ab}^{(3)}$] {};	
		\node (b1) at (-2.6,0) [circle,draw,inner sep=0.5mm,label=below:$\beta_{ab}^{(1)}$] {};		
		\node (b2) at (-0.6,0) [circle,draw,inner sep=0.5mm,label=below:$\beta_{ab}^{(2)}$] {};		
		\node (b3) at (2.6,0) [circle,draw,inner sep=0.5mm,label=below:$\beta_{ab}^{(3)}$] {};	
		\draw [postaction={decorate}] (n) to node[above]{$s_{ab}^{(1)-}$} (l1);		
		\draw [postaction={decorate}] (n) to node[above]{$s_{ab}^{(2)-}$} (l2);		
		\draw [postaction={decorate}] (l3) to node[above]{$s_{ab}^{(3)-}$} (n);			
		\draw [postaction={decorate}] (l1) to node[below]{$s_{ab}^{(1)+}$} (m);		
		\draw [postaction={decorate}] (l2) to node[below]{$s_{ab}^{(2)+}$} (m);		
		\draw [postaction={decorate}] (m) to node[below]{$s_{ab}^{(3)+}$} (l3);						
		\draw [-] (l1) to node[above]{$\Lambda_{ab}^{(1)}$} (b1);					
		\draw [-] (l2) to node[above]{$\Lambda_{ab}^{(2)}$} (b2);					
		\draw [-] (l3) to node[above]{$\Lambda_{ab}^{(3)}$} (b3);					
		\draw [-] (b1) to node[left]{$\bn_{ab}^{(1)}$} +(-0.3,0);					
		\draw [-] (b2) to +(-0.3,0);					
		\draw [-] (b3) to node[right]{$\bn_{ab}^{(3)}$} +(0.3,0);	
	}		
\end{equation}
Finally, we couple $\Lambda_{ab}^{(1)} \otimes \Lambda_{ab}^{(2)} \rightarrow \Lambda_{ab}^{(3)}$ by inserting the following sum
\begin{equation}
\begin{split}
	&
	\frac{1}{n_{ab}^{(1)}!n_{ab}^{(2)}!} \sum_{\gamma_1,\gamma_2}		
	\mytikz{		
		\node (n) at (0,2) [circle,draw,inner sep=0.5mm,label=above:$\mu_{ab}^-$] {};				
		\node (m) at (0,-2) [circle,draw,inner sep=0.5mm,label=below:$\mu_{ab}^+$] {};				
		\node (l1) at (-2,0) [circle,fill,inner sep=0.5mm] {};		
		\node (l2) at (0,0) [circle,fill,inner sep=0.5mm] {};		
		\node (l3) at (2,0) [circle,fill,inner sep=0.5mm] {};	
		\node (g1) at (-2.6,0) [rectangle,draw,label=above:$\Lambda_{ab}^{(1)}$] {$\gamma_1$};	
		\node (g2) at (-0.6,0) [rectangle,draw,label=above:$\Lambda_{ab}^{(2)}$] {$\gamma_2$};
		\node (g3) at (3.5,0) [rectangle,draw,label=above:$\Lambda_{ab}^{(3)}$] {$(\gamma_1 \circ \gamma_2)^{-1}$};		
		\node (b1) at (-3.3,0) [circle,draw,inner sep=0.5mm] {};		
		\node (b2) at (-1.3,0) [circle,draw,inner sep=0.5mm] {};		
		\node (b3) at (5,0) [circle,draw,inner sep=0.5mm] {};
		\draw [postaction={decorate}] (n) to node[above]{$s_{ab}^{(1)-}$} (l1);		
		\draw [postaction={decorate}] (n) to node[above]{$s_{ab}^{(2)-}$} (l2);		
		\draw [postaction={decorate}] (l3) to node[above]{$s_{ab}^{(3)-}$} (n);			
		\draw [postaction={decorate}] (l1) to node[below]{$s_{ab}^{(1)+}$} (m);		
		\draw [postaction={decorate}] (l2) to node[below]{$s_{ab}^{(2)+}$} (m);		
		\draw [postaction={decorate}] (m) to node[below]{$s_{ab}^{(3)+}$} (l3);						
		\draw [-] (l1) to node[above]{} (g1);					
		\draw [-] (l2) to node[above]{} (g2);					
		\draw [-] (l3) to node[above]{} (g3);					
		\draw [-] (g1) to node[above]{} (b1);					
		\draw [-] (g2) to node[above]{} (b2);					
		\draw [-] (g3) to node[above]{} (b3);					
		\draw [-] (b1) to +(-0.3,0);					
		\draw [-] (b2) to +(-0.3,0);					
		\draw [-] (b3) to +(0.3,0);	
	}		
\\ &
	= 
	\frac{1}{d(\Lambda_{ab}^{(1)})d(\Lambda_{ab}^{(2)})} \sum_{\mu_{ab}^\Lambda}
	%%%%%%%%%%%%%%%%%%%%%%%%%%%%	
	\mytikz{		
		\node (np1) at (-1.5,0) [circle,fill,inner sep=0.5mm,label=left:$\tau_{ab}^{(1)}$] {};
		\node (np2) at (0,0) [circle,fill,inner sep=0.5mm,label=right:$\tau_{ab}^{(2)}$] {};
		\node (np3) at (1.5,0) [circle,fill,inner sep=0.5mm,label=right:$\tau_{ab}^{(3)}$] {};
		\node (n) at (0,1.5) [circle,draw,inner sep=0.5mm,label=above:$\mu_{ab}^{-}$] {};
		\node (m) at (0,-1.5) [circle,draw,inner sep=0.5mm,label=below:$\mu_{ab}^{+}$] {};	
		\node (nl) at (2.5,1.5) [circle,draw,inner sep=0.5mm,label=above:$\mu_{ab}^{\Lambda}$] {};
		\draw [postaction={decorate}] (n) to node[above left]{$s_{ab}^{(1)-}$} (np1);		
		\draw [postaction={decorate}] (n) to node{$s_{ab}^{(2)-}$} (np2);		
		\draw [postaction={decorate}] (np3) to node[above]{$s_{ab}^{(3)-}$} (n);		
		\draw [postaction={decorate}] (np1) to node[below left]{$s_{ab}^{(1)+}$} (m);		
		\draw [postaction={decorate}] (np2) to node{$s_{ab}^{(2)+}$} (m);		
		\draw [postaction={decorate}] (m) to node[below right]{$s_{ab}^{(3)+}$} (np3);		
		\draw [-] (np1) to (nl);		
		\draw [-] (np2) to (nl);		
		\draw [-] (np3) to node[right]{$\Lambda_{ab}^{(3)}$} (nl);	
		\node (l1) at (1.7,1.6) {$\Lambda_{ab}^{(1)}$};	
		\node (l2) at (1.5,0.8) {$\Lambda_{ab}^{(2)}$};	
	}	
%%%%%%%%%%%%%%%%%%%%%%%%%%%%	
	\mytikz{				
		\node (n) at (0,0) [circle,draw,inner sep=0.5mm,label=above:$\mu_{ab}^{\Lambda}$] {};
		\node (b1) at (2,1.5) [circle,draw,inner sep=0.5mm,label=above:$\beta_{ab}^{(1)}$] {};
		\node (b2) at (2,0) [circle,draw,inner sep=0.5mm,label=above:$\beta_{ab}^{(2)}$] {};
		\node (b3) at (2,-1.5) [circle,draw,inner sep=0.5mm,label=above:$\beta_{ab}^{(3)}$] {};
		\node (n1) at (3,1.5) {$\bn_{ab}^{(1)}$};	
		\node (n2) at (3,0) {$\bn_{ab}^{(2)}$};	
		\node (n3) at (3,-1.5) {$\bn_{ab}^{(3)}$};			
		\draw [-] (n) to node[above]{$\Lambda_{ab}^{(1)}$} (b1);		
		\draw [-] (n) to node[above]{$\Lambda_{ab}^{(2)}$} (b2);		
		\draw [-] (n) to node[above]{$\Lambda_{ab}^{(3)}$} (b3);		
		\draw [-] (b1) to (n1);		
		\draw [-] (b2) to (n2);		
		\draw [-] (b3) to (n3);		
	}		
%%%%%%%%%%%%%%%%%%%%%%%%%%%%
\end{split}
\end{equation}
The diagram on the left hand side with inserted $\gamma_1,\gamma_2$ is equal to (\ref{eq:GKKK_deriv1}), due to the property of Clebsch-Gordan coefficients (\ref{eq:CG_gamma_pull}), which allows to pull $\gamma$'s through, and then cancel via $\mu_{ab}^-$ and $\mu_{ab}^+$ branching coefficients using (\ref{eq:B_gamma_pull}). Then applying (\ref{eq:g1g2_factoriz}) again we get the right hand side. Plugging this in (\ref{eq:GKKK_deriv2}) gives the final answer (\ref{eq:GKKK_diag}).

%%%%%%%%%%%%%%%%%%%%%%%%%%%%%%%%%%%%%%%%%%%%%%%%%%%%%%%%%%%%%%%%%%%%%
%%%%%%%%%%%%%%%%%%%%%%%%%%%%%%%%%%%%%%%%%%%%%%%%%%%%%%%%%%%%%%%%%%%%%

\section{Quivers and topological field theories on   Riemann surfaces }
\label{sec:surfaces}

The formulae for counting, two-point functions and chiral ring fusion coefficients derived 
in the previous sections have all been given in terms of permutations. For getting an orthogonal 
basis of operators, it has been convenient to Fourier transform from permutations 
to Young diagrams, which allow easy coding of finite $N$  relations. In this section we will
primarily focus on the large $N$  limit, where  $ n_a \le N_a  $. 
We will find that all the combinatoric data of counting and correlators we have considered so far
can be expressed neatly in terms of topological field theory on a Riemann surface 
obtained by thickening the quiver. The topological field theory (TFT) 
will be  a lattice topological field theory 
based on $S_n$. The choice of $n$  will depend on  the $ \{  n_{ab;\a}  \} $ specifying 
the numbers of fields of type $\alpha $ corresponding to each of the arrows in the quiver  starting from $a$ and ending at $b$.
A more elegant mathematical language might be to work with $S_{\infty} $  defined through an inductive limit, 
but in this paper  we will stick with a down to earth
presentation based on $S_n$, bearing in mind that the $n$ can be arbitrarily large, 
depending on the numbers of fields in the quiver gauge theory observables being considered. 
 These lattice topological 
field theories have been discussed in connection with Chern Simons theory \cite{dijkwit} 
and in connection with the large $N$ limit of two-dimensional Yang Mills in \cite{dadaprovero}. 
We will give a brief  review in the next subsection, and introduce some defect observables associated with subgroups of $S_n$. We will then show how the counting and 
correlators of large $N$ quiver gauge theories can be expressed 
with these TFTs on Riemann surfaces obtained by thickening the quiver.

\subsection{\texorpdfstring{$S_n$}{Sn} topological lattice gauge theory and defect operators} 

The partition function on a surface is defined by starting with a triangulation on the surface, or a more generally
 a cell-decomposition
where the 2-cells can be polygons. We associate $S_n$  group variables with each edge, and a weight function for each 
cell (or plaquette). The weight is the $  \delta ( \sigma_P ) $, where $\sigma_P$ is 
the product of group elements around the plaquette. We will call this the plaquette weight 
$Z_{P} $ 
\begin{equation}\begin{split}\label{ZedP}  
Z_{P}  ( \sigma_P )  =    \delta ( \sigma_P )  = \sum_{ R \vdash n }  { d ( R) \over n! } \chi_R ( \sigma_P ) 
\end{split}\end{equation}
The sum is over all Young diagrams with $n$ boxes, equivalently all irreps of $S_n$. 
The partition function of the manifold is defined as 
\begin{equation}\begin{split}\label{defpartfn}  
Z = { 1 \over (n!)^V  } \sum_{ \sigma_1 , \cdots , \sigma_E \in S_n } \prod_P  Z_{P}  ( \sigma_P ) 
\end{split}\end{equation}
where $V$ is the number of vertices in the triangulation. 
The topological property follows from an invariance under refinement, or conversely coarsening, 
of the cell decomposition. If we sum over an edge variable between two cells, as shown in Figure~\ref{fig:top1}, 
thus eliminating the 
edge and fusing the two cells $P_1$ and $P_2$  into a single cell $P$, we have
\begin{equation}
\begin{split}
 \sum_{ \sigma } Z_{P_1} Z_{P_2} &= \sum_{ \gamma   } \delta ( \sigma_1 \gamma \sigma_5 \sigma_6 )
 \delta ( \sigma_2 \sigma_3  \sigma_4 \gamma^{-1} ) \\
&  = \delta ( \sigma_1  \sigma_2 \sigma_3  \sigma_4 \sigma_5 \sigma_6 ) \\ 
& = Z_P 
\end{split}
\end{equation}
We denote this topological invariance property TOP1. 

\begin{figure}[h]
\begin{center}
$
\mytikz{
	\node (v1) at (-1,1) [circle,fill,inner sep=0.5mm] {};
	\node (v2) at (1,1) [circle,fill,inner sep=0.5mm] {};
	\node (v3) at (-1,0) [circle,fill,inner sep=0.5mm] {};
	\node (v4) at (1,0) [circle,fill,inner sep=0.5mm] {};
	\node (v5) at (-1,-1) [circle,fill,inner sep=0.5mm] {};
	\node (v6) at (1,-1) [circle,fill,inner sep=0.5mm] {};	
	\node (p2) at (0,0.5) {$P_2$};	
	\node (p1) at (0,-0.5) {$P_1$};
	\draw[postaction={decorate}] (v1) to node[above]{$\s_3$} (v2);
	\draw[postaction={decorate}] (v2) to node[right]{$\s_4$} (v4);
	\draw[postaction={decorate}] (v4) to node[right]{$\s_5$} (v6);
	\draw[postaction={decorate}] (v6) to node[below]{$\s_6$} (v5);
	\draw[postaction={decorate}] (v5) to node[left]{$\s_1$} (v3);
	\draw[postaction={decorate}] (v3) to node[left]{$\s_2$} (v1);
	\draw[postaction={decorate}] (v3) to node[right]{$\gamma$} (v4);
	\draw[-] (v1) to +(-0.5,0.5);
	\draw[-] (v2) to +(0.5,0.5);
	\draw[-] (v6) to +(0.5,-0.5);
	\draw[-] (v5) to +(-0.5,-0.5);
}
\quad \rightarrow \quad
\mytikz{
	\node (v1) at (-1,1) [circle,fill,inner sep=0.5mm] {};
	\node (v2) at (1,1) [circle,fill,inner sep=0.5mm] {};
	\node (v3) at (-1,0) [circle,fill,inner sep=0.5mm] {};
	\node (v4) at (1,0) [circle,fill,inner sep=0.5mm] {};
	\node (v5) at (-1,-1) [circle,fill,inner sep=0.5mm] {};
	\node (v6) at (1,-1) [circle,fill,inner sep=0.5mm] {};	
	\node (p1) at (0,0) {$P$};
	\draw[postaction={decorate}] (v1) to node[above]{$\s_3$} (v2);
	\draw[postaction={decorate}] (v2) to node[right]{$\s_4$} (v4);
	\draw[postaction={decorate}] (v4) to node[right]{$\s_5$} (v6);
	\draw[postaction={decorate}] (v6) to node[below]{$\s_6$} (v5);
	\draw[postaction={decorate}] (v5) to node[left]{$\s_1$} (v3);
	\draw[postaction={decorate}] (v3) to node[left]{$\s_2$} (v1);
	\draw[-] (v1) to +(-0.5,0.5);
	\draw[-] (v2) to +(0.5,0.5);
	\draw[-] (v6) to +(0.5,-0.5);
	\draw[-] (v5) to +(-0.5,-0.5);
}
$
\end{center}
\caption{TOP1}
\label{fig:top1}
\end{figure}

In a configuration such as shown in Figure~\ref{fig:top2}, where a bivalent vertex has holonomies $ \s_1 , \s_2$ 
on the two sides, the only combination appearing in the partition function is  $ \sigma_1 \sigma_2$. Hence  
we can rename $ \sigma_1  \rightarrow \sigma_1 \sigma_2$, do the sum over $\sigma_2 $ to get rid of 
a factor of $1/n!$, corresponding to the removal of the vertex. This leaves the partition function (\ref{defpartfn})
 invariant. We will call this topological invariance property TOP2.  
 
\begin{figure}[h]
\begin{center}
$
\mytikz{
	\node (v1) at (-1,0) [circle,fill,inner sep=0.5mm] {};
	\node (v2) at (0,0) [circle,fill,inner sep=0.5mm] {};
	\node (v3) at (1,0) [circle,fill,inner sep=0.5mm] {};	
	\draw[postaction={decorate}] (v1) to node[above]{$\s_1$} (v2);
	\draw[postaction={decorate}] (v2) to node[above]{$\s_2$} (v3);	
	\draw[-] (v1) to +(-0.5,0.5);
	\draw[-] (v3) to +(0.5,0.5);
	\draw[-] (v3) to +(0.5,-0.5);
	\draw[-] (v1) to +(-0.5,-0.5);
}
\quad \rightarrow \quad
\mytikz{
	\node (v1) at (-1,0) [circle,fill,inner sep=0.5mm] {};
	\node (v3) at (1,0) [circle,fill,inner sep=0.5mm] {};	
	\draw[postaction={decorate}] (v1) to node[above]{$\s=\s_1 \s_2$} (v3);	
	\draw[-] (v1) to +(-0.5,0.5);
	\draw[-] (v3) to +(0.5,0.5);
	\draw[-] (v3) to +(0.5,-0.5);
	\draw[-] (v1) to +(-0.5,-0.5);
}
$
\end{center}
\caption{TOP2}
\label{fig:top2}
\end{figure}

 Given a cell decomposition of a Riemann surface of genus $G$, we can use the topological invariance 
 (TOP1 and TOP2)   to  coarsen it to a single cell with $2G$ edges, and a single vertex. 
 The partition function is thus 
 \begin{equation}\begin{split} 
 Z ( \Sigma_G ) = { 1 \over n! } \sum_{ s_1 , t_1 , \cdots , s_G , t_G \in S_n } \delta ( s_1 t_1 s_1^{-1} t_1^{-1} s_2 t_2 s_2^{-1} t_2^{-1}  \cdots s_G t_G s_G^{-1} t_G^{-1}  ) 
 \end{split}\end{equation}
Manipulations which are familiar  in the large N expansion of $U(N)$  2dYM (see \cite{grta} or the review \cite{cmr})
 show that this can be expressed in terms 
of characters 
\begin{equation}\begin{split} 
Z ( \Sigma_G )  = \sum_{ R \vdash n } \left( {  d (R )  \over n! }\right)^{ 2 - 2G } 
\end{split}\end{equation}
 This formula for $S_n$ topological field theory can also be arrived at by considering the Frobenius algebra
 of conjugacy classes of $S_n$ (see e.g \cite{FHK}) or by building up  the Riemann surface from pants diagrams 
 \cite{witten2dgauge}.)  It has an interpretation in terms of covers of $\Sigma_G$ summed 
with inverse automorphism factor.

The case with  boundaries will be of particular interest. We will express the construction of
the boundary observables in a way which allows some generalized defect observables which we will need. 
For a genus $G$ surface with $B$  boundaries, we will choose a base point on each boundary and  associate  
permutations $ \s_i$ to that boundary. We require the cell-decomposition to include the boundary vertices (0-cells) 
among its vertices, and the boundary circles among its 1-cells.  For the case of the three-holed sphere we can choose another base-point in  the middle and extend edges to the boundary vertices. 
Cutting along these edges  gives a contractible 2-cell. Denoting the permutations associated with the edges as 
$\gamma_1 , \gamma_2 , \gamma_3$ we get the partition function 
\begin{equation}\begin{split}\label{B33gam}
Z ( G = 0  , B=3 ) & = { 1 \over n! } \sum_{ \g_1 , \g_2 , \g_3 \in S_n } \delta ( \gamma_1 \s_1 \gamma_1^{-1}  \gamma_2 \s_2 \gamma_2^{-1}  \gamma_3 \s_3 \gamma_3^{-1}  )  \\  
& = \sum_{ R \vdash  n } { \chi_R ( \s_1 ) \chi_R ( \s_2  ) \chi_R ( \s_3 )  \over d_R } 
\end{split}\end{equation}
The factor of $1/n! $ is due to the interior vertex, no such factors are introduced for the vertices used in the definition of the boundary observables. We could also have chosen not to include an interior vertex, and just draw two lines joining 
the vertex from one boundary circle to the vertices on the other boundary circles. Then, the definition (\ref{defpartfn})
leads us to write 
\begin{equation}\begin{split}\label{B32gam}  
Z ( G = 0  , B=3 ) & =  \sum_{ \g_1 , \g_2  \in S_n } \delta ( \gamma_1 \s_1 \gamma_1^{-1}  \gamma_2 \s_2 \gamma_2^{-1} \s_3  )
\end{split}\end{equation}
This is the same as the expression above (\ref{B33gam}), since the $\gamma_3$  can be absorbed into a  re-defintion of 
the summation variables $ \g_1 , \g_2$. 
The expression in terms of characters is  a special case of the standard result 
\begin{equation}\begin{split} 
Z ( G , B, \s_1 , \cdots , \s_B  ) & = { 1 \over n! } \sum_{ \g_1 , \cdots , \g_B  \in S_n }
 \sum_{ s_i , t_i  } \delta ( \prod_{ i=1}^G  s_i t_i s_i^{-1} t_i^{-1} \cdot \prod_{i=1}^B \g_i \s_i \g_i^{-1}  ) \\ 
 & = \sum_{ R \vdash n  }  \left( {  d( R ) \over n! } \right)^{ 2 - 2G - B  } \prod_{ i=1}^B \chi_R ( \s_i ) 
  \end{split}\end{equation}
in the normalization that appears naturally from the large $N$ expansion 
of two dimensional Yang Mills theory \cite{cmr} and which has a natural covering space interpretation.

We will now describe some more general observables in 2D  $S_n$ lattice gauge theory. 
A {\it closed $H$-defect}  will be a closed non-self-intersecting loop on the surface, equipped with the choice of a point 
on the loop. The insertion of the defect in the partition function amounts to constraining the permutation 
sums in  (\ref{defpartfn}) to require that the permutation associated with the loop is in the subgroup $H  \subset S_n$, 
and the sum over elements in $H$ is weighted by ${ 1 \over | H | } $, the inverse order of the subgroup. 
In the presence of the defect,  the cell decomposition  used to calculate the partition function must include the 
loop among its 1-cells (possibly as a composite of smaller 1-cells), and the point on the loop should be among the vertices 
of the cell decomposition.  As a  simple example, consider the circle shown in Figure~\ref{fig:torusH}. The topological properties TOP1 and TOP2 of the 
partition function ensure that we can choose a very simple cell decomposition compatible with the insertion of the defect. This is shown in 
Figure~\ref{fig:torusHcell}.  There is one 2-cell bounded by 1-cells carrying the permutations $ \gamma \sigma \gamma^{-1} \sigma^{-1}$. 
This gives 
\begin{equation}\begin{split}\label{DH} 
Z ( T^2 ;  D_H ) =  { 1 \over |H| } \sum_{ \gamma \in H } \sum_{ \sigma \in S_n } \delta ( \gamma \sigma \gamma^{-1} \sigma^{-1} )
\end{split}\end{equation}

\begin{figure}[h]
\centering
\begin{minipage}{3.0in}
	\centering
	\mytikz{
	\begin{scope}[scale=0.7]
		\draw (-3.5,0) .. controls (-3.5,1.5) and (-1.5,2) .. (0,2);
		\draw[xscale=-1] (-3.5,0) .. controls (-3.5,1.5) and (-1.5,2) .. (0,2);
		\draw[scale=-1] (-3.5,0) .. controls (-3.5,1.5) and (-1.5,2) .. (0,2);
		\draw[yscale=-1] (-3.5,0) .. controls (-3.5,1.5) and (-1.5,2) .. (0,2);
		\draw (-2,0.2) .. controls (-1.5,-0.3) and (-1,-0.5) .. (0,-0.5) .. controls (1,-0.5) and (1.5,-0.3) .. (2,0.2);
		\draw (-1.75,0) .. controls (-1.5,0.3) and (-1,0.5) .. (0,0.5) .. controls (1,0.5) and (1.5,0.3) .. (1.75,0);
		\draw (0,-0.5) .. controls +(-0.5,0) and +(-0.5,0) .. node[left]{$H$} (0,-2);
		\draw[dashed] (0,-0.5) .. controls +(0.5,0) and +(0.5,0) .. (0,-2);
		\node (v1) at (-0.37,-1.38) [circle,fill,inner sep=0.5mm] {};
	\end{scope}
	}
	\caption{Torus with defect $H$}
	\label{fig:torusH}
\end{minipage}
\begin{minipage}{3.0in}
	\centering
	\mytikz{
	\begin{scope}[scale=0.7]
	\draw (-3.5,0) .. controls (-3.5,1.5) and (-1.5,2) .. (0,2);
	\draw[xscale=-1] (-3.5,0) .. controls (-3.5,1.5) and (-1.5,2) .. (0,2);
	\draw[rotate=180] (-3.5,0) .. controls (-3.5,1.5) and (-1.5,2) .. (0,2);
	\draw[yscale=-1] (-3.5,0) .. controls (-3.5,1.5) and (-1.5,2) .. (0,2);
	\draw (-2,0.2) .. controls (-1.5,-0.3) and (-1,-0.5) .. (0,-0.5) .. controls (1,-0.5) and (1.5,-0.3) .. (2,0.2);
	\draw (-1.75,0) .. controls (-1.5,0.3) and (-1,0.5) .. (0,0.5) .. controls (1,0.5) and (1.5,0.3) .. (1.75,0);
	\draw (0,-0.5) .. controls +(-0.5,0) and +(-0.5,0) .. node[above left]{$\g$} (0,-2);
	\draw[dashed,postaction={decorate}] (0,-0.5) .. controls +(0.5,0) and +(0.5,0) .. (0,-2);
	\draw[xscale=0.8,yscale=0.7] (-3.5,0) .. controls (-3.5,1.5) and (-1.5,2) .. (0,2);
	\draw[xscale=-0.8,yscale=0.7,postaction={decorate}] (-3.5,0) .. controls (-3.5,1.5) and (-1.5,2) .. (0,2);
	\draw[xscale=-0.8,yscale=-0.7] (-3.5,0) .. controls (-3.5,1.5) and (-1.5,2) .. node[below]{$\s$} (0,2);
	\draw[xscale=0.8,yscale=-0.7] (-3.5,0) .. controls (-3.5,1.5) and (-1.5,2) .. (0,2);
	\node (v1) at (-0.37,-1.38) [circle,fill,inner sep=0.5mm] {};
	\end{scope}
	}
	\caption{ Cell decomposition : one $2$-cell}
	\label{fig:torusHcell}
\end{minipage}
\end{figure}

\begin{figure}[h]
\centering
\begin{minipage}{3.0in}
	\centering
	\mytikz{
	\begin{scope}[scale=0.7]
	% torus outer
	\draw (-3.5,0) .. controls (-3.5,1.5) and (-1.5,2) .. (0,2);
	\draw[xscale=-1] (-3.5,0) .. controls (-3.5,1.5) and (-1.5,2) .. (0,2);
	\draw[rotate=180] (-3.5,0) .. controls (-3.5,1.5) and (-1.5,2) .. (0,2);
	\draw[yscale=-1] (-3.5,0) .. controls (-3.5,1.5) and (-1.5,2) .. (0,2);
	% torus inner
	\draw (-2,0.2) .. controls (-1.5,-0.3) and (-1,-0.5) .. (0,-0.5) .. controls (1,-0.5) and (1.5,-0.3) .. (2,0.2);
	\draw (-1.75,0) .. controls (-1.5,0.3) and (-1,0.5) .. (0,0.5) .. controls (1,0.5) and (1.5,0.3) .. (1.75,0);
	% H1
	\draw (0,-0.5) .. controls +(-0.5,0) and +(-0.5,0) .. node[left]{$H_1$} (0,-2);
	\draw[dashed] (0,-0.5) .. controls +(0.5,0) and +(0.5,0) .. (0,-2);
	% H2
	\draw (0,0.5) .. controls +(-0.5,0) and +(-0.5,0) .. node[left]{$H_2$} (0,2);
	\draw[dashed] (0,0.5) .. controls +(0.5,0) and +(0.5,0) .. (0,2);
	\node (v1) at (-0.37,-1.38) [circle,fill,inner sep=0.5mm] {};
	\node (v2) at (-0.37,1.38) [circle,fill,inner sep=0.5mm] {};
	\end{scope}
	}
	\caption{Torus with defects $H_1, H_2$}
	\label{fig:torusH1H2}
\end{minipage}
\begin{minipage}{3.0in}
	\centering
	\mytikz{
	\begin{scope}[scale=0.7]
	% torus outer
	\draw (-3.5,0) .. controls (-3.5,1.5) and (-1.5,2) .. (0,2);
	\draw[xscale=-1] (-3.5,0) .. controls (-3.5,1.5) and (-1.5,2) .. (0,2);
	\draw[rotate=180] (-3.5,0) .. controls (-3.5,1.5) and (-1.5,2) .. (0,2);
	\draw[yscale=-1] (-3.5,0) .. controls (-3.5,1.5) and (-1.5,2) .. (0,2);
	% torus inner
	\draw (-2,0.2) .. controls (-1.5,-0.3) and (-1,-0.5) .. (0,-0.5) .. controls (1,-0.5) and (1.5,-0.3) .. (2,0.2);
	\draw (-1.75,0) .. controls (-1.5,0.3) and (-1,0.5) .. (0,0.5) .. controls (1,0.5) and (1.5,0.3) .. (1.75,0);
	% g1
	\draw (0,-0.5) .. controls +(-0.5,0) and +(-0.5,0) .. node[above left]{$\g_1$} (0,-2);
	\draw[dashed,postaction={decorate}] (0,-0.5) .. controls +(0.5,0) and +(0.5,0) .. (0,-2);
	% g2
	\draw (0,0.5) .. controls +(-0.5,0) and +(-0.5,0) .. node[below left]{$\g_2$} (0,2);
	\draw[dashed,postaction={decorate}] (0,0.5) .. controls +(0.5,0) and +(0.5,0) .. (0,2);
	% sigma circle
	\draw[xscale=0.8,yscale=0.7] (-3.5,0) .. controls (-3.5,1.5) and (-1.5,2) .. (0,2);
	\draw[xscale=-0.8,yscale=0.7,postaction={decorate}] (-3.5,0) .. controls (-3.5,1.5) and (-1.5,2) .. (0,2);
	\draw[xscale=-0.8,yscale=-0.7] (-3.5,0) .. controls (-3.5,1.5) and (-1.5,2) .. node[below]{$\s_1$} (0,2);
	\draw[xscale=0.8,yscale=-0.7,postaction={decorate}] (-3.5,0) .. controls (-3.5,1.5) and (-1.5,2) .. node[below]{$\s_2$} (0,2);
	% nodes
	\node (v1) at (-0.37,-1.38) [circle,fill,inner sep=0.5mm] {};
	\node (v2) at (-0.37,1.38) [circle,fill,inner sep=0.5mm] {};
	\end{scope}
	}
	\caption{ Cell decomposition: $2$ $2$-cells}
	\label{fig:torusH1H2cell}
\end{minipage}
\end{figure}

Refinements of the cell decomposition will give the same answer.
We will see shortly how this partition function comes up in counting BPS operators in $\cN =4 $ SYM. 
If we have two of these closed $H$-defects, with subgroups $H_1, H_2 $, along 
parallel circles on a torus (Figure~\ref{fig:torusH1H2}), we can compute the partition function 
by  introducing one circle transverse to the two defects (Figure~\ref{fig:torusH1H2cell}). 
There are now two 2-cells in the cell decomposition. The partition function with these defect insertions is 
\begin{equation}\begin{split}\label{DH1DH2}  
Z ( T^2  ; D_{H_1} , D_{H_2}  ) = { 1 \over |H_1|    | H_2| }  \sum_{ \gamma_1 \in H_1 } \sum_{ \gamma_2 \in H_2 }  \sum_{ \sigma_1 ,  \sigma_2 \in S_n } \delta ( \gamma_1 \sigma_1 \gamma_2^{-1} \sigma_1^{-1} ) 
                              \delta ( \gamma_2 \sigma_2 \gamma_1^{-1} \sigma_2^{-1} )  
\end{split}\end{equation}
We will shortly see applications of this formula  to the counting of chiral operators for the conifold. 

These closed $H$-defects can also be inserted at the boundaries of Riemann surfaces. 
If we insert an $H_1$-defect at one end of a cylinder ($ S^1 \times I $) and an $H_2$ defect at the other end, 
the partition function is 
\begin{equation}\begin{split} 
Z ( S^1 \times I ; D_{H_1} , D_{H_2} ) = { 1 \over |H_1 | |H_2| } \sum_{ \gamma_1 \in H_1 } \sum_{ \gamma_2 \in H_2 } 
\sum_{ \sigma \in S_n } 
\delta ( \gamma_1 \sigma \gamma_2 \sigma^{-1} ) 
\end{split}\end{equation}
By choosing $H_1$ and $H_2$ to be appropriate wreath products, this partition function was shown to count Feynman graphs
in  \cite{feynmanstrings}. This formula arose because Feynman graphs can be out in one-to-one correspondence with
points in a double coset, which consists of  permutations $\sigma$, subject to equivalences defined by left and right multiplication 
by permutations $ \gamma_1 , \gamma_2$  in $H_1 , H_2$ respectively.  The symmetry factor of a fixed Feynman diagram 
was computed by fixing the permutation $\sigma$ along a line joining the two ends of the cylinder.

\begin{figure}[h]
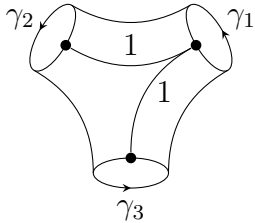

\centering
$
\mytikz{
	\def\R{1.2}
	\def\rx{0.5}
	\def\ry{0.2}
	\draw[rotate=90,postaction={decorate}] (-\R,0) ellipse ({\ry} and {\rx});
	\draw[rotate=210,postaction={decorate}] (-\R,0) ellipse ({\ry} and {\rx});
	\draw[rotate=330,postaction={decorate}] (-\R,0) ellipse ({\ry} and {\rx});
	\draw ($(0,0)+(30:\R)+(-60:\rx)$) to [bend right=30] ($(0,0)+(-90:\R)+(-180:-\rx)$);
	\draw ($(0,0)+(150:\R)+(60:\rx)$) to [bend right=30] ($(0,0)+(30:\R)+(-60:-\rx)$);
	\draw ($(0,0)+(-90:\R)+(-180:\rx)$) to [bend right=30] ($(0,0)+(150:\R)+(60:-\rx)$);
	\node (v1) at ($(0,0)+(30:\R)+(-150:\ry)$) [circle,fill,inner sep=0.5mm] {};
	\node (v2) at ($(0,0)+(150:\R)+(-30:\ry)$) [circle,fill,inner sep=0.5mm] {};
	\node (v3) at ($(0,-\R)+(0,\ry)$) [circle,fill,inner sep=0.5mm] {};
	\draw (v1) to [bend left=30] node[above]{$1$} (v2);
	\draw (v3) to [bend left=30] node[right]{$1$} (v1);
	\node (s2) at ($(0,0)+(30:\R)+(30:0.5)$) {$\g_1$};
	\node (s1) at ($(0,0)+(150:\R)+(150:0.5)$) {$\g_2$};
	\node (s3) at ($(0,0)+(270:\R)+(270:0.5)$) {$\g_3$};
}
$
\caption{3-holed sphere with defect, imposing $\gamma_3^{-1} = \g_1\g_2$}
\label{fig:3sphere}
\end{figure}

This introduces us to the second type of defect we will need here. It is a line joining two distinct points, with the 
associated permutation fixed. We will call this a {\it open Wilson line} defect. 
 In the applications to chiral operator counting for quiver thories, we will often use
 open-line defects, where the permutation is fixed to be the identity, which we can call
 {\it unit  open Wilson line} defects.
In particular we will consider a
3-holed sphere, with the permutation associated with line,  shown on the left,  
constrained to be the identity (Figure~\ref{fig:3sphere}). 
If we have permutations $ \g_1 , \g_2 , \g_3 $ at the boundaries, all measured according to the orientation induced 
on the boundaries by the orientation of the surface  and we introduce   the 
unit open Wilson lines shown  (left Figure in \ref{fig:3sphere}), then the partition function
 (equation \ref{defpartfn}) is non-vanishing  provided 
\begin{equation}\begin{split} 
\g_1 \g_2  \g_3   =1 
\end{split}\end{equation}
so that $ \g_3^{-1}  =  ( \g_1 \g_2  )  $. 
This is to be contrasted with the 3-holed sphere partition 
function without these defect insertions  (\ref{B33gam}) where 
$\g_3 $ can be any permutation that appears when taking the product of a permutation in the conjugacy class of $\g_2 $ 
and a permutation in the conjugacy class of $\g_1$.  An obvious generalization is to consider a $k+1$ holed sphere
with unit open Wilson lines joining a point on the $k+1$'th boundary to points on the $k$ boundaries. Then we will have 
\begin{equation}\begin{split} 
\g_{k+1}^{-1}  =  ( \prod_{i=1}^k \gamma_{i} )     
\end{split}\end{equation}

\noindent
This type of 3-holed vertex shows up in $G$-equivariant TFT for $G=S_n$ 
\cite{turaev,mooseg}. We will comment more on this in Section \ref{sec:funcov}. 

\subsection{Counting, correlators and defects in \texorpdfstring{$S_n$}{Sn} TFT }

\begin{figure}[h]
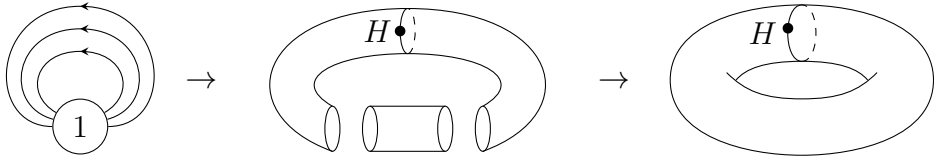

\centering
$
\mytikz{ 
	\node (s1) at (0,0) [circle,draw] {$1$};							
	\draw [postaction={decorate}] (s1.30) .. controls +(30:0.5) and +(0.6,0) .. (0,1)
		.. controls +(-0.6,0) and +(150:0.5) .. (s1.150);
	\draw [postaction={decorate}] (s1.15) .. controls +(15:0.7) and +(0.9,0) .. (0,1.3)
		.. controls +(-0.9,0) and +(165:0.7) .. (s1.165);
	\draw [postaction={decorate}] (s1.0) .. controls +(0:0.9) and +(1.2,0) .. (0,1.6)
		.. controls +(-1.2,0) and +(180:0.9) .. (s1.180);		
}
\rightarrow
\mytikz{
	\draw (-1,0) ellipse (0.1 and 0.3);
	\draw (1,0) ellipse (0.1 and 0.3);
	\draw (-1,0.3) .. controls +(150:0.7) and +(180:1) .. (0,1);
	\draw (-1,-0.3) .. controls +(160:1.5) and +(180:2) .. (0,1.6);
	\draw[xscale=-1] (-1,0.3) .. controls +(150:0.7) and +(180:1) .. (0,1);
	\draw[xscale=-1] (-1,-0.3) .. controls +(160:1.5) and +(180:2) .. (0,1.6);
	\draw (-0.5,0) ellipse (0.1 and 0.3);
	\draw (0.5,0) ellipse (0.1 and 0.3);
	\draw (-0.5,0.3) to (0.5,0.3);
	\draw (-0.5,-0.3) to (0.5,-0.3);
	\draw (0,1.6) arc (90:270:0.1 and 0.3);	
	\draw[dashed] (0,1) arc (-90:90:0.1 and 0.3);
	\node at (-0.4,1.3) {$H$};
	\node at (-0.1,1.3) [circle,fill,inner sep=0.5mm] {};
}
\rightarrow \quad
\mytikz{
	\begin{scope}[scale=0.5]
	% torus outer
	\draw (-3.5,0) .. controls (-3.5,1.5) and (-1.5,2) .. (0,2);
	\draw[xscale=-1] (-3.5,0) .. controls (-3.5,1.5) and (-1.5,2) .. (0,2);
	\draw[rotate=180] (-3.5,0) .. controls (-3.5,1.5) and (-1.5,2) .. (0,2);
	\draw[yscale=-1] (-3.5,0) .. controls (-3.5,1.5) and (-1.5,2) .. (0,2);
	% torus inner
	\draw (-2,0.2) .. controls (-1.5,-0.3) and (-1,-0.5) .. (0,-0.5) .. controls (1,-0.5) and (1.5,-0.3) .. (2,0.2);
	\draw (-1.75,0) .. controls (-1.5,0.3) and (-1,0.5) .. (0,0.5) .. controls (1,0.5) and (1.5,0.3) .. (1.75,0);
	% H2
	\draw (0,0.5) .. controls +(-0.5,0) and +(-0.5,0) .. node[left]{$H$} (0,2);
	\draw[dashed] (0,0.5) .. controls +(0.5,0) and +(0.5,0) .. (0,2);
	\node (v2) at (-0.37,1.38) [circle,fill,inner sep=0.5mm] {};
	\end{scope}
}
$
\caption{Transforming $\mC^3$ quiver into a Riemann surface}
\label{fig:c3gluing}
\end{figure}

Consider the number of chiral operators for $\mC^3$.  We specify  $ (n_1 , n_2 , n_3 )$ as the number of 
$X , Y , Z $.  We will work in the region $ n =  n_1 + n_2 + n_3 \le N$. The operators are parameterized by the 
permutation $ \sigma $ which relates the upper and lower indices of these chiral fields, transforming in the 
fundamental and anti-fundamental of $U(N)$.  Two permutations $\sigma, \sigma'  \in S_n $  give the same operator 
if there is some $ \gamma \in S_{n_1} \times S_{n_2} \times S_{n_3} $ relating  them  as $ \sigma' = \gamma  \sigma \gamma^{-1} $. 
In the region $ n \le N$, there are no additional finite $N$ relations. The finite $N$ relations  were considered in Section~\ref{sec:constops}
and solved using a representation theoretic basis involving Young diagrams and branching coefficients. Here we focus 
on the large $N$ limit and associated geometry.  An easy way to count the permutation $\sigma $ subject 
to the specified equivalence is to use the Burnside Lemma for  group actions (see e.g. \cite{Cameron}). This gives the 
number of orbits as the average number of fixed points of the group action. Applied to the case at hand, we have  
\begin{equation}\begin{split} 
\cN_{\mC^3}  ( n_1 , n_2 , n_3 ) = { 1 \over n_1! n_2! n_3 ! } 
 \sum_{ \gamma \in S_{n_1} \times S_{n_2} \times S_{n_3 } }\sum_{ \sigma \in S_n  } \delta ( \gamma \sigma \gamma^{-1} \sigma^{-1}  )
\end{split}\end{equation} 
This is of the form  (\ref{DH}) with $  H = S_{n_1} \times S_{n_2} \times S_{n_3} $ and $n = n_1 + n_2 + n_3 $
\begin{equation}\begin{split}
\cN_{\mC^3}  ( n_1 , n_2 , n_3 )  = Z ( T^2 ; D_H ) \qquad \qquad H = S_{n_1} \times S_{n_2} \times S_{n_3} 
\end{split}\end{equation}
There is a simple relation between the quiver of $\mC^3$ and the $T^2$  Riemann surface, with defect, that we have ended up with. 
Take the three edges of the quiver and collapse them to a single one. In general, we will use the operation of collapsing
all the edges having the same start and end-points to a single one. Take a cylinder corresponding to the node and a cylinder 
corresponding to the edge of the quiver. Insert  an $H$- defect around the cylinder corresponding to the edge. Glue the cylinders
 together. This is illustrated in Figure~\ref{fig:c3gluing}.

Using (\ref{eq:OQ_sig_2pt}) (alternatively see \cite{BHR1}) 
The two-point function for the $\mC^3 $ case is 
\begin{equation}\begin{split} 
\langle \cO_{ \sigma } \cO_{\tilde \sigma }^{\dagger}  \rangle = \sum_{\tau  \in S_n } 
 \sum_{ \gamma \in H } \delta ( \gamma \sigma  \gamma^{-1}  \tilde \sigma^{-1}\tau  )  N^{ C_{\tau} }  
\end{split}\end{equation}
We can modify the surface in Figure~\ref{fig:torusHcell}  to arrive 
at the surface, where the  $S_n$ TFT computes the correlator. 
Replace the loop labelled $\sigma$  by a pair of loops related to the original loop by deforming slightly away in opposite directions. 
Cut out the region  between them. Label these $\s  , \tilde \s $. Cut out another hole based at the same point and insert 
the sum $\sum_{\tau \in S_n } N^{C_{\tau} } \tau$ at the boundary of the hole. 
This is shown in Figure~\ref{fig:c3holes}. Traversing round the 2-cell gives the contractible path associated with $\gamma \sigma  \gamma^{-1}  \tilde \sigma \tau$. Hence the two-point function is the $S_n$-TFT partition functions 
associated with the surface shown in Figure~\ref{fig:c3holes}. The cut-out regions are shaded 
and there is  an observable $ \sum_{ \tau } N^{C_{\tau } }  \tau $ 
inserted at the loop labelled $\tau $.

\begin{figure}[h]
\centering
$
\mytikz{ 		
	\fill [black!20!white] (-1,-1) to [bend left=20](1,-1);
	\fill [black!20!white] (-1,1) to [bend right=20](1,1);
	\fill [black!20!white] (v1) .. controls +(-30:1.5) and +(-80:1.5) .. (v1);
	\node (v1) at (-1,1) [circle,fill,inner sep=0.5mm] {};
	\node (v2) at (1,1) [circle,fill,inner sep=0.5mm] {};
	\node (v3) at (1,-1) [circle,fill,inner sep=0.5mm] {};
	\node (v4) at (-1,-1) [circle,fill,inner sep=0.5mm] {};		
	\draw[dashed] (v1) to (v2);
	\draw[dashed] (v3) to (v4);	
	\draw [postaction={decorate}] (v2) to node[right]{$\gamma$} (v3);
	\draw [postaction={decorate}] (v1) to node[left]{$\gamma$} (v4);
	\draw [postaction={decorate}] (v4) to [bend left=20] node[above right]{$\sigma$} (v3);
	\draw [postaction={decorate}] (v1) to [bend right=20] node[below right]{$\tl\sigma$} (v2);	
	\draw [postaction={decorate}] (v1) .. controls +(-30:1.5) and +(-80:1.5) .. node[below]{$\tau $} (v1);	
}
$
\caption{$\mC^3$ torus with holes}
\label{fig:c3holes}
\end{figure}

The three-point function for the $\mC^3$ case (see equation \ref{prodops})  is 
\begin{equation}\begin{split} 
\langle \cO_{ \sigma^{(1)}  } \cO_{\sigma^{(2)}  }  ( \cO_{\tilde \sigma} )^{\dagger} \rangle 
= \sum_{\tau  \in S_n } \sum_{ \gamma \in H }  \delta ( \gamma  ( \sigma^{(1)}   \circ \sigma^{(2)}  ) 
\gamma^{-1} \tilde \sigma^{-1}\tau  )  N^{ C_{\tau } }  
\end{split}\end{equation}
Here $\s_1 \in S_{n^{(1)} } , \s_2 \in S_{n^{(2)}} , \tilde \s \in S_{n^{(1)} + n^{(2)} } $
and we defined $n = n^{(1)} + n^{(2)}$. This is computed by $S_n$ TFT on the same surface 
shown in Figure~\ref{fig:c3holes}, but with the  boundary permutation $\sigma $ replaced by $ \sigma^{(1)}  \circ \sigma^{(2)}  $. 

\begin{figure}[h]
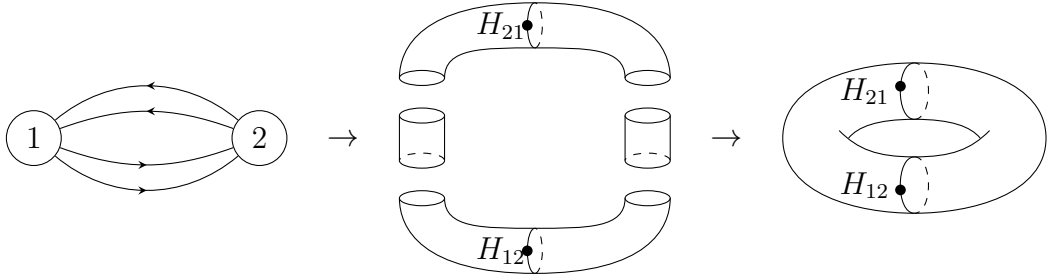

\centering
$
\mytikz{
	\node (s1) at (-1.5,0) [circle,draw] {$1$};
	\node (s2) at (1.5,0) [circle,draw] {$2$};
	\draw [postaction={decorate}] (s1) to [bend right=20] (s2);
	\draw [postaction={decorate}] (s1) to [bend right=40] (s2);
	\draw [postaction={decorate}] (s2) to [bend right=20] (s1);
	\draw [postaction={decorate}] (s2) to [bend right=40] (s1);
}
\quad \rightarrow \quad
\mytikz{
% top tube
	\draw (-1.5,0.8) ellipse (0.3 and 0.1);
	\draw (1.5,0.8) ellipse (0.3 and 0.1);
	\draw (-1.2,0.8) .. controls +(90:0.4) and +(180:0.7) .. (0,1.2);
	\draw (-1.8,0.8) .. controls +(90:0.4) and +(180:1.5) .. (0,1.8);
	\draw[xscale=-1] (-1.2,0.8) .. controls +(90:0.4) and +(180:0.7) .. (0,1.2);
	\draw[xscale=-1] (-1.8,0.8) .. controls +(90:0.4) and +(180:1.5) .. (0,1.8);	
% H_21	
	\draw (0,1.8) arc (90:270:0.1 and 0.3);	
	\draw[dashed] (0,1.2) arc (-90:90:0.1 and 0.3);
	\node at (-0.45,1.5) {$H_{21}$};			
% bottom tube
	\draw (-1.5,-0.8) ellipse (0.3 and 0.1);	
	\draw (1.5,-0.8) ellipse (0.3 and 0.1);	
	\draw (-1.2,-0.8) .. controls +(-90:0.4) and +(180:0.7) .. (0,-1.2);
	\draw (-1.8,-0.8) .. controls +(-90:0.4) and +(180:1.5) .. (0,-1.8);
	\draw[xscale=-1] (-1.2,-0.8) .. controls +(-90:0.4) and +(180:0.7) .. (0,-1.2);
	\draw[xscale=-1] (-1.8,-0.8) .. controls +(-90:0.4) and +(180:1.5) .. (0,-1.8);	
% H_12
	\draw (0,-1.2) arc (90:270:0.1 and 0.3);	
	\draw[dashed] (0,-1.8) arc (-90:90:0.1 and 0.3);
	\node at (-0.45,-1.5) {$H_{12}$};				
% left cylinder 
	\draw (-1.5,0.3) ellipse (0.3 and 0.1);
	\draw[dashed] (-1.2,-0.3) arc (0:180:0.3 and 0.1);	
	\draw (-1.2,-0.3) arc (360:180:0.3 and 0.1);
	\draw (-1.8,0.3) to (-1.8,-0.3);
	\draw (-1.2,0.3) to (-1.2,-0.3);
% right cylinder 
	\draw (1.5,0.3) ellipse (0.3 and 0.1);
	\draw[dashed] (1.8,-0.3) arc (0:180:0.3 and 0.1);	
	\draw (1.8,-0.3) arc (360:180:0.3 and 0.1);
	\draw (1.8,0.3) to (1.8,-0.3);
	\draw (1.2,0.3) to (1.2,-0.3);	
	\node at (-0.1,1.5) [circle,fill,inner sep=0.5mm] {};
	\node at (-0.1,-1.5) [circle,fill,inner sep=0.5mm] {};
}
\quad \rightarrow \quad
\mytikz{
	\begin{scope}[scale=0.5]
	% torus outer
	\draw (-3.5,0) .. controls (-3.5,1.5) and (-1.5,2) .. (0,2);
	\draw[xscale=-1] (-3.5,0) .. controls (-3.5,1.5) and (-1.5,2) .. (0,2);
	\draw[rotate=180] (-3.5,0) .. controls (-3.5,1.5) and (-1.5,2) .. (0,2);
	\draw[yscale=-1] (-3.5,0) .. controls (-3.5,1.5) and (-1.5,2) .. (0,2);
	% torus inner
	\draw (-2,0.2) .. controls (-1.5,-0.3) and (-1,-0.5) .. (0,-0.5) .. controls (1,-0.5) and (1.5,-0.3) .. (2,0.2);
	\draw (-1.75,0) .. controls (-1.5,0.3) and (-1,0.5) .. (0,0.5) .. controls (1,0.5) and (1.5,0.3) .. (1.75,0);
	% H1
	\draw (0,-0.5) .. controls +(-0.5,0) and +(-0.5,0) .. node[left]{$H_{12}$} (0,-2);
	\draw[dashed] (0,-0.5) .. controls +(0.5,0) and +(0.5,0) .. (0,-2);
	% H2
	\draw (0,0.5) .. controls +(-0.5,0) and +(-0.5,0) .. node[left]{$H_{21}$} (0,2);
	\draw[dashed] (0,0.5) .. controls +(0.5,0) and +(0.5,0) .. (0,2);
	\node (v1) at (-0.37,-1.38) [circle,fill,inner sep=0.5mm] {};
	\node (v2) at (-0.37,1.38) [circle,fill,inner sep=0.5mm] {};
	\end{scope}
}
$
\caption{Transforming conifold quiver into a Riemann surface}
\label{fig:congluing}
\end{figure}

For the case of the conifold,  we fix the numbers $n_{12}^{(1)} , n_{12}^{(2)} $ of operators 
in the $ ( N , \bar N )$ representation, and $n_{21}^{(1)} , n_{21}^{(2)} $ in the representation $(\bar N , N )$.
We have  $n =   n_{12}^{(1)} + n_{12}^{(2)} =  n_{21}^{(1)} + n_{21}^{(2)}  $. We 
define $H_{12}  = S_{n_{12}^{(1)} } \times S_{n_{12}^{(2)}} $  and 
$H_{21} = S_{n_{21}^{(1)} } \times S_{ n_{21}^{(2)}}$. The operators 
are described by two permutations $ ( \s_1 , \s_2 )$ 
each in $S_n$, which relate fundamental and anti-fundamental indices 
for the two gauge groups. There are equivalences 
\begin{equation}\begin{split} 
( \s_1 , \s_2 ) \sim ( \gamma_{21} \s_1 \gamma_{12}^{-1}  , \gamma_{12} \s_2 \gamma_{21}^{-1}  ) 
\end{split}\end{equation}
Restricting to $ n \le N$, where there are no further finite $N$ constraints, and using the  Burnside Lemma 
\begin{equation}\begin{split} 
\cN_{\cC } ( n_{12}^{(1)} , n_{12}^{(2)} , n_{21}^{(1)} , n_{21}^{(2)} ) 
= { 1 \over |H_1|  |H_2|  } \sum_{\gamma_{12} \in H_{12}    }  \sum_{ \gamma_{21} \in H_{21} } 
    \delta ( \gamma_{21} \sigma_1 \gamma_{12}^{-1} \sigma_{1}^{-1}   ) 
     \delta (     \gamma_{12} \sigma_2 \gamma_{21}^{-1} \sigma_{2}^{-1}   )  
\end{split}\end{equation}
Comparing to (\ref{DH1DH2}) we see that 
\begin{equation}\begin{split} 
\cN_{\cC} ( n_{12}^{(1)} , n_{12}^{(2)} , n_{21}^{(1)} , n_{21}^{(2)} )  = Z ( T^2  ; D_{H_{12} } , D_{H_{21} }  ) 
\end{split}\end{equation}
The procedure for going from the quiver to the TFT data is the same as in the case of $\mC^3$. 
Collapse multiple edges with 
 the same start and end-points to a single edge. Take a cylinder for each gauge group
and a cylinder for each edge. This is essentially the operation of
 thickening the quiver diagram into a surface. 
Equivalently, we can describe this as forming  a  split-node version of the quiver 
where multiple edges have been replaced with single edges, and then thickening all the 
edges of this quiver.  The cylinders for the matter edges are equipped with closed $H$-defects. 
Glue the cylinders 
together. This is illustrated in Figure~\ref{fig:congluing}.
The TFT partition function can be computed using a simple cell decomposition 
with two 2-cells, summing over additional permutations $\s_1 , \s_2 $ extending
 along lines located along the thickened tubes for each gauge group as in 
Figure~\ref{fig:torusH1H2cell}.  
 
\begin{figure}[h]
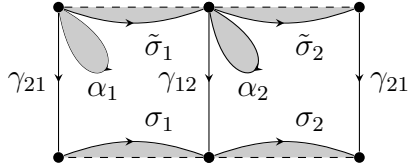

\centering
$
\mytikz{ 	
\begin{scope}	
	\fill [black!20!white] (-1,-1) to [bend left=20](1,-1);
	\fill [black!20!white] (-1,1) to [bend right=20](1,1);
	\fill [black!20!white] (v1) .. controls +(-30:1.5) and +(-80:1.5) .. (v1);
	\node (v1) at (-1,1) [circle,fill,inner sep=0.5mm] {};
	\node (v2) at (1,1) [circle,fill,inner sep=0.5mm] {};
	\node (v3) at (1,-1) [circle,fill,inner sep=0.5mm] {};
	\node (v4) at (-1,-1) [circle,fill,inner sep=0.5mm] {};		
	\draw[dashed] (v1) to (v2);
	\draw[dashed] (v3) to (v4);	
	\draw [postaction={decorate}] (v2) to node[left]{$\gamma_{12}$} (v3);
	\draw [postaction={decorate}] (v1) to node[left]{$\gamma_{21}$} (v4);
	\draw [postaction={decorate}] (v4) to [bend left=20] node[above right]{$\sigma_1$} (v3);
	\draw [postaction={decorate}] (v1) to [bend right=20] node[below right]{$\tl\sigma_1$} (v2);	
	\draw [postaction={decorate}] (v1) .. controls +(-30:1.5) and +(-80:1.5) .. node[below]{$\alpha_1$} (v1);	
\end{scope}
\begin{scope}[xshift=2cm]
	\fill [black!20!white] (-1,-1) to [bend left=20](1,-1);
	\fill [black!20!white] (-1,1) to [bend right=20](1,1);
	\fill [black!20!white] (v1) .. controls +(-30:1.5) and +(-80:1.5) .. (v1);
	\node (v1) at (-1,1) [circle,fill,inner sep=0.5mm] {};
	\node (v2) at (1,1) [circle,fill,inner sep=0.5mm] {};
	\node (v3) at (1,-1) [circle,fill,inner sep=0.5mm] {};
	\node (v4) at (-1,-1) [circle,fill,inner sep=0.5mm] {};		
	\draw[dashed] (v1) to (v2);
	\draw[dashed] (v3) to (v4);	
	\draw [postaction={decorate}] (v2) to node[right]{$\gamma_{21}$} (v3);
	\draw [postaction={decorate}] (v4) to [bend left=20] node[above right]{$\sigma_2$} (v3);
	\draw [postaction={decorate}] (v1) to [bend right=20] node[below right]{$\tl\sigma_2$} (v2);	
	\draw [postaction={decorate}] (v1) .. controls +(-30:1.5) and +(-80:1.5) .. node[below]{$\alpha_2$} (v1);	
\end{scope}
}
$
\caption{Conifold torus with holes}
\label{fig:conholes}
\end{figure}

\noindent
The two-point functions for the conifold, in the permutation basis for the operators, are 
\begin{equation}\begin{split}\label{2ptfuncon} 
& \langle \cO_{ \s_1  , \s_2 }  (\cO_{\tilde \s_1 , \tilde  \s_2 })^{\dagger}  \rangle = \sum_{\gamma_{12} \in H_{12}   } \sum_{ \gamma_{21} \in H_{21}} 
\sum_{\tau_1  \in S_{n} } \sum_{\tau_2 \in S_{n}} N^{C_{\tau_1} } N^{C_{\tau_2} } 
 \delta (      \gamma_{21} \sigma_1 \gamma_{12}^{-1} \tilde \sigma_{1}^{-1}\tau_1      )
 \delta (       \gamma_{12} \sigma_2 \gamma_{21}^{-1} \tilde \sigma_{2}^{-1}\tau_2                  ) 
\end{split}\end{equation}
To obtain this as a partition function in TFT, we use Figure~\ref{fig:conholes}. 
The lines associated with $\sigma_1$ and $\sigma_2$, extending along the lines for 
each  gauge group,  have been cut to separate $\sigma_1 , \sigma_2 $ from $ \tilde \sigma_1 , \tilde \sigma_2 $. 
And we have inserted on the 2-cells previously associated to each gauge group 
additional boundaries carrying permutations $\sum_{\tau_1} N^{C_{\tau_1}}\tau_1 $  and
 $\sum_{\tau_2} N^{C_{\tau_2}}\tau_2 $. The two contractible 2-cells, 
  associated to one gauge group each, give following (\ref{defpartfn}),  the 
  correct delta functions in (\ref{2ptfuncon}).

\noindent 
The three point functions  $ \langle \cO_{ \s_1^{(1) } , \s_2^{(1)}  } \cO_{ \s_1^{(2)}  , \s_2^{(2)} } (\cO_{ \tilde \s_1 , \tilde \s_2 })^{\dagger} \rangle
$ are  obtained by the  replacement $ \sigma_a \rightarrow \s_a^{(1)}  \circ \s_{a}^{(2)}  $ in (\ref{2ptfuncon}), which is a simple 
replacement in the  TFT defect operators  of Figure~\ref{fig:conholes}. 

For $ \mC^3 / \mZ_2$, with specified numbers $n_{11} , n_{12}^{(1)} , n_{12}^{(2)} , n_{21}^{(1)} , n_{21}^{(2)} , n_{22} $, 
 we have the counting 
\begin{equation}\begin{split}\label{countC3Z2}  
& \cN_{ \mC^3/Z_2 }   ( n_{11} , n_{12}^{(1)} , n_{12}^{(2)} , n_{21}^{(1)} , n_{21}^{(2)} , n_{22} )  \\ 
&  =  
{ 1 \over |H_{12} | | H_{21} |  | H_{11} | | H_{22} | } 
\sum_{ \gamma_{11} \in S_{n_{11}} } \sum_{ \gamma_{22} \in S_{n_{22}} } 
\sum_{ \gamma_{12} \in H_{12} } \sum_{ \gamma_{21} \in H_{21} }
 \sum_{ \s_1 \in S_{n_1} }  
 \sum_{ \s_2  \in S_{n_2} } \\ 
& \delta ( (  \g_{11} \circ \g_{21} )     \s_{1}        (  \g_{11}^{-1}  \circ \g_{12}^{-1}  )  \s_{1}^{-1}                               )
       \delta (                   (  \g_{22} \circ \g_{12} )     \s_{2}        (  \g_{22}^{-1}  \circ \g_{21}^{-1}  )  \s_{2}^{-1}  ) 
\end{split}\end{equation}
with $H_{12} = S_{n_{12}^{(1)} } \times S_{n_{12}^{(2)} }$,  $H_{21} = S_{n_{21}^{(1)} } \times S_{n_{21}^{(2)} }$
and 
\begin{equation}\begin{split} 
&  n_1 = n_{21}^{(1)} + n_{21}^{(2)}  + n_{11} = n_{11} + n_{12}^{(1)} + n_{12}^{(2)}   \\ 
 &  n_{2} = n_{12}^{(1)} + n_{12}^{(2)}  + n_{22} =  n_{12}^{(1)} + n_{12}^{(2)}  + n_{22} 
\end{split}\end{equation}
Note that $ n_{21}^{(1)} + n_{21}^{(2)} = n_{12}^{(1)} + n_{12}^{(2)}$. 
The total number of distinct indices being permuted is $n_{11} + n_{22} + n_{12}^{(1)} + n_{12}^{(2)}$ , 
so this will be related to $S_n$  TFT with $n = n_{11} + n_{22} + n_{12}^{(1)} + n_{12}^{(2)}$. 
The relations between $S_n$ and its different subgroups is best expressed with the diagram in 
\ref{quivertrace}.  
The counting is  reproduced as a TFT partition function on a genus two surface. 
To describe this surface, and the associated defects, we first replace multiple edges
with same start and end points by single edges, then we form the  split-node version of this 
quiver. This has edges for the gauge group and for the matter fields. Build the surface by 
taking a tube for each edge. The vertices become  three-holed spheres. Insert closed $H$-defects
on the matter tubes, so that the permutations around these loops are constrained as 
$\gamma_{ab} \in S_{n_{ab}} \subset S_n$.  We introduce  unit open Wilson 
lines connecting the $H$-defects on the 3-holed spheres,  as  in Figure~\ref{fig:3sphere}. 
The construction of the genus two Riemann surface is shown in Figure~\ref{fig:C3Z2surface}.
This ensures that the holonomies are  
$\gamma_{11} \circ \gamma_{21} $ and  $ \gamma_{22}^{-1}  \circ \gamma_{21}^{-1}  $ 
at the left and right upper ends of the gauge group  cylinders ; 
they are $ \gamma_{11}^{-1}  \circ \gamma_{12}^{-1}  $  and $ \gamma_{22 } \circ \gamma_{12 }  $ 
at the left and right lower circles of the two cylinders. There are line defects 
with holonomies $\s_1$ constrained to be in $S_{n_1} \in S_n$, 
and $\s_2 \in S_{n_2} \in S_n $. The TFT partition function in the presence of these defects 
leads to the delta functions in (\ref{countC3Z2}), coming from the two 2-cells 
of the gauge-group cylinders. 
 So we 
can state that 
\begin{equation}\begin{split}
 \cN_{\mC^3/Z_2 }  ( n_{11} , n_{12}^{(1)} , n_{12}^{(2)} , n_{21}^{(1)} , n_{21}^{(2)} , n_{22} ) 
 = Z ( \Sigma_{ G=2} ; D_{H_{11} } , D_{H_{21} } , D_{ H_{12} } ,  D_{ H_{21} }  ; W   ) 
\end{split}\end{equation}
$H_{ab} $ are the groups $S_{n_{ab}} $. 
$W$ stands for the set of unit open Wilson lines on the three-holed spheres
and the constraints requiring the $\s_1, \s_2$ to live in in the subgroups $S_{n_1} , S_{n_2}$. 

\begin{figure}[h]
\centering
$
\mytikz{
	\node (s1) at (-1.5,0) [circle,draw] {$1$};
	\node (s2) at (1.5,0) [circle,draw] {$2$};
	\draw [postaction={decorate}] (s1.-120) .. controls +(-120:0.5) and +(0,-0.6) .. (-2.5,0)
		.. controls +(0,0.6) and +(120:0.5) .. (s1.120);
	\draw [postaction={decorate}] (s2.60) .. controls +(60:0.5) and +(0,0.6) .. (2.5,0)
		.. controls +(0,-0.6) and +(-60:0.5) .. (s2.-60);		
	\draw [postaction={decorate}] (s1) to [bend right=20] (s2);
	\draw [postaction={decorate}] (s1) to [bend right=40] (s2);
	\draw [postaction={decorate}] (s2) to [bend right=20] (s1);
	\draw [postaction={decorate}] (s2) to [bend right=40] (s1);	
}
\quad \rightarrow \quad
\mytikz{
% top tube
	\draw (-1.5,0.8) ellipse (0.3 and 0.1);
	\draw (1.5,0.8) ellipse (0.3 and 0.1);
	\draw (-1.2,0.8) .. controls +(90:0.4) and +(180:0.7) .. (0,1.2);
	\draw[xscale=-1] (-1.2,0.8) .. controls +(90:0.4) and +(180:0.7) .. (0,1.2);
	\draw (0,1.8) arc (90:270:0.1 and 0.3);	
	\draw[dashed] (0,1.2) arc (-90:90:0.1 and 0.3) node[above]{$H_{21}$};			
% bottom tube
	\draw (-1.5,-0.8) ellipse (0.3 and 0.1);	
	\draw (1.5,-0.8) ellipse (0.3 and 0.1);	
	\draw (-1.2,-0.8) .. controls +(-90:0.4) and +(180:0.7) .. (0,-1.2);
	\draw[xscale=-1] (-1.2,-0.8) .. controls +(-90:0.4) and +(180:0.7) .. (0,-1.2);
	\draw (0,-1.2) arc (90:270:0.1 and 0.3) node[below]{$H_{12}$};	
	\draw[dashed] (0,-1.8) arc (-90:90:0.1 and 0.3);
% left tube	
\begin{scope}[xshift=-1.5cm,rotate=90]
	\draw (-1.2,0.8) .. controls +(90:0.4) and +(180:0.7) .. (0,1.2);
	\draw (-1.8,0.8) .. controls +(90:0.4) and +(180:1.5) .. (0,1.8);
	\draw[xscale=-1] (-1.2,0.8) .. controls +(90:0.4) and +(180:0.7) .. (0,1.2);
	\draw[xscale=-1] (-1.8,0.8) .. controls +(90:0.4) and +(180:1.5) .. (0,1.8);	
	\draw (0,1.8) arc (90:270:0.1 and 0.3);	
	\draw[dashed] (0,1.2) arc (-90:90:0.1 and 0.3) node[left]{$H_{11}$};		
\end{scope}
% right tube
\begin{scope}[xshift=1.5cm,rotate=90]
	\draw (-1.2,-0.8) .. controls +(-90:0.4) and +(180:0.7) .. (0,-1.2);
	\draw (-1.8,-0.8) .. controls +(-90:0.4) and +(180:1.5) .. (0,-1.8);
	\draw[xscale=-1] (-1.2,-0.8) .. controls +(-90:0.4) and +(180:0.7) .. (0,-1.2);
	\draw[xscale=-1] (-1.8,-0.8) .. controls +(-90:0.4) and +(180:1.5) .. (0,-1.8);	
	\draw (0,-1.2) arc (90:270:0.1 and 0.3) node[right]{$H_{22}$};	
	\draw[dashed] (0,-1.8) arc (-90:90:0.1 and 0.3);
\end{scope}			
% left cylinder 
	\draw (-1.5,0.3) ellipse (0.3 and 0.1);
	\draw[dashed] (-1.2,-0.3) arc (0:180:0.3 and 0.1);	
	\draw (-1.2,-0.3) arc (360:180:0.3 and 0.1);
	\draw (-1.8,0.3) to (-1.8,-0.3);
	\draw (-1.2,0.3) to (-1.2,-0.3);	
% right cylinder 
	\draw (1.5,0.3) ellipse (0.3 and 0.1);
	\draw[dashed] (1.8,-0.3) arc (0:180:0.3 and 0.1);	
	\draw (1.8,-0.3) arc (360:180:0.3 and 0.1);
	\draw (1.8,0.3) to (1.8,-0.3);
	\draw (1.2,0.3) to (1.2,-0.3);	
% top connectors
	\draw (-2.3,1.8) to (2.3,1.8);
	\draw (-2.3,1.2) to [bend left=45] (-1.8,0.8);	
	\draw (2.3,1.2) to [bend right=45] (1.8,0.8);
% bottom connectors
	\draw (-2.3,-1.8) to (2.3,-1.8);
	\draw (-2.3,-1.2) to [bend right=45] (-1.8,-0.8);	
	\draw (2.3,-1.2) to [bend left=45] (1.8,-0.8);	
% nodes and "1" lines
	\node (v12) at (-0.1,-1.5) [circle,fill,inner sep=0.5mm] {};
	\node (v21) at (-0.1,1.5) [circle,fill,inner sep=0.5mm] {};
	\node (v11) at (-3,-0.1) [circle,fill,inner sep=0.5mm] {};
	\node (v22) at (3,-0.1) [circle,fill,inner sep=0.5mm] {};
	\draw (v11) .. controls (-3,1.5) .. (v21);
	\draw (v22) .. controls (3,1.5) .. (v21);
	\draw (v11) .. controls (-3,-1.5) .. (v12);
	\draw (v22) .. controls (3,-1.5) .. (v12);
	\node at (-2.4,1.6) {$1$};
	\node at (2.4,1.6) {$1$};
	\node at (-2.4,-1.6) {$1$};
	\node at (2.4,-1.6) {$1$};
% extra nodes and lines
	\node (vl1) at (-1.5,0.9) [circle,fill,inner sep=0.5mm] {};
	\node (vl2) at (-1.5,0.2) [circle,fill,inner sep=0.5mm] {};
	\node (vl3) at (-1.5,-0.4) [circle,fill,inner sep=0.5mm] {};	
	\node (vl4) at (-1.5,-0.9) [circle,fill,inner sep=0.5mm] {};
	\node (vr1) at (1.5,0.9) [circle,fill,inner sep=0.5mm] {};
	\node (vr2) at (1.5,0.2) [circle,fill,inner sep=0.5mm] {};
	\node (vr3) at (1.5,-0.4) [circle,fill,inner sep=0.5mm] {};	
	\node (vr4) at (1.5,-0.9) [circle,fill,inner sep=0.5mm] {};
	\draw (vl1) .. controls (-1.3,1.3) .. node[left]{$1$} (v21);
	\draw (vl4) .. controls (-1.3,-1.3) .. node[left]{$1$} (v12);
	\draw (vr1) .. controls (1.3,1.3) .. node[right]{$1$} (v21);
	\draw (vr4) .. controls (1.3,-1.3) .. node[right]{$1$} (v12);	
	\draw (vl2) to (vl3);	
	\draw (vr2) to (vr3);		
	\node (s1) at (-0.5,0.4){$\sigma_1$};
	\node (s2) at (0.5,0.4){$\sigma_2$};
	\draw [dotted,->] (s1) to (-1.4,0);
	\draw [dotted,->] (s2) to (1.4,0);
}
$
\caption{Transforming $\mC^3/\mZ_2$ quiver into a Riemann surface}
\label{fig:C3Z2surface}
\end{figure}
\begin{figure}[h]
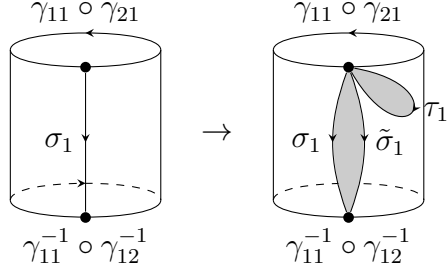

\centering
$
\mytikz{
	\def\x{1};
	\def\y{1};
	\def\dy{0.3};	
	\draw[postaction={decorate}] (\x,\y) .. controls +(0,\dy) and +(0,\dy) .. node[above]{$\g_{11}\circ\g_{21}$} (-\x,\y);	
	\draw (-\x,\y) .. controls +(0,-\dy) and +(0,-\dy) .. (\x,\y);
	\draw (\x,-\y) .. controls +(0,-\dy) and +(0,-\dy) .. node[below]{$\g_{11}^{-1}\circ \g_{12}^{-1}$} (-\x,-\y);	
	\draw[dashed,postaction={decorate}] (-\x,-\y) .. controls +(0,\dy) and +(0,\dy) .. (\x,-\y);
	\draw (-\x,\y) to (-\x,-\y);
	\draw (\x,\y) to (\x,-\y);	
	\node (v1) at (0,0.77) [circle,fill,inner sep=0.5mm] {};
	\node (v2) at (0,-1.23) [circle,fill,inner sep=0.5mm] {};
	\draw[postaction={decorate}] (v1) to node[left]{$\s_1$} (v2);
}
\quad\rightarrow\quad
\mytikz{
	\def\x{1};
	\def\y{1};
	\def\dy{0.3};	
	\draw[postaction={decorate}] (\x,\y) .. controls +(0,\dy) and +(0,\dy) .. node[above]{$\g_{11}\circ\g_{21}$} (-\x,\y);	
	\draw (-\x,\y) .. controls +(0,-\dy) and +(0,-\dy) .. (\x,\y);
	\draw (\x,-\y) .. controls +(0,-\dy) and +(0,-\dy) .. node[below]{$\g_{11}^{-1}\circ \g_{12}^{-1}$} (-\x,-\y);	
	\draw[dashed,postaction={decorate}] (-\x,-\y) .. controls +(0,\dy) and +(0,\dy) .. (\x,-\y);
	\draw (-\x,\y) to (-\x,-\y);
	\draw (\x,\y) to (\x,-\y);	
	\fill [black!20!white] (0,0.77) to [bend right=20] (0,-1.23) to [bend right=20] (0,0.77);
	\fill [black!20!white] (v1) .. controls +(-10:1.5) and +(-60:1.5) .. (v1);
	\node (v1) at (0,0.77) [circle,fill,inner sep=0.5mm] {};
	\node (v2) at (0,-1.23) [circle,fill,inner sep=0.5mm] {};
	\draw[postaction={decorate}] (v1) to [bend right=20] node[left]{$\s_1$} (v2);
	\draw[postaction={decorate}] (v1) to [bend left=20] node[right]{$\tl\s_1$} (v2);
	\draw[postaction={decorate}] (v1) .. controls +(-10:1.5) and +(-60:1.5) .. node[right]{$\tau_1$} (v1);
}
$
\caption{Splitting $\sigma_a$ in gauge group cylinder to go from counting to 2-point function.}
\label{fig:cylinder_split}
\end{figure}

 
\noindent 
Again,  going from counting to 2-point functions, is a simple step 
in the $S_n$ TFT, as shown in Figure~\ref{fig:cylinder_split}.
The $\sigma_1$-edge on the first  cylinder   is split into 
two edges joined to form a circle surrounding a hole in the surface, now carrying fixed permutations 
$ \sigma_1, \tilde \sigma_1$ from the two chosen operators. 
Likewise the  
$\sigma_{2} $ edge is split into a pair of edges carrying $ \sigma_2 , \tilde \sigma_2$ 
permutations.  An additional hole in each gauge group cylinder carries $ \sum_{\tau_a \in S_{n_a}} N^{ C_{\tau_a} }
\tau_a$.
\begin{equation}\begin{split} 
& \langle \cO_{ \sigma_1 , \sigma_2 }  ( \cO_{\tilde  \sigma_1 , \tilde \sigma_2 }) ^{\dagger} \rangle \\ 
& \qquad =
\sum_{ \gamma_{11} \in S_{n_{11}} } \sum_{ \gamma_{22} \in S_{n_{22}} } 
\sum_{ \gamma_{12} \in H_{12} } \sum_{ \gamma_{21} \in H_{21} }
 \sum_{ \tau_1 \in S_{n_1} }  
 \sum_{ \tau_2  \in S_{n_2} } \\ 
& \qquad N^{C_{\tau_1} +C_{\tau_2}  } ~ \delta ( (  \g_{11} \circ \g_{21} )     \s_{1}        
(  \g_{11}^{-1}  \circ \g_{12}^{-1}  )  \tilde \s_{1}^{-1}    \tau_1       )  
       \delta (                   (  \g_{22} \circ \g_{12} )     \s_{2}        (  \g_{22}^{-1}  \circ \g_{21}^{-1}  ) \tilde  \s_{2}^{-1} \tau_2 ) 
\end{split}\end{equation}
We have the equality 
\begin{equation}\begin{split} 
 \langle \cO_{ \sigma_1 , \sigma_2 }  ( \cO_{\tilde  \sigma_1 , \tilde \sigma_2 }) ^{\dagger}  \rangle 
 =  Z ( \Sigma_{ G=2 , B=2 } ;  \sigma_1 , \sigma_2 , \tilde \sigma_1 , \tilde \sigma_2 ; 
  D_{H_{11} } , D_{H_{11} } , D_{ H_{12} } ,  D_{ H_{21} } ; W  , S ) 
\end{split}\end{equation}
Again the quiver correlator to $S_n $ TFT correspondence generalizes simply from two to three-point functions
\begin{equation}\begin{split}
 & \langle \cO_{ \sigma_1^{(1)}  , \sigma_2^{(1)}  }  \cO_{ \sigma_1^{(2)}  , \sigma_2^{(2)}  } 
  ( \cO_{\tilde  \sigma_1 , \tilde \sigma_2 }) ^{\dagger}  \rangle 
\\ & \quad = 
Z ( \Sigma_{ G=2 , B=2 } ;  \sigma_1^{(1)} \circ  \sigma_1^{(2)} , \sigma_2^{(1)} \circ \sigma_{2}^{(2)}  , \tilde \sigma_1 , \tilde \sigma_2 ; 
  D_{H_{11} } , D_{H_{22} } , D_{ H_{12} } ,  D_{ H_{21} } ; W ) 
\end{split}\end{equation}
In this case, $n = \tilde n_{11} + \tilde n_{22} + \tilde n_{12}^{(1)}  + \tilde n_{12}^{(2)} $. 
And we have selection rules 
$ n_{ab}^{(1)}  + n_{ab}^{(2)} =  \tilde n_{ab}   $ .

\begin{figure}[h]
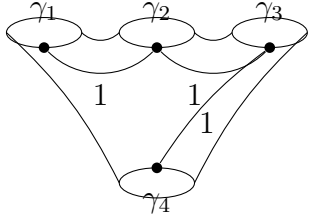

\centering
$
\mytikz{
	\def\R{1.2}
	\def\rx{0.5}
	\def\ry{0.2}
	\draw (0,-1) ellipse ({\rx} and {\ry}) node[below]{$\g_4$};
	\draw (-1.5,1) ellipse ({\rx} and {\ry}) node[above]{$\g_1$};
	\draw (0,1) ellipse ({\rx} and {\ry}) node[above]{$\g_2$};
	\draw (1.5,1) ellipse ({\rx} and {\ry}) node[above]{$\g_3$};
	\draw ($(0,-1)+(-\rx,0)$) to [bend right=10] ($(-1.5,1)+(-\rx,0)$);
	\draw ($(0,-1)+(\rx,0)$) to [bend left=10] ($(1.5,1)+(\rx,0)$);
	\draw ($(-1.5,1)+(\rx,0)$) to [bend right=45] ($(0,1)+(-\rx,0)$);
	\draw ($(0,1)+(\rx,0)$) to [bend right=45] ($(1.5,1)+(-\rx,0)$);
	\node (v1) at ($(-1.5,1)+(0,-\ry)$) [circle,fill,inner sep=0.5mm] {};
	\node (v2) at ($(0,1)+(0,-\ry)$) [circle,fill,inner sep=0.5mm] {};
	\node (v3) at ($(1.5,1)+(0,-\ry)$) [circle,fill,inner sep=0.5mm] {};
	\node (v4) at ($(0,-1)+(0,\ry)$) [circle,fill,inner sep=0.5mm] {};
	\draw (v1) to [bend right=45] node[below]{$1$} (v2);
	\draw (v2) to [bend right=45] node[below left]{$1$} (v3);
	\draw (v3) to [bend right=10] node[below]{$1$} (v4);
}
$
\caption{4-holed sphere with defects, imposing $\gamma_4^{-1} =  \g_1 \g_2 \g_3 $}
\label{fig:4sphere}
\end{figure}

The generalization of the above constructions to an arbitary quiver 
should be clear from the above examples. Having chosen
$n_{ab}^{\alpha}$ and $n_{ab} = \sum_{ \alpha } n_{ab}^{\alpha} $, the counting  will be given 
in terms of gauge theory with an $S_n$ group which contains all the $S_{n_{ab}}$ and $S_{n_a}$. 
The way these are embedded in $S_n$  can be drawn with a diagram such as Figure~(\ref{quivertrace}). 
There are constraints 
\bea\label{nanab} 
n_a = \sum_{ b } n_{ba} = \sum_{ b} n_{ab}
\eea
and groups $H_{ab} = \times_{\alpha } S_{n_{ab}^{\alpha} } $. 
There are subsets $\bS (n_{ab}) $ of the integers $\{ 1 , \cdots , n \}$ corresponding to strands 
in the diagram of the type  ~(\ref{quivertrace}). There are subsets $ \bS ( n_a )$ 
related to $\bS ( n_{ab} ) $ by equations reflecting (\ref{nanab}) but in terms of subset embeddings :  
\bea 
 \bS ( n_a ) = \bigcup_{b } \bS ( n_{ba} ) = \bigcup_b   \bS ( n_{ab} )  
\eea
The integer $n$ is given by 
\bea\label{nnanab}  
n = \sum_{a} n_{aa} + \sum_{ a < b } n_{ab} 
\eea
Correspondingly the set $\bS ( n ) = \{ 1 , \cdots , n  \} $ is a union reflecting  (\ref{nnanab})
\bea 
\bS ( n ) = \bigcup_{ a } \bS ( n_{aa} )  ~ \bigcup_{ a < b }  \bS ( n_{ba} )
\eea
To express the counting in terms of TFT, we use $S_n$-TFT with $n$ given above. 
To get the surface, we collapse 
all the directed edges from a fixed $a$ to $b$ into a single directed edge. We form the split node quiver, 
thicken it by introducing cylinders for the matter edges and the gauge groups, multi-holed spheres at the incoming and outgoing nodes. We insert  closed-$S_{n_{ab}}$ defects on the $a \rightarrow b $ cylinders. 
The multi-holed spheres have appropriate unit-Wilson lines as in  Figure~\ref{fig:4sphere}. 
The cylinder for gauge 
group $a$ has a Wilson line constrained to have holonomy $\sigma_a$  in $S_{n_a} \subset S_{n}$. 
To go from counting of operators  to 2-point and 3-point correlators involves 
the same steps as above,  applied separately to each cylinder.

\subsection{Fundamental groups, covering spaces  and worldsheets}
\label{sec:funcov} 

We have emphasized the interpretation of the quiver counting and correlators 
in terms of 2D  $S_n$ lattice TFT on Riemann surfaces $ \Sigma_G$ equipped with defects, 
since the latter is a concrete computable physical model. There are other fascinating
geometry constructions that should  link to quiver free field observables via the the $S_n$-TFT.  
These will be interesting research  avenues for the future. 
The simplest constructions in lattice $S_n$ TFT on a Riemann surface can be 
interpreted in terms of covering spaces of the Riemann surface, of degree $n$.
This is indeed crucial to the string interpretation of the large N expansion of
two dimensional Yang Mills theory \cite{grta,dadaprovero,cmr}. 
The $S_n$  holonomies of $S_n$ gauge theory on $ \Sigma_G$  are interpreted in terms 
of permutations of  the covering sheets induced by lifting paths in $\Sigma_G$ to 
covers of $\Sigma_G$.  The presence of  closed $H$-defects, where $H$ takes the form of product subgroups
such as $S_{n_{11}} \times S_{n_{21}}  \times \cdots $, can be interpreted in terms of covers involving multiple types 
of sheets.  Variations on the standard covering 
space mathematics occur  in the context of the large N expansion of 
two dimensional Yang Mills \cite{grta,cmr1,SRwil}, particularly when 
Wilson loops (possibly intersecting ones) are introduced. Another setting 
for permutation defects is in  2D conformal field theory  \cite{permbranes}.  
From an  AdS/CFT perspective, the appearance of covering spaces 
of a two dimensional space for a large class of quiver gauge theories
suggests the interpretation of the covering spaces as string worldsheets, 
and the two dimensional base-space of the TFT  as a part of the dual  spacetime. 
 Can this interpretation be developed in terms of the Sasaki-Einstein 
duals of the gauge theory at non-zero coupling \cite{BFHMSVW1,*BFHMSVW2} ? 

A systematic account of the relation between cutting  and gluing of Riemann surfaces 
and constructions of 2D TFT connects with the description of TFT as a functor between 
a category of 1-dimensional objects and two dimensional cobordisms on the one hand 
and a category of Frobenius algebras on the other. These constructions \cite{atiyah,FHK} 
have been generalized  \cite{turaev,mooseg,barsch,dkr} to the 
equivariant case, which should be relevant here.  The paper \cite{turaev}
includes  lattice constructions similar to what we have used in describing 
the $H$-defects.  To get the counting and correlators of quiver theories,  we need 
to specialize the general $G$-equivariant discussion to $S_n$, but allow $n$ to be arbitrary 
as part of an inductive $S_{\infty}$ construction. This type 
of $S_{\infty} $ TFT (in  the non-equivariant setting) has already been discussed \cite{mmn}.

Many developments in 2D TFT treat the base space of the TFT as string worldsheet. 
Here, as emphasized through the analogies to large $N$ 2dYM, the base space
of the TFT should be considered as the target space of strings. 
The covering spaces are string worldsheets. 
This is also a feature of Matrix strings where $S_n$ orbifold CFTs (which are related to 
$S_n$ TFTs) are treated as spacetime CFTs \cite{MatSt}. There should also be a TFT 
on the worldsheets, with $1/N$ palying the role of string coupling, with different 
regions of the worldsheets mapping to different spacetime regions (cut-out by the defects
in the spacetime TFT) being chracterized by distinct worldsheet phases. It will be very 
interesting to infer the systematics of this worldsheet TFT, by using the link 
to the  spacetime $S_n$ TFT provided by covering space theory.

\subsection{Young diagram basis and TFT constructions} 

\begin{figure}[h]
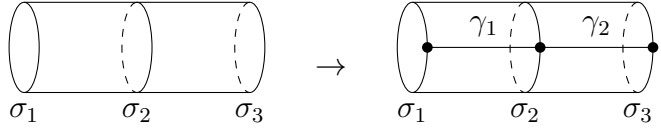

\centering
$
\mytikz{
	\draw (-1.5,0.6) arc (90:270:0.2 and 0.6) node[below]{$\sigma_1$};	
	\draw (-1.5,-0.6) arc (-90:90:0.2 and 0.6);
	\draw[dashed] (0,0.6) arc (90:270:0.2 and 0.6) node[below]{$\sigma_2$};	
	\draw (0,-0.6) arc (-90:90:0.2 and 0.6);
	\draw[dashed] (1.5,0.6) arc (90:270:0.2 and 0.6) node[below]{$\sigma_3$};	
	\draw (1.5,-0.6) arc (-90:90:0.2 and 0.6);
	\draw (-1.5,0.6) to (1.5,0.6);
	\draw (-1.5,-0.6) to (1.5,-0.6);	
}
\quad\rightarrow\quad
\mytikz{
	\draw (-1.5,0.6) arc (90:270:0.2 and 0.6) node[below]{$\sigma_1$};
	\draw (-1.5,-0.6) arc (-90:90:0.2 and 0.6);
	\draw[dashed] (0,0.6) arc (90:270:0.2 and 0.6) node[below]{$\sigma_2$};
	\draw (0,-0.6) arc (-90:90:0.2 and 0.6);
	\draw[dashed] (1.5,0.6) arc (90:270:0.2 and 0.6) node[below]{$\sigma_3$};
	\draw (1.5,-0.6) arc (-90:90:0.2 and 0.6);
	\draw (-1.5,0.6) to (1.5,0.6);
	\draw (-1.5,-0.6) to (1.5,-0.6);	
	\draw (-1.3,0) to node[above]{$\gamma_1$} (0.2,0);
	\draw (0.2,0) to node[above]{$\gamma_2$} (1.7,0);
	\node (v1) at (-1.3,0) [circle,fill,inner sep=0.5mm] {};	
	\node (v1) at (0.2,0) [circle,fill,inner sep=0.5mm] {};	
	\node (v1) at (1.7,0) [circle,fill,inner sep=0.5mm] {};	
}
$
\caption{Cylinder with $\s_1,\s_2,\s_3$ insertions}
\label{fig:cyls1s2s3}
\end{figure}

We have not so far expressed 
our construction of orthogonal bases at finite $N$ 
in terms of 2D TFT. We expect this should be possible by Fourier transforming 
from the permutation basis to representation bases. 
An encouraging hint is that the basic quantities
entering the counting, namely the Littlewood-Richardson coefficients  $ g ( R_1, R_2 , R_3 ) $
as well as the Kronecker product coefficients $ C ( R_1 , R_2 , R_3 )$  
can be constructed in  $S_n$ TFT using the kind of defects we have considered. 
Consider the partition function  $ Z ( \s_1 , \s_2 , \s_3 ) $  of the  3-holed sphere shown in Figure~\ref{fig:3sphere}
\begin{equation}\begin{split} 
Z ( \Sigma_{ G=0 , B=3} ;  \s_1 , \s_2 , \s_3 ) 
= \delta ( \s_1\s_2 \s_3 ) 
\end{split}\end{equation}
Sum over  $\sigma_1 \in S_{n_1} \subset S_{n = n_1 + n_2 } $ 
and $\sigma_2 \in S_{n_2} \subset S_{n = n_1 + n_2 } $ 
with the normalization $ { 1 \over n_1 ! n_2! } $.  Multiply by $ \chi_{R_1} ( \s_1 )  \chi_{R_2} ( \s_2 ) \chi_{R_3} ( \s_3 ) $ 
and sum over $R_1 \vdash n_1 , R_2 \vdash n_2 , R_3 \vdash n$ to get 
\begin{equation}\begin{split} 
& Z ( \Sigma_{ G=0 , B=3} ;  R_1  , R_2 , R_3 )
\\
& = { 1 \over n_1 ! n_2! } \sum_{ \s_1 \in S_{n_1} }   \sum_{ \s_2 \in S_{n_2 } } 
\chi_{R_1} ( \s_1 )  \chi_{R_2} ( \s_2 ) \chi_{R_3} ( \s_3 ) Z ( \Sigma_{ G=0 , B=3} ;  \s_1 , \s_2 , \s_3 ) \\ 
& = { 1 \over n_1 ! n_2! }   \sum_{ \s_1 , \s_2 } \chi_{R_1} ( \s_1 )  \chi_{R_2} ( \s_2 ) \chi_{R_3} ( \s_1 \circ \s_2  ) \\ 
& = g ( R_1 , R_2 , R_3 ) 
\end{split}\end{equation} 
To get $ C ( R_1  , R_2 , R_3 ) $  take a cylinder, with boundary permutations 
$ \s_1 , \s_3 $ and insert a closed defect in the middle with permutation $\sigma_2$
 (see Figure~\ref{fig:cyls1s2s3}).  
All three are in $S_n$. The partition function is 
\begin{equation}\begin{split} 
Z ( S^1 \times I ; \sigma_1 , \sigma_2 , \sigma_3 ) 
= \sum_{ \gamma_1 , \gamma_2 \in S_n } 
  \delta (\sigma_1 \gamma_1 \sigma_2^{-1} \gamma_1^{-1}  ) 
    \delta (\sigma_2 \gamma_2 \sigma_1^{-1} \gamma_2^{-1}  ) 
\end{split}\end{equation}
Sum over permutations, weighted by characters to get a representation basis partition functions 
\begin{equation}\begin{split} 
& Z ( S^1 \times I  ;  R_1 , R_2 , R_3 ) 
\\
& = { 1 \over (n!)^3  }  \sum_{ R_1 , R_2 , R_3 \vdash n }
   \chi_{R_1} ( \s_1 ) \chi_{R_2}  ( \s_2 ) \chi_{R_3}  ( \s_3 ) 
    Z ( S^1 \times I ; \sigma_1 , \sigma_2 , \sigma_3 ) \\ 
 &     = { 1 \over n! } \sum_{\s_1 \in S_n } \chi_{ R_1} ( \s_1 ) \chi_{ R_3 } ( \s_1 ) \chi_{ R_3} ( \s_1 ) \\ 
 & = C ( R_1 , R_2 , R_3 ) 
\end{split}\end{equation}

%%%%%%%%%%%%%%%%%%%%%%%%%%%%%%%%%%%%%%%%%%%%%%%%%%%%%%%%%%

\section{Interacting chiral ring}
\label{sec:int-chir-ring}

So far we have constructed the finite-$N$ chiral ring of the \emph{free} quiver theories, that is, with zero superpotential. In the context of AdS/CFT theories without superpotential 
arise at special points of a moduli space of CFTs, generically with non-zero superpotential. 
For the generic CFTs,  the free chiral ring gets modified by identifying F-terms with zero.

In general, there are physical arguments that the mesonic chiral ring of the gauge theory on $D3$ branes at a Calabi-Yau singularity $Y^6$ is the coordinate ring of the symmetric product space $\Sym^N(Y^6)$. In other words, the partition function of the chiral ring is counting the states of $N$ identical bosons on $Y^6$. In some cases this can be argued by using the geometric invariant theory \cite{Luty:1995sd}.

Such arguments, however, work at the level of counting, and do not provide an explicit construction of the operators, which could be identified with dual BPS states in AdS. Here we make the first steps in the construction of the interacting chiral ring at finite $N$, using the free orthogonal basis derived in the previous sections.

%%%%%%%%%%%%%%%%%%%%%%%%%%%%%%%%%%%%%%%%%%%%%%%%%%%%%%%%%%%%%%%%%%%%%

\subsection{Review of the chiral ring}

As an example we take the theory on D3 branes at a conifold singularity. It has the quiver shown in Figure~\ref{fig:quivercon} that we have analyzed before, but at a generic fixed point there is a non-zero superpotential
\begin{equation}
	W_{\cC} = \tr(A_1 B_1 A_2 B_2) - \tr(A_1 B_2 A_2 B_1)
\end{equation}
In general, such quiver-superpotential pairs (for $\mC^3$, $\C3Z2$, $dP_0$ and many others) can be constructed using the technology of brane tilings \cite{tilings1,tilings2}.

When we have $W$, the chiral ring gets modified compared to the free theory: the operators have to be identified up to F-terms
\begin{equation}
	F = \left\{ \frac{\partial W}{\partial \Phi_{ab;\a}} \right\}
\end{equation}
For example, the conifold F-terms are
\begin{equation}
\label{eq:Fterms_conifold}
\begin{split}
	F_{\cC} = \{ 
		& B_1 A_2 B_2 - B_2 A_2 B_1, \quad B_2 A_1 B_1 - B_1 A_1 B_2, \\
		& A_2 B_2 A_1 - A_1 B_2 A_2, \quad A_1 B_1 A_2 - A_2 B_1 A_1 \}
\end{split}	
\end{equation}
In the interacting chiral ring they are identified with 0
\begin{equation}
	F \sim 0
\end{equation}
For the conifold expressions (\ref{eq:Fterms_conifold}) implies that within the chiral ring we can commute $A$'s through $B$'s and vice versa. The resulting mesonic chiral ring at large $N$ is thus spanned by 
\begin{equation}
\label{eq:str_conifold}
	S^{i_1 i_2 \ldots i_n} S^{j_1 j_2 \ldots j_n} \, \tr (A_{i_1} B_{j_1} A_{i_2} B_{j_2} \ldots A_{i_n} B_{j_n} )
\end{equation}
where $S$ is a symmetric tensor, and products of such symmetrized traces.

To get the interacting chiral ring at finite $N$ we have to enforce both finite $N$ constraints, and F-terms. Note that they might not be independent, for example, at $N=1$ the F-terms $F_\cC$ themselves vanish, so the free and interacting chiral rings are the same. In order to clarify the situation, let us define the construction more rigorously.

Let $V^{(\infty)}$ be the ring of chiral gauge invariant operators of the free theory at $N=\infty$, that is, treating operators as formal products of traces, without any finite $N$ identifications. The basis can be labelled by $\bL$ or $\bK$ as constructed in the previous sections 
\begin{equation}
	V^{(\infty)} = \{ \cO(\bL) \}
\end{equation}
At finite $N$ some operators in $V^{(\infty)}$ vanish -- they form an ideal\footnote{
$V_N$ is an ideal of $V^{(\infty)}$ because a product of vanishing operator and any other operator is also vanishing
} 
$V_N \subset V^{(\infty)}$
\begin{equation}
\label{eq:VN_defn}
	V_N = \{ \cO(\bL) \, | \, l(R_a) > N \}
\end{equation}
The quotient is the free chiral ring at finite $N$
\begin{equation}
	V^{(N)} = V^{(\infty)} / V_N
\end{equation}
which is spanned by operators with $l(R_a) \le N$. Now, let $V_F$ be the space of all gauge invariant operators at $N=\infty$, which are identified with zero by F-terms. It is spanned by all operators containing an F-term anywhere within a trace
\begin{equation}
	V_F = \{ \tr(f \, \Phi_{i_1 j_1;\alpha_1} \Phi_{i_2 j_2;\alpha_2} \ldots) \cO(\bL) \, | \, f \in F \}
\end{equation}
$V_F$ is also an ideal of $V^{(\infty)}$. The $N=\infty$ interacting chiral ring is then the quotient
\begin{equation}
\label{eq:Vint_from_Vfree}
	V_{\rm int}^{(\infty)} = V^{(\infty)} / V_F
\end{equation}
It is spanned by products of symmetrized traces as in (\ref{eq:str_conifold}). Finally, the finite $N$ interacting chiral ring is
\begin{equation}
	V_{\rm int}^{(N)} = V^{(\infty)} / (V_F \cup V_N)
\end{equation}
that is, we identify operators in $V^{(\infty)}$ if they differ by $V_F$ \emph{or} $V_N$.
This quotient can be implemented explicitly using computational algebraic geometry \cite{bfhvz}. 
This is practical at small $N$ but becomes computationally prohibitive at large $N$.

To illustrate the different spaces $(V^{(\infty)}, V^{(N)}, V_{\rm int}^{(\infty)}, V_{\rm int}^{(N)}, V_F, V_N)$ we list the corresponding partition functions for the conifold theory. The operators in $V^{(\infty)}$ are counted by (\ref{eq:counting_con_inf}):
\begin{equation}
	Z^{(\infty)} = \prod_{k=1}^\infty \frac{1}{( 1 - a_1^{k} b_1^{k} - a_1^{k} b_2^{k} - a_2^{k} b_1^{k} - a_2^{k} b_2^{k})}
\end{equation}
The finite $N$ free chiral ring $V^{(N)}$ counting is given explicitly by our construction (\ref{eq:counting_con})
\begin{equation}
\begin{split}
	Z^{(N)} &= 
	\sum_{\substack{R_1,R_2 \\ l(R_a) \le N}}	
	\sum_{\substack{r_{A_1},r_{A_2} \\ r_{B_1},r_{B_2}}}  
	a_1^{|r_{A_1}|} \, a_2^{|r_{A_2}|} \, b_1^{|r_{B_1}|} \, b_2^{|r_{B_2}|} \,
\\	& \times
	g(r_{A_1},r_{A_2};R_1) g(r_{B_1},r_{B_2};R_1) g(r_{A_1},r_{A_2};R_2) g(r_{B_1},r_{B_2};R_2)
\end{split}	
\end{equation}
The size of $V_N$ is the difference
\begin{equation}
\label{eq:Z_N}
	Z_N = Z^{(\infty)} - Z^{(N)}
\end{equation}
The partition function of $V_{\rm int}^{(\infty)}$ can be written from first principles, by counting products of symmetrized traces, containing equal number of $A$'s and $B$'s: 
\begin{equation}
\label{eq:Z_int}
	Z_{\rm int}^{(\infty)} = \prod_{n=1}^\infty \prod_{n_1=0}^{n} \prod_{m_1 = 0}^{n} \frac{1}{(1 - a_1^{n_1} b_1^{m_1} a_2^{n-n_1} b_2^{n-n_2})}
\end{equation}
which also gives us $V_F$ via (\ref{eq:Vint_from_Vfree}):
\begin{equation}
	Z_F = Z^{(\infty)} - Z_{\rm int}^{(\infty)}
\end{equation}
Finally, according to the argument that $V^{(N)}_{\rm int}$ is the coordinate ring of $\Sym^N(\cC)$, the partition function for it, using the technology of plethystics \cite{plethystics}, is
\begin{equation}
\label{eq:Z_intN_bosons}
	Z_{\rm int}^{(N)} = \left[ \prod_{n=0}^\infty \prod_{n_1=0}^{n} \prod_{m_1 = 0}^{n} \frac{1}{(1 - \nu \, a_1^{n_1} b_1^{m_1} a_2^{n-n_1} b_2^{n-n_2})} \right]_{\nu^N}
\end{equation}
Here $[\ldots]_{\nu^N}$ denotes taking the coefficient of $\nu^N$ term. This allows to calculate the size of the union $V_F \cup V_N$
\begin{equation}
	Z_{F \cup N} = Z^{(\infty)} - Z_{\rm int}^{(N)}
\end{equation}

%%%%%%%%%%%%%%%%%%%%%%%%%%%%%%%%%%%%%%%%%%%%%%%%%%%%%%%%%%%%%%%%%%%%%

\subsection{Chiral ring from operators}

In this section we apply our technology to explicitly derive the size of the conifold finite $N$ chiral ring $V^{(N)}_{\rm int}$ for a simple choice of charges
\begin{equation}
\label{eq:sample_charges}
	n_1 = m_1 = 1, \quad n_2 = m_2 = N
\end{equation}
That is, we look at operators of the form $(A_1 B_1) (A_2 B_2)^N$.
This is restrictive but is valid for any $N$. Any direct computational method of tackling this runs into having to deal with $N^2$ variables, which is not viable at large $N$. 

Specifically, the goal is to calculate $[Z^{(N)}_{\rm int}]_{a_1 b_1 a_2^N b_2^N}$ \emph{without} using the $N$-boson counting (\ref{eq:Z_intN_bosons}), but instead relying on the explicit description
\begin{equation}
	V_{\rm int}^{(N)} = V^{(\infty)} / (V_F \cup V_N) 
\end{equation}
We can write the corresponding partition function as
\begin{equation}
\label{eq:ZintN_expanded}
\begin{split}
	Z_{\rm int}^{(N)} &= Z^{(\infty)} - (Z_N + Z_F - Z_{N\cap F}) \\
	&= Z^{(\infty)}_{\rm int} - (Z_N - Z_{N\cap F})
\end{split}
\end{equation}
In other words, to get the finite $N$ ring counting $Z_{\rm int}^{(N)}$ we can take the ring spanned by symmetrized traces $Z^{(\infty)}_{\rm int}$ and subtract the number of finite $N$ constraints $Z_N$, but excluding $Z_{N\cap F}$ compensating for the fact that some $V_N$ is already included in $V_F$. We know $Z^{(\infty)}_{\rm int}$ (\ref{eq:Z_int}) and $Z_N$ (\ref{eq:Z_N}), but the calculation of $Z_{N \cap F}$ is non-trivial. In fact, the choice of charges (\ref{eq:sample_charges}) is the first example where the intersection $V_F \cap V_N$ is non-empty.

First, let us calculate the expected result using the $N$-boson description. The combination
\begin{equation}
	Z^{(\infty)}_{\rm int} - Z^{(N)}_{\rm int} = Z_N - Z_{N \cap F}
\end{equation}
counts the number of boson states with more than $N$ bosons. With our choice of charges there are just two such states, involving $N+1$ bosons:
\begin{equation}
	|(A_1 B_1), (A_2 B_2)^{\otimes N}\rangle \quad |(A_1 B_2), (A_2 B_1), (A_2 B_2)^{\otimes (N-1)} \rangle
\end{equation}
Thus, we expect
\begin{equation}\label{ExpVint}
	Z_N - Z_{N \cap F} = 2
\end{equation}
We confirm this with the explicit description of operators forming $V^{\infty} , V_N , V_F $ in Appendix \ref{Appsec:ExpIntChi}. 

%%%%%%%%%%%%%%%%%%%%%%%%%%%%%%%%%%%%%%%%%%%%%%%%%%%%%%%%%%%%%%%%%%%%%

\subsection{Giant gravitons in the conifold}

In this section we find dual operators to certain giant graviton states, following the identification of \cite{Beasley,Butti:2006au,Forcella:2007wk,hanany07}. Related discussions 
in the context of the conifold or ABJM theory appear in \cite{ABJMcor1,*ABJMcor2,*ABJMcor3,*ABJMcor4,*ABJMcor5,GntsConCP1,*GntsConCP2,*GntsConCP3}. The main 
purpose here is to identify the operators previously considered in connection with giants among 
those spanning the complete orthogonal bases we have described. 

In general, D3 branes wrapping non-contractible cycles in geometry are identified with baryonic operators. In the conifold theory we have $AdS_5 \times T^{1,1}$ geometry in the bulk, where $T^{1,1}$ is the base of the conifold cone. In homogeneous coordinates the conifold is given by an identification
\begin{equation}
\cC: \quad (a_1,a_2,b_1,b_2) \sim (\lambda a_1, \lambda a_2, \lambda^{-1} b_1, \lambda^{-1} b_2), \quad \lambda \in \mC
\end{equation}
The $T^{1,1}$ base can be expressed as:
\begin{equation}
\begin{split}
T^{1,1}:
\quad &|a_1|^2+|a_2|^2 = |b_1|^2 + |b_2|^2 = 1, 
\\
&(a_1,a_2,b_1,b_2) \sim (e^{i\a} a_1, e^{i\a} a_2, e^{-i\a} b_1, e^{-i\a} b_2), 
\quad \alpha \in \mR
\end{split}
\end{equation}
Minimal non-contractible D3 branes in $T^{1,1}$, using Mikhailov's \cite{mikhailov} construction, are given by
\begin{equation}
a_i = 0 \quad (B=1), \qquad b_i = 0 \quad (B=-1)
\end{equation}
The baryon number $(B=\pm 1)$ corresponds to the homology class of $T^{1,1}$. The dual operators in the chiral ring are determinants
\begin{equation}
a_i = 0 \;\leftrightarrow\; \cO=\det(A_i), \qquad b_i = 0 \;\leftrightarrow\; \cO=\det(B_i)
\end{equation}
In the basis constructed in this paper we deal only mesonic operators, that is, $B=0$ sector. We can identify composite giant configurations with total $B=0$, for example
\begin{equation}
a_1 b_1 = 0 \quad \leftrightarrow \quad \cO_{11} = \det(A_1) \det(B_1)
\end{equation}
The dual operator is mesonic and, in fact, can be expressed nicely in our basis
\begin{equation}
\cO_{11} = \det(A_1 B_1) = \cO\left(
\mytikz{	
	\node (m1) at (-1.5,0.5) [circle,draw,inner sep=0.5mm] {};
	\node (n1) at (-1.5,-0.5) [circle,draw,inner sep=0.5mm] {};
	\node (m2) at (1.5,-0.5) [circle,draw,inner sep=0.5mm] {};
	\node (n2) at (1.5,0.5) [circle,draw,inner sep=0.5mm] {};	
	\draw[postaction={decorate}] (m1) to node[left]{$R_1 = {\tiny \yng(1,1,1,1)}\;$} (n1);
	\draw[postaction={decorate}] (m2) to node[right]{$R_2 = {\tiny \yng(1,1,1,1)}$} (n2);
	\draw [postaction={decorate}] (n1) to [bend right=15] node[above]{$r_{A_2} = \emptyset$} (m2);
	\draw [postaction={decorate}] (n1) to [bend right=40] node[below]{$r_{A_1} = {\tiny \yng(1,1,1,1)}$} (m2);
	\draw [postaction={decorate}] (n2) to [bend right=15] node[below]{$r_{B_2} = \emptyset$} (m1);
	\draw [postaction={decorate}] (n2) to [bend right=40] node[above]{$r_{B_1} = {\tiny \yng(1,1,1,1)}$} (m1);			
}
\right)
\end{equation}
Here the meaning of the diagram is just a convenient visualization of the labels $\bL$, while the actual operator is $\cO_\cC(\bL)$ as defined in (\ref{eq:OL_con}). The single-column Young diagrams are understood to have $N$ boxes. Note because of the single-column diagrams there are no multiplicities at the white nodes, which makes them particularly nice examples.
Next, if we fix $R_1=R_2=[1^N]$, but allow different numbers $(n_{A_1},n_{A_2},n_{B_1},n_{B_2})$ of fields, subject to restriction $n_{A_1}+n_{A_2}=n_{B_1}+n_{B_2}=N$, we get one operator for each choice of charges
\begin{equation}
\cO\left(
\mytikz{	
	\node (m1) at (-1.5,0.5) [circle,draw,inner sep=0.5mm] {};
	\node (n1) at (-1.5,-0.5) [circle,draw,inner sep=0.5mm] {};
	\node (m2) at (1.5,-0.5) [circle,draw,inner sep=0.5mm] {};
	\node (n2) at (1.5,0.5) [circle,draw,inner sep=0.5mm] {};	
	\draw[postaction={decorate}] (m1) to node[left]{$R_1 = {\tiny \yng(1,1,1,1)}\;$} (n1);
	\draw[postaction={decorate}] (m2) to node[right]{$R_2 = {\tiny \yng(1,1,1,1)}$} (n2);
	\draw [postaction={decorate}] (n1) to [bend right=15] node[above]{$r_{A_2} = {\tiny \yng(1,1)}$} (m2);
	\draw [postaction={decorate}] (n1) to [bend right=40] node[below]{$r_{A_1} = {\tiny \yng(1,1)}$} (m2);
	\draw [postaction={decorate}] (n2) to [bend right=15] node[below]{$r_{B_2} = {\tiny \yng(1)}$} (m1);
	\draw [postaction={decorate}] (n2) to [bend right=40] node[above]{$r_{B_1} = {\tiny \yng(1,1,1)}$} (m1);			
}
\right)
\end{equation}
This is because $r_{A_i}, r_{B_i}$ diagrams are forced to be single-column by Littlewood-Richardson rule.
In fact, that is precisely what is needed to fill out the $(N+1,N+1)$ representation of $SU(2)\times SU(2)$ of which $\det(A_1 B_1)$ is the highest weight state. Using covariant basis we can label the whole representation by  $R_1,R_2,\Lambda_A,\Lambda_B$ as in Figure~\ref{fig:covlabel_column}.
\begin{figure}[h]
\centering
 \mytikz{	
	\node (m1) at (-1.5,0.5) [circle,draw,inner sep=0.5mm] {};
	\node (n1) at (-1.5,-0.5) [circle,draw,inner sep=0.5mm] {};
	\node (m2) at (1.5,-0.5) [circle,draw,inner sep=0.5mm] {};
	\node (n2) at (1.5,0.5) [circle,draw,inner sep=0.5mm] {};	
	\node (t12) at (0,-0.7) [circle,fill,inner sep=0.5mm] {};	
	\node (b12) at (0,-1.5) {$\Lambda_{A}={\tiny \yng(4)}$};
	\node (t21) at (0,0.7) [circle,fill,inner sep=0.5mm] {};
	\node (b21) at (0,1.5) {$\Lambda_{B}={\tiny \yng(4)}$};
	\draw[postaction={decorate}] (m1) to node[left]{$R_1 = {\tiny \yng(1,1,1,1)}\;$} (n1);
	\draw[postaction={decorate}] (m2) to node[right]{$R_2 = {\tiny \yng(1,1,1,1)}$} (n2);
	\draw [postaction={decorate}] (n1) to [bend right=5] (t12);
	\draw [postaction={decorate}] (t12) to [bend right=5] (m2);
	\draw [postaction={decorate}] (n2) to [bend right=5] (t21);
	\draw [postaction={decorate}] (t21) to [bend right=5] (m1);
	\draw[-] (t12) to (b12);
	\draw[-] (t21) to (b21);
}
\caption{Representation containing giant gravitons expanding in $T^{1,1}$}
\label{fig:covlabel_column}
\end{figure} 

\begin{figure}[h]
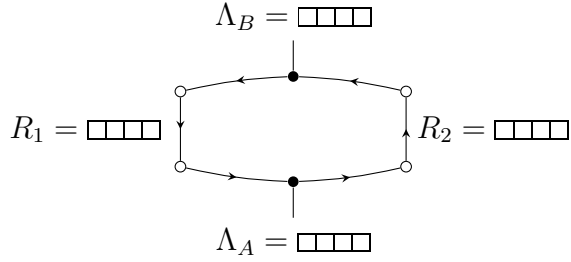

\centering
 \mytikz{	
	\node (m1) at (-1.5,0.5) [circle,draw,inner sep=0.5mm] {};
	\node (n1) at (-1.5,-0.5) [circle,draw,inner sep=0.5mm] {};
	\node (m2) at (1.5,-0.5) [circle,draw,inner sep=0.5mm] {};
	\node (n2) at (1.5,0.5) [circle,draw,inner sep=0.5mm] {};	
	\node (t12) at (0,-0.7) [circle,fill,inner sep=0.5mm] {};	
	\node (b12) at (0,-1.5) {$\Lambda_{A}={\tiny \yng(4)}$};
	\node (t21) at (0,0.7) [circle,fill,inner sep=0.5mm] {};
	\node (b21) at (0,1.5) {$\Lambda_{B}={\tiny \yng(4)}$};
	\draw[postaction={decorate}] (m1) to node[left]{$R_1 ={\tiny \yng(4)}\;$} (n1);
	\draw[postaction={decorate}] (m2) to node[right]{$R_2 ={\tiny \yng(4)}$} (n2);
	\draw [postaction={decorate}] (n1) to [bend right=5] (t12);
	\draw [postaction={decorate}] (t12) to [bend right=5] (m2);
	\draw [postaction={decorate}] (n2) to [bend right=5] (t21);
	\draw [postaction={decorate}] (t21) to [bend right=5] (m1);
	\draw[-] (t12) to (b12);
	\draw[-] (t21) to (b21);
}
\caption{Representation containing $AdS$ giants}
\label{fig:covlabel_row}
\end{figure}

In analogy with half-BPS states in $\mC^3$, it is natural to suggest that if single-column operators $R_1=R_2=[1^N]$ correspond to giants expanding in the compact $T^{1,1}$, then single-row operators $R_1=R_2=[n]$ would be dual giants, expanding in $AdS_5$ and point-like in $T^{1,1}$. These states live in the $SU(2)\times SU(2)$ representation $(N+1,N+1)$ labelled as in Figure~\ref{fig:covlabel_row}.

\begin{figure}[h]
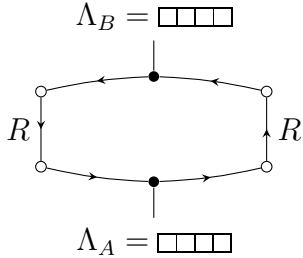

\centering
 \mytikz{	
	\node (m1) at (-1.5,0.5) [circle,draw,inner sep=0.5mm] {};
	\node (n1) at (-1.5,-0.5) [circle,draw,inner sep=0.5mm] {};
	\node (m2) at (1.5,-0.5) [circle,draw,inner sep=0.5mm] {};
	\node (n2) at (1.5,0.5) [circle,draw,inner sep=0.5mm] {};	
	\node (t12) at (0,-0.7) [circle,fill,inner sep=0.5mm] {};	
	\node (b12) at (0,-1.5) {$\Lambda_{A}={\tiny \yng(4)}$};
	\node (t21) at (0,0.7) [circle,fill,inner sep=0.5mm] {};
	\node (b21) at (0,1.5) {$\Lambda_{B}={\tiny \yng(4)}$};
	\draw[postaction={decorate}] (m1) to node[left]{$R$} (n1);
	\draw[postaction={decorate}] (m2) to node[right]{$R$} (n2);
	\draw [postaction={decorate}] (n1) to [bend right=5] (t12);
	\draw [postaction={decorate}] (t12) to [bend right=5] (m2);
	\draw [postaction={decorate}] (n2) to [bend right=5] (t21);
	\draw [postaction={decorate}] (t21) to [bend right=5] (m1);
	\draw[-] (t12) to (b12);
	\draw[-] (t21) to (b21);
}
\caption{Protected representation}
\label{fig:covlabel_R}
\end{figure}

It is important to note that, in principle, in order to match with the D3 brane states on the bulk side, we need to use the interacting chiral ring. The reason we can rely on the free chiral ring operators in these examples, is that the highest weight state involves only $A_1,B_1$ and no $A_2,B_2$. The F-terms only identify operators by symmetrizing $A_1,A_2$ and $B_1,B_2$, so for an operator like $\det(A_1 B_1)$ there are no F-term identifications. In other words, the elements of the interacting chiral ring are equivalence classes up to F-terms, but an operator only involving $A_1,B_1$ is the unique operator in its equivalence class. Therefore, these operators are protected from mixing, in a similar way like half-BPS operators in $\mC^3$. We can identify all such protected operators: in order to have a highest weight state with only $A_1,B_1$, the $SU(2)\times SU(2)$ representation must be $(\Lambda_A,\Lambda_B) = ([n], [n])$, where $\Lambda$'s are single-row. This is analogous to half-BPS operators in $\mC^3$ having $\Lambda=[n]$. Then $R_1=R_2$ can be anything, but are forced to be equal, in order to have $[n]$ in their product. Thus we have a class of operators in the chiral ring labelled by $R \vdash n$ for any $n$ as in Figure~\ref{fig:covlabel_R}.

The highest weight operator in this representation can be expressed as
\begin{equation}
\cO_\cC(R) = \frac{1}{n!} \sum_{\sigma} \chi_R(\sigma) \trVN{n}(\sigma \, (A_1 B_1)^{\otimes n})
\end{equation}

%%%%%%%%%%%%%%%%%%%%%%%%%%%%%%%%%%%%%%%%%%%%%%%%%%%%%%%%%%%%%%%%%%%%%
%%%%%%%%%%%%%%%%%%%%%%%%%%%%%%%%%%%%%%%%%%%%%%%%%%%%%%%%%%%%%%%%%%%%%

\section{The case with fundamental matter} 
\label{sec:fundmatt} 

In this section we sketch how our techniques can be extended to quivers involving fundamental and anti-fundamental matter. These are represented by different kind of nodes, with only incoming or only outgoing arrows.

The simplest example is SQCD. The counting of chiral operators
has been studied in connection with  the moduli space for SQCD in papers 
\cite{chenmek,hanmek,ghhjm,JJV}.  
These papers have made a connection between Schur polynomials and this counting. 
Here we will construct a basis for these operators which diagonalizes the (Zamolodchikov)  inner product 
given by the 2-point functions in the  free field theory at finite $N$.
The enumeration of states in the basis will agree with the previous counting results.

\begin{figure}[h!]
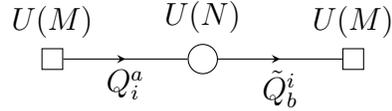

\centering
\mytikz{
	\node (n) at (0,0) [draw,circle,label=above:$U(N)$] {};
	\node (m1) at (-2,0) [draw,rectangle,label=above:$U(M)$] {};
	\node (m2) at (2,0) [draw,rectangle,label=above:$U(M)$] {};
	\draw [postaction={decorate}] (m1) to node[below]{$Q^a_i$} (n);
	\draw [postaction={decorate}] (n) to node[below]{$\tl Q^i_b$} (m2);
}
\caption{SQCD quiver for gauge group $U(N)$, $M$ quarks $Q^a_i$ and $M$ antiquarks $\tl Q^i_b$}
\label{fig:quiver-sqcd}
\end{figure}

Consider $U(N)$ gauge theory with $M$ chiral multiplets in a fundamental representation (quarks) and $M$ in anti-fundamental (anti-quarks), the quiver is shown in Figure~\ref{fig:quiver-sqcd}.
Gauge invariant operators can be written like in analogy with (\ref{eq:Osig_C3Z2}), but now with open flavour indices
\begin{equation}
\begin{split} 
\cO( \sigma , \vec a , \vec b ) &= 
(Q^{a_1}_{j_1  }  \cdots Q^{  a_n }_{ j_n })
(\delta_{ i_{ \s (  1 )  }}^{j_1}  \cdots \delta_{ i_{\s ( n ) } }^{j_n})
(\tQ^{i_1 }_{ b_1} \cdots \tQ^{i_n }_{ b_n })
\\
&\equiv  
\quad
\mytikz {
	\node (v) at (0,0.7) {};
	\node (v) at (0,-0.7) {};
	\node (a) at (-3,0) {$\vec a$};
	\node (q1) at (-1.5,0) [draw,rectangle] {$Q^{\otimes n}$};
	\node (s) at (0,0) [draw,rectangle] {$\sigma$};
	\node (q2) at (1.5,0) [draw,rectangle] {$\tl Q^{\otimes n}$};
	\node (b) at (3,0) {$\vec b$};
	\draw [postaction={decorate}] (a) to (q1);
	\draw [postaction={decorate}] (q1) to (s);
	\draw [postaction={decorate}] (s) to (q2);
	\draw [postaction={decorate}] (q2) to (b);
} 
\end{split}
\end{equation}
We observe, as in (\ref{eq:OQ_adj}), there is an invariance
\begin{equation}\begin{split} 
\cO( \gamma_1  \sigma \gamma_2 ; \vec a , \vec b )
= \cO( \sigma ; \gamma_1 ( \vec a )  ,  \gamma_2^{-1}  ( \vec b ) ),
\qquad \g_1,\g_2 \in S_M
\end{split}\end{equation}
Define the Fourier transform and use the constraint
\begin{equation}\begin{split} 
& \cO^{R , S , T }_{ ij ; M_S , M_T ; k l   } = \sum_{ \sigma \in S_n  } { d_R }   D^R_{ij} ( \sigma  )C_{\vec a }^{ S , M_S , k } C^{\vec b }_{ T , M_T , l } \cO ( \s ; \vec a , \vec b ) \\ 
& = { 1 \over n!^2 } \sum_{ \g_1 , \g_2 , \s } D^R_{ij} ( \g_1 \s \g_2 ) C^{ S , M_S , k }_{ \gamma_1 ( \vec a ) }  C^{ \gamma_2^{-1}  ( \vec b )  }_{ T , M_T , l } \cO ( \s ; \vec a , \vec b ) \\ 
& = { 1 \over n!^2 } \sum_{ \g_1 , \g_2 , \s }   D_{  k k'}^S   ( \g_1) D^{T}_{l' l } ( \g_2 )  D^R_{ij} ( \g_1 \s \g_2 )  C^{ S , M_S , k' }_{  \vec a  }C^{ \vec b  }_{ T , M_T , l' } \cO ( \s ; \vec a , \vec b ) \\ 
& = \delta_{ R , S } \delta_{ R, T } \delta_{ik} \delta_{jl} \cO^{ R , S , T}_{ k' l' ; M_S , M_T ; k' l' } 
\end{split}\end{equation}
This means that  we can define 
\begin{equation}\begin{split}\label{oplabels} 
\cO^R_{M_R, M'_R} = \sum_{ i , j } \cO^{R , R , R }_{ ij ; M_R , M_R' ; ij    } 
\end{split}\end{equation}
which solve  the constraint. Since the representation $R$  of $S_n$ comes from matrix elements of a 
permutation which permutes indices $ i_1 , \cdots , i_M$ which range over $ 1 \cdots N$, there is a 
cutoff $ l(R) \le N$. Anti-symmetrizations of more than $N$ copies of such indices always give zero.

Note the labelling of the operators follows the split-node quiver pattern, shown in Figure~\ref{fig:split-quiver-sqcd}, the new ingredient is the open lines carrying flavour indices. In this simple case the branching nodes causes all irreps to be the same, but it is easy to see, how to generalize this to more complicated quivers, with non-trivial branchings.
\begin{figure}[h!]
\centering
$$
\cO^R_{M_R, M'_R}
\quad \sim \quad
\mytikz{
	\node (v) at (0,0.7) {};
	\node (v) at (0,-0.7) {};
	\node (a) at (-2,0) {$M_R$};
	\node (n1) at (-0.7,0) [circle,draw,inner sep=0.5mm] {};
	\node (n2) at (0.7,0) [circle,draw,inner sep=0.5mm] {};	
	\node (b) at (2,0) {$\tl M_R$};
	\draw [postaction={decorate}] (a) to node[above]{$R$} (n1);
	\draw [postaction={decorate}] (n1) to node[above]{$R$} (n2);
	\draw [postaction={decorate}] (n2) to node[above]{$R$} (b);		
}
$$
\caption{Labelled split-node SQCD quiver, with flavour state labels $M_R,\tl M_R$}
\label{fig:split-quiver-sqcd}
\end{figure}

This has implications for counting. Let $ \cN ( n_1 , \cdots , n_M ; \tilde n_1 , \cdots , \tilde n_M   )$ be the number of states
with $n_i $ copies of  $ Q_i$ and $\tilde n_i $ copies of $  \tilde Q_i$. They diagonalize the generators of 
two copies of $U(M)$ which are $ H_i \equiv \sum_{a=1}^N   Q_i^a { \partial \over \partial  Q^a_i } $ and  $ \tilde  H_i \equiv \sum_{a=1}^N   \tilde Q_i^a  {  \partial \over \partial  \tilde Q^a_i } $ 
The generating function for these numbers can be defined as  
\begin{equation}\begin{split} 
& \cN ( t_1 , \cdots , t_M ; \tilde t_1 , \cdots , \tilde t_M )  =
\sum_{ n_i , \tilde n_i   } \cN ( n_1 , \cdots , n_M ; \tilde n_1 , \cdots , \tilde n_M   ) t_1^{n_1 } t_2^{n_2} \cdots t_M^{ n_M } \tilde t_1^{\tilde n_1 } \tilde t_2^{ \tilde n_2} \cdots \tilde t_M^{\tilde  n_M } 
\end{split}\end{equation}
where the powers of $t_i , \tilde t_i $ give  the eigenvalues of  $H_i , \bar H_i$. This can be expressed in terms of the characters 
of $ U(M) \times U(M) $ 
\begin{equation}
 \chi_S ( \vec t ) = tr_S \prod_i t_i^{H_i}  \qquad 
 \chi_T ( \vec \tilde t  )  = tr_T \prod_i \tilde t_i^{ \tilde  H_i  } 
\end{equation} 
So the counting function  is 
\begin{equation}\begin{split} 
 \cN ( t_1 , \cdots , t_M ; \tilde t_1 , \cdots , \tilde t_M )  = \sum_{ S , T } M ( S , T )  \chi_S ( t_1 , \cdots , t_M ) \chi_T ( \tilde t_1 , \cdots , t_M ) 
\end{split}\end{equation}
where $M( S , T ) $ is the number of times irreps $S  \times T$ of $U(M) \times U(M)$ appear in the space of chiral operators. 
According to (\ref{oplabels}), we have $ M( S , T ) = \delta_{ S , T }   $ when $ l( S ) = l( T )  \le N $   and zero otherwise. 
So we conclude that 
\begin{equation}\begin{split}\label{countsqcd}  
 \cN ( t_1 , \cdots , t_M ; \tilde t_1 , \cdots , \tilde t_M )  = \sum_{ \substack{ S \in Rep  (U(M) )  \\ l( S ) \le N } } \chi_S ( \vec t ) \chi_S ( \vec { \tilde t}   ) 
\end{split}\end{equation}
When $M < N $, the $l(S) \le N $ constraint is automatically satisfied by irreps of $U(M)$ so we have 
\begin{equation}\begin{split} 
\cN ( t_1 , \cdots , t_M ; \tilde t_1 , \cdots , \tilde t_M )  = \sum_{  S  \in Rep  (U(M) )  } \chi_S ( \vec t ) \chi_S ( \vec { \tilde t}   )  
= \prod_{  i , j  =1}^{M }   {1  \over ( 1 - t_i \tilde t_j )  }  
\end{split}\end{equation}
In the last line, we used the Cauchy-Liitlewood formula. See similar discussion in \cite{JJV} and earlier papers 
\cite{chenmek,hanmek,ghhjm}.

Now we will see that the Fourier basis diagonalizes the two point function in the free theory
\begin{equation}\begin{split}
< Q^a_i Q^{b\dagger}_j > = \delta^{ab} \delta_{ij}, \qquad
< \tl Q_a^i \tl Q_{b}^{j\dagger} > = \delta_{ab} \delta^{ij}
\end{split}\end{equation}

The first steps are written diagrammatically as:
\begin{equation}
\begin{split}
\la O^R_{M_R, M'_R} O^{\dagger S}_{M_S, M'_S} \ra 
&=
D^R_{i_1 j_1}(\s_1) D^S_{i_2 j_2}(\s_2) 
\la
\mytikz{
	\node (a) at (0,2.5) {$R, M_R, j_1$};
	\node (n1) at (0,1.8) [circle,draw,inner sep=0.5mm] {};
	\node (q1) at (0,1) [draw,rectangle] {$Q^{\otimes n}$};
	\node (s) at (0,0) [draw,rectangle] {$\sigma_1$};
	\node (q2) at (0,-1) [draw,rectangle] {$\tl Q^{\otimes n}$};
	\node (n2) at (0,-1.8) [circle,draw,inner sep=0.5mm] {};
	\node (b) at (0,-2.5) {$R, M'_R, i_1$};
	\draw [postaction={decorate}] (a) to (n1);
	\draw [postaction={decorate}] (n1) to (q1);
	\draw [postaction={decorate}] (q1) to (s);
	\draw [postaction={decorate}] (s) to (q2);
	\draw [postaction={decorate}] (q2) to (n2);
	\draw [postaction={decorate}] (n2) to (b);
}
\mytikz{
	\node (a) at (0,2.5) {$S, M_S, i_2$};
	\node (n1) at (0,1.8) [circle,draw,inner sep=0.5mm] {};
	\node (q1) at (0,1) [draw,rectangle] {$Q^{\dagger \otimes n}$};
	\node (s) at (0,0) [draw,rectangle] {$\sigma_2$};
	\node (q2) at (0,-1) [draw,rectangle] {$\tl Q^{\dagger \otimes n}$};
	\node (n2) at (0,-1.8) [circle,draw,inner sep=0.5mm] {};
	\node (b) at (0,-2.5) {$S, M'_S, j_2$};
	\draw [postaction={decorate}] (n1) to (a);
	\draw [postaction={decorate}] (q1) to (n1);
	\draw [postaction={decorate}] (s) to (q1);
	\draw [postaction={decorate}] (q2) to (s);
	\draw [postaction={decorate}] (n2) to (q2);
	\draw [postaction={decorate}] (b) to (n2);
}
\ra
\\
& =
\sum_{\g_1,\g_2} D^R_{i_1 j_1}(\s_1) D^S_{i_2 j_2}(\s_2) 
\la
\mytikz{
	\def\x{1};
	\node (a) at (-\x,2.5) {$R, M_R, j_1$};
	\node (n1) at (-\x,1.8) [circle,draw,inner sep=0.5mm] {};
	\node (s) at (-\x,0) [draw,rectangle] {$\sigma_1$};
	\node (n2) at (-\x,-1.8) [circle,draw,inner sep=0.5mm] {};
	\node (b) at (-\x,-2.5) {$R, M'_R, i_1$};
	\draw [postaction={decorate}] (a) to (n1);
	\draw [postaction={decorate}] (n2) to (b);
	\node (a2) at (\x,2.5) {$S, M_S, i_2$};
	\node (n12) at (\x,1.8) [circle,draw,inner sep=0.5mm] {};
	\node (s2) at (\x,0) [draw,rectangle] {$\sigma_2$};
	\node (n22) at (\x,-1.8) [circle,draw,inner sep=0.5mm] {};
	\node (b2) at (\x,-2.5) {$S, M'_S, j_2$};
	\draw [postaction={decorate}] (n12) to (a2);
	\draw [postaction={decorate}] (b2) to (n22);
	\node (g1) at (0,1.5) [draw,rectangle] {$\g_1$};
	\node (g1i) at (0,0.5) [draw,rectangle] {$\g_1^{-1}$};
	\node (g2) at (0,-1.5) [draw,rectangle] {$\g_2$};
	\node (g2i) at (0,-0.5) [draw,rectangle] {$\g_2^{-1}$};
	\draw [postaction={decorate}] (n1) .. controls (-\x,1.5) .. (g1);
	\draw [postaction={decorate}] (g1) .. controls (\x,1.5) .. (n12);
	\draw [postaction={decorate}] (s2) .. controls (\x,0.5) .. (g1i);
	\draw [postaction={decorate}] (g1i) .. controls (-\x,0.5) .. (s);
	\draw [postaction={decorate}] (n22) .. controls (\x,-1.5) .. (g2);
	\draw [postaction={decorate}] (g2) .. controls (-\x,-1.5) .. (n2);
	\draw [postaction={decorate}] (s) .. controls (-\x,-0.5) .. (g2i);
	\draw [postaction={decorate}] (g2i) .. controls (\x,-0.5) .. (s2);
}
\ra
\end{split}
\end{equation}
This leads to 
\begin{equation}\begin{split} 
\langle \cO^{R}_{ M_R , M'_R} \cO^{\dagger S }_{M_S , M'_S } \rangle 
& = \sum_{ \gamma_1 , \gamma_2 } \delta_{R, S } \delta_{ M_R , M_S } \delta_{ M'_R , M'_S } 
 D^R_{ j_1 i_2 } ( \gamma_1 ) D^R_{j_2 i_1} ( \gamma_2 ) D^R_{ i_1 j_1} ( \sigma_1 )  D^S_{i_2 j_2 } ( \s_2 ) \\ 
& \times\sum_T Dim T \chi_T (  \s_1 \g_2^{-1} \s_2 \g_1^{-1} )  
\end{split}\end{equation}
Expansing the character in terms of matrix elements, and using orthogonality of 
elements gives 
\begin{equation}\begin{split}
{ (n!)^4 Dim R \over d_R^2 } \delta^{ R , S  } \delta_{ M_R , M'_S } \delta_{ M_S , M'_R } 
\end{split}\end{equation}
This shows that the basis is orthogonal. 
The Clebsch's $ C_{\vec a}^{ R, M_R , i } $ can be written  by labelling the states $M_R$ using the  numbers $n_i$ of the 
different fkavours of $Q$, along with branching coefficients  for the decomposition of the representation  $R$ of $S_n$ 
to the invarint representation of $\prod_i S_{n_i}$ (as we have done elsewhere).

%%%%%%%%%%%%%%%%%%%%%%%%%%%%%%%%%%%%%%%%%%%%%%%%%%%%%%%%%%%%%%%%%%%%%
%%%%%%%%%%%%%%%%%%%%%%%%%%%%%%%%%%%%%%%%%%%%%%%%%%%%%%%%%%%%%%%%%%%%%

\section{Discussion and future avenues }
\label{sec:discussion} 

\subsection{IR fixed point}
\label{sec:IRfxdpoint}

Let us discuss in some more detail the RG flow of the conifold theory (see \cite{Strassler:2005qs} for a good review). The superpotential is
\begin{equation}
W = h \left( \tr(A_1 B_1 A_2 B_2) - \tr(A_1 B_2 A_2 B_1) \right)
\end{equation}
where we have reinstated the coupling constant $h$. Let us define a dimensionless coupling constant $\eta = h \mu$, with energy scale $\mu$. The dimensionless couplings of the theory are then $(g_1,g_2,\eta)$, where $(g_1,g_2)$ are the gauge couplings of the two group factors. We focus on the case where $g_1=g_2\equiv g$. 

\begin{figure}[h]
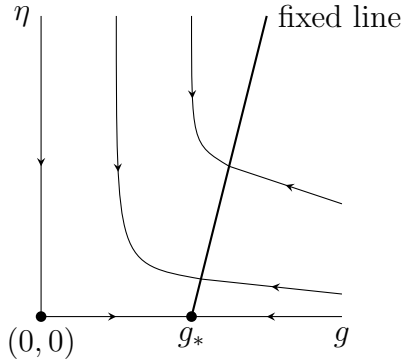

\centering
\mytikz{	
	\draw[postaction={decorate}] (0,0) to (2,0);
	\draw[postaction={decorate}] (4,0) to (2,0);
	\draw (4,0) node[below]{$g$};
	\draw[postaction={decorate}] (0,4) to (0,0);	
	\draw (0,4) node[left]{$\eta$};
	\node (g0) at (0,0) [circle,fill,inner sep=0.5mm] {};
	\node (gs) at (2,0) [circle,fill,inner sep=0.5mm] {};
	\draw (0,0) node[below]{$(0,0)$};
	\draw (2,0) node[below]{$g_*$};
	\draw[thick] (2,0) to (3,4) node[right]{fixed line};
	\draw[postaction={decorate}] (1,4) .. controls (1,0.7) .. (2.125,0.5);
	\draw[postaction={decorate}] (4,0.3) to (2.125,0.5);
	\draw[postaction={decorate}] (2,4) .. controls (2,2.3) .. (2.5,2);	
	\draw[postaction={decorate}] (4,1.5) to (2.5,2);
}
\caption{RG flow diagram of the conifold theory}
\label{fig:conifoldRG}
\end{figure}

The theory is asymptotically free, so perturbatively $g$ increases in the IR. The coupling $\eta$ classically scales like $\mu$ and vanishes in the IR, corresponding to the fact that $W$ is classically irrelevant. The full non-perturbative RG flow diagram, however, looks like in Figure~\ref{fig:conifoldRG}. There is a line of fixed points in the $(g,\eta)$ plane, which originates at the $(g,\eta) = (g_*, 0)$ point and extends up towards the strongly coupled regime. This means the theory in the IR has a marginal coupling, which controls the position on the fixed line. In the context of AdS/CFT correspondence this marginal coupling is related to $\alpha'$ in the bulk, and the supergravity regime corresponds to strong coupling, that is, being far up along the line.

Note there is also a trivial free fixed point, disconnected from the line, at $(g,\eta)=(0,0)$. Let us focus on the RG flow from this UV fixed point $(0,0)$ to the IR fixed point $(g_*,0)$. The theory in the IR is a strongly coupled CFT, but with zero superpotential. This fixed point is similar to the usual Seiberg fixed point in a $N_f = 2 N_c$ SQCD, and is qualitatively different from the rest of the fixed line. With $W=0$ the F-terms vanish, and the chiral ring is much larger compared to $\eta\neq 0$ theory.

Our main observation here, is that many of the results in this paper regarding ``free'' theory are valid not only in the UV free fixed point, but also in the IR interacting fixed point $g_*$. Consider the basis of operators $\cO(\bL)$ or $\cO(\bK)$. Part of the motivation for this particular basis is that it diagonalizes the free field metric (\ref{eq:free_metric}), valid in the UV. This will get modified along the flow, and $\cO(\bL)$ will likely no longer be orthogonal in the IR, using the CFT two-point function. However, it is known that chiral ring itself is not changed along the flow \cite{cdsw}, so our basis will still be a \emph{complete linearly independent finite $N$ basis} for the chiral ring in the IR. From this perspective, the free two-point function could be seen as a particular inner product on the chiral ring states, which allows to solve the finite $N$ constraints. Therefore, one of the key results of our paper, the chiral ring structure constants of the ``free'' operators (\ref{eq:GLLL_diag}) and (\ref{eq:GKKK_diag}), which depend only on the holomorphic information and not on the two-point function, are valid in the interacting fixed point $g_*$.

\subsection{Directions for the future} 

We outline some applications, extensions and questions arising from this work. 

\begin{itemize}

\item The counting formula \ref{Infprod}   which we derived for quivers describing general bifundamental fields (including adjoints)
should admit a generalization to the case with arbitrary number of fundamentals 
as well as bi-fundamentals. 
A derivation following the methods here should be possible. 
Some  external nodes of the quiver will have only incoming arrows 
or outgoing arrows.
The counting by splitting the internal nodes and  associating Young diagrams 
to all the edges, and Littlewood-Richardson coefficients  to the  nodes, 
should continue to work. The construction of orthogonal  
operators should proceed by similar methods, with quiver chracters 
having permutations inserted between splittings of the internal nodes, 
and branching coefficients at vertices. A first step focusing on SQCD has been 
taken in Section \ref{sec:fundmatt}.

\item  We have considered the 2-point functions and multiplication of  operators constructed from 
scalar bi-fundamentals. In the case of $\cN =1 $ SUSY, the results for fermions in a chiral multiplet 
can be obtained by applying the supersymmetry algebra. 
Deriving the formulae for the case of fermionic operators directly 
from the Wick contractions would be interesting, with applicability extending to non-supersymmetric theories
such as those that play a a role in dimensional deconstruction \cite{ahcg}.

\item The counting and chiral ring struture constants have been computed for 
standard quiver theories. It will be interesting to see how far we can 
apply the 
present methods to compute these quantities in theories described by 
generalized quivers \cite{gaiotto}. 

\item 
Generalize the present  results to the case where the $U(N_a)$ gauge groups 
are replaced by other classical groups.  An immediate question is to 
extend the discussion to include baryonic vertices which are important 
for $SU(N_a)$ gauge groups. For the one-matrix problem, corresponding to 
the quiver with one node and one edge, there has been progress on the 
 $O(N)$ case \cite{AABF,CDD}.

\item 
The appearance of emergent riemann surfaces  in connection 
with gauge theory here  is reminiscent of \cite{WittenM5,gaiotto}. The present story involves correlators (of arbitrarily high dimension operators)  at a 
fixed point of moduli space without vevs being turned on, while the one of \cite{WittenM5,gaiotto} is 
looking at the non-perturbative moduli space of vacua and the 6D origin of 4D theories.
In the present story, we encounter sums over covers of the Riemann surface, 
while \cite{gaiotto} involves 
a distinguished covering of the UV curve by the IR curve. 
Despite these differences, it is tempting to ask if there is some 
unified story that explains these two apperances of 
Riemann surfaces correlated with gauge group and matter content. 

\item 
The association of Young diagrams to quiver gauge theory data we have encountered here 
is also reminiscent of the topological vertex \cite{AKMV03}. For quiver theories 
arising from branes at 
toric singularities, the latter uses toric diagrams of the CY which appears in the moduli space 
of the gauge theory at non-zero superpotential. Our constructions have been in the limit of 
vanishing superpotential. Further work along the lines 
of Section \ref{sec:int-chir-ring} may help 
in exploring the relation of the present constructions to \cite{AKMV03}. 

\item 
Our results are relevant for free quiver gauge theories in any dimension. 
As such they are also relevant to  matrix models and  matrix quantum mechanics, 
associated with quivers. This type of quantum mechanics has been useful in the context of black holes in $N=2$ compactifications (e.g. \cite{BBdEV12}). It would be interesting to explore potential applications of  
the quiver characcters  and $S_n$ TFTs we find here in this black hole context \cite{BBdEV12}.

\item
The focus in this paper has been on explicit computations at the free
 field point, of quantities such as the CFT inner product and fusion coefficients. 
In cases with enough supersymmetry, there are differential 
equations on the moduli space of CFTs for the inner product \cite{papttstar}. 
The free field results can serve as boundary conditions for solving 
these equations. 

\item 
The constraints on the permutations $\sigma_a$ define a double coset, although 
we have not used this language. This double coset is 
\bea 
\prod_a \left ( ( \prod_{b,\alpha}    S_{n_{ba;\alpha} }       \times \prod_{b,\alpha}   S_{n_{ab;\alpha} } )  \setminus   ( S_{n_a} \times S_{n_a} ) / Diag ( S_{n_a} ) \right)  
\eea
Double cosets are known to admit products. It would be interesting to investigate the meaning of these products in the context of the present gauge theory applications.

\end{itemize}

\vskip.3in

{ \Large 
{ \centerline { \bf Acknowledgements }  } } 

\vskip.1in 
We are grateful  to Diego Rodriguez-Gomez for very helpful discussions and 
collaboration in the early stages of this work. We thank Ofer Aharony, Robert de Mello Koch, 
David Garner, Ami Hanany, Rodolfo Russo, Gabriele Travaglini, Brian Wecht  for discussions. 
We are grateful to the Isaac Newton Institute, Cambridge and the 
Institute for Advanced Study, Princeton for hospitality where part of this work was done. 
SR  is supported by STFC Grant ST/J000469/1, String theory, gauge theory, and duality. 
JP is supported by a Queen Mary, University of London studentship.

\vskip.5in

\begin{appendix}

%%%%%%%%%%%%%%%%%%%%%%%%%%%%%%%%%%%%%%%%%%%%%%%%%%%%%%%%%%%%%%%%%%%%%
%%%%%%%%%%%%%%%%%%%%%%%%%%%%%%%%%%%%%%%%%%%%%%%%%%%%%%%%%%%%%%%%%%%%%

\section{Symmetric group formulae}
\label{appsec:symgpform} 

\subsection{General}

$ R \vdash n $ will denote a Young diagram with $n$ boxes, associated with an irreducible representaton (irrep) of $S_n$. 
 A Young diagram $R$ is also 
associated with an irrep of $U(N)$, when the length  of the first column $l(R)$ obeys the constraint 
$l(R) \le N$.  $\Dim_N(R)$ denotes  the dimension of $U(N)$ irrep $R$. $d(R)$ is the dimension of $S_n$ irrep $R$.
\begin{equation}
	\Dim_N(R) = \frac{f_N(R)}{h(R)}, \quad d(R) = \frac{n!}{h(R)}
\end{equation}
$\Dim_N(R)$ is the dimension of $U(N)$ irrep $R$. $d(R)$ is the dimension of $S_n$ irrep $R$.
$f_N(R)$ is the ($N$-dependent) product of weights of boxes in the Young diagram. $h(R)$ is the product of hook
lengths. Describing the boxes of a Young diagram with coordinates $(i,j)$ running along 
rows and columns respectively, with $r_i$ being the row lengths and $c_j$ the column lengths 
\begin{equation}
\begin{split}
& f_N ( R ) = \prod_{i,j} ( N -i + j ) \\
& h(R ) = \prod_{i,j} ( r_i + c_j -i - j +  1 ) 
\end{split}
\end{equation}

The Kronecker Delta over the symmetric group $\delta(\sigma)$, defined to be $1$ if the argument is $1$ and 
zero otherwise. It is also defined, by linearity,  
over formal sums of group elements with complex coefficients
(the group algebra) by picking the coefficient of the identity permutation. 
It has an expansion in characters.
 There is a  related character orthogonality relation, 
obtained by summing over irreps
\begin{align}
\label{eq:chiR_sum}
	& \sum_{R \vdash n} 	\frac{d(R)}{n!} \chi_R(\sigma) = \delta(\sigma)	
\\
	& \sum_{R \vdash n} \chi_R(\sigma) \chi_R(\tau) = \sum_{\gamma \in S_n} \delta(\gamma \sigma \gamma^{-1} \tau^{-1})	
\end{align}

The characters are traces of matrix elements $ \chi_R ( \sigma )  = \sum_i  D^R_{ii}(\s)$. The matrix elements 
satisfy $D^R_{ij} ( \sigma ) = D^R_{ji} ( \sigma^{-1} )$ ). 
Orthogonality relations from summing over $\sigma$ are 
\begin{align}
\label{eq:D_sum_sigma}
	&\sum_{\s\in S_n} D^R_{ij}(\s) D^S_{kl}(\s) = \frac{n!}{d(R)} \delta_{RS} \delta_{ik} \delta_{jl} 
\\
& \sum_{\s\in S_n} \chi_R (\s) \chi_S (\s \tau) = \frac{n!}{d(R)} \delta_{RS} \chi_R ( \tau ) 
\\
& \sum_{\s\in S_n} \chi_R (\s) \chi_S (\s) = n! \, \delta_{RS} 
\\ 
	& \sum_{\sigma \in S_n} D^R_{ij}(\sigma) N^{c(\sigma)} = \delta_{ij} f_N(R)
\\
	& \sum_{\sigma \in S_n} \chi_R(\sigma) N^{c(\sigma)} = d(R) f_N(R) = n! \, \Dim_N(R)
\\
\label{eq:DP_tr_sum}
	&\sum_{\s\in S_n} \tr\left(P^{\nu^-,\nu^+}_{R\rightarrow \br } D^R(\s) \right) N^{c(\sigma)} = \delta^{\nu^-\nu^+} d(\br) \, f_N(R)
\end{align}
The last equation involves generalized projectors (intertwining operators) 
linking different copies (labelled by $\nu^+ , \nu^- $)  of the irrep $\br$ of  a subgroup $H \subset S_n $.
We will describe these subgroup reduction in more detail in the next subsection. For derivations of the above identities, the reader may consult e.g. 
\cite{Hamermesh}. 

\subsection{Branching coefficients}
\label{app:branching}

Consider a subgroup $H \subset S_n$. In this paper $H$ will be of the form 
\begin{equation}
\begin{split}
	H &= S_{n_1} \times S_{n_2} \times \ldots
\end{split}
\end{equation}
An irrep $R$  of $S_n$ can be decomposed into irreps $\br = (r_1,r_2,\ldots)$ of $H$
\begin{equation}
\begin{split}\label{SnHreduction} 
	V_R^{(S_n)} &= \bigoplus_{\substack{r_1 \vdash n_1 \\ r_2 \vdash n_2}}  V_{r_1}^{(S_{n_1})} \otimes V_{r_2}^{(S_{n_2})} \otimes V^{r_1 r_2}_R \\
	| V^{r_1 r_2}_R | &= g(r_1,r_2;R)
\end{split}	
\end{equation}
The states in $R$ are spanned by the basis $| R ; \br, \nu, \bl \rangle$ where $\br,\nu$ labels the irrep of $H$ ($\nu$ is the multiplicity label, if $\br$ appears multiple times in the decomposition), and $\bl=(l_1,l_2,\ldots)$ is a state in $\br=(r_1,r_2,\ldots)$. Branching coefficients $B^{R\rightarrow \br,\nu}_{i\rightarrow \bl}$ are defined to be the components of the vector $| R ; \br, \nu, \bl \rangle$ in terms of 
any orthogonal basis for $R$.   
\begin{equation}
\label{eq:Bil_matrix}
	B^{R\rightarrow \br,\nu}_{i\rightarrow \bl} = \langle R; i |R ; \br, \nu, \bl \rangle = \langle R ; \br, \nu, \bl | R; i \rangle
\end{equation}
Since the representations of $S_n$ can be chosen to be real, branching coefficients are real $(B^{R\rightarrow \br,\nu}_{i\rightarrow \bl})^* = B^{R\rightarrow \br,\nu}_{i\rightarrow \bl}$.

The multiplicities  $g(r_1,r_2;R)$ are given by the Littlewood-Richardson rule, 
which instructs us to put together the boxes of $r_2$ alongside those of $r_1$, subject to 
some conditions (see e.g \cite{fulhar}). These are usually first encountered in physics in the context of irreps of $U(N)$
but the present description in terms of reduction $S_n \rightarrow H$ is related to that by Schur-Weyl duality. 
Some times we will informally write
\begin{equation}
	r_1 \otimes r_2 = \bigoplus_{R} g(r_1,r_2;R) \, R
\end{equation}
in place of the more accurate (\ref{SnHreduction}). 

We use the following diagrammatic notation for the branching coefficients
\begin{equation}
\label{eq:Bil_diagram}
	B^{R\rightarrow (r_1,r_2,r_3),\nu}_{i\rightarrow (l_1,l_2,l_3)} \equiv \;
	\mytikz{
		\node (i) at (-1,0) {$i$};		
		\node (n) at (0,0) [circle,draw,inner sep=0.5mm,label=below:$\nu$] {};
		\node (l1) at (1.5,1) {$l_1$};
		\node (l2) at (1.5,0) {$l_2$};
		\node (l3) at (1.5,-1) {$l_3$};
		\draw [postaction={decorate}] (i) to node[above]{$R$} (n);
		\draw [postaction={decorate}] (n) to node[above]{$r_1$} (l1);
		\draw [postaction={decorate}] (n) to node[above right]{$r_2$} (l2);
		\draw [postaction={decorate}] (n) to node[below]{$r_3$} (l3);
	}
\end{equation}
Because of reality, the diagram with arrows reversed is equal.

Here we list the properties of branching coefficients in the diagrammatic notation, followed by the corresponding equations. 
 For illustration we  take the  subgroup $H=S_{n_1}\times S_{n_2}$, with the generalization to more factors being straightforward.
\begin{align}
\label{eq:B_gamma_pull}
	\mytikz{
		\node (m) at (0,0) [circle,draw,inner sep=0.5mm,label=below right:$\nu$] {};	
		\node (g1) at (-1,0.7) [rectangle,draw] {$\gamma_1$};
		\node (g2) at (-1,-0.7) [rectangle,draw] {$\gamma_2$};
		\draw [postaction={decorate}] (m) to node[above]{$R$} +(1,0);	
		\draw [postaction={decorate}] (g1) to node[above]{$r_1$} (m);
		\draw [postaction={decorate}] (g2) to node[below]{$r_2$} (m);		
		\draw [postaction={decorate}] ($(g1)+(-1.0,0)$) to (g1);
		\draw [postaction={decorate}] ($(g2)+(-1.0,0)$) to (g2);
	}
\; &= \;
	\mytikz{
		\node (s) at (0.5,0) [rectangle,draw] {$\gamma_1 \circ \gamma_2$};		
		\node (m) at (-1.5,0) [circle,draw,inner sep=0.5mm,label=below right:$\nu$] {};	
		\draw [postaction={decorate}] (s) to +(1.2,0);	
		\draw [postaction={decorate}] (m) to node[above]{$R$} (s);		
		\draw [postaction={decorate}] ($(m)+(-1,0.7)$) to node[above]{$r_1$} (m);
		\draw [postaction={decorate}] ($(m)+(-1,-0.7)$) to node[below]{$r_2$} (m);
	}
\\	
\label{eq:BB_R_delta}
	\mytikz{				
		\node (n) at (-0.7,0) [circle,draw,inner sep=0.5mm,label=below:$\nu$] {};	
		\node (tn) at (0.7,0) [circle,draw,inner sep=0.5mm,label=below:$\tl\nu$] {};	
		\node (i1) at (-2,0.7) {};		
		\node (i2) at (-2,-0.7) {};
		\node (j1) at (2,0.7) {};		
		\node (j2) at (2,-0.7) {};				
		\draw [postaction={decorate}] (i1) to node[above]{$r_1$} (n);
		\draw [postaction={decorate}] (i2) to node[below]{$r_2$} (n);
		\draw [postaction={decorate}] (n) to node[above]{$R$} (tn);		
		\draw [postaction={decorate}] (tn) to node[above]{$\tl r_1$} (j1);
		\draw [postaction={decorate}] (tn) to node[below]{$\tl r_2$} (j2);
	}
\; &= \;	
	\mytikz{
		\node (i1) at (-1,0.5) {};		
		\node (i2) at (-1,-0.5) {};
		\node (j1) at (1,0.5) {};		
		\node (j2) at (1,-0.5) {};
		\draw [postaction={decorate}] (i1) to node[above]{$r_1$} (j1);
		\draw [postaction={decorate}] (i2) to node[below]{$r_2$} (j2);
	}
	\times 
	\delta_{r_1 \tl r_1} \delta_{r_2 \tl r_2} \delta_{\nu \tl\nu}
\\	
\label{eq:BB_rr_sum}
	\sum_{r_1,r_2,\nu}
	\mytikz{				
		\node (n) at (-0.7,0) [circle,draw,inner sep=0.5mm,label=below:$\nu$] {};	
		\node (tn) at (0.7,0) [circle,draw,inner sep=0.5mm,label=below:$\nu$] {};	
		\node (i) at (-2,0) {};				
		\node (j) at (2,0) {};							
		\draw [postaction={decorate}] (n) to [bend left=45] node[above]{$r_1$} (tn);
		\draw [postaction={decorate}] (n) to [bend right=45] node[below]{$r_2$} (tn);		
		\draw [postaction={decorate}] (i) to node[above]{$R$} (n);
		\draw [postaction={decorate}] (tn) to node[above]{$R$} (j);		
	}
\; &= \;	
	\mytikz{
		\node (i) at (-1,0) {};				
		\node (j) at (1,0) {};	
		\draw [postaction={decorate}] (i) to node[above]{$R$} (j);
	}	
\\	
\label{eq:BB_g1g2}
	\sum_{r_1,r_2,\nu}
	\mytikz{				
		\node (n) at (-1.2,0) [circle,draw,inner sep=0.5mm,label=below:$\nu$] {};	
		\node (tn) at (1.2,0) [circle,draw,inner sep=0.5mm,label=below:$\nu$] {};	
		\node (s1) at (0,0.5) [rectangle,draw] {$\gamma_1$};
		\node (s2) at (0,-0.5) [rectangle,draw] {$\gamma_2$};
		\node (i) at (-2.2,0) {};				
		\node (j) at (2.2,0) {};							
		\draw [postaction={decorate}] (n) to [bend left=20] node[above]{$r_1$} (s1);
		\draw [postaction={decorate}] (n) to [bend right=20] node[below]{$r_2$} (s2);		
		\draw [postaction={decorate}] (s1) to [bend left=20] (tn);
		\draw [postaction={decorate}] (s2) to [bend right=20] (tn);
		\draw [postaction={decorate}] (i) to node[above]{$R$} (n);
		\draw [postaction={decorate}] (tn) to node[above]{$R$} (j);		
	}		
\; &= \;	
	\mytikz{
		\node (i) at (-1.7,0) {};			
		\node (s) at (0,0) [rectangle,draw] {$\gamma_1 \circ \gamma_2$};		
		\node (j) at (1.7,0) {};	
		\draw [postaction={decorate}] (i) to node[above]{$R$} (s);
		\draw [postaction={decorate}] (s) to (j);
	}	
\end{align}
The equations can be read off by assiging some state labels to each edge and branching 
coefficients for each white node. As usual with index notation, we need free indices  matching 
on both sides of the equation for the open ends of lines, and repeated  indices appearing 
on internal legs are assumed to be summed : 
\begin{align}
  D^{r_1}_{ i_1 j_1}  ( \gamma_1 )  D^{r_2}_{ i_2  j_2}  ( \gamma_2 )   B^{R\rightarrow  ( r_1 r_2  )  ,\nu}_{~ k \rightarrow j_1 , j_2}
 &=  B^{R\rightarrow (  r_1 r_2 )  ,\nu}_{~j  \rightarrow i_1 , i_2} D^{R}_{ k j }  ( \gamma_1 \circ \gamma_2 ) \\
 B^{ R \rightarrow (r_1,r_2) ; \nu }_{ ~ k \rightarrow  i_1 , i_2 }   B^{ R \rightarrow (\tl r_1,\tl r_2) ; \tl \nu }_{ ~ k \rightarrow  j_1 , j_2 } 
  &= \delta_{i_1 j_1 }\delta_{i_2 j_2 }   \delta_{ \nu  \tl \nu } \delta_{ r_1  \tl r_1  } \delta_{r_2  \tl r_2 } \\ 
  \sum_{ r_1 , r_2 , \nu }   B^{R\rightarrow (  r_1 r_2 )  ,\nu}_{~i  \rightarrow k_1 , k_2} 
  B^{R\rightarrow (  r_1 r_2 )  ,\nu}_{~ j  \rightarrow k_1 , k_2} &= \delta_{ i j } \\
   \sum_{ r_1 , r_2 , \nu }   B^{R\rightarrow (  r_1 r_2 )  ,\nu}_{~i  \rightarrow j_1 , j_2} D^{R_1}_{j_1 k_1} ( \gamma_1 ) 
  D^{R_2}_{j_2 k_2} ( \gamma_2 )   B^{R\rightarrow (  r_1 r_2 )  ,\nu}_{~j   \rightarrow k_1 , k_2}
 & = D^{R}_{ij} ( \gamma_1 \circ \gamma_2 )  
\end{align} 
As an example of the generalization to $ H = \times_b S_{n_b }$ with an arbitrary finite number of factors, the second 
equation above becomes : 
\bea 
B^{ R \rightarrow \cup_{b} r_b ; \nu }_{ ~ k \rightarrow \cup_{b}  i_b  } 
  B^{ R \rightarrow \cup_b  \tl r_b  ; \tl \nu }_{ ~ k \rightarrow  \cup_b j_b } 
  &=   \delta_{ \nu , \tl \nu } \prod_{b } \delta_{ r_b , \tilde  r_b }  \delta_{i_b  j_b }
\eea

Another useful identity is
\begin{equation}
\chi_R(\gamma_1 \circ \gamma_2) = \sum_{r_1,r_2} g(r_1,r_2;R) \chi_{r_1}(\gamma_1)\chi_{r_2}(\gamma_2)
\end{equation}
which we get by taking the trace of (\ref{eq:BB_g1g2}).

\subsection{Clebsch-Gordan coefficients}
\label{app:clebsch}

The standard tensor product of $S_n$ irreps, where we take a tensor 
product of two irreps $R, S $ of $S_n$ and then decompose into 
irreps $T$ of $S_n$ with multiplicities  $C(R,S,T)$, also plays a key  role in this paper. 
\begin{equation}
\begin{split}
	V_R^{(S_n)} \otimes V_S^{(S_n)} &= \bigoplus_{T \vdash n} V_T \otimes V_{RS}^T \\
	| V_{RS}^T | &= C(R,S,T)
\end{split}	
\end{equation}
To distinguish the coupling of irreps $(r_1 , r_2, \cdots )$  of $H = S_{n_1} \times S_{n_2} \cdots $ into irreps $R$ 
of $S_n$  (with $\sum_b n_b = n $) with the present decomposition 
relating three irreps of $S_n$, the former are sometimes called outer products of symmetric group 
irreps. while the latter are called Kronecker products. The Kronecker products are also called inner products sometimes 
but we will avoid that terminology, to avoid confusion with the scalar product of states within an irrep, which we will 
freely call inner product. 

The diagrammatic notation for the Clebsch-Gordan coefficient will be a black node:
\begin{equation}
\label{eq:CG_diag}
S^{R_1 R_2,\Lambda\,\tau}_{\,i_1\;i_2,\,m}
\; = \;
\mytikz{	
	\node (t) at (0,0) [circle,fill,inner sep=0.5mm,label=below:$\tau$] {};	
	\node (i1) at (-1.5,0.7) {$i_1$};
	\node (i2) at (-1.5,-0.7) {$i_2$};
	\node (m) at (1.5,0) {$m$};
	\draw [postaction={decorate}] (t) to node[above]{$\Lambda$} (m);		
	\draw [postaction={decorate}] (i1) to node[above]{$R_1$} (t);
	\draw [postaction={decorate}] (i2) to node[below]{$R_2$} (t);
}
\end{equation}

It obeys the following identities:
\begin{align}
\label{eq:CG_gamma_pull}
	\mytikz{
		\node (m) at (0,0) [circle,fill,inner sep=0.5mm,label=below:$\tau$] {};	
		\node (g1) at (-1,0.7) [rectangle,draw] {$\gamma$};
		\node (g2) at (-1,-0.7) [rectangle,draw] {$\gamma$};
		\draw [postaction={decorate}] (m) to node[above]{$\Lambda$} +(1,0);	
		\draw [postaction={decorate}] (g1) to node[above]{$R_1$} (m);
		\draw [postaction={decorate}] (g2) to node[below]{$R_2$} (m);		
		\draw [postaction={decorate}] ($(g1)+(-1.0,0)$) to (g1);
		\draw [postaction={decorate}] ($(g2)+(-1.0,0)$) to (g2);
	}
\; &= \;
	\mytikz{
		\node (s) at (-0.2,0) [rectangle,draw] {$\gamma$};		
		\node (m) at (-1.5,0) [circle,fill,inner sep=0.5mm,label=below:$\tau$] {};	
		\draw [postaction={decorate}] (s) to +(0.7,0);	
		\draw [postaction={decorate}] (m) to node[above]{$\Lambda$} (s);		
		\draw [postaction={decorate}] ($(m)+(-1,0.7)$) to node[above]{$R_1$} (m);
		\draw [postaction={decorate}] ($(m)+(-1,-0.7)$) to node[below]{$R_2$} (m);
	}
\\		
	\mytikz{				
		\node (n) at (-0.7,0) [circle,fill,inner sep=0.5mm,label=below:$\tau$] {};	
		\node (tn) at (0.7,0) [circle,fill,inner sep=0.5mm,label=below:$\tl \tau$] {};	
		\node (i) at (-2,0) {};				
		\node (j) at (2,0) {};							
		\draw [postaction={decorate}] (n) to [bend left=45] node[above]{$R_1$} (tn);
		\draw [postaction={decorate}] (n) to [bend right=45] node[below]{$R_2$} (tn);		
		\draw [postaction={decorate}] (i) to node[above]{$\Lambda$} (n);
		\draw [postaction={decorate}] (tn) to node[above]{$\tl \Lambda$} (j);		
	}
\; &= \;	
	\mytikz{
		\node (i) at (-1,0) {};				
		\node (j) at (1,0) {};	
		\draw [postaction={decorate}] (i) to node[above]{$\Lambda$} (j);
	}
	\times 
	\delta_{\Lambda \tl \Lambda} \delta_{\tau \tl \tau}
\\	
	\sum_{\Lambda,\tau}
	\mytikz{				
		\node (n) at (-0.7,0) [circle,fill,inner sep=0.5mm,label=below:$\tau$] {};	
		\node (tn) at (0.7,0) [circle,fill,inner sep=0.5mm,label=below:$\tau$] {};	
		\node (i1) at (-2,0.7) {};		
		\node (i2) at (-2,-0.7) {};
		\node (j1) at (2,0.7) {};		
		\node (j2) at (2,-0.7) {};				
		\draw [postaction={decorate}] (i1) to node[above]{$R_1$} (n);
		\draw [postaction={decorate}] (i2) to node[below]{$R_2$} (n);
		\draw [postaction={decorate}] (n) to node[above]{$\Lambda$} (tn);		
		\draw [postaction={decorate}] (tn) to node[above]{$R_1$} (j1);
		\draw [postaction={decorate}] (tn) to node[below]{$R_2$} (j2);
	}
\; &= \;	
	\mytikz{
		\node (i1) at (-1,0.5) {};		
		\node (i2) at (-1,-0.5) {};
		\node (j1) at (1,0.5) {};		
		\node (j2) at (1,-0.5) {};
		\draw [postaction={decorate}] (i1) to node[above]{$R_1$} (j1);
		\draw [postaction={decorate}] (i2) to node[below]{$R_2$} (j2);
	}		
\\	
	\sum_{\Lambda,\tau}
	\mytikz{				
		\node (n) at (-1,0) [circle,fill,inner sep=0.5mm,label=below:$\tau$] {};
		\node (s) at (0,0) [rectangle,draw] {$\gamma$};		
		\node (tn) at (1,0) [circle,fill,inner sep=0.5mm,label=below:$\tau$] {};	
		\node (i1) at (-2,0.7) {};		
		\node (i2) at (-2,-0.7) {};
		\node (j1) at (2,0.7) {};		
		\node (j2) at (2,-0.7) {};				
		\draw [postaction={decorate}] (i1) to node[above]{$R_1$} (n);
		\draw [postaction={decorate}] (i2) to node[below]{$R_2$} (n);
		\draw [postaction={decorate}] (n) to node[above]{$\Lambda$} (s);		
		\draw [postaction={decorate}] (s) to (tn);		
		\draw [postaction={decorate}] (tn) to node[above]{$R_1$} (j1);
		\draw [postaction={decorate}] (tn) to node[below]{$R_2$} (j2);
	}
\; &= \;	
	\mytikz{
		\node (i1) at (-1.2,0.5) {};		
		\node (i2) at (-1.2,-0.5) {};
		\node (j1) at (1.2,0.5) {};		
		\node (j2) at (1.2,-0.5) {};
		\node (s1) at (0,0.5) [rectangle,draw] {$\gamma$};
		\node (s2) at (0,-0.5) [rectangle,draw] {$\gamma$};
		\draw [postaction={decorate}] (i1) to node[above]{$R_1$} (s1);
		\draw [postaction={decorate}] (s1) to (j1);
		\draw [postaction={decorate}] (i2) to node[below]{$R_2$} (s2);
		\draw [postaction={decorate}] (s2) to (j2);
	}			
\end{align}

The corresponding equations are:
\begin{align}
D^{R_1}_{i_1 j_1}(\g) D^{R_2}_{i_2 j_2}(\g) 
S^{R_1 R_2,\Lambda\,\tau}_{\,j_1\;j_2,\,m}
&=
S^{R_1 R_2,\Lambda\,\tau}_{\,i_1\;i_2,\,l}
D^{\Lambda}_{l m}(\g) 
\\
S^{R_1 R_2,\Lambda\,\tau}_{\,i_1\;i_2,\,l}
S^{R_1 R_2,\tl\Lambda\,\tl\tau}_{\,i_1\;i_2,\,m}
&=
\delta_{\Lambda \tl \Lambda} \delta_{\tau \tl \tau} \delta_{lm}
\\
\sum_{\Lambda,\tau}
S^{R_1 R_2,\Lambda\,\tau}_{\,i_1\;i_2,\,m}
S^{R_1 R_2,\Lambda\,\tau}_{\,j_1\;j_2,\,m}
&=
\delta_{i_1 j_1} \delta_{i_2 j_2}
\\
\sum_{\Lambda,\tau}
S^{R_1 R_2,\Lambda\,\tau}_{\,i_1\;i_2,\,l}
D^\Lambda_{lm}(\g)
S^{R_1 R_2,\Lambda\,\tau}_{\,j_1\;j_2,\,m}
&=
D^{R_1}_{i_1 j_1}(\g) D^{R_2}_{i_2 j_2}(\g)
\end{align}

%%%%%%%%%%%%%%%%%%%%%%%%%%%%%%%%%%%%%%%%%%%%%%%%%%%%%%%%%%%%%%%%%%%%%

\subsection{Multiplicities}

Here we collect identities involving multiplicities $g(r_1,r_2;R)$ and $C(R_1,R_2,\Lambda)$. 

Using (\ref{eq:BB_g1g2}) and (\ref{eq:BB_R_delta}) leads to:
\begin{equation}
	\chi_R(\sigma_1 \circ \sigma_2) = \sum_{r_1\vdash n_1} \sum_{r_2 \vdash n_2} g(r_1,r_2;R) \chi_{r_1}(\sigma_1) \chi_{r_2}(\sigma_2)
\end{equation}
From this, Littlewood-Richardson multiplicity can be calculated as
\begin{equation}
g(r_1,r_2;R) = \frac{1}{n_1! n_2!} \sum_{\sigma_1 \in S_{n_1}} \sum_{\sigma_2 \in S_{n_2}} \chi_{r_1}(\sigma_1) \chi_{r_2}(\sigma_2) \chi_R(\sigma_1 \circ \sigma_2)
\end{equation}

Analogously, for Clebsch-Gordan coefficients:
\begin{equation}
	\chi_{R_1}(\sigma) \chi_{R_2}(\sigma) = \sum_{\Lambda \vdash n} C(R_1,R_2,\Lambda) \chi_{\Lambda}(\sigma)
\end{equation}
and
\begin{equation}
C(R_1,R_2,\Lambda) = \frac{1}{n!} \sum_{\sigma \in S_n} \chi_{R_1}(\sigma) \chi_{R_2}(\sigma) \chi_{\Lambda}(\sigma)
\end{equation}

Combining the above we find:
\begin{equation}
\label{eq:gg_Cg}
\begin{split}
	&\frac{1}{n_1!n_2!}\sum_{\sigma_1 \in n_1}\sum_{\sigma_2 \in n_2} \chi_{R_1}(\sigma_1 \circ \sigma_2) \chi_{R_2}(\sigma_1 \circ \sigma_2)
\\
	& \qquad=  \sum_{r_1 \vdash n_1} \sum_{r_2 \vdash n_2} g(r_1,r_2;R_1) g(r_1,r_2;R_2)
\\
	& \qquad= \sum_{\Lambda \vdash n} C(R_1,R_2,\Lambda) g([n_1],[n_2];\Lambda)
\end{split}
\end{equation}
where $[n_1]$ and $[n_2]$ are trivial representations for the corresponding groups, arising from $\frac{1}{n_1!n_2!} \sum_{\sigma_1,\sigma_2} \chi_\Lambda(\sigma_1\circ\sigma_2)$.

%%%%%%%%%%%%%%%%%%%%%%%%%%%%%%%%%%%%%%%%%%%%%%%%%%%%%%%%%%%%%%%%%%%%%
%%%%%%%%%%%%%%%%%%%%%%%%%%%%%%%%%%%%%%%%%%%%%%%%%%%%%%%%%%%%%%%%%%%%%

\section{Quiver characters}

\subsection{Symmetric group characters}
\label{app:identities_characters}

The usual symmetric group characters $\chi_R(\sigma) \equiv \tr(D^R(\sigma))$ obey the following identities
\begin{align}
\label{eq:chi_inv}
\chi_{R} ( \sigma ) & =  \chi_R  ( \sigma^{-1}  ) \\  
\label{eq:chi_adj}
\chi_R ( \sigma  )  &=  \chi_R (\gamma \sigma \gamma^{-1} ) \\  
\label{eq:chi_sumsig}
\frac{1}{n!} \sum_{ \sigma \in S_n } \chi_R ( \sigma ) \chi_S ( \sigma )  &=  \delta_{ RS } \\  
\label{eq:chi_sumR}
\sum_{R \vdash n} \chi_R ( \sigma ) \chi_R ( \tau  )  & =    \sum_{\gamma \in S_n } \delta ( \sigma \g \tau \g^{-1} ) 
\end{align}
They could be summarized as: invariance under inversion (\ref{eq:chi_inv}); invariance under conjugation (\ref{eq:chi_adj}); orthogonality in representation labels (\ref{eq:chi_sumsig}); orthogonality in conjugacy classes (\ref{eq:chi_sumR}). There is also a useful generalization of (\ref{eq:chi_sumsig})
\begin{equation}
\label{eq:chi_sumsig_tau}
\frac{1}{n!} \sum_{ \sigma \in S_n } \chi_R ( \sigma ) \chi_S ( \sigma \tau )  =   \delta_{ RS } \frac{\chi_R ( \tau )}{d(R)}
\end{equation}

\subsection{Restricted quiver characters}
\label{app:identities_restricted}

Restricted quiver character is defined as 
\begin{equation}
\begin{split}
	\chi_Q(\bL,\bsig) &= 
	\prod_a 
	D^{R_a}_{i_a j_a} (\sigma_a)
	B^{R_a \rightarrow \bigcup_{b,\alpha} r_{ab;\a}, \nu^-_a}_{j_a \rightarrow \bigcup_{b,\alpha} l_{ab;\a}}
	B^{R_a \rightarrow \bigcup_{b,\alpha} r_{ba;\a}, \nu^+_a}_{i_a \rightarrow \bigcup_{b,\alpha} l_{ba;\a}}
\end{split}
\end{equation}
with
\begin{equation}
	\bL \equiv \{ R_a, r_{ab;\a}, \nu^-_a, \nu^+_a \},
	\qquad
	\bsig \equiv \{ \sigma_a \}
\end{equation}
Diagrammatically, for the case $\mC^3/Z_2$, 
\begin{equation}
\label{eq:chiL_C3Z2_app}
	\chi_{\C3Z2}(\bL,\bsig) = \;
	\mytikz{
	%
	% Quiver with R,r,mu,nu labels
	%
		\node (s1) at (0,0) [rectangle,draw] {$\sigma_1$};		
		\node (m1) at (0,1) [circle,draw,inner sep=0.5mm,label=above:$\nu^+_1$] {};
		\node (n1) at (0,-1) [circle,draw,inner sep=0.5mm,label=below:$\nu^-_1$] {};
		\node (s2) at (3,0) [rectangle,draw] {$\sigma_2$};
		\node (n2) at (3,1) [circle,draw,inner sep=0.5mm,label=above:$\nu^-_2$] {};
		\node (m2) at (3,-1) [circle,draw,inner sep=0.5mm,label=below:$\nu^+_2$] {};
		\draw [postaction={decorate}] (m1) to node[auto]{$R_1$} (s1);
		\draw [postaction={decorate}] (s1) to node[auto]{} (n1);
		\draw [postaction={decorate}] (n1) to [bend left=90] node[auto]{$r_{11}$} (m1);
		\draw [postaction={decorate}] (n1) to [bend left=10] node[above]{$r_{12;1}$} (m2);
		\draw [postaction={decorate}] (n1) to [bend right=10] node[below]{$r_{12;2}$} (m2);
		\draw [postaction={decorate}] (m2) to node[auto]{$R_2$} (s2);
		\draw [postaction={decorate}] (s2) to node[auto]{} (n2);
		\draw [postaction={decorate}] (n2) to [bend left=90] node[auto]{$r_{22}$} (m2);
		\draw [postaction={decorate}] (n2) to [bend left=10] node[below]{$r_{21;1}$} (m1);
		\draw [postaction={decorate}] (n2) to [bend right=10] node[above]{$r_{21;2}$} (m1);		
	}
\end{equation}
Note that for the case of a trivial quiver with a single node and a single field, the quiver character is precisely the symmetric group character.

They obey analogous identities to (\ref{eq:chi_inv}), (\ref{eq:chi_adj}), (\ref{eq:chi_sumsig}), (\ref{eq:chi_sumR}):
\begin{align}
\chi_Q ( \bL, \bsig  )  &=  \chi_Q ( \bL, \bsig^{-1} ) 
\\
\label{eq:chiQ_inv}
\chi_Q ( \bL, \bsig  )  &=  \chi_Q ( \bL, \Adj_{\bgam} (\bsig) ) 
\\
\label{eq:chiQ_sig_sum}
\frac{1}{\prod_a n_a!}
\sum_{\bsig} 
\frac{\prod_{a} d(R_a)}{ \prod_{a,b,\alpha} d(r_{ab;\a}) }  
\chi_Q(\bL,\bsig) \chi_Q(\tl\bL,\bsig) 
&=	\delta_{\bR\tl\bR}\delta_{\br\tl\br}\delta_{\bnu^+\tl\bnu^+}\delta_{\bnu^-\tl\bnu^-} 
\\
\label{eq:chiQ_L_sum}
\sum_{\bL} 
\frac{\prod_{a} d(R_a)}{ \prod_{a,b,\alpha} d(r_{ab;\a}) }  
\chi_Q(\bL, \bsig) \chi_Q(\bL, \btau) 
&= 
\frac{\prod_a n_a!}{\prod_{a,b,\alpha} n_{ab;\a} !}
\sum_{\bgam} \prod_a \delta(\Adj_{\bgam}(\sigma_a) \tau_a^{-1}  ) 
\end{align}
For the proofs see Appendix~\ref{appsec:proofs-gen-char-ids}.

The generalization of (\ref{eq:chi_sumsig_tau}) is
\begin{equation}
\label{eq:CQ_sigsig_sum}
	\sum_{\bsig} \chi_Q(\bL,\btau\bsig) \chi_Q(\tl\bL,\bsig) =
	\delta_{\bR\tl\bR}\delta_{\br\tl\br}\delta_{\bnu^-\tl\bnu^-}
	\prod_a \frac{n_a!}{d(R_a)} \tr\left(D^{R_a}(\tau_a) P^{\nu^+_a\tl\nu^+_a}_{R_a\rightarrow \bigcup_{b,\alpha} r_{ba;\a}} \right)
\end{equation}
where 
\begin{equation}
	(P^{\nu_a^+\tl\nu_a^+}_{R_a\rightarrow \bigcup_{b,\alpha} r_{ba;\a}})_{i_a \tl i_a} \equiv 
	B^{R_a \rightarrow \bigcup_{b,\alpha} r_{ba;\a}, \nu^+_a}_{i_a \rightarrow \bigcup_{b,\alpha} l_{ba;\a}}	
	B^{R_a \rightarrow \bigcup_{b,\alpha} r_{ba;\a}, \tl \nu^+_a}_{\tl i_a \rightarrow \bigcup_{b,\alpha} l_{ba;\a}}
\end{equation}

\subsection{Covariant quiver characters}
\label{app:identities_covariant}

The covariant quiver characters are defined as
\begin{equation}
\begin{split}
	\chi_Q(\bK,\bsig) &= 
	\left(
	\prod_a 
	D^{R_a}_{i_a j_a} (\sigma_a)
	B^{R_a \rightarrow \bigcup_{b} s_{ab}^-, \nu^-_a}_{j_a \rightarrow \bigcup_{b} l_{ab}^-}
	B^{R_a \rightarrow \bigcup_{b} s_{ba}^+, \nu^+_a}_{i_a \rightarrow \bigcup_{b} l_{ba}^+}
	\right)
\left(
	\prod_{a,b} 
	B^{\Lambda_{ab} \rightarrow [\bn_{ab}],\beta_{ab}}_{l_{ab}} 
	S^{\; s_{ab}^+ \; s_{ab}^-,\, \Lambda_{ab}\tau_{ab}}_{\;l_{ab}^+\;\tl l_{ab}^-,\,l_{ab}}
\right)
\end{split}
\end{equation}
with
\begin{equation}
	\bK = \{ R_a, s_{ab}^+, s_{ab}^-, \nu_a^+, \nu_a^-, \Lambda_{ab}, \tau_{ab}, n_{ab;\a}, \beta_{ab} \}, 
	\qquad
	\bsig = \{ \sigma_a \}
\end{equation}
Diagrammatically, for the $\mC^3/\mZ_2$ case, 
\begin{equation}
	\chi_{\C3Z2}(\bK,\bsig) = \;
	\mytikz{
	%
	% Quiver with R,r,mu,nu labels
	%
		\node (s1) at (0,0) [rectangle,draw] {$\sigma_1$};		
		\node (m1) at (0,1) [circle,draw,inner sep=0.5mm,label=above:$\nu^+_1$] {};
		\node (n1) at (0,-1) [circle,draw,inner sep=0.5mm,label=below:$\nu^-_1$] {};
		\node (s2) at (3,0) [rectangle,draw] {$\sigma_2$};
		\node (n2) at (3,1) [circle,draw,inner sep=0.5mm,label=above:$\nu^-_2$] {};
		\node (m2) at (3,-1) [circle,draw,inner sep=0.5mm,label=below:$\nu^+_2$] {};
		\node (t12) at (1.5,-1) [circle,fill,inner sep=0.5mm,label=above:$\tau_{12}$] {};
		\node (t21) at (1.5,1) [circle,fill,inner sep=0.5mm,label=below:$\tau_{21}$] {};
		\node (t11) at (-1,0) [circle,fill,inner sep=0.5mm] {};
		\node (t22) at (4,0) [circle,fill,inner sep=0.5mm] {};
		\node (n12) at (1.5,-2.5) {};
		\node (n21) at (1.5,2.5) {};
		\node (b12) at (1.5,-2) [circle,draw,inner sep=0.5mm,label=left:$\beta_{12}$] {};
		\node (b21) at (1.5,2) [circle,draw,inner sep=0.5mm,label=right:$\beta_{21}$] {};
		\draw [postaction={decorate}] (m1) to node[left]{$R_1$} (s1);
		\draw [postaction={decorate}] (s1) to node[left]{} (n1);
		\draw [postaction={decorate}] (n1) to [bend left=45] node[left]{$s_{11}$} (t11);
		\draw [postaction={decorate}] (t11) to [bend left=45] node[left]{$s_{11}$} (m1);
		\draw [postaction={decorate}] (n1) to node[above]{$s_{12}^-$} (t12);
		\draw [postaction={decorate}] (t12) to node[above]{$s_{12}^+$} (m2);
		\draw [-] (t12) to node[right]{$\Lambda_{12}$} (b12);
		\draw [-] (b12) to node[right]{$\bn_{12}$} (n12);
		\draw [postaction={decorate}] (m2) to node[right]{$R_2$} (s2);
		\draw [postaction={decorate}] (s2) to node[right]{} (n2);
		\draw [postaction={decorate}] (n2) to [bend left=45] node[right]{$s_{22}$} (t22);
		\draw [postaction={decorate}] (t22) to [bend left=45] node[right]{$s_{22}$} (m2);
		\draw [postaction={decorate}] (n2) to node[below]{$s_{21}^-$} (t21);
		\draw [postaction={decorate}] (t21) to node[below]{$s_{21}^+$} (m1);		
		\draw [-] (t21) to node[left]{$\Lambda_{21}$} (b21);
		\draw [-] (b21) to node[left]{$\bn_{12}$} (n21);
	}
\end{equation}

They obey identities:
\begin{align}
\chi_Q ( \bK, \bsig  )  &=  \chi_Q ( \bK, \bsig^{-1} ) 
\\
\label{eq:chiQK_inv}
\chi_Q ( \bK, \bsig  )  &=  \chi_Q ( \bK, \Adj_{\bgam} (\bsig) ) 
\\
\frac{1}{\prod_a n_a!}
\sum_{\bsig} \left( \prod_{a} d(R_a) \right)
\chi_Q(\bK,\bsig) \chi_Q(\tl\bK,\bsig) &=
	\delta_{\bK\tl\bK}
\\
\label{eq:chiQK_K_sum} 
\sum_{\bK} 
\left( \prod_{a} d(R_a) \right)
\chi_Q(\bK, \bsig) \chi_Q(\bK, \btau) 
&= 
\frac{\prod_a n_a!}{\prod_{a,b,\alpha} n_{ab;\a} !}  \,
\sum_{\bgam} \prod_a \delta(\Adj_{\bgam}(\sigma_a) \tau_a^{-1}  ) 
\end{align}	
	
And
\begin{equation}
\begin{split}
\label{eq:chiQK_sigsig_sum}
	&\sum_{\bsig} \chi_Q(\bK,\btau\bsig) \chi_Q(\tl\bK,\bsig)  =
	\delta_{\bR\tl\bR}\delta_{\bs^-\tl\bs^-}\delta_{\bnu^-\tl\bnu^-} 
\\ & \quad
	\times \prod_a \left( \frac{n_a!}{d(R_a)} 
	D^{R_a}_{i_a \tl i_a}(\tau_a) 
	B^{R_a \rightarrow \bigcup_{b} s_{ba}^+, \nu^+_a}_{i_a \rightarrow \bigcup_{b} l_{ba} }
	B^{R_a \rightarrow \bigcup_{b} \tl s_{ba}^+, \tl\nu^+_a}_{\tl i_a \rightarrow \bigcup_{b} \tl l_{ba} }	
	\prod_b 
		S_{l_{ba} k_{ba}}^{s_{ba}^+ s_{ba}^- ; \Lambda_{ba} \tau_{ba} \beta_{ba} \bn_{ba}}
		S_{\tl l_{ba} k_{ba}}^{\tl s_{ba}^+ s_{ba}^- ; \tl\Lambda_{ba} \tl\tau_{ba} \tl\beta_{ba} \bn_{ba}}
\right)
\end{split}
\end{equation}
\begin{equation}
\sum_{\bsig,\btau} \chi_Q(\bK,\btau\bsig) \chi_Q(\tl\bK,\bsig) \prod_a N^{c(\tau_a)}  =
\delta_{\bK\tl\bK} \prod_a \frac{n_a!}{d(R_a)} f_N(R_a)
\end{equation}

%%%%%%%%%%%%%%%%%%%%%%%%%%%%%%%%%%%%%%%%%%%%%%%%%%%%%%%%%%%%%%%%%%%%%
%%%%%%%%%%%%%%%%%%%%%%%%%%%%%%%%%%%%%%%%%%%%%%%%%%%%%%%%%%%%%%%%%%%%%

\section{General basis from invariance}
\label{app:basis_from_invariance}

Here we want to show how solving the invariance (\ref{eq:OQ_adj})
\begin{equation}
	\cO_Q (\bn,\bsig) = \cO_Q (\bn, \Adj_{\bgam}(\bsig))
\end{equation}
naturally leads to the complete bases (\ref{eq:OL_defn})
\begin{equation}
	\cO_Q(\bL) = \frac{1}{\prod n_a!}\sqrt{\frac{\prod d(R_a)}{\prod d(r_{ab;\a})}} \; \sum_{\bsig} \chi_Q(\bL,\bsig) \cO_Q(\bn,\bsig)
\end{equation}
and (\ref{eq:OK_defn})
\begin{equation}
	\cO_Q(\bK) = \frac{\sqrt{\prod d(R_a)}}{\prod n_a!} \; \sum_{\bsig} \chi_Q(\bK,\bsig) \cO_Q(\bn,\bsig)
\end{equation}

\subsection{Review of \texorpdfstring{$\mC$}{C}}

First, let us start with the familiar example of half-BPS operators. Those are described by a trivial quiver $\mC$, with single node and single field $\Phi_{11}$. The operators obey invariance
\begin{equation}
	\cO_\mC(\sigma) = \cO_\mC(\gamma \sigma \gamma^{-1}), \quad \gamma \in S_n
\end{equation}
We want to express this as a projection to the invariant subspace
\begin{equation}
\begin{split}
	\cO_\mC(\sigma) &= \frac{1}{n!} \sum_{\gamma \in S_n} \cO_\mC(\gamma \sigma \gamma^{-1}) = \sum_{\rho \in S_n} \left( \frac{1}{n!} \sum_{\gamma \in S_n} \delta(\gamma \sigma \gamma^{-1} \rho^{-1}) \right) \cO_\mC(\rho)
\end{split}	
\end{equation}
Now
\begin{equation}
	P(\sigma,\rho) = \frac{1}{n!} \sum_{\gamma \in S_n} \delta(\gamma \sigma \gamma^{-1} \rho^{-1}) 
\end{equation}
is a projector, and we want to find the explicit basis that it projects to. That amounts to being able to write $P(\sigma,\rho) = \sum_L \Psi_L(\sigma) \Psi^*_L(\rho)$ for some labels $L$ and wavefunctions $\Psi_L(\sigma)$. In order to do that, we rewrite $\delta(\sigma)$ using (\ref{eq:chiR_sum})
\begin{equation}
\label{eq:Pst_1}
\begin{split}
	P(\sigma,\rho) &= \sum_{R \vdash n} \frac{d(R)}{(n!)^2} \sum_{\gamma} \chi_R(\gamma \sigma \gamma^{-1} \rho^{-1}) \\
	& = \sum_{R \vdash n} \frac{d(R)}{(n!)^2} \sum_{\gamma} D^R_{ij}(\gamma) D^R_{jk}( \sigma) D^R_{kl}(\gamma^{-1}) D^R_{li}(\rho^{-1})
\end{split}
\end{equation}
This allows us to perform $\gamma$ sum using (\ref{eq:D_sum_sigma}), resulting in
\begin{equation}
	P(\sigma,\rho) = \frac{1}{n!}\sum_{R \vdash n} \chi_R(\sigma) \chi_R(\rho)
\end{equation}
which is the desired explicit form for the projector. It leads to the complete basis (up to normalization, chosen for convenience) -- Schur polynomial basis
\begin{equation}
	\cO_\mC(R) = \frac{1}{n!} \sum_{\sigma} \chi_R(\sigma) \cO_\mC(\sigma)
\end{equation}

\subsection{Review of \texorpdfstring{$\mC^3$}{C3}}

Now let us see how the same procedure is applied to $\mC^3$. The operators are invariant under (\ref{eq:Osig_C3_conj})
\begin{equation}
	\cO_{\mC^3}(\bn,\gamma \sigma \gamma^{-1}) = \cO_{\mC^3}(\bn,\sigma), \quad \gamma \in S_{n_1} \times S_{n_2} \times S_{n_3} \equiv H \subset S_n
\end{equation}
This leads to a projection
\begin{equation}
	\cO_{\mC^3}(\bn,\sigma) = \sum_{\rho \in S_n} P(\sigma,\rho) \cO_{\mC^3}(\bn,\rho)
\end{equation} 
with
\begin{equation}
\label{eq:Pst_0}
	P(\sigma,\rho) = \frac{1}{|H|} \sum_{\gamma \in H} \delta(\gamma \sigma \gamma^{-1} \rho^{-1}) 
\end{equation}
Again introducing sum over $R$ we get the same as (\ref{eq:Pst_1})
\begin{equation}
\label{eq:Pst_2}
	P(\sigma,\rho) 
	= \sum_{R \vdash n} \frac{d(R)}{|H|\,n!} \sum_{\gamma \in H} D^R_{ij}(\gamma) D^R_{jk}( \sigma) D^R_{km}(\gamma^{-1}) D^R_{mi}(\rho^{-1})
\end{equation}
with a key difference that now the sum 
\begin{equation}
\label{eq:DD_sum_gamma}
	\sum_{\gamma \in S_{n_1} \times S_{n_2} \times S_{n_3}} D^R_{ij}(\gamma) D^R_{km}(\gamma^{-1}) 
\end{equation}
is only over the subgroup of $S_n$.

There are two ways to evaluate (\ref{eq:DD_sum_gamma}), eventually leading to the two different bases (\ref{eq:OL_C3}) and (\ref{eq:OK_C3}). For the first one, we introduce explicit representations for the subgroup $S_{n_1} \times S_{n_2} \times S_{n_3}$ by inserting a delta function as a sum over projectors (\ref{eq:Pij_Bil})
\begin{equation}
	\delta_{i j} = \sum_{r_1,r_2,r_3,\nu} (P^{\nu,\nu}_{R\rightarrow r_1,r_2,r_3})_{i j } =  
	\sum_{r_1,r_2,r_3,\nu}
	B^{R\rightarrow \br,\nu}_{i\rightarrow \bl} B^{R\rightarrow \br,\nu}_{j \rightarrow \bl}
\end{equation}
When $\gamma \in S_{n_1} \times S_{n_2} \times S_{n_3}$, $D^R(\gamma)$ can be moved through the branching coefficients, which allows us to express
\begin{equation}
	D^R_{ij}(\gamma_1 \circ \gamma_2 \circ \gamma_3) = \sum_{r_1,r_2,r_3,\nu} 
		B^{R\rightarrow \br,\nu}_{i\rightarrow \bl}
		D^{r_1}_{l_1 \tl l_1}(\gamma_1) D^{r_2}_{l_2 \tl l_2}(\gamma_2) D^{r_3}_{l_3 \tl l_3}(\gamma_3)
	 B^{R\rightarrow \br,\nu}_{j \rightarrow \tl \bl}
\end{equation}
Applying this to both terms in (\ref{eq:DD_sum_gamma}) we get
\begin{equation}
\label{eq:DD_deriv_C3_1}
\begin{split}
	\sum_{\gamma\in H} D^R_{ij}(\gamma) D^R_{km}(\gamma^{-1}) 
	&= \sum_{\br^+,\nu^+}\sum_{\br^-,\nu^-} \sum_{\gamma_1,\gamma_2,\gamma_3}
	B^{R\rightarrow \br^+,\nu^+}_{i\rightarrow \bl^+}
		D^{r^+_1}_{l^+_1 \tl l^+_1}(\gamma_1) D^{r^+_2}_{l^+_2 \tl l^+_2}(\gamma_2) D^{r^+_3}_{l^+_3 \tl l^+_3}(\gamma_3)
	 B^{R\rightarrow \br^+,\nu^+}_{j \rightarrow \tl \bl^+}
\\
	& ~~~~~~~~~~~~ \times 
		B^{R\rightarrow \br^-,\nu^-}_{k\rightarrow \bl^-}
		D^{r^-_1}_{l^-_1 \tl l^-_1}(\gamma_1^{-1}) D^{r^-_2}_{l^-_2 \tl l^-_2}(\gamma_2^{-1}) D^{r^-_3}_{l^-_3 \tl l^-_3}(\gamma_3^{-1})
	 B^{R\rightarrow \br^-,\nu^-}_{m \rightarrow \tl \bl^-}
\end{split}
\end{equation}
Now the $\gamma_1,\gamma_2,\gamma_3$ sums give $(\delta^{r_1^+ r_1^-}\delta_{l_1^+\tl l_1^-}\delta_{l_1^- \tl l_1^+})$ etc, which reconnect the branching coefficients. The final answer for (\ref{eq:DD_sum_gamma}) is thus
\begin{equation}
\label{eq:DD_deriv_C3_2}
\begin{split}
	\sum_{\gamma\in H} D^R_{ij}(\gamma) D^R_{km}(\gamma^{-1}) 
	&=	 
	\sum_{\br,\nu^+,\nu^-} \frac{n_1!n_2!n_3!}{d(r_1)d(r_2)d(r_3)}
	B^{R\rightarrow \br,\nu^-}_{m \rightarrow \bl}
	B^{R\rightarrow \br,\nu^+}_{i\rightarrow \bl}	
	B^{R\rightarrow \br,\nu^-}_{k\rightarrow \tl\bl}
	B^{R\rightarrow \br,\nu^+}_{j \rightarrow \tl\bl}
\\
	&= 	
	\sum_{\br,\nu^+,\nu^-} \frac{n_1!n_2!n_3!}{d(r_1)d(r_2)d(r_3)}
	(P_{R\rightarrow \br}^{\nu^-,\nu^+})_{mi}
	(P_{R\rightarrow \br}^{\nu^-,\nu^+})_{kj}
\end{split}
\end{equation}
The projector (\ref{eq:Pst_2}) is thus
\begin{equation}
	P(\sigma,\rho) = \frac{1}{n!} \sum_{R,\br,\nu^+,\nu^-}
	\frac{d(R)}{d(r_1)d(r_2)d(r_3)}
	\,
	\tr(P_{R\rightarrow \br}^{\nu^-,\nu^+} D^R(\sigma))
	\,
	\tr(P_{R\rightarrow \br}^{\nu^-,\nu^+} D^R(\rho))
\end{equation}
This is again of the form $\sum_L \Psi_L(\sigma)\Psi^*_L(\rho)$, with labels $\bL=\{R,r_1,r_2,r_3,\nu^+,\nu^-\}$, thus we conclude that the complete basis is (\ref{eq:OL_C3})
\begin{equation}
	\cO_{\mC^3}(\bL) \sim \sum_{\sigma} \tr(P_{R\rightarrow \br}^{\nu^-,\nu^+} D^R(\sigma)) \, \cO_{\mC^3}(\bn,\sigma)
\end{equation}
up to a normalization coefficient.

An alternative way to evaluate the sum (\ref{eq:DD_sum_gamma}) is to observe that $D^R_{ij}(\gamma)D^R_{mk}(\gamma)$ is a representation matrix of $\gamma$ in the tensor product $R \otimes R$ rep. We can decompose this into irreps using $S_n$ Clebsch-Gordan coefficients
\begin{equation}
\label{eq:DD_to_SS_deriv}
	D^R_{ij}(\gamma)D^R_{mk}(\gamma) = \sum_{\L,\tau} D^\Lambda_{l \tl l}(\gamma) S^{R R,\Lambda \tau}_{\,i\,m,\,l} S^{R R,\Lambda \tau}_{\,j\,k,\,\tl l}
\end{equation}
Now the $\gamma$ only appears in a single $D(\gamma)$, and the sum over $\gamma \in S_{n_1}\times S_{n_2}\times S_{n_3}$ is simply a projection to invariants under the subgroup
\begin{equation}
\label{eq:D_to_BB_deriv}
	\sum_{\gamma \in S_{n_1}\times S_{n_2}\times S_{n_3}} D^\Lambda_{l\tl l}(\gamma) =
	n_1!n_2!n_3! \sum_{\beta=1}^{g([\bn];\Lambda)} B^{\Lambda\rightarrow [\bn],\beta}_{l} B^{\Lambda\rightarrow [\bn],\beta}_{\tl l}
\end{equation}
The branching coefficients have the same meaning as before: $[\bn]$ denotes three single-row Young diagrams of length $n_1,n_2,n_3$, which is the trivial one-dimensional representation of $ S_{n_1}\times S_{n_2}\times S_{n_3}$. Since it is one-dimensional, we suppress the index for it. $\beta$ is the multiplicity for how many times $[\bn]$ appears in $\Lambda$. Branching coefficient $B^{\Lambda\rightarrow [\bn],\beta}_{l}$ itself is a vector in $\Lambda$, which is invariant under $ S_{n_1}\times S_{n_2}\times S_{n_3}$, labelled by $\beta$. Note the number of invariants is $g([n_1],[n_2],[n_3];\Lambda)$, that is, how many ways there are to combine three single-row diagrams into $\Lambda$ using Littlewood-Richardson rule. It vanishes if $\Lambda$ has more than three rows, so we have a constraint
\begin{equation}
	l(\Lambda) \le 3
\end{equation}
$\Lambda$ is a representation of the global symmetry $U(3)$. The full sum (\ref{eq:DD_sum_gamma}) is thus
\begin{equation}
\label{eq:Pst_3}
	\sum_{\gamma \in H} D^R_{ij}(\gamma) D^R_{km}(\gamma^{-1}) 
	=
	n_1!n_2!n_3!
	\sum_{\Lambda,\tau,\beta}
	 \left(B^{\Lambda\rightarrow [\bn],\beta}_{l} S^{R R,\Lambda \tau}_{\,i\,m,\,l}\right)
	 \left(B^{\Lambda\rightarrow [\bn],\beta}_{\tl l} S^{R R,\Lambda \tau}_{\,j\,k,\,\tl l}\right)
\end{equation}
and the projector (\ref{eq:Pst_2}) evaluates to
\begin{equation}
	P(\sigma,\rho) = \sum_{R, \Lambda,\tau,\beta} \frac{d(R)}{n!} 	
	 \left(B^{\Lambda\rightarrow [\bn],\beta}_{l} S^{R R,\Lambda \tau}_{\,i\,m,\,l} D^R_{im}(\rho)  \right)
	 \left(B^{\Lambda\rightarrow [\bn],\beta}_{\tl l} S^{R R,\Lambda \tau}_{\,j\,k,\,\tl l} D^R_{jk}(\sigma)  \right)
\end{equation}
This leads to the basis (\ref{eq:OK_C3})
\begin{equation}
	\cO(\bK) \sim \sum_{\sigma\in S_n} B^{\Lambda\rightarrow [\bn],\beta}_{m} S^{R R,\Lambda \tau}_{\,i\,j,\,m} D^R_{ij}(\sigma) \, \cO(\bn, \sigma)
\end{equation}
up to normalization.

\subsection{General quiver}

Now let us extend this derivation for a general quiver. We need to solve the invariance (\ref{eq:OQ_adj})
\begin{equation}	
	\cO_Q (\bn,\bsig) = \cO_Q (\bn, \Adj_{\bgam}(\bsig))
\end{equation}
that is, to evaluate the projector 
\begin{equation}
\label{eq:Pst_general}
\begin{split}
	P(\bsig,\brho) &= \frac{1}{|H|} \sum_{\bgam \in H} \delta(\Adj_{\bgam}(\bsig) \brho^{-1})	\\ &=
	\frac{1}{|H|} \sum_{\bgam \in H} \prod_a \delta(\Adj_{\bgam}(\sigma_a) \rho^{-1}_a)
\end{split}
\end{equation}
in analogy with (\ref{eq:Pst_0}). The invariance group is
\begin{equation}
	H = \bigotimes_{a,b,\alpha} S_{n_{ab;\a}}, \quad\quad |H| = \prod_{a,b,\alpha} n_{ab;\a} !
\end{equation}
Note beforehand, that $\chi_Q(\bL,\bsig)$  obeys exactly the required itentity (\ref{eq:chiQ_L_sum}),  which allows to write (\ref{eq:Pst_general}) like $\sum_{\bL} \chi_Q(\bL,\bsig) \chi_Q(\bL,\brho)$ , leading to the $\cO_Q(\bL)$ basis. The same is true of $\chi_Q(\bK,\bsig)$, which obeys (\ref{eq:chiQK_K_sum}), leading to $\cO_Q(\bK)$ basis. The purpose here, however, is to constructively \emph{derive} $\chi_Q(\bL,\bsig)$ and $\chi_Q(\bK,\bsig)$ as the basis diagonalizing $P(\bsig,\brho)$. 

Like before, we expand the deltas in terms of characters
\begin{equation}
\label{eq:Pst_Dgab}
\begin{split}
	P(\bsig,\brho) &=
	\frac{1}{|H|} \sum_{\bR} \sum_{\bgam \in H}  \prod_a \frac{d(R_a)}{n_a!} \chi_{R_a}(\Adj_{\bgam}(\sigma_a) \rho^{-1}_a)
\\
	&= \frac{1}{|H|} \sum_{\bR} \sum_{\bgam \in H}  \prod_a \frac{d(R_a)}{n_a!}
	D^{R_a}_{i_aj_a}(\otimes_{b,\alpha} \gamma_{ba;\a} ) \, D^{R_a}_{j_ak_a}(\sigma_a) \, D^{R_a}_{k_am_a}(\otimes_{b,\alpha} \gamma_{ab;\a}^{-1} ) \,  D^{R_a}_{m_ai_a}( \rho^{-1}_a) 
\end{split}
\end{equation}
The question is, again, how to perform the $\gamma_{ab;\a}$ sum
\begin{equation}
\label{eq:DD_sum_problem}
	\sum_{\bgam\in H} \prod_{a} 
	D^{R_a}_{i_aj_a}(\otimes_{b,\alpha} \gamma_{ba;\a} )
	D^{R_a}_{k_am_a}(\otimes_{b,\alpha} \gamma_{ab;\a}^{-1} )
\end{equation}
Note each $\gamma_{ab;\a}$ and $\gamma_{ab;\a}^{-1}$ occurs exactly once.

One way, in analogy to the restricted Schur basis, is to insert the branching coefficients around $\gamma$'s:
\begin{equation}
	D^{R_a}_{i_aj_a}(\otimes_{b,\alpha} \gamma_{ba;\a} ) =	
	\sum_{\bigcup_{b,\alpha}r_{ba;\a}} \sum_{\nu}
		B^{R_a \rightarrow \bigcup_{b,\alpha}r_{ba;\a},\nu_a}_{i_a \rightarrow \bigcup_{b,\alpha}l_{ba;\a}}
		B^{R_a \rightarrow \bigcup_{b,\alpha}r_{ba;\a},\nu_a}_{j_a \rightarrow \bigcup_{b,\alpha}\tl l_{ba;\a}}
		\prod_{b,\alpha}
		D^{r_{ba;\a}}_{l_{ba;\a} \tl l_{ba;\a}}( \gamma_{ba;\a} ) 
\end{equation}
Replacing all $D(\gamma)$ and $D(\gamma^{-1})$ we get analogous expansion to (\ref{eq:DD_deriv_C3_1}), which allows us to perform $\gamma_{ab;\a}$ sums. They generate delta functions which contract the branching coefficients in analogy to (\ref{eq:DD_deriv_C3_1}) as follows:
\begin{equation}
\label{eq:DD_gammaab_sum}
\begin{split}
	&\sum_{\bgam\in H} \prod_{a} 
	D^{R_a}_{i_aj_a}(\otimes_{b,\alpha} \gamma_{ba;\a} )
	D^{R_a}_{k_am_a}(\otimes_{b,\alpha} \gamma_{ab;\a}^{-1} )
\\
	&=	 
	\sum_{\{r_{ab;\a}\}} \sum_{\{\nu_a^+\}}\sum_{\{\nu_a^-\}}
	\frac{\prod n_{ab;\a}!}{\prod d(r_{ab;\a})}
	\prod_a
	\left(
	B^{R_a \rightarrow \bigcup_{b,\alpha}r_{ab;\a},\nu^-_a}_{m_a \rightarrow \bigcup_{b,\alpha}l_{ab;\a}}
	B^{R_a \rightarrow \bigcup_{b,\alpha}r_{ba;\a},\nu^+_a}_{i_a \rightarrow \bigcup_{b,\alpha}l_{ba;\a}}
	\right)
\\
	& ~~~~~~~~~~~~~~~~~~~~~~~~~~~~~~~~~~ \times	
\left(
	B^{R_a \rightarrow \bigcup_{b,\alpha}r_{ab;\a},\nu^-_a}_{k_a \rightarrow \bigcup_{b,\alpha}\tl l_{ab;\a}}
	B^{R_a \rightarrow \bigcup_{b,\alpha}r_{ba;\a},\nu^+_a}_{j_a \rightarrow \bigcup_{b,\alpha}\tl l_{ba;\a}}
	\right)	
\end{split}
\end{equation}
This leads to
\begin{equation}
	P(\bsig,\brho) = \frac{1}{\prod n_a!} \sum_{\bR,\br,\bnu^+,\bnu^-}
	\frac{\prod d(R_a)}{\prod d(r_{ab;\a})}
	\chi_Q(\bL,\bsig) \chi_Q(\bL,\brho)
\end{equation}
with $\chi_Q(\bL,\bsig)$ defined as in (\ref{eq:chiQ_defn}) and thus the basis
\begin{equation}
	\cO_Q(\bL) = \frac{1}{\prod n_a!} \sqrt{\frac{\prod d(R_a)}{ \prod d(r_{ab;\a})} } \; \sum_{\bsig} \chi_Q(\bL,\bsig) \cO_Q(\bsig) .
\end{equation}

An alternative way of evaluating (\ref{eq:DD_sum_problem}) is to use Clebsch-Gordan coefficients, leading to the covariant basis. In order to apply (\ref{eq:Pst_3})  we need a term  $D(\gamma)D(\gamma^{-1})$ with $\gamma$ in some subgroup of $S_n$. In general, however, (\ref{eq:DD_sum_problem}) does not have that form, because $D(\otimes_{b,\alpha} \gamma_{ba;\a})$ contains permutations corresponding to fields incoming to $a$, and $D(\otimes_{b,\alpha} \gamma_{ab;\a}^{-1})$ contains outgoing. Therefore, first we have to insert branching coefficients to separate fields between different quiver nodes
\begin{equation}
\label{eq:DD_sum_partial}
\begin{split}
	&\sum_{\bgam\in H} \prod_{a} 
	D^{R_a}_{i_aj_a}(\otimes_{b,\alpha} \gamma_{ba;\a} )
	D^{R_a}_{k_am_a}(\otimes_{b,\alpha} \gamma_{ab;\a}^{-1} )
\\
	&=
	\sum_{\bgam\in H} \prod_{a} 
	\left(
	\sum_{\bigcup_b s_{ba}^+}\sum_{\nu_a^+}	
	\,
	B^{R_a\rightarrow \bigcup_b s_{ba}^+,\nu_a^+}_{i_a\rightarrow \bigcup_b l_{ba}^+}
	B^{R_a\rightarrow \bigcup_b s_{ba}^+,\nu_a^+}_{j_a\rightarrow \bigcup_b \tl l_{ba}^+}
	\prod_b
	D^{s_{ba}^+}_{l_{ba}^+ \tl l_{ba}^+}(\otimes_{\alpha} \gamma_{ba;\a})		
	\right)
\\
	& ~~~~~~~~ \times	
	\left(	
	\sum_{\bigcup_b s_{ab}^-}\sum_{\nu_a^-} 
	\,
	B^{R_a\rightarrow \bigcup_b s_{ab}^-,\nu_a^-}_{k_a\rightarrow \bigcup_b l_{ab}^-}
	B^{R_a\rightarrow \bigcup_b s_{ab}^-,\nu_a^-}_{m_a\rightarrow \bigcup_b \tl l_{ab}^-}
	\prod_b
	D^{s_{ab}^-}_{l_{ab}^- \tl l_{ab}^-}(\otimes_{\alpha} \gamma_{ab;\a}^{-1})		
	\right)
\end{split}
\end{equation}
Now for each ordered pair of quiver nodes $(a,b)$, where we have $M_{ab}$ fields labelled by $\alpha$, we can apply (\ref{eq:Pst_3})
\begin{equation}
\begin{split}
	& \sum_{\bigcup_{\alpha} \gamma_{ab;\a}}
	D^{s_{ab}^+}_{l_{ab}^+ \tl l_{ab}^+}(\otimes_{\alpha} \gamma_{ab;\a})
	D^{s_{ab}^-}_{\tl l_{ab}^- l_{ab}^- }(\otimes_{\alpha} \gamma_{ab;\a})	
\\
	& \quad =
	(\prod_\alpha	n_{ab;\a}!)
	\sum_{\Lambda_{ab},\tau_{ab},\beta_{ab}}
	\left(
	 B^{\Lambda_{ab} \rightarrow [\bn_{ab}],\beta_{ab}}_{l_{ab}} 
	 S^{ s_{ab}^+ \; s_{ab}^-,\Lambda\tau_{ab}}_{\,l_{ab}^+\,\tl l_{ab}^-,\,l_{ab}}
	 \right)
	\left(
	 B^{\Lambda_{ab} \rightarrow [\bn_{ab}],\beta_{ab}}_{\tl l_{ab}} 
	 S^{s_{ab}^+ \; s_{ab}^-,\Lambda\tau_{ab}}_{\,\tl l_{ab}^+\,l_{ab}^-,\,\tl l_{ab}}
	 \right)	 
\end{split}
\end{equation}
Note that the effect on (\ref{eq:DD_sum_partial}) is to reconnect $l_{ab}^+$ with $\tl l_{ab}^-$ via the Clebsch-Gordan coefficient $S^{s_{ab}^+ \; s_{ab}^-,\Lambda \tau_{ab}}_{\,l_{ab}^+\,\tl l_{ab}^-,\,l_{ab}}$, and the same for $\tl l_{ab}^+$ with $l_{ab}^-$. This produces the right structure where the branching coefficients factor into two quivers. The end result, putting everything back into (\ref{eq:Pst_Dgab}) is
\begin{equation}
	P(\bsig,\brho) = \frac{1}{\prod n_a!} \sum_{\bK}
	\left( \prod_a d(R_a) \right)
	\chi_Q(\bK,\bsig) \chi_Q(\bK,\brho)
\end{equation}
where the label $\bK$ includes
\begin{equation}
	\bK = \{ R_a, s_{ab}^+, s_{ab}^-, \nu_a^+, \nu_a^-, \Lambda_{ab}, \tau_{ab}, \beta_{ab}, n_{ab;\a} \}
\end{equation}
and $\chi_Q(\bK,\bsig)$ is as in (\ref{eq:chiK_defn}).
The basis is then
\begin{equation}
\cO_Q(\bK) = \frac{\sqrt{\prod d(R_a)}}{\prod n_a!} \; \sum_{\bsig} \chi_Q(\bK,\bsig) \cO_Q(\bn,\bsig)	
\end{equation}

%%%%%%%%%%%%%%%%%%%%%%%%%%%%%%%%%%%%%%%%%%%%%%%%%%%%%%%%%%%%%%%%%%%%%
%%%%%%%%%%%%%%%%%%%%%%%%%%%%%%%%%%%%%%%%%%%%%%%%%%%%%%%%%%%%%%%%%%%%%

\section{Proofs}

%%%%%%%%%%%%%%%%%%%%%%%%%%%%%%%%%%%%%%%%%%%%%%%%%%%%%%%%%%
%%%%%%%%%%%%%%%%%%%%%%%%%%%%%%%%%%%%%%%%%%%%%%%%%%%%%%%%%%

\subsection{Proof of large \texorpdfstring{$N$}{N} counting}\label{App:dercount} 

We need to do some sums over $R_a,  S^{\pm }_{ab} $ in order to arrive at the 
(\ref{deltamuaforcount}) starting  from (\ref{covcountgen}).  We apply the identity 
\bea 
\sum_{ R  \vdash  n } \chi_{ R } ( \sigma_1 ) \chi_R ( \sigma_2 ) = \sum_{ \gamma  \in S_n} \delta ( \gamma \sigma_1 \gamma^{-1} \sigma_2 )  
\eea
to the quantity $\cN (\{  t_{ab;\a} \}   ; \{  M_{ab} \}    )  $ in  (\ref{deltamuaforcount}) to obtain 
\bea 
&& \cN (  \{  t_{ab;\a} \}   ,\{  M_{ab} \}   )  \cr 
&& =  \sum_{ R_a \vdash n_a } \sum_{ \L_{ab} \vdash n_{ab} } \sum_{ S^{\pm}_{ab} \vdash n_{ab} }
  \sum_{\{ \sigma_{ab}^+ \in S_{n_{ab}} \} }   \sum_{\{  \s_{ab}^- \in S_{n_{ab}}\} } 
\prod_a \chi_{R_a} ( \prod_b^{\circ} \s_{ba}^{+} ) \chi_{R_a} ( \prod_b^{\circ} \s_{ba}^{- } )\cr 
&&  \qquad \qquad  \qquad \qquad  \prod_{a,b}  { \chi_{ S^+_{ab}} ( \s_{ab}^+ ) \over  n_{ab}!  }  {  \chi_{ S^-_{ab}} ( \s_{ab}^- ) \over n_{ab}!    } 
\chi_{ S^+_{ab}} ( \tau_{ab} ) \chi_{S^-_{ab} } ( \tau_{ab}  ) \chi_{ \Lambda_{ab}} ( \tau_{ab} ) \chi_{ \L_{ab}} 
\left ( \mT_{ab} \right ) \cr 
&&=  \prod_a  \sum_{ \gamma_{ab}^{ \pm}  \vdash S_{n_{ab}} } 
     \delta_{ S_{n_{ab}} }\left ( \prod_b \s_{ba}^+ ~  \cdot ~ \mu_a  ~\cdot ~  \prod_{b}^{\circ} \s_{ab}^- ~ \cdot ~ \mu_a^{-1} \right )
 \cr 
&&  ~~~~   \prod_{a,b} { 1 \over ( n_{ab}! )^2 }  \delta_{ S_{n_{ab} } }  \left(  \g^{+}_{ab} \s^+_{ab} (\g^+_{ab})^{-1}   \tau_{ab}  \right) 
 \delta_{ S_{n_{ab} } }  \left (  \g^{- }_{ab} \s^-_{ab} (\g^-_{ab})^{-1}   \tau_{ab}  \right ) \tr \left ( \mT_{ab} \tau_{ab} \right )     \cr 
&& 
\eea 
We can use the delta functions to 
solve for $\tau_{ab} $   as $\g^{+}_{ab} (  \s^+_{ab})^{-1}  ( \g^{+}_{ab})^{-1} ) ) $. 
There is the  invariance 
\bea 
 \tr ( \mT_{ab} \g^{+}_{ab}  (\s_{ab}^+)^{-1}  ( \g^{+}_{ab})^{-1} ) )  = 
 \tr ( \mT_{ab} ( \s^+_{ab} )^{-1}  ) 
\eea
of the trace in $V_{M_{ab}}^{\otimes n_{ab}} $. 
The sum over $ \g^{-}_{ab}$ is invariant under left multiplication  by $( \g^+_{ab})^{-1} $. 
Hence we obtain 
\bea 
&& \cN ( \{  t_{ab;\a} \}   ;  \{ M_{ab} \}   )  = \prod_a \sum_{ \g_a} \delta_{ S_{n_a} } \left ( \prod_a^{\circ} \s_{ba}^+ ~ \cdot ~ \g_a ~ \cdot \prod_a^{\circ} \s_{ba}^- ~ \cdot ~ \g_a^{-1} \right )    \cr
&& \qquad \qquad   \prod_{a,b} { 1 \over  n_{ab} !  } \sum_{ \g^-_{ab} }
 \delta_{S_{n_ab}} \left (    \g_{ab}^- \s_{ab}^-    ( \g_{ab}^- )^{-1}  (\s_{ab}^+)^{-1} \right ) \tr  \left ( \mT_{ab} ( \s_{ab}^+  )^{-1} \right )
\eea
Now we can solve for  $\s_{ab}^-$, use invariance of the trace under conjugation of $\mT_{ab}$ by $ \g^-_{ab}$, 
as well as invariance of the sums over $\g_a \in S_{n_a} $ under right multiplication by $\gamma_{ab}^- \in S_{n_{ab}} \subset S_{n_a} $
to arrive at 
\bea 
 && \cN ( \{  t_{ab;\a} \}   ; \{  M_{ab} \}   )  = \prod_{a} \sum_{ \g_a} \delta_{S_{n_a} } \left  ( \prod_b \s_{ab}^+ ~ \cdot ~  \g_a ~ \cdot ~ 
 \prod_b \s_{ab}^+ ~ \cdot ~ \g_a^{-1}  \right ) \prod_{a,b}  \tr  \left( \mT_{ab} ( \s_{ab}^+ )^{-1} \right) \cr 
 && 
\eea
Renaming $ \s^+_{ab} \rightarrow \s_{ab} $ we arrive at  (\ref{deltamuaforcount}) 

%%%%%%%%%%%%%%%%%%%%%%%%%%%%%%%%%%%%%%%%%%%%%%%%%%%%%%%%%%
%%%%%%%%%%%%%%%%%%%%%%%%%%%%%%%%%%%%%%%%%%%%%%%%%%%%%%%%%%

\subsection{Proofs of quiver character identities} 
\label{appsec:proofs-gen-char-ids} 

Here we prove the identities (\ref{eq:chiQ_inv}), (\ref{eq:chiQ_sig_sum}), (\ref{eq:chiQ_L_sum}), (\ref{eq:CQ_sigsig_sum}) obeyed by the restricted quiver characters $\chi_Q(\bL,\bsig)$.

\paragraph{Invariance of $\chi_Q(\bL,\bsig)$}$\,$

Here we show that restricted quiver characters $\chi_Q(\bL,\bsig)$ obey (\ref{eq:chiQ_inv}), invariance under $\bsig \rightarrow \Adj_{\bgam}(\bsig)$. 

It is easiest to see from a diagram. For example, if we take simplified version of (\ref{eq:chiL_C3Z2_app}) with only single flavor, we have:
\begin{equation}
\begin{split}
\chi_Q(\bL,\Adj_{\bgam}(\bsig)) &\sim 
	\mytikz{
	%
	% Quiver with R,r,mu,nu labels
	%
		\node (s1) at (0,0) [rectangle,draw] {$\sigma_1$};
		\node (g1) at (0,0.7) [rectangle,draw] {$\gamma_{11}\circ\gamma_{21}$};		
		\node (g1i)at (0,-0.7) [rectangle,draw] {$\gamma_{11}^{-1}\circ\gamma_{12}^{-1}$};		
		\node (m1) at (0,1.7) [circle,draw,inner sep=0.5mm,label=above:$\nu^+_1$] {};		
		\node (n1) at (0,-1.7) [circle,draw,inner sep=0.5mm,label=below:$\nu^-_1$] {};
		\node (s2) at (3,0) [rectangle,draw] {$\sigma_2$};
		\node (g2) at (3,-0.7) [rectangle,draw] {$\gamma_{22}\circ\gamma_{12}$};		
		\node (g2i)at (3,0.7) [rectangle,draw] {$\gamma_{22}^{-1}\circ\gamma_{21}^{-1}$};
		\node (m2) at (3,-1.7) [circle,draw,inner sep=0.5mm,label=below:$\nu^+_2$] {};
		\node (n2) at (3,1.7) [circle,draw,inner sep=0.5mm,label=above:$\nu^-_2$] {};		
		\draw [postaction={decorate}] (m1) to node[right]{$R_1$} (g1);
		\draw [postaction={decorate}] (g1) to (s1);
		\draw [postaction={decorate}] (s1) to (g1i);
		\draw [postaction={decorate}] (g1i)to node[right]{$R_1$} (n1);
		\draw [postaction={decorate}] (n1) to [bend left=90] node[left]{$r_{11}$} (m1);		
		\draw [postaction={decorate}] (n1) to [bend right=0] node[below]{$r_{12}$} (m2);
		\draw [postaction={decorate}] (m2) to node[left]{$R_2$} (g2);
		\draw [postaction={decorate}] (g2) to (s2);
		\draw [postaction={decorate}] (s2) to (g2i);
		\draw [postaction={decorate}] (g2i) to node[left]{$R_2$} (n2);
		\draw [postaction={decorate}] (n2) to [bend left=90] node[right]{$r_{22}$} (m2);		
		\draw [postaction={decorate}] (n2) to [bend right=0] node[above]{$r_{21}$} (m1);		
	}
=
	\mytikz{
	%
	% Quiver with R,r,mu,nu labels
	%
		\node (s1) at (0,0) [rectangle,draw] {$\sigma_1$};				
		\node (m1) at (0,1.4) [circle,draw,inner sep=0.5mm,label=above:$\nu^+_1$] {};		
		\node (n1) at (0,-1.4) [circle,draw,inner sep=0.5mm,label=below:$\nu^-_1$] {};
		\node (g11) at (-0.7,0.7) [rectangle,draw] {$\gamma_{11}$};		
		\node (g11i)at (-0.7,-0.7) [rectangle,draw] {$\gamma_{11}^{-1}$};
		\node (g21) at (0.7,1.4) [rectangle,draw] {$\gamma_{21}$};		
		\node (g12i)at (0.7,-1.4) [rectangle,draw] {$\gamma_{12}^{-1}$};
		\node (s2) at (3,0) [rectangle,draw] {$\sigma_2$};		
		\node (m2) at (3,-1.4) [circle,draw,inner sep=0.5mm,label=below:$\nu^+_2$] {};
		\node (n2) at (3,1.4) [circle,draw,inner sep=0.5mm,label=above:$\nu^-_2$] {};		
		\node (g22) at (3.7,-0.7) [rectangle,draw] {$\gamma_{22}$};		
		\node (g22i)at (3.7,0.7) [rectangle,draw] {$\gamma_{22}^{-1}$};
		\node (g12) at (2.3,-1.4) [rectangle,draw] {$\gamma_{12}$};		
		\node (g21i)at (2.3,1.4) [rectangle,draw] {$\gamma_{21}^{-1}$};
		\draw [postaction={decorate}] (m1) to node[right]{$R_1$} (s1);		
		\draw [postaction={decorate}] (s1)to node[right]{$R_1$} (n1);
		\draw [postaction={decorate}] (n1) to [bend left=45] (g11i);
		\draw [postaction={decorate}] (g11i) to node[left]{$r_{11}$} (g11);
		\draw [postaction={decorate}] (g11) to [bend left=45] (m1);
		\draw [postaction={decorate}] (n1) to (g12i);
		\draw [postaction={decorate}] (g12i) to node[below]{$r_{12}$} (g12);
		\draw [postaction={decorate}] (g12) to (m2);
		\draw [postaction={decorate}] (m2) to node[left]{$R_2$} (s2);		
		\draw [postaction={decorate}] (s2) to node[left]{$R_2$} (n2);			
		\draw [postaction={decorate}] (n2) to [bend left=45] (g22i);
		\draw [postaction={decorate}] (g22i) to node[right]{$r_{22}$} (g22);
		\draw [postaction={decorate}] (g22) to [bend left=45]  (m2);		
		\draw [postaction={decorate}] (n2) to (g21i);
		\draw [postaction={decorate}] (g21i) to node[above]{$r_{21}$} (g21);		
		\draw [postaction={decorate}] (g21) to (m1);
	}
\\ &= 
	\chi_Q(\bL,\bsig)
\end{split}
\end{equation}
This follows from the property (\ref{eq:B_gamma_pull}) of the branching coefficients, which allows to pull $\gamma$'s through and cancel with each other

This procedure can be written explicitly for the general case (\ref{eq:chiQ_defn}):
\begin{equation}
\begin{split}
	\chi_Q(\bL,\Adj_{\bgam}(\bsig)) &= 
	\prod_a 	
		D^{R_a}_{i_a i'_a} (\otimes_{b,\alpha} \gamma_{ba;\a} )
		D^{R_a}_{i'_a j'_a} (\sigma_a)
		D^{R_a}_{j'_a j_a} (\otimes_{b,\alpha} \gamma_{ab;\a}^{-1} )	
\\
	&\quad\quad\times 
	B^{R_a \rightarrow \bigcup_{b,\alpha} r_{ba;\a}, \nu^+_a}_{i_a \rightarrow \bigcup_{b,\alpha} l_{ba;\a}}
	B^{R_a \rightarrow \bigcup_{b,\alpha} r_{ab;\a}, \nu^-_a}_{j_a \rightarrow \bigcup_{b,\alpha} l_{ab;\a}}	
\\
	&= 
	\prod_a 			
		D^{R_a}_{i_a j_a} (\sigma_a)
		B^{R_a \rightarrow \bigcup_{b,\alpha} r_{ba;\a}, \nu^+_a}_{i_a \rightarrow \bigcup_{b,\alpha} {l'}_{ba;\a}}
	B^{R_a \rightarrow \bigcup_{b,\alpha} r_{ab;\a}, \nu^-_a}_{j_a \rightarrow \bigcup_{b,\alpha} {l''}_{ab;\a}}	
\\
	&\quad\quad\times 
	  \left( \prod_{b,\alpha} D^{r_{ba;\a}}_{l_{ba;\a} {l'}_{ba;\a}} (\gamma_{ba;\a} )	\right)
		\left( \prod_{b,\alpha} D^{r_{ab;\a}}_{{l''}_{ab;\a}l_{ab;\a}} (\gamma_{ab;\a}^{-1} )	\right)
\\ 
  &= \prod_a 			
		D^{R_a}_{i_a j_a} (\sigma_a)
		B^{R_a \rightarrow \bigcup_{b,\alpha} r_{ba;\a}, \nu^+_a}_{i_a \rightarrow \bigcup_{b,\alpha} {l'}_{ba;\a}}
	B^{R_a \rightarrow \bigcup_{b,\alpha} r_{ab;\a}, \nu^-_a}_{j_a \rightarrow \bigcup_{b,\alpha} {l''}_{ab;\a}}	
\\
& ~~~
	   \prod_{a , b,\alpha} D^{r_{ab ;\a}}_{l_{a b ;\a} {l'}_{ab ;\a}} (\gamma_{ab ; \a} )	
	 D^{r_{ab;\a}}_{{l''}_{ab;\a}l_{ab;\a}} (\gamma_{ab;\a}^{-1} )	
\\ 
	&= 
	\prod_a 			
	D^{R_a}_{i_a j_a} (\sigma_a)		
	B^{R_a \rightarrow \bigcup_{b,\alpha} r_{ba;\a}, \nu^+_a}_{i_a \rightarrow \bigcup_{b,\alpha} l_{ba;\a}}
	B^{R_a \rightarrow \bigcup_{b,\alpha} r_{ab;\a}, \nu^-_a}_{j_a \rightarrow \bigcup_{b,\alpha} l_{ab;\a}}	
\\ 
   & = \chi_Q ( \bL , \bsig ) 
\end{split}
\end{equation}

\paragraph{Proof of orthogonality in $\bL$ of  $\chi_Q(\bL,\bsig)$    }$\,$

Here we will prove (\ref{eq:CQ_sigsig_sum}), of which (\ref{eq:chiQ_sig_sum}) is a special case. Expanding the definition of $\chi_Q(\bL,\bsig)$:
\begin{equation}
\label{eq:CQ_sisig_deriv1}
\begin{split}
	& \sum_{\tl\bsig} \chi_Q(\bL,\bsig \tl\bsig) \chi_Q(\tl\bL,\tl\bsig) 
= 		
	\sum_{\tl\bsig}
	\prod_a 
	D^{R_a}_{i_a j_a} (\sigma_a \tl\sigma_a)
	D^{\tl R_a}_{\tl i_a \tl j_a} (\tl\sigma_a)
\\ & \quad\quad\quad\quad\quad\quad \times
	B^{R_a \rightarrow \bigcup_{b,\alpha} r_{ba;\a}, \nu^+_a}_{i_a \rightarrow \bigcup_{b,\alpha} l_{ba;\a}}
	B^{R_a \rightarrow \bigcup_{b,\alpha} r_{ab;\a}, \nu^-_a}_{j_a \rightarrow \bigcup_{b,\alpha} l_{ab;\a}}
	B^{\tl R_a \rightarrow \bigcup_{b,\alpha} \tl r_{ba;\a}, \tl \nu^+_a}_{\tl i_a \rightarrow \bigcup_{b,\alpha} \tl l_{ba;\a}}
	B^{\tl R_a \rightarrow \bigcup_{b,\alpha} \tl r_{ab;\a}, \tl \nu^-_a}_{\tl j_a \rightarrow \bigcup_{b,\alpha} \tl l_{ab;\a}}
\end{split}
\end{equation}
We apply identity (\ref{eq:D_sum_sigma})
to do the sum in each product term
\begin{equation}	
	\sum_{\tl\sigma_a} D^{R_a}_{i_a j_a} (\sigma_a \tl\sigma_a) D^{\tl R_a}_{\tl i_a \tl j_a} (\tl\sigma_a) 
	= \frac{n_a!}{d(R_a)} \delta_{R_a \tl R_a} D^{R_a}_{i_a \tl i_a}(\sigma_a) \delta_{j_a \tl j_a} 
\end{equation}
Now contract a pair of branching coefficients with $\delta_{j_a \tl j_a}$, applying (\ref{eq:BB_R_delta})
\begin{equation}
	B^{R_a \rightarrow \bigcup_{b,\alpha} r_{ab;\a}, \nu^-_a}_{j_a \rightarrow \bigcup_{b,\alpha} l_{ab;\a}}
	B^{R_a \rightarrow \bigcup_{b,\alpha} \tl r_{ab;\a}, \tl\nu^-_a}_{j_a \rightarrow \bigcup_{b,\alpha} \tl l_{ab;\a}}
	=
	\delta_{\nu^-_a \tl\nu^-_a} \prod_{b,\alpha} \delta_{r_{ab;\a} \tl r_{ab;\a}} \delta_{l_{ab;\a} \tl l_{ab;\a}}
\end{equation}
Since this appears in (\ref{eq:CQ_sisig_deriv1}) under $\prod_a$, we effectively get a delta on all $\nu^-_a, r_{ab;\a}, l_{ab;\a}$. So the sum is
\begin{equation}
\label{eq:CQ_sisig_deriv2}
\begin{split}
	\sum_{\tl\bsig} \chi_Q(\bL,\bsig \tl\bsig) \chi_Q(\tl\bL,\tl\bsig) 
&= 	
	\delta_{\bR\tl\bR}\delta_{\br\tl\br}\delta_{\bnu^-\tl\bnu^-} 
\prod_a \frac{n_a!}{d(R_a)}
	D^{R_a}_{i_a \tl i_a} (\sigma_a)	
	B^{R_a \rightarrow \bigcup_{b,\alpha} r_{ba;\a}, \nu^+_a}_{i_a \rightarrow \bigcup_{b,\alpha} l_{ba;\a}}	
	B^{R_a \rightarrow \bigcup_{b,\alpha} r_{ba;\a}, \tl \nu^+_a}_{\tl i_a \rightarrow \bigcup_{b,\alpha} l_{ba;\a}}
\end{split}
\end{equation}
which is (\ref{eq:CQ_sigsig_sum}).

\paragraph{Proof of orthogonality in $\bsig$ conjugacy class  of  $\chi_Q(\bL,\bsig)$ }$\,$

Here we show (\ref{eq:chiQ_L_sum}).

Consider the product of quiver characters $\chi_Q(\bL,\bsig) \chi_Q(\bL,\btau)$  :
\begin{equation}
\begin{split} 
	&\chi_Q(\bL,\bsig) \chi_Q(\bL,\btau) 
\\
&= 	
	\prod_a 
		D^{R_a}_{i_a j_a} (\sigma_a)
		B^{R_a \rightarrow \bigcup_{b,\alpha} r_{ab;\a}, \nu^-_a}_{j_a \rightarrow \bigcup_{b,\alpha} l_{ab;\a}}
		B^{R_a \rightarrow \bigcup_{b,\alpha} r_{ba;\a}, \nu^+_a}_{i_a \rightarrow \bigcup_{b,\alpha} l_{ba;\a}}	
		D^{R_a}_{\tl j_a \tl i_a} (\tau_a^{-1})
		B^{R_a \rightarrow \bigcup_{b,\alpha} r_{ab;\a}, \nu^-_a}_{\tl j_a \rightarrow \bigcup_{b,\alpha} \tl l_{ab;\a}}
		B^{R_a \rightarrow \bigcup_{b,\alpha} r_{ba;\a}, \nu^+_a}_{\tl i_a \rightarrow \bigcup_{b,\alpha} \tl l_{ba;\a}}	
\end{split}
\end{equation}
We flipped $D^R_{ij}(\tau)=D^R_{ji}(\tau^{-1})$ in the second character for later convenience.
Each index $l_{ab;\a}$ appears once in a branching coefficient with $\nu^+$ and once with $\nu^-$, which are contracted together (and same for $\tl l_{ab;\a}$). Next we ``reconnect'' the branching coefficients by inserting
\begin{equation}
	\delta_{i_{ab;\a} j_{ab;\a}}
	\delta_{\tl i_{ab;\a} \tl j_{ab;\a}} =
	\frac{d(r_{ab;\a})}{n_{ab;\a}!}
	\sum_{\gamma_{ab;\a}} 		
		D^{r_{ab;\a}}_{\tl i_{ab;\a} i_{ab;\a}}(\gamma_{ab;\a})
		D^{r_{ab;\a}}_{j_{ab;\a} \tl j_{ab;\a}}(\gamma_{ab;\a}^{-1})
\end{equation}
for each $l_{ab;\a}, \tl l_{ab;\a}$:
\begin{equation}
\label{eq:chiQ_Lderiv_2}
\begin{split} 
	&\chi_Q(\bL,\bsig) \chi_Q(\bL,\btau) 
\\ & =	
	\prod_a  				
		D^{R_a}_{i_a j_a} (\sigma_a)
		B^{R_a \rightarrow \bigcup_{b,\alpha} r_{ab;\a}, \nu^-_a}_{j_a \rightarrow \bigcup_{b,\alpha} j_{ab;\a}}
		B^{R_a \rightarrow \bigcup_{b,\alpha} r_{ba;\a}, \nu^+_a}_{i_a \rightarrow \bigcup_{b,\alpha} i_{ba;\a}}			
		D^{R_a}_{\tl j_a \tl i_a} (\tau_a^{-1})
		B^{R_a \rightarrow \bigcup_{b,\alpha} r_{ab;\a}, \nu^-_a}_{\tl j_a \rightarrow \bigcup_{b,\alpha} \tl j_{ab;\a}}
		B^{R_a \rightarrow \bigcup_{b,\alpha} r_{ba;\a}, \nu^+_a}_{\tl i_a \rightarrow \bigcup_{b,\alpha} \tl i_{ba;\a}}			
\\ & \quad
	\times 
	\left(		
		\prod_{a,b,\alpha} 
		\frac{d(r_{ab;\a})}{n_{ab;\a}!}
		\sum_{\gamma_{ab;\a}} 		
		D^{r_{ab;\a}}_{\tl i_{ab;\a} i_{ab;\a}}(\gamma_{ab;\a})
		D^{r_{ab;\a}}_{j_{ab;\a} \tl j_{ab;\a}}(\gamma_{ab;\a}^{-1})
	\right)
\end{split}
\end{equation}
After this, $i_{ba;\a} , \ti_{ba;\a} $ appear in a matrix element of $ \gamma_{ba;\a}$, 
hence they link, via branching coefficients, to $ \sigma_a , \tau_a^{-1}  $. Likewise   
 $j_{ba;\a} , \tj_{ba;\a} $ appear in a matrix element of $ (  \gamma_{ba;\a})^{-1} $ and, via branching coeffients, 
link $ \sigma_{b} , \tau_b^{-1} $. 
This reconnection step can be understood diagrammatically, for each $r_{ab;\a}$:
\begin{equation}
	\mytikz{
		\node (sa) at (-2.8,1) [rectangle,draw] {$\sigma_a$};		
		\node (na1) at (-1,1) [circle,draw,inner sep=0.5mm,label=below:$\nu_a^-$] {};
		\node (nb1) at (1,1) [circle,draw,inner sep=0.5mm,label=below:$\nu_b^+$] {};
		\node (sb) at (2.8,1) [rectangle,draw] {$\sigma_b$};		
		\node (ta) at (-2.8,-1) [rectangle,draw] {$\tau_a^{-1}$};		
		\node (na2) at (-1,-1) [circle,draw,inner sep=0.5mm,label=below:$\nu_a^-$] {};
		\node (nb2) at (1,-1) [circle,draw,inner sep=0.5mm,label=below:$\nu_b^+$] {};
		\node (tb) at (2.8,-1) [rectangle,draw] {$\tau_b^{-1}$};		
		\draw [postaction={decorate}] (sa) to node[below]{$R_a$} (na1);
		\draw [postaction={decorate}] (na1) to node[below]{$r_{ab;\a}$} (nb1);
		\draw [postaction={decorate}] (nb1) to node[below]{$R_a$} (sb);
		\draw [postaction={decorate}] (na1) to +(0,0.5);
		\draw [postaction={decorate}] ($(nb1)+(0,0.5)$) to (nb1);
		\draw [postaction={decorate}] (na2) to node[below]{$R_a$} (ta);
		\draw [postaction={decorate}] (nb2) to node[below]{$r_{ab;\a}$} (na2);
		\draw [postaction={decorate}] (tb) to node[below]{$R_a$} (nb2);
		\draw [postaction={decorate}] ($(na2)+(0,0.5)$) to (na2);
		\draw [postaction={decorate}] (nb2) to +(0,0.5);
	}
= \frac{d(r_{ab;\a})}{n_{ab;\a}!} \sum_{\gamma}
	\mytikz{
		\node (sa) at (-2.8,1) [rectangle,draw] {$\sigma_a$};		
		\node (na1) at (-1.2,1) [circle,draw,inner sep=0.5mm,label=below:$\nu_a^-$] {};
		\node (nb1) at (1.2,1) [circle,draw,inner sep=0.5mm,label=below:$\nu_b^+$] {};
		\node (sb) at (2.8,1) [rectangle,draw] {$\sigma_b$};		
		\node (ta) at (-2.8,-1) [rectangle,draw] {$\tau_a^{-1}$};
		\node (na2) at (-1.2,-1) [circle,draw,inner sep=0.5mm,label=below:$\nu_a^-$] {};
		\node (nb2) at (1.2,-1) [circle,draw,inner sep=0.5mm,label=below:$\nu_b^+$] {};
		\node (tb) at (2.8,-1) [rectangle,draw] {$\tau_b^{-1}$};
		\node (g1) at (-0.5,0) [rectangle,draw] {$\gamma^{-1}$};	
		\node (g2) at (0.5,0) [rectangle,draw] {$\,\gamma^{\phantom{1}}$};		
		\draw [postaction={decorate}] (sa) to node[below]{$R_a$} (na1);
		\draw [postaction={decorate}] (na1) to [bend left=45] node[right]{$r_{ab;\a}$} (g1);
		\draw [postaction={decorate}] (g2) to [bend left=45] (nb1);
		\draw [postaction={decorate}] (nb1) to node[below]{$R_a$} (sb);
		\draw [postaction={decorate}] (na1) to +(0,0.5);
		\draw [postaction={decorate}] ($(nb1)+(0,0.5)$) to (nb1);
		\draw [postaction={decorate}] (na2) to node[below]{$R_a$} (ta);		
		\draw [postaction={decorate}] (g1) to [bend left=45] (na2);
		\draw [postaction={decorate}] (nb2) to [bend left=45] node[left]{$r_{ab;\a}$} (g2);
		\draw [postaction={decorate}] (tb) to node[below]{$R_a$} (nb2);
		\draw [postaction={decorate}] ($(na2)+(0,0.5)$) to (na2);
		\draw [postaction={decorate}] (nb2) to +(0,0.5);
	}
\end{equation}
Performing reconnection for all legs, the group factors disconnect into pieces like
\begin{equation}
	\mytikz{
		\def\x{1.8};
		\node (s1) at (-\x,0) [rectangle,draw] {$\,\sigma_a^{\phantom{1}}$};		
		\node (s2) at (\x,0) [rectangle,draw] {$\tau_a^{-1}$};				
		\node (np1) at (-\x,1.3) [circle,draw,inner sep=0.5mm,label=left:$\nu_a^+$] {};
		\node (np2) at (\x,1.3) [circle,draw,inner sep=0.5mm,label=right:$\nu_a^+$] {};
		\node (nm1) at (-\x,-1.3) [circle,draw,inner sep=0.5mm,label=left:$\nu_a^-$] {};
		\node (nm2) at (\x,-1.3) [circle,draw,inner sep=0.5mm,label=right:$\nu_a^-$] {};
		\node (gab1) at (0,-1.5) [rectangle,draw] {$\gamma_{ab_3;\a}^{-1}$};
		\node (gab2) at (0,-2.5) [rectangle,draw] {$\gamma_{ab_4;\a}^{-1}$};
		\node (gba1) at (0,1.5) [rectangle,draw] {$\,\gamma_{b_2a;\a}^{\phantom{1}}$};
		\node (gba2) at (0,2.5) [rectangle,draw] {$\,\gamma_{b_1a;\a}^{\phantom{1}}$};
		\draw [postaction={decorate}] (np1) to node[left]{} (s1);	
		\draw [postaction={decorate}] (s1) to node[left]{$R_a$} (nm1);	
		\draw [postaction={decorate}] (s2) to node[right]{$R_a$} (np2);	
		\draw [postaction={decorate}] (nm2) to node[right]{} (s2);			
		\draw [postaction={decorate}] (nm1) to [bend right=20] node[above]{} (gab1);
		\draw [postaction={decorate}] (nm1) to [bend right=40] node[below]{} (gab2);
		\draw [postaction={decorate}] (gab1) to [bend right=20] node[above]{$r_{ab_3;\a}$} (nm2);
		\draw [postaction={decorate}] (gab2) to [bend right=40] node[below right]{$r_{ab_4;\a}$} (nm2);
		\draw [postaction={decorate}] (np2) to [bend right=20] node[below]{} (gba1);
		\draw [postaction={decorate}] (np2) to [bend right=40] node[above]{} (gba2);
		\draw [postaction={decorate}] (gba1) to [bend right=20] node[below]{$r_{b_2a;\a}$} (np1);
		\draw [postaction={decorate}] (gba2) to [bend right=40] node[above left]{$r_{b_1a;\a}$} (np1);
	}
\end{equation}
Here $r_{b_1a;\a}$, $r_{b_2a;\a}$ represent fields incoming to $a$, and $r_{ab_3;\a}$, $r_{ab_4;\a}$ represent fields outgoing from $a$. The full expression (\ref{eq:chiQ_Lderiv_2}) is just a product of such factors over $a$.

We can move $D(\gamma)$ and $D(\gamma^{-1})$ through branching coefficients next to $D(\sigma)$
\begin{equation}
\begin{split} 
	&\chi_Q(\bL,\bsig) \chi_Q(\bL,\btau) 
\\ & =	
	\frac{\prod d(r_{ab;\a})}{\prod n_{ab;\a}!}
	\sum_{\bgam} \prod_a  				
		D^{R_a}_{i_a j_a} ( (\otimes_{b,\alpha} \gamma_{ba;\a})  \sigma_a (\otimes_{b,\alpha} \gamma_{ab;\a}^{-1}))
		D^{R_a}_{\tl j_a \tl i_a } (\tau_a^{-1})		
\\ & ~~~~~~~~~~~~~~~~~~~ \times
		B^{R_a \rightarrow \bigcup_{b,\alpha} r_{ab;\a}, \nu^-_a}_{j_a \rightarrow \bigcup_{b,\alpha} j_{ab;\a}}
		B^{R_a \rightarrow \bigcup_{b,\alpha} r_{ab;\a}, \nu^-_a}_{\tl j_a \rightarrow \bigcup_{b,\alpha} j_{ab;\a}}
		B^{R_a \rightarrow \bigcup_{b,\alpha} r_{ba;\a}, \nu^+_a}_{\tl i_a \rightarrow \bigcup_{b,\alpha} i_{ba;\a}}					
		B^{R_a \rightarrow \bigcup_{b,\alpha} r_{ba;\a}, \nu^+_a}_{i_a \rightarrow \bigcup_{b,\alpha} i_{ba;\a}}		
\end{split}
\end{equation}
Now the branching coefficients are contracted in a way to make projectors, which we can sum over, using (\ref{eq:BB_rr_sum})
\begin{equation}
	\sum_{\{ r_{ab,\a} \},\nu_a^-} 
		B^{R_a \rightarrow \bigcup_{b,\alpha} r_{ab;\a}, \nu^-_a}_{j_a \rightarrow \bigcup_{b,\alpha} j_{ab;\a}}
		B^{R_a \rightarrow \bigcup_{b,\alpha} r_{ab;\a}, \nu^-_a}_{\tl j_a \rightarrow \bigcup_{b,\alpha} j_{ab;\a}}
	= 
	\sum_{r_{ab},\nu_a^-} P^{R_a \rightarrow \bigcup_{b,\alpha} r_{ab;\a}, \nu^-_a}_{j_a \tl j_a}
	= \delta_{j_a \tl j_a}
\end{equation}
Performing this for both pairs of branching coefficients we arrive at
\begin{equation}
	\sum_{\bL} \frac{\prod n_{ab;\a}!}{\prod d(r_{ab;\a})}
	\chi_Q(\bL,\bsig) \chi_Q(\bL,\btau)
	=
		\sum_{R_a} \sum_{\bgam} \prod_a 
		\chi_{R_a}( (\otimes_{b,\alpha} \gamma_{ba;\a})  \sigma_a (\otimes_{b,\alpha} \gamma_{ab;\a}^{-1}) \tau_a^{-1})	
\end{equation}
Finally, the sum over $R_a$ can be done for each group factor using (\ref{eq:chiR_sum}), if we include a factor $\frac{d(R_a)}{n_a!}$
\begin{equation}
\begin{split}
	\sum_{\bL} 
	\frac{\prod n_{ab;\a}!}{\prod d(r_{ab;\a})}
	\frac{\prod d(R_a)}{\prod n_a!}
	\chi_Q(\bL,\bsig) \chi_Q(\bL,\btau)
	&=
		\sum_{\bgam} \prod_a \sum_{R_a}
		\frac{\prod d(R_a)}{\prod n_a!}
		\chi_{R_a}( (\otimes_{b,\alpha} \gamma_{ba;\a})  \sigma_a (\otimes_{b,\alpha} \gamma_{ab;\a}^{-1}) \tau_a^{-1} )	
\\
	&=
		\sum_{\bgam} \prod_a		
		\delta( (\otimes_{b,\alpha} \gamma_{ba;\a})  \sigma_a (\otimes_{b,\alpha} \gamma_{ab;\a}^{-1}) \tau_a^{-1} )			
\end{split}		
\end{equation}
Thus we arrive at (\ref{eq:chiQ_L_sum}).

\subsection{Derivation of two-point function}\label{appsec:2ptfunction} 

Here we show (\ref{eq:OQ_sig_2pt}), the two-point function of operators $\cO_Q(\bn,\bsig)$ defined in (\ref{eq:OQ_sig_defn}), which is used to show the orthogonality of restricted basis in Section~\ref{sec:twopoint}. 

The conjugated operator is:
\begin{equation}
\label{eq:OQ_sig_conj}
\begin{split}
	\cO_Q(\bn,\bsig)^\dagger &= 
	\prod_{a,b,\alpha} \left( \bar{\Phi}_{ab;\a}^{\otimes n_{ab;\a}} \right)^{\bm{I}_{ab;\a}}_{\bm{J}_{ab;\a}}
	\prod_a \left( \sigma_a \right)^{ \bigcup_{b,\alpha} \bm{J}_{ba;\a}}_{ \bigcup_{b,\alpha} \bm{I}_{ab;\a}} \\
	&=
	\prod_{a,b,\alpha} \left( \Phi_{ab;\a}^{\dagger \; \otimes n_{ab;\a}} \right)_{\bm{I}_{ab;\a}}^{\bm{J}_{ab;\a}}
	\prod_a \left( \sigma_a^{-1} \right)_{ \bigcup_{b,\alpha} \bm{J}_{ba;\a}}^{ \bigcup_{b,\alpha} \bm{I}_{ab;\a}}
\end{split}
\end{equation}
In the first line, since $\cO_Q$ is a scalar, conjugation is simply a complex conjugation of the fields $\bar\Phi$. In the second line we convert it to Hermitian conjugate by transposing both $(\bar\Phi)^i_j = (\Phi^\dagger)^j_i$ and $(\sigma)^i_j = (\sigma^{-1})^j_i$. The appearance of $\sigma^{-1}$ indicates reversal of cyclic order, so that e.g. $\tr(XYZ)^\dagger = \tr(Z^\dagger Y^\dagger X^\dagger)$. The two point function for two fields is
\begin{equation}
	\la (\Phi_{ab;\a})^i_j (\Phi^{\dagger\,\alpha}_{ab})^k_l \ra = \delta^i_l \delta^k_j
\end{equation}
The Wick contraction between $n_{ab;\a}$ fields generate
\begin{equation}
\label{eq:phi_n_wick}
	\la 
	\left( \Phi_{ab;\a}^{\otimes n_{ab;\a}}  \right)^{ \bm{I}_{ab;\a}}_{ \bm{J}_{ab;\a}}
	\left( \Phi_{ab;\a}^{\dagger \; \otimes n_{ab;\a}} \right)_{ \bm{\tl I}_{ab;\a}}^{ \bm{\tl J}_{ab;\a}}
	\ra 
	= \sum_{\gamma \in S_{n_{ab;\a}} } \delta^{ \gamma(\bm{I}_{ab;\a}) }_{\bm{\tl I}_{ab;\a}} \, \delta^{ \bm{\tl J}_{ab;\a}}_{ \gamma(\bm{J}_{ab;\a}) }	
	= \sum_{\gamma \in S_{n_{ab;\a}} } (\gamma^{-1})^{ \bm{I}_{ab;\a}}_{ \bm{\tl I}_{ab;\a}} \, (\gamma)^{ \bm{\tl J}_{ab;\a}}_{ \bm{J}_{ab;\a}}
\end{equation}
So the two point function, combining (\ref{eq:OQ_sig_defn}), (\ref{eq:OQ_sig_conj}) and (\ref{eq:phi_n_wick}):
\begin{equation}
\label{eq:OQ_sig_2pt_deriv}
\begin{split}
	\la \cO_Q(\bn,\bsig)\cO_Q(\bn,\tl\bsig)^\dagger \ra 
&=
	\sum_{\bgam}
	\prod_{a,b,\alpha}
	(\gamma_{ab;\a}^{-1})^{ \bm{I}_{ab;\a}}_{ \bm{\tl I}_{ab;\a}} \, (\gamma_{ab;\a})^{ \bm{\tl J}_{ab;\a}}_{ \bm{J}_{ab;\a}}
	\prod_a 
	\left( \sigma_a \right)^{ \bigcup_{b,\alpha} \bm{J}_{ba;\a}}_{ \bigcup_{b,\alpha} \bm{I}_{ab;\a}}
	\left( \tl\sigma_a^{-1} \right)_{ \bigcup_{b,\alpha} \bm{\tl J}_{ba;\a}}^{ \bigcup_{b,\alpha} \bm{\tl I}_{ab;\a}}
\\
&= \sum_{\bgam} \prod_{a} 
	\tr\left(
		\sigma_a \,
		(\otimes_{b,\alpha} \gamma_{ab;\a}^{-1} ) \, 
		\tl\sigma_{a}^{-1} \, 
		(\otimes_{b,\alpha} \gamma_{ba;\a} )
	\right)
\\
&\equiv \sum_{\bgam} \prod_{a} 
	\tr\left(
		\Adj_{\bgam}(\sigma_a)
		\tl\sigma_a^{-1}
	\right)
\end{split}	
\end{equation}
which gives (\ref{eq:OQ_sig_2pt}).

This calculation can also be understood diagrammatically. As an example let us take a simplified $\C3Z2$ quiver, with only a single flavor of $\Phi_{12}$ and $\Phi_{21}$
\begin{equation}
\label{eq:O2_sig_diag}
	\cO(\bn,\bsig) = \;
	\mytikz{
		\node (s1) at (0,0) [rectangle,draw] {$\sigma_1$};				
		\node (s2) at (3,0) [rectangle,draw] {$\sigma_2$};					
		\draw [postaction={decorate}] (s1.-120) .. controls +(-135:1.5) and +(135:1.5) .. node[left]{$\Phi_{11}$} (s1.120);
		\draw [postaction={decorate}] (s1.-60) to [bend right=30] node[below]{$\Phi_{12}$} (s2.-120);			
		\draw [postaction={decorate}] (s2.60) .. controls +(45:1.5) and +(-45:1.5) .. node[right]{$\Phi_{22}$} (s2.-60);
		\draw [postaction={decorate}] (s2.120) to [bend right=30] node[above]{$\Phi_{21}$} (s1.60);		
	}
\end{equation}
Conjugate operator (\ref{eq:OQ_sig_conj}) is represented by
\begin{equation}
\label{eq:O2_sig_diag_conj}
	\cO(\bn,\bsig)^\dagger = \;
	\mytikz{
		\node (s1) at (0,0) [rectangle,draw] {$\sigma_1^{-1}$};				
		\node (s2) at (3,0) [rectangle,draw] {$\sigma_2^{-1}$};					
		\draw [postaction={decorate}] (s1.120) .. controls +(135:1.5) and +(-135:1.5) .. node[left]{$\Phi^\dagger_{11}$} (s1.-120);
		\draw [postaction={decorate}] (s2.-120) to [bend left=30] node[below]{$\Phi^\dagger_{12}$} (s1.-60);			
		\draw [postaction={decorate}] (s2.-60) .. controls +(-45:1.5) and +(45:1.5) .. node[right]{$\Phi^\dagger_{22}$} (s2.60);
		\draw [postaction={decorate}] (s1.60) to [bend left=30] node[above]{$\Phi^\dagger_{21}$} (s2.120);		
	}
\end{equation}
Our convention is that outgoing arrow corresponds to lower index, and incoming to upper index, so the reversed arrows indicate transposed indices in the second line of (\ref{eq:OQ_sig_conj}). The Wick contraction between blocks of conjugate fields (\ref{eq:phi_n_wick}) is, diagrammatically
\begin{equation}
\la
	\mytikz{
		\node (p) at (0,0) [rectangle,draw] {$(\Phi_{ab;\a})^{\otimes n}$};				
		\node (pd) at (2,0) [rectangle,draw] {$(\Phi^{\dagger\,\alpha}_{ab})^{\otimes n}$};
		\draw [postaction={decorate}] ($(p)+(0,1)$) to (p);
		\draw [postaction={decorate}] (p) to +(0,-1);
		\draw [postaction={decorate}] ($(pd)+(0,1)$) to (pd);
		\draw [postaction={decorate}] (pd) to +(0,-1);
	}
\ra 
\; = \;
	\sum_{\gamma\in S_n} \;
	\mytikz{
		\node (p) at (0,0) [rectangle,draw] {$\gamma^{-1}$};				
		\node (pd) at (1.5,0) [rectangle,draw] {$\gamma$};
		\draw [postaction={decorate}] ($(p)+(0,1)$) to (p);
		\draw [postaction={decorate}] (p.-90) .. controls +(0,-0.5) and +(0,0.5) .. +(1.5,-1);
		\draw [postaction={decorate}] ($(pd)+(0,1)$) to (pd);
		\draw [postaction={decorate}] (pd.-90) .. controls +(0,-0.5) and +(0,0.5) .. +(-1.5,-1);
	}
\end{equation}
Applying this rule to the product of diagrams (\ref{eq:O2_sig_diag}) and (\ref{eq:O2_sig_diag_conj}) we find
\begin{equation}
	\la \cO(\bn,\bsig) \cO(\bn,\tl\bsig)^\dagger \ra = \sum_{\bgam} \;
	\mytikz{
		\node (g12i) at (-0.7,0) [rectangle,draw] {$\gamma_{12}^{-1}$};
		\node (g11i) at (-1.7,0) [rectangle,draw] {$\gamma_{11}^{-1}$};
		\node (g11) at (-3,0) [rectangle,draw] {$\gamma_{11}$};
		\node (g21) at (-4,0) [rectangle,draw] {$\gamma_{21}$};
		\node (s1) at (-2.35,1) [rectangle,draw] {$\sigma_1$};
		\node (s1i) at (-2.35,-1) [rectangle,draw] {$\tl\sigma_1^{-1}$};
		\draw [postaction={decorate}] (s1) to [bend left=30] (g11i);
		\draw [postaction={decorate}] (s1) to [bend left=30] (g12i);
		\draw [postaction={decorate}] (g11i) to [bend left=30] (s1i);
		\draw [postaction={decorate}] (g12i) to [bend left=30] (s1i);
		\draw [postaction={decorate}] (s1i) to [bend left=30] (g11);
		\draw [postaction={decorate}] (s1i) to [bend left=30] (g21);
		\draw [postaction={decorate}] (g11) to [bend left=30] (s1);
		\draw [postaction={decorate}] (g21) to [bend left=30] (s1);
		\node (g12) at (0.7,0) [rectangle,draw] {$\gamma_{12}$};
		\node (g22) at (1.7,0) [rectangle,draw] {$\gamma_{22}$};
		\node (g22i) at (3,0) [rectangle,draw] {$\gamma_{22}^{-1}$};
		\node (g21i) at (4,0) [rectangle,draw] {$\gamma_{21}^{-1}$};
		\node (s2) at (2.35,1) [rectangle,draw] {$\sigma_2$};
		\node (s2i) at (2.35,-1) [rectangle,draw] {$\tl\sigma_2^{-1}$};
		\draw [postaction={decorate}] (s2) to [bend left=30] (g22i);
		\draw [postaction={decorate}] (s2) to [bend left=30] (g21i);
		\draw [postaction={decorate}] (g22i) to [bend left=30] (s2i);
		\draw [postaction={decorate}] (g21i) to [bend left=30] (s2i);
		\draw [postaction={decorate}] (s2i) to [bend left=30] (g22);
		\draw [postaction={decorate}] (s2i) to [bend left=30] (g12);
		\draw [postaction={decorate}] (g22) to [bend left=30] (s2);
		\draw [postaction={decorate}] (g12) to [bend left=30] (s2);
	}
\end{equation}
It is easy to see that in general quivers will break up into separate factors for each group, with $\sigma_a$ and $\tl\sigma_a^{-1}$ connected by $\gamma_{ab;\a}^{-1}$ and $\gamma_{ba;\a}$. This reproduces (\ref{eq:OQ_sig_2pt_deriv}).

%%%%%%%%%%%%%%%%%%%%%%%%%%%%%%%%%%%%%%%%%%%%%%%%%%%%%%%%%%%%%%%%%%%%%
%%%%%%%%%%%%%%%%%%%%%%%%%%%%%%%%%%%%%%%%%%%%%%%%%%%%%%%%%%%%%%%%%%%%%

\subsection{Derivation of chiral ring structure constants} 
\label{der:ChirStConst} 

Here we explain the formulae corresponding to the diagrammatic derivation of (\ref{eq:GLLL_result})
given in section \ref{sec:restschurfus}. 

We can write (\ref{formula-chiring}) as 
\begin{equation}\begin{split} 
 G ( \bL^{(1)} , \bL^{(2)} ; \bL^{(3)} ) 
&= 
	\tl f_{\bL^{(1)}\bL^{(2)}}^{\bL^{(3)}}	
	\frac{1}{\prod_a n_a^{(1)}!n_a^{(2)}!}
\sum_{ \bsig^{(1)}  , \bsig^{(2)}   } \\
& \qquad \prod_a   \left (  \prod_{p=1}^3 B^{R_a^{(p)} \rightarrow \cup_{b ,\alpha}  r_{ba;\a}   }_{ i_{a}^{(p)} \rightarrow    \cup_{b ,\alpha}  l_{ba}^{(p)\alpha}   }   \right ) 
 \left (  \prod_{p=1}^3 B^{R_a^{(p)} \rightarrow \cup_{b ,\alpha}  r_{ab;\a}   }_{ i_{a}^{(p)} \rightarrow    \cup_{b ,\alpha}  l_{ab}^{(p)\alpha}   }   \right ) 
 \\
&\qquad \times D^{R_a^{(1)} }_{ i_a^{(1)} j_a^{(1)}     } ( \sigma_a^{(1)} )D^{R_a^{(2)} }_{ i_a^{(2)} j_a^{(2)}     } ( \sigma_a^{(2)} )
D^{R_a^{(3)} }_{ i_a^{(3)} j_a^{(2)}     } (  \sigma_a^{(1)} \circ \sigma_a^{(2)} ) \\
&= 
	\tl f_{\bL^{(1)}\bL^{(2)}}^{\bL^{(3)}}	
	\frac{1}{\prod_a n_a^{(1)}!n_a^{(2)}!}
\sum_{ \bsig^{(1)}  , \bsig^{(2)}   } \\
& \qquad \prod_a   \left (  \prod_{p=1}^3 B^{R_a^{(p)} \rightarrow \cup_{b ,\alpha}  r_{ba;\a}   }_{ i_{a}^{(p)} \rightarrow    \cup_{b ,\alpha}  l_{ba}^{(p)\alpha}   }   \right ) 
 \left (  \prod_{p=1}^3 B^{R_a^{(p)} \rightarrow \cup_{b ,\alpha}  r_{ab;\a}   }_{ i_{a}^{(p)} \rightarrow    \cup_{b ,\alpha}  l_{ab}^{(p)\alpha}   }   \right ) 
\\ 
& \qquad \times D^{R_a^{(1)} }_{ i_a^{(1)} j_a^{(1)}     } ( \sigma_a^{(1)} )D^{R_a^{(2)} }_{ i_a^{(2)} j_a^{(2)}     } ( \sigma_a^{(2)} )
D^{R_a^{(3)} }_{ j_a^{(3)}  i_a^{(3)}    } (  ( \sigma_a^{(1)} )^{-1}  \circ ( \sigma_a^{(2)})^{-1}  )
\end{split}\end{equation}

Next we do the sum over the $ \sigma_{a}^{(1)} , \sigma_{a}^{(2)} $, expressing the answer in terms of branching coefficients
as in (\ref{eq:g1g2_factoriz}). 

\begin{equation}\begin{split} 
& \sum_{ \sigma^{(1)} \in S_{n^{(1)} } } \sum_{  \sigma^{(2)} \in S_{n^{(2)}  } }
D^{R^{(1)} }_{ i_1 j_1 } ( \sigma^{(1)} ) D^{R^{(2)} }_{ i_2 j_2 } ( \sigma^{(2)} )
D^{ R^{(3)}  }_{i_3 j_3  } ( \sigma^{(1)} \circ \sigma^{(2)} )  \\ 
& =  \sum_{ \sigma^{(1)} , \sigma^{(2)} } 
\sum_{ S^{(1)} , S^{(2)} } \sum_{\nu} 
 D^{R^{(1)} }_{ i_1 j_1 } ( \sigma^{(1)} ) D^{R^{(2)} }_{ i_2 j_2 } ( \sigma^{(2)} )
B^{R^{(3)}  \rightarrow S^{(1)} , S^{(2)} ; \nu  }_{ i_3 \rightarrow k_1 , k_2     }  D^{S^{(1)}  }_{ k_1 m_1 }  ( \sigma^{(1)} ) 
  D^{S^{(2)}  }_{ k_2 m_2 }  ( \sigma^{(2)} ) B^{R^{(3) } \rightarrow S^{(1)} , S^{(2)} ; \nu }_{ j_3 \rightarrow m_1 , m_2     }\\ 
& =\sum_{ S^{(1)} , S^{(2)} } \sum_{\nu}   {n^{(1)} ! \over d( R^{(1)} )  } {n^{(2)} ! \over d( R^{(2)} )  }\delta_{ R^{(1)} , S^{(1)}  }\delta_{ R^{(2)} , S^{(2)}  }   \delta_{i_1 k_1 } \delta_{j_1 m_1 } \delta_{ i_2 k_2 } \delta_{ j_2 m_2} B^{R^{(3)}  \rightarrow S^{(1)} , S^{(2)} ; \nu  }_{ i_3 \rightarrow k_1 , k_2     }
 B^{R^{(3)}  \rightarrow S^{(1)} , S^{(2)} ; \nu  }_{ j_3 \rightarrow m_1 , m_2     } \\ 
& = {n^{(1)} ! \over d( R^{(1)} )  } {n^{(2)} ! \over d( R^{(2)} )  } 
\sum_{\nu}  B^{R^{(3)}  \rightarrow R^{(1)} , R^{(2)} ; \nu  }_{ i_3 \rightarrow i_1 , i_2     } B^{R^{(3) } \rightarrow S^{(1)} , S^{(2)} ; \nu }_{ j_3 \rightarrow j_1 , j_2     }
\end{split}\end{equation}
Applying this at each node, gives two extra branching coefficients at each node of the quiver $Q$, leading to:
\begin{equation}\begin{split}\label{quiver-fusion-coeffs} 
& G ( \bL^{(1)} , \bL^{(2)} ; \bL^{(3)} )  = 
\frac{\tl f_{\bL^{(1)}\bL^{(2)}}^{\bL^{(3)}} }{ \prod_a d( R_a^{(1)} ) d( R_a^{(2)} ) }
\sum_{\{\nu_a\}} \\ & \prod_{a}
 B^{R_a^{(1)} \rightarrow \cup_{b,\alpha} r_{ba;\a}^{(1)} ;~  \nu_{a}^{(1) + }}_{~  i^{(1)}_{ a  } \rightarrow \cup_{ b ,\alpha} l_{ba;\a}^{(1)}  }      B^{R_a^{(2)} \rightarrow \cup_{b,\alpha} r_{ba;\a}^{(2)} ;~  \nu_{a}^{(2) + }}_{~  i^{(2)}_{ a  } \rightarrow \cup_{ b ,\alpha} l_{ba;\a}^{(2)}  }   B^{ R^{(3)}_a \rightarrow R^{(1)}_a , R^{(2)}_a  ; \nu^{+}_a }_{i^{(3)}_a \rightarrow i^{(1)}_a  , i^{(2)}_{a} } 
B^{R^{(3)  }_a   \rightarrow  \cup_{ b ,\alpha}   r_{ba;\a}^{(3)}   ;  \nu^{ (3) + }_a   }_{ i^{(3)}_{a} \rightarrow   \cup_{ b ,\alpha}   l_{ba;\a}^{(3)}  } \\ 
& \times
B^{R^{(3)}_a   \rightarrow  \cup_{ b ,\alpha}   r_{ab;\a}^{(3)}   ;  \nu^{ (3) - }_a   }_{ j^{(3)}_{a} \rightarrow   \cup_{ b ,\alpha}   l_{ab;\a}^{(3)}  }  B^{ R^{(3)}_a \rightarrow R^{(1)}_a , R^{(2)}_a  ;   \nu^{-}_a  }_{j^{(3)}_a \rightarrow j^{(1)}_a  , j^{(2)}_{a} } B^{R_a^{(1)} \rightarrow \cup_{b,\alpha} r_{ab;\a}^{(1)} ;~  \nu_{a}^{(1) - }}_{~  j^{(1)}_{ a  } \rightarrow \cup_{ b ,\alpha} l_{ab;\a}^{(1)}  }      B^{R_a^{(2)} \rightarrow \cup_{b,\alpha} r_{ab;\a}^{(2)} ;~  \nu_{a}^{(2) - }}_{~  j^{(2)}_{ a  } \rightarrow \cup_{ b ,\alpha} l_{ab;\a}^{(2)}  } \\ 
& = \frac{\tl f_{\bL^{(1)}\bL^{(2)}}^{\bL^{(3)}} }{ \prod_a d( R_a^{(1)} ) d( R_a^{(2)} ) }
\sum_{\{\nu_a\}}  \prod_{a} B^{ R^{(3)}_a \rightarrow R^{(1)}_a , R^{(2)}_a  ; \nu^{+}_a }_{i^{(3)}_a \rightarrow i^{(1)}_a  , i^{(2)}_{a} }  B^{ R^{(3)}_a \rightarrow R^{(1)}_a , R^{(2)}_a  ;   \nu^{-}_a  }_{j^{(3)}_a \rightarrow j^{(1)}_a  , j^{(2)}_{a} } 
\\ & 
\qquad \qquad \qquad \left ( \prod_{p=1}^3  B^{R_a^{(p)} \rightarrow \cup_{b,\alpha} r_{ba}^{(p)\alpha} ;~  \nu_{a}^{(p) + }}_{~  i^{(p)}_{ a  } \rightarrow \cup_{ b ,\alpha} l^{(p )\alpha}_{ba}  }   \right ) 
\left ( \prod_{p=1}^3   B^{R^{(p)}_a   \rightarrow  \cup_{ b ,\alpha}   r^{(p)\alpha}_{ab}   ;  \nu^{ (3) - }_a   }_{ j^{(3)}_{a} \rightarrow   \cup_{ b ,\alpha}   l_{ab}^{(p)\alpha}  } \right ) 
\end{split}\end{equation}

The label  $\nu_a $ is summed over the Littlewood-Richardson multiplicity 
$g ( R_a^{(1)}   , R_a^{(2)}   ; R^{(3)}_a  ) $  for the reduction of the irrep $R^{(3)}_a$ 
of $S_{n_a^{(3)} }$  to irrep $ R^{(1)} \otimes R^{ (2)}_a $ of $S_{n_a^{(1)}} \times S_{ n_a^{(2)} } $. 
By Schur-Weyl duality, this is also the multiplicity of the $U(N_a)$ representation $R_a^{(3)}$
in the tensor product of $ R_a^{(1)}  \otimes R_a^{(2)}  $.

The next step is to exploit the invariance, 
 under the action of $ \times_{ a ,  b ,\alpha} S_{n_{ab;\a}^{(1)} } \times   
S_{n_{ab;\a}^{(2)}}  $,   of the   branching coefficients
in  (\ref{quiver-fusion-coeffs}) labeled by $\nu_a^{(1) - } , \nu_a^{(2)- } \nu_a^{(3) - }$
 (we could equally well have chosen to work with the $\nu_a^{(1) + } , \nu_a^{(2) +  } \nu_a^{(3)  + }$) as indicated in (\ref{eq:GLLL_deriv1}). 
\begin{equation}
\begin{split} 
& G ( \bL^{(1)} , \bL^{(2)} ; \bL^{(3)} ) 
= 	\tl f_{\bL^{(1)}\bL^{(2)}}^{\bL^{(3)}}	
	\frac{1}{\prod_a d(R_a^{(1)}) d( R_a^{(2)}) } 	\frac{1}{\prod_{ a , b ,\alpha}  n_{ab;\a}^{(1)} ! n_{ab;\a}^{(2)} !}
\sum_{ \{ \nu_a \} } \sum_{   \gamma_{ab;\a}^{(1)} , \gamma_{ab;\a}^{(2)}  } 	
	\\ 
	& 
	 \prod_{a} B^{ R^{(3)}_a \rightarrow R^{(1)}_a , R^{(2)}_a  ; \nu^{+}_a }_{i^{(3)}_a \rightarrow i^{(1)}_a  , i^{(2)}_{a} }  B^{ R^{(3)}_a \rightarrow R^{(1)}_a , R^{(2)}_a  ;   \nu^{-}_a  }_{j^{(3)}_a \rightarrow j^{(1)}_a  , j^{(2)}_{a} } 
\qquad \qquad \qquad \left ( \prod_{p=1}^3  B^{R_a^{(p)} \rightarrow \cup_{b,\alpha} r_{ba}^{(p)\alpha} ;~  \nu_{a}^{(p) + }}_{~  i^{(p)}_{ a  } \rightarrow \cup_{ b ,\alpha} l^{(p )\alpha}_{ba}  }   \right ) \\ 
&     \left (  \prod_{a , b,\alpha}
 D^{r_{ab}^{(1)}\alpha}_{ k_{ab;\a}^{(1)}  l_{ab;\a}^{(1)}  }   ( \gamma_{ab}^{ (1)\alpha} )
 D^{r_{ab}^{(2)}\alpha}_{ k_{ab;\a}^{(2)}  l_{ab;\a}^{(2)}  }   ( \gamma_{ab}^{ (2)\alpha} )
  D^{r_{ab}^{(3)}\alpha}_{ k_{ab;\a}^{(3)}  l_{ab;\a}^{(3)}  }  
   (    (\gamma_{ab}^{ (1)\alpha})^{-1}    \circ   (\gamma_{ab}^{ (2)\alpha})^{-1 }   ) ~~
 \prod_{p=1}^3   B^{R^{(p)}_a   \rightarrow  \cup_{ b ,\alpha}   r^{(p)\alpha}_{ab}   ;  \nu^{ (3) - }_a   }_{ j^{(3)}_{a} \rightarrow   \cup_{ b ,\alpha}   k_{ab}^{(p)\alpha}  } \right ) 
\end{split} 
\end{equation}
Now we do the sum over the permutations  $ \{ \gamma_{ab;\a}^{(1)} , \gamma_{ab;\a}^{(2)}\} $
which introduces branching  coefficients for $ r_{ab;\a}^{(3)} \rightarrow r_{ab;\a}^{(1)} \otimes  r_{ab;\a}^{(2)}$

\begin{equation}
\begin{split} 
& G ( \bL^{(1)} , \bL^{(2)} ; \bL^{(3)} ) 
= 	\tl f_{\bL^{(1)}\bL^{(2)}}^{\bL^{(3)}}	
	\frac{1}{\prod_a d( R_a^{(1)} ) d( R_a^{(2)} ) } 	
	\frac{1}{\prod_{ a , b ,\alpha} d (r_{ab;\a}^{(1)}) d (r_{ab;\a}^{(2)} )}\\
	&  	 \sum_{ \{ \nu_{a} , \nu_{ab;\a}\}  }   \prod_{a}	 \left ( B^{ R^{(3)}_a \rightarrow R^{(1)}_a , R^{(2)}_a  ; \nu^{+}_a }_{i^{(3)}_a \rightarrow i^{(1)}_a  , i^{(2)}_{a} }  B^{ R^{(3)}_a \rightarrow R^{(1)}_a , R^{(2)}_a  ;   \nu^{-}_a  }_{j^{(3)}_a \rightarrow j^{(1)}_a  , j^{(2)}_{a} }
\right ) 	
	 \\
& \prod_{ a  } 
\left ( \prod_{p=1}^3  B^{R_a^{(p)} \rightarrow \cup_{b,\alpha} r_{ba}^{(p)\alpha} ;~  \nu_{a}^{(p) + }}_{~  i^{(p)}_{ a  } \rightarrow \cup_{ b ,\alpha} l^{(p )\alpha}_{ba}  }   \right ) 
\left ( \prod_{p=1}^3   B^{R^{(p)}_a   \rightarrow  \cup_{ b ,\alpha}   r^{(p)\alpha}_{ab}   ;  \nu^{ (p) - }_a   }_{ j^{(p)}_{a} \rightarrow   \cup_{ b ,\alpha}   k_{ab}^{(p)\alpha}  } \right ) \\ 
& 
\prod_{ a , b ,\alpha} 
B^{r_{ab;\a}^{(3)}  \rightarrow  r_{ab;\a}^{(1)} ,   r_{ab;\a}^{(2)} ; \nu_{ab;\a}  }_{ 
~ l_{ab;\a}^{(3)} \rightarrow  l_{ab;\a}^{(1)} ,   l_{ab;\a}^{(2)}}
B^{r_{ab;\a}^{(3)}  \rightarrow  r_{ab;\a}^{(1)} ,   r_{ab;\a}^{(2)} ; \nu_{ab;\a}  }_{ 
~ k_{ab;\a}^{(3)} \rightarrow  k_{ab;\a}^{(1)} ,   k_{ab;\a}^{(2)} }  
\end{split} 
\end{equation}
We now see that  there is a factorization between state indices for Young diagrams associated branching 
coefficients carrying  $\nu^-$ indices and those  for Young diagrams associated branching 
coefficients carrying  $\nu^-$ indices, which corresponds to the factorized form in the diagram (\ref{eq:GLLL_diag})
\begin{equation}\begin{split}
& G ( \bL^{(1)} , \bL^{(2)} ; \bL^{(3)} ) 
= 	\tl f_{\bL^{(1)}\bL^{(2)}}^{\bL^{(3)}}	
	\frac{1}{\prod_a d( R_a^{(1)} ) d( R_a^{(2)} ) } 	
	\frac{1}{\prod_{ a , b ,\alpha} d (r_{ab;\a}^{(1)}) d (r_{ab;\a}^{(2)} )}\\
& 	 \sum_{ \{ \nu_{a} , \nu_{ab;\a}\}  }  \prod_{a}	 \left ( B^{ R^{(3)}_a \rightarrow R^{(1)}_a , R^{(2)}_a  ; \nu^{+}_a }_{i^{(3)}_a \rightarrow i^{(1)}_a  , i^{(2)}_{a} }  \prod_{p=1}^3  B^{R_a^{(p)} \rightarrow \cup_{b,\alpha} r_{ba}^{(p)\alpha} ;~  \nu_{a}^{(p) + }}_{~  i^{(p)}_{ a  } \rightarrow \cup_{ b ,\alpha} l^{(p )\alpha}_{ba}  } \prod_{ b ,\alpha} B^{r_{ab;\a}^{(3)}  \rightarrow  r_{ab;\a}^{(1)} ,   r_{ab;\a}^{(2)} ; \nu_{ab;\a}  }_{ 
~ l_{ab;\a}^{(3)} \rightarrow  l_{ab;\a}^{(1)} ,   l_{ab;\a}^{(2)}} \right )  \\ 
& \prod_a \left (  B^{ R^{(3)}_a \rightarrow R^{(1)}_a , R^{(2)}_a  ;   \nu^{-}_a  }_{j^{(3)}_a \rightarrow j^{(1)}_a  , j^{(2)}_{a} }
	 \prod_{p=1}^3   B^{R^{(p)}_a   \rightarrow  \cup_{ b ,\alpha}   r^{(p)\alpha}_{ab}   ;  \nu^{ (p) - }_a   }_{ j^{(p)}_{a} \rightarrow   \cup_{ b ,\alpha}   k_{ab}^{(p)\alpha}  }  
\prod_{   b ,\alpha}  B^{r_{ab;\a}^{(3)}  \rightarrow  r_{ab;\a}^{(1)} ,   r_{ab;\a}^{(2)} ; \nu_{ab;\a}  }_{ 
~ k_{ab;\a}^{(3)} \rightarrow  k_{ab;\a}^{(1)} ,   k_{ab;\a}^{(2)} }   \right ) 
\end{split}\end{equation}
This is the factorized result, where we have a factor for each gauge group, and  for each gauge group
there is a factorization separating the $\nu^+$  branching coefficients 
from the $\nu^-$ branching coefficients. The close connection between 
the final formula and the diagrammatic moves means that we can interpret the 
process of constructing the final answer diagrammatically. Start with 
the original quiver and modify it to produce the split-node version with $R_a$ lines 
joining the  plus and minus nodes. 
 Now cut this split-node quiver along  all the  edges, 
thus separating it into a collection of nodes labelled $ \nu_a^+ , \nu_a^- $. 
 The $\nu_a^{+} $ nodes have a collection of directed   lines carrying labels $ R_{a} , r_{ba;\a}$.
 The $\nu_a^- $ nodes have outgoing directed lines labeled $ R_{a} , r_{ab;\a}$. 
 Doing this cutting procedure for the three labelled quivers, to produce 
 nodes $ ( \nu_a^{(I)+}  ,  \nu_a^{(I)-} )  $ (for $I=1,2,3$) with dangling lines
 labeled $ R_{a}^{(I)}  , r_{ba}^{(I)\alpha}$. 
 Link up the nodes $\nu_{a}^{(I) + } $ 
  using new nodes $ \mu_{a}^+   $ for $ ( R_a^{(1)} , R_a^{(2)} ) \rightarrow  R_a^{(3)} $, and new 
  nodes  $\mu_{ab;\a}$ for the
   $  r_{ba;\a}^{(1)}  ,    r_{ba;\a}^{(2)}    \rightarrow    r_{ba;\a}^{(3)}$. 
   This gives a graph  for each gauge group labelled $a$, with nodes labelled by $ \{ \cup_I \nu_a^{(I) + } , \mu_a , \mu_{ab;\a} \} $.    Repeating the same procedure for the minus nodes gives
 another set of graphs for each gauge group,    with nodes labeled 
$ \{ \cup_I \nu_a^{(I) -  } , \mu_a , \mu_{b a;\a} \} $. So the result for the chiral ring 
   structure constants can be obtained by cutting and gluing of  the split-node  quivers labeled $ \bL_1 , \bL_2 , \bL_3$. This is an illustration of the power of  quivers as calculators.

\subsection{Finite \texorpdfstring{$N$}{N} chiral ring with superpotential, using explicit operators }
\label{Appsec:ExpIntChi}   

Here we confirm the expected dimension from (\ref{ExpVint}) using 
the explicit description of operators in $V_N , V_F $. The space $V_N$ (\ref{eq:VN_defn}) is spanned by the operators, where any $R_a$ is taller than $N$. For our choice of charges there are three such operators (using the restricted Schur basis):
\begin{equation}
\begin{split}
	& \qquad V_N = \{ O_1, O_2, O_3 \}  \\
	\cO_1 &\equiv \cO(R_1=[1^{N+1}],R_2=[1^{N+1}];\br) \\
	\cO_2 &\equiv \cO(R_1=[1^{N+1}],R_2=[2,1^{N-1}];\br) \\
	\cO_3 &\equiv \cO(R_1=[2,1^{N-1}],R_2=[1^{N+1}];\br)
\end{split}
\end{equation}
with $r_{A_1} = [1], r_{B_1} = [1], r_{A_2} = [1^N], r_{B_2} = [1^N]$ for all three. 

A convenient way to characterize $V_F$ is as the kernel
\begin{equation}
	V_F = {\rm Ker}(\cP)
\end{equation}
where $\cP$ is the \emph{symmetrization operator} acting on $V^{(\infty)}$ as a linear operator. For example
\begin{equation}
	\cP \; \tr(A_1 B_1 A_2 B_2) = \frac{1}{2} \tr(A_1 B_1 A_2 B_2) + \frac{1}{2} \tr(A_1 B_2 A_1 B_2)
\end{equation}
Then $V_F$ is the subspace of $V^{(\infty)}$ annihilated by $\cP$. In order to find $V_F \cap V_N$ we need the operators annihilated by $\cP$ in $V_N$
\begin{equation}
	V_F \cap V_N = {\rm Ker}(\cP_{V_N} )
\end{equation}

The action of $\cP$ is easily written in the product-of-traces or the permutation basis $\cO(\bsig)$, but we have $V_N$ in terms of  the $\cO(\bL)$ basis. In order to calculate $\cP$ acting on $\cO_1,\cO_2,\cO_3$ we need to expand them in terms of $\cO(\bsig)$ using the definitions (\ref{eq:OL_defn}). The first operator is easy, since all representations are one-dimensional and branching coefficients are trivial:
\begin{equation}
	\cO_1 = \frac{1}{(N+1)!^2} \sum_{\sigma_1,\sigma_2 \in S_{N+1}} (-1)^{\sigma_1+\sigma_2} \, \cO(\sigma_1,\sigma_2)
\end{equation}

For $\cO_2$ and $\cO_3$ we need the branching coefficient $B^{[2,1^{N-1}] \rightarrow [1],[1^N]}_{i}$. The representation $([1],[1^N])$ of the subgroup $S_1 \times S_N$ is one-dimensional, so we do not include a label for it. Representation $[2,1^{N-1}]$ is $N$-dimensional, for which we use the Young-Yamanouchi (YY) basis. The YY-basis is labelled by Young tableaux, i.e. Young diagrams with integers $\{1, \ldots, N+1\}$ in the boxes. We use the convention where the numbers are decreasing along rows and down columns. For example, $[2,1^3]$ is spanned by:
\begin{equation}		
	\left\{
		{\tiny \young({5}{1},{4},{3},{2})} , \;
		{\tiny \young({5}{2},{4},{3},{1})} , \;
		{\tiny \young({5}{3},{4},{2},{1})} , \;
		{\tiny \young({5}{4},{3},{2},{1})}
	\right\}
\end{equation}
The YY-basis is particularly convenient for our purpose, because it is constructed using the decomposition $S_{N+1} \rightarrow S_1 \times S_N$. The state in $[2,1^{N-1}]$ which transforms according to $([1],[1^N])$ of $S_1 \times S_N$ is precisely the one which has the label 1 in the second column. Thus the branching coefficient we need is simply\footnote{Here the diagram denotes the first column of any height, with numbers from $2$ to $N+1$.}
\begin{equation}	
	B^{[2,1^{N-1}] \rightarrow [1],[1^N]}_{i} = \delta(i={\tiny \young({\,}{1},{\,},{\,},{\,})} )
\end{equation}
The operator is thus
\begin{equation}\label{eq:O2_deriv} 
\begin{split} 
&\cO_{2}  = \frac{\sqrt{N}}{(N+1)!^2}  \sum_{\substack{ \s_1  \in S_{N+1} \\ \s_2 \in S_{N+1}} }
   (-1)^{ \s_1 } \left \langle \tiny{ \young({\,}1,{\,},{\,},{\,})}     
   \right | \s_2    \left | \tiny{ \young({\,}1,{\,},{\,},{\,}) }   \right \rangle \cO( \s_1 , \s_2  ) \\
& = 
\frac{\sqrt{N}}{(N+1)!^2}
\sum_{\substack{ \s_1  \in S_{N+1} \\ \s_2 \in S_{N}} }
(-1)^{\s_1} 
\left(    
    \left \langle \tiny{ \young({\,}1,{\,},{\,},{\,})}  \right | \s_2    \left | \tiny{ \young({\,}1,{\,},{\,},{\,}) }   \right \rangle     
    \cO( \s_1 , \s_2)
	+ 
	\sum_{ k = 2}^{ N+1} 
  \left \langle   {\tiny{ \young({\,}1,{\,},{\,},{\,}) }} 
\right | \s_2 (1 k ) \left |  \tiny{ \young({\,}1,{\,},{\,},{\,}) }   \right\rangle
	\cO( \s_1 , \s_2(1k))
\right) 
\end{split} 
\end{equation} 
We have split the sum over $\sigma_2 \in S_{N+1}$ into a part where $\sigma_2 \in S_1 \times S_N$, and the rest, where first element gets permuted.
The corresponding matrix elements are:
\begin{equation}
	\left \langle \tiny{ \young({\,}1,{\,},{\,},{\,})}  \right | \s_2    \left | \tiny{ \young({\,}1,{\,},{\,},{\,}) } \right\rangle = (-1)^{\s_2}
	, \quad
	\left \langle \tiny{ \young({\,}1,{\,},{\,},{\,})}  \right | \s_2 (1k)    \left | \tiny{ \young({\,}1,{\,},{\,},{\,}) } \right\rangle = \frac{(-1)^{\s_2}}{N}
	, \quad (\sigma_2 \in S_1 \times S_N)
\end{equation}
Substituting this in (\ref{eq:O2_deriv}):
\begin{equation}
\label{eq:O2}
\begin{split}
	\cO_2 &= 	
		\frac{\sqrt{N}}{(N+1)!^2}
		\sum_{\substack{ \s_1  \in S_{N+1} \\ \s_2 \in S_{N}} }
		(-1)^{\s_1} (-1)^{\s_2}
  	\left( \cO( \s_1 , \s_2) + \frac{1}{N} \sum_{ k = 2}^{ N+1}  \cO( \s_1 , \s_2(1k)) \right) 
  \\
  &= \frac{N! \sqrt{N} }{(N+1)!^2}
  	\sum_{\s  \in S_{N+1} }
		(-1)^{\s} \left( \cO( \s , \mI) + \cO( \s , (12)) \right) 
\end{split}
\end{equation}
In the second line we performed the $\sigma_2$ sum by using invariance (\ref{eq:OQ_adj}) to set $O(\sigma_1,\sigma_2)=O(\sigma_1 \sigma_2^{-1},\mI)$ and redefining $\s=\s_1 \s_2^{-1}$. This is possible because $\sigma_2$ does not run over the full $S_{N+1}$, but only the subgroup (\ref{eq:gamma_subgroup}). Also using invariance we find $O(\sigma,(1k))=O((2k)\sigma(2k),(12))$, which allows to remove $k$ dependence. 

Analogously the final operator in $V_N$ is
\begin{equation}
\label{eq:O3}
	\cO_3 	
  = \frac{N! \sqrt{N} }{(N+1)!^2}
  	\sum_{\s  \in S_{N+1} }
		(-1)^{\s} \left( \cO( \mI , \s) + \cO( (12) , \s) \right) 
\end{equation}

Now, the question is, how many linear combinations of $\cO_1,\cO_2,\cO_3$ are annihilated by $\cP$. This will give $V_N \cap V_F$. 
First, observe that $\cO_1$ is unchanged by the symmetrization
\begin{equation}
	\cP \, \cO_1 = \cO_1
\end{equation}
because any permutation of $A$'s or $B$'s is already included in the sum
\begin{equation}
\label{eq:O1_gg}
\begin{split}
	\sum_{\sigma_1,\sigma_2 \in S_{N+1}} (-1)^{\sigma_1+\sigma_2} \, \cO(\sigma_1 , \sigma_2) &=
	\sum_{\sigma_1,\sigma_2 \in S_{N+1}} (-1)^{\sigma_1+\sigma_2} \, \cO(\sigma_1 \gamma, \gamma^{-1}\sigma_2) 
	\\
	&= 
	\sum_{\sigma_1,\sigma_2 \in S_{N+1}} (-1)^{\sigma_1+\sigma_2} \, \cO(\gamma\sigma_1 , \sigma_2\gamma^{-1}) 	
\end{split}
\end{equation}
so all permutations within a trace are already present. Note in (\ref{eq:O1_gg}) we do not use (\ref{eq:OQ_adj}), because $\gamma \notin S_1 \times S_N$, instead we absorb $\gamma$ in the sums $\sigma_1,\sigma_2$. The same relationship is not obeyed by $\cO_2,\cO_3$, because of non-trivial $\sigma_1,\sigma_2$ dependence.

Now, let us deal with $\cO_2,\cO_3$. It is useful to separate (\ref{eq:O2}), (\ref{eq:O3})
\begin{equation}
\label{eq:O23_split}
\begin{split}
	\cO_2^{\mI} &= \sum_{\sigma\in S_{N+1}} (-1)^\sigma \cO(\sigma,\mI) , \quad
	\cO_2^{(12)} = \sum_{\sigma\in S_{N+1}} (-1)^\sigma \cO(\sigma,(12)) \\
	\cO_3^{\mI} &= \sum_{\sigma\in S_{N+1}} (-1)^\sigma \cO(\mI,\sigma) , \quad
	\cO_3^{(12)} = \sum_{\sigma\in S_{N+1}} (-1)^\sigma \cO((12),\sigma)
\end{split}
\end{equation}
so that
\begin{equation}
\cO_2 \sim \cO_2^{\mI} + \cO_2^{(12)}, \quad 
\cO_3 \sim \cO_3^{\mI} + \cO_3^{(12)}
\end{equation}
We can evaluate (\ref{eq:O23_split}) explicitly
\begin{equation}
\begin{split}
	\cO_2^{\mI} &= \sum_{\sigma\in S_{N+1}} (-1)^\sigma \tr(\sigma \, (B_1 A_1) \otimes (B_2 A_2)^{\otimes N})
\\ 
	\cO_3^{\mI} &= \sum_{\sigma\in S_{N+1}} (-1)^\sigma \tr(\sigma \, (A_1 B_1) \otimes (A_2 B_2)^{\otimes N})
\\ 
	\cO_2^{(12)} &= - \sum_{\sigma\in S_{N+1}} (-1)^\sigma \tr(\sigma \, (B_2 A_1 ) \otimes (B_1 A_2) \otimes (B_2 A_2)^{\otimes N - 1})	
\\ 
	\cO_3^{(12)} &= - \sum_{\sigma\in S_{N+1}} (-1)^\sigma \tr(\sigma \, (A_1 B_2) \otimes (A_2 B_1) \otimes (A_2 B_2)^{\otimes N - 1})		
\end{split}
\end{equation}
These are ``determinant-like'' operators made from composites $A_i B_j$. It is easy to see that
\begin{equation}
	\cP \, \cO_2^{\mI} = \cP \, \cO_3^{\mI} , \quad \cP \, \cO_2^{(12)} = \cP \, \cO_3^{(12)}
\end{equation}
because $\cO_2^{\mI},\cO_3^{\mI}$ and $\cO_2^{(12)},\cO_3^{(12)}$ only differ by the ordering inside the trace. This leads to
\begin{equation}
	\cP \cO_2 = \cP \cO_3
\end{equation}
Also we can check that $\cP \cO_2 \neq \cP \cO_1$ by using an example, so these are in fact two linearly independent operators spanning image of $\cP$.

This leads, finally, to the single operator in the kernel
\begin{equation}
	V_N \cap V_F = \{ \cO_2 - \cO_3 \}
\end{equation}
which is annihilated by $\cP$. Thus we have derived the size of the interacting chiral ring (\ref{eq:ZintN_expanded}) from first principles, in agreement with $N$-boson counting (\ref{eq:Z_intN_bosons}).

%%%%%%%%%%%%%%%%%%%%%%%%%%%%%%%%%%%%%%%%%%%%%%%%%%%%%%%%%%

\end{appendix}

%%%%%%%%%%%%%%%%%%%%%%%%%%%%%%%%%%%%%%%%%%%%%%%%%%%%%%%%%%

\bibliography{QuivCal}        %or whatever your .bib file is
\bibliographystyle{utphys}   %if you use utphys.bst

\end{document}